\documentclass[a4paper,11pt]{article}
\pdfoutput=1

\usepackage{jheppub}
\usepackage[T1]{fontenc}
\usepackage[utf8]{inputenc}
\usepackage{lmodern}
\usepackage{amsmath,amssymb,amsthm,mathtools,mathrsfs,bm}
\usepackage{booktabs,array}
\usepackage{enumitem}
\usepackage{microtype}
\usepackage{orcidlink}

\setlist{itemsep=2pt,topsep=4pt,parsep=0pt,partopsep=0pt}
\allowdisplaybreaks
\AtBeginDocument{%
  \setlength{\abovedisplayskip}{7pt plus 2pt minus 2pt}%
  \setlength{\belowdisplayskip}{7pt plus 2pt minus 2pt}%
  \setlength{\abovedisplayshortskip}{5pt plus 2pt minus 2pt}%
  \setlength{\belowdisplayshortskip}{5pt plus 2pt minus 2pt}%
}
\makeatletter
\let\KrylovOriginalAddcontentsline\addcontentsline
\newif\ifKrylovWriteContentsLine
\renewcommand{\addcontentsline}[3]{%
  \begingroup
  \KrylovWriteContentsLinetrue
  \def\KrylovContentsFile{#1}%
  \def\KrylovContentsToc{toc}%
  \def\KrylovContentsLevel{#2}%
  \def\KrylovSubsubsectionLevel{subsubsection}%
  \def\KrylovParagraphLevel{paragraph}%
  \def\KrylovSubparagraphLevel{subparagraph}%
  \ifx\KrylovContentsFile\KrylovContentsToc
    \ifx\KrylovContentsLevel\KrylovSubsubsectionLevel
      \KrylovWriteContentsLinefalse
    \fi
    \ifx\KrylovContentsLevel\KrylovParagraphLevel
      \KrylovWriteContentsLinefalse
    \fi
    \ifx\KrylovContentsLevel\KrylovSubparagraphLevel
      \KrylovWriteContentsLinefalse
    \fi
  \fi
  \ifKrylovWriteContentsLine
    \endgroup
    \KrylovOriginalAddcontentsline{#1}{#2}{#3}%
  \else
    \endgroup
  \fi
}
\makeatother

\newtheorem{proposition}{Proposition}[section]
\newtheorem{corollary}[proposition]{Corollary}
\newtheorem{lemma}[proposition]{Lemma}
\theoremstyle{definition}

\newtheorem{example}[proposition]{Example}
\theoremstyle{remark}

\newcommand{\ket}[1]{|#1\rangle}
\newcommand{\bra}[1]{\langle #1|}
\newcommand{\braket}[2]{\langle #1\mid #2\rangle}
\newcommand{\dd}{\mathrm d}
\newcommand{\ii}{\mathrm i}
\newcommand{\ee}{\mathrm e}
\newcommand{\CC}{\mathbb C}
\newcommand{\RR}{\mathbb R}

\newcommand{\cH}{\mathcal H}
\newcommand{\cK}{\mathcal K}
\newcommand{\cA}{\mathcal A}
\newcommand{\cP}{\mathcal P}

\newcommand{\cF}{\mathcal F}
\newcommand{\Jac}{\mathsf J}
\newcommand{\Id}{\mathbf 1}
\newcommand{\Span}{\operatorname{span}}
\newcommand{\Dom}{\operatorname{Dom}}
\newcommand{\supp}{\operatorname{supp}}
\newcommand{\rank}{\operatorname{rank}}
\newcommand{\diag}{\operatorname{diag}}

\newcommand{\He}{\operatorname{He}}

\newcommand{\im}{\operatorname{Im}}
\newcommand{\tphi}{\widetilde\phi}
\newcommand{\thh}{\widetilde h}
\newcommand{\tK}{\widetilde{\mathbb K}}
\newcommand{\Qhat}{\widehat Q}
\newcommand{\Qsh}{Q^{\sharp}}
\newcommand{\cS}{\mathcal S}

\title{Polynomial
Initial-State Jumps and Christoffel Transforms in Krylov Complexity}
\author[a\orcidlink{0000-0003-0948-4817}]{Abhishek Chowdhury}
\emailAdd{achowdhury@iitbbs.ac.in}
\author[a\orcidlink{0009-0002-0199-9066}]{Ajit Prasad Mahapatra}
\affiliation[a]{School of Basic Sciences, Indian Institute of Technology Bhubaneswar, Argul, Odisha 752050, India}
\emailAdd{a25ph09001@iitbbs.ac.in}

\abstract{
At fixed Hamiltonian $H$, changing the initial state changes the cyclic pair and hence
generally the Lanczos basis and spread. For every normalizable polynomial preparation
$\lvert\psi_Q\rangle=N_Q^{-1/2}Q(H)\lvert K_0\rangle$, we reconstruct its state-Krylov problem
exactly from reference cyclic data. The transfer assumes no integrability and applies on finite
or infinite cyclic support.
Reweighting the reference spectral measure by $\lvert Q\rvert^2$ gives a finite-band connector
for every prepared amplitude and a finite-rank Christoffel--Darboux projection for cumulative
probabilities and spread, without rerunning ambient-space Lanczos. Fixed-degree seeds preserve the limiting Jacobi coefficients of asymptotically constant chains.
Every Charlier number-state jump obeys $K_r(t)\ge K_0(t)$, with strict inequality for $r\ge1$ away
from revivals.
At every fixed finite $r$, the Hermite endpoint has exact all-level amplitudes, whereas the
continuous-$q$-Hermite chord chain of double-scaled SYK has an all-level root-free re-Lanczos
reconstruction. Writing $D_r^{\rm H}$ and $D_r^{\mathrm{ch}}(t)$ for the mean absolute Fock and
chord displacements, respectively, the bounds are $K_r\ge D_r^{\rm H}$ and
$K_r(t)\ge D_r^{\mathrm{ch}}(t)\ge\lvert\overline n_r(t)-r\rvert$.
Thus rebuilt spread bounds physical displacement and signed chord drift. At fixed $0\le q<1$ and
time, $K_r(t)-D_r^{\mathrm{ch}}(t)\to0$ as $r\to\infty$, with both approaching the same
folded-Bessel limit. In Liouville space, the exact unnormalized gap-measure family on an open
inverse-temperature interval determines the positive transition-resolved measure, assuming the thermal
kernel is known and strictly positive and the requisite exponential moments exist.
One cyclic solution can therefore
be reused across polynomial seeds while separating physical propagation, basis response, and
rebuilt spread.}

\begin{document}
\hypersetup{pageanchor=false}
\maketitle
\hypersetup{pageanchor=true}

\section{Introduction}
\label{sec:intro}

State Krylov complexity represents unitary evolution within the cyclic subspace of a chosen state as propagation along a one-dimensional Jacobi chain. Starting from a normalized seed $\ket{K_0}$, a self-adjoint, time-independent Hamiltonian generates the orbit 
$\{\ket{K_0},H\ket{K_0},H^2\ket{K_0},\ldots\}$ and the Lanczos orthogonalization converts this orbit into an ordered basis
$\{\ket{K_0},\ket{K_1},\ldots\}$ along which the time-evolved state spreads. For state, or spread, complexity, this ordering is intrinsic to the construction. It is selected by the pair $(H,\ket{K_0})$ and is distinguished by the nested short-time minimization underlying spread complexity \cite{Balasubramanian:2022tpr}. The relevant dynamical datum is therefore not the Hamiltonian alone, but the Hamiltonian together with its cyclic seed.

Our central question is whether the Krylov dynamics produced by changing the initial state within the reference cyclic subspace at fixed $H$ can be recovered from the original Lanczos problem, or whether the construction must be repeated from the beginning. Although two complete orthonormal bases of a common subspace are related by a unitary transformation, the Krylov ordering is reconstructed from the orbit of the chosen seed under $H$ and is not preserved by a generic change of basis. A linear change of seed therefore does not, in general, induce a linear transformation of either the Lanczos coefficients or the ordered Krylov basis. The ordered Krylov basis must be reconstructed even when the new state belongs to the reference Krylov span, and in finite dimension its orbit may have smaller spectral support and terminate earlier. Variations of initial conditions, Hamiltonians and Hilbert-space dimension have recently been studied from several complementary viewpoints \cite{KunduMalvimatSinha2023,SreeramKannanModakAravinda2025,Balasubramanian:2025variations}, among them \emph{Koherence}, the entropy of coherence between the perturbed and unperturbed Krylov bases \cite{Balasubramanian:2025variations}. We isolate the fixed-$H$ problem and use the term ``initial-state jump'' for a change of cyclic seed, with no quench or discontinuity in the time evolution.

In this paper, we solve this relative problem for polynomially related seeds,
\begin{equation}
   \ket{\psi_Q}\propto Q(H)\ket{K_0}.
   \label{eq:intro-seed}
\end{equation}
This family is already intrinsic to the reference Krylov construction. Every reference vector has the form
$\ket{K_r}=P_r(H)\ket{K_0}/\sqrt{h_r}$, where $P_r$ is the $r$-th reference monic orthogonal polynomial and $h_r$ is its squared norm \cite{MuckYang:2022}. More generally, a normalized finite superposition
$\sum_{r=0}^{R}c_r\ket{K_r}$, with $\sum_r|c_r|^2=1$, is generated by
\begin{equation}
   Q=\sum_{r=0}^{R}c_r\frac{P_r}{\sqrt{h_r}}\,,
   \qquad N_Q=1\,.
\end{equation}
Thus, individual reference jumps and arbitrary finite superpositions belong to the same polynomial class. In the solvable chains developed here, this includes number-state jumps in the Heisenberg--Weyl oscillator, fixed Fock-number preparations at its Hermite endpoint, fixed-$H$ localized-site jumps in constant-coefficient tight-binding models, and fixed-chord preparations in double-scaled SYK. Polynomial filtering also occurs in quantum-query problems, where the object of interest is the cost of preparing the filtered state rather than its subsequent Krylov dynamics \cite{Adhikari2026Query}.

\paragraph{Results.}
Five consequences illustrate complementary aspects of the transfer machinery. In these statements $K_r$ denotes the
spread complexity of the jump $\ket{K_0}\to\ket{K_r}$ to the $r$-th reference Krylov vector,
$\lambda$ is the Poisson parameter of the oscillator's spectral measure, $\tau$ the dimensionless
oscillator time ($\tau_{\rm H}$ its rescaling in the Hermite limit), and $\widetilde a_n^{(r)}$,
$\widetilde\beta_n^{(r)}$ are the diagonal and squared off-diagonal Lanczos coefficients of the
jumped chain.  For the continuous-$q$-Hermite chord
chain, $0\le q<1$ is the DSSYK chord parameter, $\overline n_r$ is the mean chord number,
$D_r^{\rm ch}$ is the mean absolute displacement from the prepared chord, and $D_r^{\rm H}$ is
the analogous Fock-number displacement at the Hermite endpoint.

\begin{enumerate}[label=\emph{(\roman*)},leftmargin=2.2em,itemsep=4pt,topsep=4pt,parsep=0pt]
\item \emph{A vacuum lower bound for oscillator jumps.}
The inequality is proved for the
oscillator chain in corollary~\ref{cor:H1-monotonicity}. No number-state jump lowers the oscillator complexity,
\begin{equation}
   K_r(\tau)\ \ge\ K_0(\tau)=2\lambda(1-\cos\tau),
   \label{eq:intro-vacuum-bound}
\end{equation}
strictly for $r\ge1$ away from the revival times at which $K_0$ vanishes. The restriction to the
oscillator is necessary, because on finite support the same first jump can strictly \emph{lower}
the complexity, by deleting a spectral atom and re-timing the revivals
(eq.~\eqref{eq:three-level-decrease}).
The oscillator inequality has a direct physical origin.  Displacement evolves the number
state $\ket r$ with reference-basis mean level $r+K_0(\tau)$, while the degree-$r$ filtration can
subtract at most the initial offset $r$.  The probability that returns below level $r$
contributes an
additional nonnegative term.  Subsection~\ref{subsec:HW} turns this argument into an exact identity.

\item \emph{An exact
continuous-support amplitude closure.}
At the Hermite endpoint, proposition~\ref{prop:Hermite-all-r-complexity-app} gives a root-free
amplitude closure for every fixed finite $r$. The amplitude at level $n$ is a Gaussian times
$(-\ii\tau_{\rm H})^n$ and a degree-$r$ polynomial generated by a scalar differential
recurrence.
The same result gives an exact physical Fock-distance observable in the original number basis, defined as the mean absolute displacement $|m-r|$ from the prepared level $r$.  With $x=\tau_{\rm H}^2$,
\begin{equation}
   K_r(\tau_{\rm H})\ge D_r^{\rm H}(x)
   \ge x+2\ee^{-x}\frac{x^r}{(r-1)!}
   >x=K_0(\tau_{\rm H}),
   \qquad r\ge1,\quad x>0\,,
   \label{eq:intro-Hermite-radial-bound}
\end{equation}
and
$K_r-D_r^{\rm H}=r(r+1)\tau_{\rm H}^6/[2(2r+1)]
+O_r(\tau_{\rm H}^8)$.  Thus physical Fock displacement and rebuilt spread agree through fourth
order but are not identical.  The generalized-Hermite parity reduction at $r=1$ further sums the
complexity to
\begin{equation}
   K_1(\tau_{\rm H})=\tau_{\rm H}^2+1-\ee^{-2\tau_{\rm H}^2}\,,
   \qquad
   K_0(\tau_{\rm H})=\tau_{\rm H}^2 \,,
   \label{eq:intro-hermite-endpoint}
\end{equation}
so the first jump has a one-unit late-time excess.

\item \emph{Bulk rigidity, fixed-chord closure and a controlled deep-seed fold.}
If the reference measure belongs to the Nevai class $M(a,b)$, every nonzero polynomial $Q$ of
fixed degree produces a shifted measure in the same class.  Thus
$\widetilde a_n^{[Q]}\to a$ and $\widetilde b_n^{[Q]}\to b$ without any restriction on the roots
of $Q$.  Real roots may lie in the limiting band, at visible atoms or outside the band, and
nonreal roots are also allowed.  The proposition 
leaves finite rows, convergence rates and subleading coefficients free to retain preparation
dependence.  The unbounded Charlier chain makes that surviving dependence explicit,
\begin{equation}
   \widetilde a_n^{(r)}=n+s_r(\lambda)+O(n^{-1}),
   \qquad
   \widetilde\beta_n^{(r)}
   =\lambda\bigl[n+2r-s_r(\lambda)\bigr]+O(n^{-1}),
   \label{eq:intro-slopes-intercepts}
\end{equation}
where $s_r(\lambda)$ counts the deleted Poisson atoms.  The leading slopes are rigid.  The two subleading intercepts have fixed sum $2r$. Each deleted Poisson atom
increases the diagonal intercept by one and decreases the intercept of
$\widetilde\beta_n^{(r)}/\lambda$ by one.  This intercept behavior provides an analytic realization of the broader preparation dependence discussed in refs.~\cite{SreeramKannanModakAravinda2025,Balasubramanian:2025variations}.

For the continuous-$q$-Hermite chain of double-scaled SYK,
corollary~\ref{cor:qhermite-root-free-app} reconstructs the complete re-Lanczos problem for every
fixed finite chord $r$ and every shifted level from an $r\times r$ Gram matrix built from polynomial-remainder data.  The
normalized chord-number motion and rebuilt Krylov spread obey the exact two-basis comparison
\begin{equation}
   K_r(t)\ge D_r^{\rm ch}(t)\ge
   \bigl|\overline n_r(t)-r\bigr|.
   \label{eq:intro-chord-radial-bound}
\end{equation}
This is an all-level reconstruction with the seed chord $r$ held fixed.  It does not, however,
assert uniformity as $r$ grows.

Write $p_r=P_r/\sqrt{h_r}$, so that
$\ket{K_r}=p_r(H)\ket{K_0}$.  The complementary deep-seed limit follows this preparation as
$r\to\infty$.  At every fixed shifted level, the chain induced by $\ket{K_r}$ from a reference Jacobi chain in $M(a,b)$ approaches the folded arcsine
Jacobi chain with first hopping $\sqrt2\,b$ and all later hoppings $b$.  For the $q$-Hermite chain
of double-scaled SYK at fixed $q<1$ and fixed time, the physical chord distribution becomes a
two-sided Bessel-function profile with vanishing signed mean displacement, while the re-Lanczos chain folds the two
directions and retains their mean absolute displacement.  The physical mean absolute chord displacement and
the rebuilt complexity then have the same folded limit, so their nonnegative difference
vanishes.  Section~\ref{sec:rigidity} proves both
limits and keeps their fixed-degree and growing-degree hypotheses separate.

\item \emph{Arithmetic support loss on a finite lattice.}
 For the
constant-coefficient tight-binding chain of $N_{\rm lat}$ sites, moving the initially occupied site
to $\ket k$ deletes exactly
\begin{equation}
   r_{\rm zero}=\gcd(k+1,N_{\rm lat}+1)-1
   \label{eq:intro-gcd-support-loss}
\end{equation}
spectral levels from the shifted problem, so the launch site and the chain length together decide
which sine modes are invisible to the preparation.
 The shifted cyclic subspace has dimension
$N_{\rm lat}-r_{\rm zero}$, and its last positive-norm Krylov vector is at level
$N_{\rm lat}-r_{\rm zero}-1$. Subsection~\ref{subsec:tight-binding} and appendix~\ref{app:finite} obtain
this alongside every localized-site connector explicitly, through the terminal edge.

\item \emph{Basis response, response geometry and transition tomography.}
For a polynomial preparation, the amplitude connector factorizes as
$\mathsf T_Q=\mathsf U_Q^\dagger Q(\Jac)/\sqrt{N_Q}$, where $\Jac$ is the reference Jacobi matrix.  The overlap matrix $\mathsf U_Q$
determines Koherence and measures how the ordered Krylov basis changes, whereas $\mathsf T_Q$
transports the prepared dynamics.  The two objects therefore answer different questions.  On a
real parameter domain where the Toda normalizer $Z_Q(\boldsymbol s)$ is finite and twice differentiable under the integral, its response
satisfies
\begin{equation}
   \partial_{s_j}\partial_{s_k}\log Z_Q
   =\operatorname{Cov}_{Q,\boldsymbol s}(H^j,H^k)
   =F^{\rm Q}_{jk}\,,
   \qquad
   g^{\rm FS}_{jk}=\frac14F^{\rm Q}_{jk}\,.
\end{equation}
This is an exact information geometry of the prepared pure states.  Its derivation uses only the commutativity of the functions of $H$ and does not assume a thermodynamic ensemble.  Interpreting the parameters as generalized inverse temperatures
additionally requires the chosen powers to be physically admissible conserved charges, typically
independent extensive or quasilocal quantities in the thermodynamic limit.

On bounded support, polynomial approximation controls the normalized filtered state and its
compact-time Krylov dynamics.  Quantum signal processing realizes suitable bounded and
parity-compatible approximants only after specifying a block-encoding or qubitized access model,
rescaling the spectrum to $[-1,1]$, and accounting for polynomial degree, target precision,
normalization, success probability and any amplitude-amplification overhead.

A separate transition-tomography result arises in Liouville space.  A single-temperature gap measure discards the source energy $F=E_b$ while retaining only the gap $\omega=E_a-E_b$.  After division by the known nonzero
thermal kernel and multiplication by the known partition function, the unnormalized family known on an open interval of inverse temperatures is a measure-valued bilateral Laplace transform in the source-energy variable.  Finite
exponential moments and Laplace uniqueness then determine the positive transition-resolved
measure, but not the phases, eigenvectors or higher correlations.  Coarse-grained ETH
\cite{Srednicki1999ETH} gives an additional interpretation of the recovered density rather than
an input to the exact inversion.  The Fisher information lost by projecting to the gap is the
expected conditional source-energy variance.

\end{enumerate}

\paragraph{Scope and mechanism.}
A degree-$r$ polynomial produces a state in
$\Span\{\ket{K_0},\ldots,\ket{K_r}\}$ before terminal reduction.
In a finite cyclic space of dimension $d$, the Lanczos chain ends at level $d-1$ and the
degree-$d$ terminal polynomial $T_d$ annihilates the seed.  A polynomial preparation therefore acts
through its class modulo $T_d$, as developed in section~\ref{sec:setup}.
Fixed degree gives a controlled notion of proximity in the reference Krylov filtration, not in real
space.  A many-body operator $Q(H)$ is generally global, although it reaches only the first few
reference Krylov levels.  The formalism remains exact at every finite degree.  The interpretation
as a nearby preparation, and the rigidity statements proved below, instead hold with the degree
fixed before the large-system or large-depth limit. 
This regime also makes a finite-depth reference calculation reusable.  Once its recurrence and
Fourier moments are known, finite connectors and projected kernels generate a family of prepared
dynamics for the same $H$.  Related filtering and restarted-Lanczos algorithms reuse reference
data to form modified static problems \cite{Sorensen1992,LiXiVecharynski2016,GumerovRiggSlevinsky2026}.
Here the output is dynamical and consists of prepared amplitudes, cumulative probabilities and
spread complexity.
The same distinction applies in operator space.  For a positive inner product under which
$\mathcal L=[H,\cdot]$ is self-adjoint on the cyclic domain, polynomial seeds
$Q(\mathcal L)O$ are finite nested-commutator descendants
\cite{Parker2019,Nandy2024,Rabinovici2021Edge,Rabinovici2025}.  They filter Liouvillian gaps and
need not represent spatial translations.  Matrix-valued seed families, scalar mixed-state
functionals and physical operator evolution are separated in section~\ref{sec:matrix} and
appendices~\ref{app:mixed-seeds}--\ref{app:operator-krylov}.

The reason polynomial jumps are tractable is most transparent in the spectral representation. Let
$\dd\mu(E)=\bra{K_0}\Pi(\dd E)\ket{K_0}$ be the scalar spectral measure of the reference seed.\footnote{Here ``scalar'' refers to the one-component spectral measure and Jacobi problem obtained after a seed direction has been chosen, in contrast with the matrix-valued parent measure of section~\ref{sec:matrix}. It does not distinguish a jump to one reference Krylov vector from a finite superposition. Both define scalar problems.}
The normalized state $N_Q^{-1/2}Q(H)\ket{K_0}$ carries the measure
\begin{equation}
   \dd\nu_Q(E)=\frac{|Q(E)|^2}{N_Q}\dd\mu(E),
   \qquad
   N_Q=\int |Q(E)|^2\dd\mu(E).
   \label{eq:intro-measure}
\end{equation}
Thus, changing the seed produces a positive polynomial, or Christoffel-type, modification of the reference measure. For real $Q$ the multiplier is $Q^2$ and may be treated through iterated or confluent Christoffel transformations. For complex coefficients it is
$|Q|^2=\Qsh Q$ on the real spectrum, where
$\Qsh(z)=\overline{Q(\bar z)}$. If a zero of $Q$ lies at a spectral atom, meaning a point $E_\alpha$ carrying strictly positive
weight $\mu(\{E_\alpha\})>0$, an eigenvalue of $H$ visible to the reference seed,
then that atom is removed, and in finite dimension the cyclic dimension drops accordingly.
In the spectral-theory terminology used below, an \emph{atom} is such a point of nonzero
spectral weight, and that weight is conventionally called its \emph{mass}.  When no ambiguity is
possible we use the phrase ``spectral weight''.

A root at a visible eigenvalue
removes the reference seed's cyclic component at that eigenvalue from the new seed's spectral
support.  The Hamiltonian, its eigenspace and its spectrum remain unchanged, while
$Q(H)\ket{K_0}$ has zero projection onto the reference cyclic component.  This is algebraically analogous to
state-deleting Darboux transformations such as the Krein--Adler construction
\cite{OdakeSasaki2013KreinAdler}, which instead deforms the operator itself.  The distinction is
essential here because the fixed-$H$ problem asks how the lost spectral direction propagates to
amplitudes, complexity and the finite-support inverse calculus of appendix~\ref{app:finite}.

The problem is therefore to propagate a measure transformation to the time-dependent amplitudes and the spread complexity.  Rerunning Lanczos would solve each shifted problem separately, but it would not provide the relative transfer from the resolved reference spectral data developed here.
The shifted orthogonal family is determined by the reference spectral data through closed finite
formulas. The distinction matters and is worth stating once. Iterative Lanczos generates
each new basis vector from the previous ones by an orthogonalization step whose cost and numerical
behavior depend on the ambient dimension. What replaces it here is a finite linear system, of size
set by $\deg Q$ alone, whose solution gives the connector row at level $n$ directly. The general root system is eq.~\eqref{eq:general-root-system}. In the Charlier chain, the jet/remainder form is eq.~\eqref{eq:H1-jet-connector-app},
together with a three-term recurrence in $n$ for the rows themselves,
eq.~\eqref{eq:H1-G-recurrence-app}. In these direct root, jet and remainder realizations, no
orthogonalization against previously computed shifted vectors is required.

The first consequence is an exact finite-band transfer of the amplitudes. Let $P_n$ and $R_n^{[Q]}$ denote the reference and shifted monic orthogonal polynomials, with squared norms $h_n$ and $\thh_n^{[Q]}$, respectively. If
$W_Q=\Qsh Q/N_Q$ has degree $L=2\deg Q$, monicity and orthogonality give
\begin{equation}
   S_n^{[Q]}(E)=W_Q(E)R_n^{[Q]}(E)
   =\sum_{m=n}^{n+L}\Gamma_{n,m}^{[Q]}P_m(E)
   \label{eq:intro-finite-band}
\end{equation}
at every nonterminal level of an infinite chain. The corresponding amplitude is
\begin{equation}
   \tphi_n^{[Q]}(t)=
   \frac{1}{\sqrt{\thh_n^{[Q]}}}
   \sum_{m=n}^{n+L}\Gamma_{n,m}^{[Q]}I_m(t),
   \qquad
   I_m(t)=\sqrt{h_m}\,\phi_m^{(0)}(t)
   =\int\ee^{-\ii Et}P_m(E)\dd\mu(E).
   \label{eq:intro-amplitude}
\end{equation}
The connector $\Gamma_{n,m}^{[Q]}$ carries the seed dependence, while the time dependence is inherited from the reference Fourier--orthogonal-polynomial moments $I_m(t)$. At each shifted level, only a finite window of reference moments is required, with width fixed by the degree of the seed. In a finite cyclic space of dimension $d$, the product $W_QR_n^{[Q]}$ is first reduced modulo the terminal polynomial $T_d$ and then expanded in the physical basis $P_0,\ldots,P_{d-1}$. The corresponding amplitude sum terminates at $m=d-1$ rather than following the infinite-chain range in eq.~\eqref{eq:intro-amplitude}.

Individual amplitudes are naturally described by the connector, whereas cumulative probabilities are more directly obtained from reproducing kernels \cite{SimonCDK}. Multiplication by a degree-$r$ polynomial embeds the shifted polynomial space of degree at most $\ell$ into the reference space of degree at most $\ell+r$. Its image is characterized by the divisibility constraints at the zeros of $Q$, so the shifted Christoffel--Darboux kernel is the reference kernel projected away from the corresponding constraint directions. For a real polynomial with simple real roots $\xi_a$,
\begin{equation}
\begin{gathered}
   \frac{Q(E)Q(F)}{N_Q}\tK_\ell^{[Q]}(E,F)
   =\mathbb K_{\ell+r}^{\mu}(E,F)
    -\mathbf k_\ell(E)^T G_\ell^{-1}\mathbf k_\ell(F),\\[2pt]
   (\mathbf k_\ell(E))_a=\mathbb K_{\ell+r}^{\mu}(E,\xi_a),
   \qquad
   (G_\ell)_{ab}=\mathbb K_{\ell+r}^{\mu}(\xi_a,\xi_b).
\end{gathered}
   \label{eq:intro-projected-kernel}
\end{equation}

Repeated roots replace point evaluations by derivative jets and their jet Gram matrix.\footnote{The kernel is set in blackboard bold, $\mathbb K$, to separate it from the spread
complexity $K_Q(t)$ and from the Krylov vectors $\ket{K_n}$, which are the other objects denoted by $K$ in this paper. The superscript $\mu$ on $\mathbb K_M^\mu$ identifies the reference measure and is not a
spacetime or component index. Also, $K_r$ abbreviates $K_Q$ for the
reference-state jump $Q=P_r/\sqrt{h_r}$ throughout the paper.} For complex seed coefficients, the left-hand side is
$\Qhat(E)\overline{\Qhat(F)}\,\tK_\ell^{[Q]}(E,F)$, where
$\Qhat=Q/\sqrt{N_Q}$, and the corresponding coefficient-space projector is Hermitian. For finite support, the root or jet evaluation formula applies without modification while $\ell+r\le d-1$. At and beyond the terminal edge, the exact object is the quotient image projector rather than an extension of the reproducing-kernel sum. After insertion into the spectral double integral, the projected kernel gives
\begin{equation}
   \Phi_\ell^{[Q]}(t)
   =\sum_{n=0}^{\ell}|\tphi_n^{[Q]}(t)|^2,
   \qquad
   K_Q(t)=\sum_{\ell\ge0}\bigl(1-\Phi_\ell^{[Q]}(t)\bigr).
\end{equation}
For an infinite chain the second identity is understood in $[0,\infty]$ until finiteness is established, whereas in finite dimension the tail sum terminates. Thus, cumulative probabilities and spread complexity are obtained without first constructing every
shifted Krylov vector. Throughout, $K_0(t)$ denotes the spread complexity of the reference pair
$(H,\ket{K_0})$ and $K_Q(t)$ that of the jumped pair $(H,\ket{\psi_Q})$, both being the mean
Krylov level of the evolved state in the respective Lanczos basis.

Once the reference moments $I_m(t)$ and the finite connector are known, the shifted survival amplitude, the Mandelstam--Tamm variance, the currents across Krylov
cuts and the instantaneous variation of $K_Q(t)$ all follow from the same reference problem.
These quantities enter standard quantum and Krylov speed bounds
\cite{MandelstamTamm1945,DeffnerCampbell2017,Hornedal2022,GillSarkar2024}.  The result used here is
their exact transfer from the reference pair to the polynomially prepared pair.

The solvable chains make the two global issues explicit.  In the Heisenberg--Weyl/Charlier
oscillator, the factorization $I_n(\tau)=\cS(\tau)q(\tau)^n$ in eq.~\eqref{eq:H1-old-moments-main} reduces amplitudes and finite kernel
cuts to finite algebra.  Root-free remainder recurrences then determine every number-state jump,
including the large-index intercepts, the finite-time tail bound and the Hermite endpoint.  Roots
on the Poisson support delete visible atoms.  They change the subleading intercepts but not the
linear slopes.
In the constant-hopping chain, a polynomial jump moves the initially occupied site without
changing the lattice Hamiltonian.  On a finite chain, a truncated Chebyshev product and its
product-Gram factorization continue the connector through support loss and the terminal edge, in
agreement with the sine-transform dynamics \cite{Balasubramanian:2025variations}.  The two models
therefore isolate the distinct infinite-tail and finite-terminal completions of the same transfer
construction.  Changes of representation or Hilbert-space dimension remain outside the fixed-$H$ problem.  The Hermite endpoint and the continuous-$q$-Hermite chord chain add two continuous-support completions.  The first gives root-free all-level amplitudes and a Fock-distance bound, while the second gives fixed-chord all-level re-Lanczos data and exposes the folded deep-chord limit in DSSYK.

Explicit special-function formulas for the reference polynomials are useful but not required.  Recurrence data $a_n,b_n$, squared norms $h_n$ and
Fourier moments $I_n(t)$ to a given depth determine every fixed-degree connector available at
that depth.  A degree-$r$ seed needs reference data through level $n+2r$ to reconstruct shifted
level $n$, with terminal reduction in finite dimension.  The operation is a finite linear solve
whose size depends on $r$, not an ambient-space Lanczos recursion for the new seed.  Outside the
solvable families the closed expressions disappear, but the relative construction remains exact
at every level supported by the reference data and the required domains.  Appendix
\ref{app:numerics} separates connector, rank, evolution and tail errors in this implementation.

Let
$V_0=(\ket{v_1},\ldots,\ket{v_p}):\CC^p\to\mathcal H$ be the isometry whose orthonormal columns
span a finite family of pure seed directions.  This family has the positive matrix-valued parent
measure $\dd M=V_0^\dagger\Pi(\dd E)V_0$.  It retains the spectral
interference among those directions and is also the input for block Lanczos, subject to the usual
rank-deflation qualifications \cite{CrapsEvninPascuzzi2024}.  It does not, however, select a
common scalar Lanczos basis or a unique scalar spread complexity.  Choosing a unit direction $c$
produces the scalar compression $c^\dagger\dd M c$ and hence a distinct scalar Jacobi problem.
Mixedness enters only after a density operator has been chosen.  Restriction to the commutative
algebra generated by $H$ gives the scalar energy measure
$\dd\mu_\rho(E)=\operatorname{Tr}[\rho\Pi(\dd E)]$.  If $\rho$ is supported on a finite reference
Krylov window, this measure is a trace contraction of $\dd M$.  Its density with respect to
the reference measure is a globally nonnegative polynomial and therefore admits scalar spectral factorization, making the contraction
measure-equivalent to an effective pure polynomial seed.  This equivalence holds only at the level
of the scalar energy distribution.  The preparation remains mixed, and its scalar complexity is
neither an average of pure-seed complexities nor physical density-matrix evolution.

Physical density-matrix evolution is instead a Liouville-space problem.  For a fixed positive
operator inner product under which $\mathcal L$ is self-adjoint on the relevant cyclic domain,
polynomial descendants $Q(\mathcal L)O$ filter energy gaps.  At fixed
temperature, some changes of inner-product kernel reduce to scalar filters of the gap measure, but
a generic temperature change reweights transitions with the same gap according to their source
energies and therefore requires the joint transition data of
appendix~\ref{app:operator-krylov}, where $\omega=E_a-E_b$ is the transition gap and $F=E_b$ the source energy.  After removal of the known thermal kernel and restoration of
the known partition-function factor, the exact unnormalized family known on an open interval of inverse temperatures is a measure-valued bilateral Laplace transform in $F$.  Under the stated
exponential-moment and Laplace-uniqueness hypotheses it determines the positive transition-resolved measure, but not phases,
eigenvectors, degenerate transition labels or higher correlations.   Section~\ref{sec:context}
summarizes the resulting tomography and Fisher-information statements.

The resulting relative calculus brings the spectral theory of Jacobi operators,
modified measures and orthogonal polynomials
\cite{Chihara1978,Szego1975,SimonOPRL1,Uvarov1969,KautskyGolub1983},
into direct contact with the Fourier--orthogonal-polynomial representation of
Krylov dynamics and spread complexity
\cite{Balasubramanian:2022tpr,MuckYang:2022,Nandy2024,Qu2025LanczosOP}.
Christoffel/Darboux transformations determine
the shifted Jacobi problem from the reference data, finite-band
connectors transfer the time-dependent amplitudes, and finite-rank
kernel projections reconstruct cumulative probabilities and spread
complexity. These ingredients define an exact fixed-$H$ dynamical
calculus for polynomially related seeds, including complex
superpositions, confluent constraints, support loss and finite
terminal closure.

\paragraph{Guide to the paper.}
Section~\ref{sec:setup} formulates polynomial preparation in
the cyclic spectral problem and establishes composition, support and resolvent properties.
Section~\ref{sec:finiteband} derives the finite-band amplitude transfer, while
section~\ref{sec:kernels} reconstructs cumulative probabilities and spread complexity through
finite-rank Christoffel--Darboux projections.  Section~\ref{sec:examples} develops the Charlier
and constant-hopping chains together with the Hermite endpoint.  Section~\ref{sec:matrix}
organizes finite families of seed directions through a matrix-valued parent measure.
Section~\ref{sec:rigidity} proves all-root invariance of the limiting Nevai bulk, derives the
degree-resolved finite-time ceiling, identifies the subleading memory of atom deletion and obtains
the folded deep-seed limit.  
Section~\ref{sec:context} extends the physical applications in two directions, with particular emphasis on chord dynamics in double-scaled SYK.  The
chord-dynamics part separates physical chord motion from re-Lanczos spread,
proves the fixed-chord physical-distance comparison and its deep-seed saturation, and factors dynamical
transport from Koherence.  The spectral-response part identifies the Toda response with
pure-state quantum Fisher geometry and connects the preparation calculus to determinant
insertions, filtered Loschmidt and thermofield-double amplitudes, bounded-filter algorithms and
temperature tomography.  Section~\ref{sec:discussion} sets the scope of these results and the
open joint seed-depth, observed-level and time-scaling problems.

Appendices~\ref{app:lanczos}--\ref{app:projected-kernel} establish the Lanczos, connector and
projection identities used in the main text.  Appendix~\ref{app:christoffel} contains the
all-root Nevai proof, the root-Gram fixed-level formulas and their continuous-$q$-Hermite
corollary, the controlled deep-seed proposition and the compact-filter closure.
Appendix~\ref{app:H1} develops Charlier number-state jumps from root-free remainder and jet
recurrences through the large-index, tail and operator-domain estimates that control the
unbounded chain.  It then derives the fixed-seed, all-level Hermite amplitude closure, Fock-distance filtration
and short-time defect.  This separates the finite algebra needed at each rebuilt level from the
global tail estimates that prove finiteness of the Krylov-index first moment on the infinite chain.
Appendix~\ref{app:finite} completes the finite-chain analysis through terminal folding and
support loss, while appendix~\ref{app:numerics} gives reference-space implementations and diagnostics.
Appendices~\ref{app:mixed-seeds} and~\ref{app:operator-krylov} treat scalar mixed-state
contractions and transition-resolved Liouville-space dynamics.

\section{Polynomial seed jumps in the cyclic spectral problem}
\label{sec:setup}

Comparing different seeds at fixed $H$ requires a common cyclic spectral description. We first fix the reference Jacobi problem and its finite terminal closure, then identify polynomial seed changes with positive modifications of the scalar measure, determine how relative filters compose and when they lose spectral support, and reconstruct the local shifted Jacobi data from resolvents and moments. Appendix~\ref{app:lanczos} derives the Lanczos--orthogonal-polynomial dictionary, finite closure and Weyl-function identities used in this section.

\subsection{Reference cyclic data and polynomial seed measures}
\label{subsec:reference-seed-measures}
Let $H$ be a time-independent self-adjoint operator and $\ket{K_0}$ a normalized seed. The polynomial cyclic subspace of the reference pair is
\begin{equation}
   \cK(H,\ket{K_0})
   =
   \operatorname{cl}_{\mathcal H}
   \Span\{p(H)\ket{K_0}:p\in\CC[E],\ p(H)\ket{K_0}\text{ exists}\}.
   \label{eq:cyclic-subspace-main}
\end{equation}
The closure in eq.~\eqref{eq:cyclic-subspace-main} need not be the full Hilbert space. Even when the spectral variable is continuous, the Krylov basis remains the countable orthonormal-polynomial basis indexed by $n=0,1,2,\ldots$. The continuum labels the spectral representation rather than the Krylov chain.

For a complete infinite chain, we assume that
$\ket{K_0}\in\bigcap_{n\ge0}\Dom(H^n)$ and that the polynomials are dense in $L^2(\mu)$, where
$\mu$ is the seed spectral measure constructed in eq.~\eqref{eq:spectral-measure} below. These
regularity assumptions are automatic for compact spectral support, with finite support producing a
terminal chain, and they hold for the determinate measures used below, a measure being
determinate when it is uniquely fixed by its power moments. For noncompact support, a standard
sufficient condition for determinacy and polynomial density is Carleman's criterion,
$
   \sum_{n\ge1}b_n^{-1}=\infty,
$
stated in terms of the off-diagonal Lanczos coefficients $b_n$ of the three-term recurrence,
eq.~\eqref{eq:base-recurrence} below. It also implies essential self-adjointness of the
corresponding minimal Jacobi operator \cite{SimonMomentProblem1998}, where the moment problem
is developed precisely in this operator form. For every infinite shifted problem we likewise assume that the polynomials are dense in $L^2(\nu_Q)$.  This is automatic on compact support and, for the explicit unbounded models below, follows from the shifted linear Jacobi bound proved in proposition~\ref{prop:finite-time-control-main}. 
The proposition derives the shifted hopping bound from the reference coefficients by degree filtration, without assuming shifted polynomial density. Unless the shifted chain has already terminated, that bound makes the shifted Carleman series diverge; determinacy, polynomial density and essential self-adjointness of the shifted minimal Jacobi operator then follow. Under the domain and density assumptions above, the polynomial cyclic subspace coincides with the
full spectral cyclic subspace of $\ket{K_0}$. We use two counting facts repeatedly.
The reference orthogonal-polynomial construction through any nonterminal degree $M$ requires only
the moments through order $2M$, and after a seed modification by a polynomial of degree $r$ (the operation defined in eq.~\eqref{eq:seed-normalization} below), shifted orthogonality through a level $n$ before shifted termination requires the reference moments through order $2(n+r)$. The spectral theorem identifies $H|_{\cK}$ with multiplication by $E$ on $L^2(\mu)$, where
\begin{equation}
   \dd\mu(E)=\bra{K_0}\Pi(\dd E)\ket{K_0},
   \qquad
   \bra{K_0}f(H)\ket{K_0}=\int f(E)\dd\mu(E)\,.
   \label{eq:spectral-measure}
\end{equation}
The second identity holds for bounded Borel functions $f$ and extends whenever the matrix element and the integral both exist. The measure $\mu$ may be discrete, absolutely continuous, singular-continuous, or of mixed type. Complete infinite-chain statements are always understood within the polynomial closure in $L^2(\mu)$.

The real monic orthogonal polynomials $P_n$ for this spectral measure, together with their squared norms, are defined by
\begin{equation}
   \int P_n(E)P_m(E)\dd\mu(E)=h_n\delta_{nm}\,,
   \qquad
   P_0=1,\qquad h_0=1\,,
   \label{eq:base-orthogonality}
\end{equation}
and the reference Krylov basis is
$
   \ket{K_n}=P_n(H)\ket{K_0}/\sqrt{h_n}\,.
$
Equivalently, for every nonterminal Krylov index $n$, there is a unique orthonormal polynomial $p_n$ of degree exactly $n$ such that
\begin{equation}
   \ket{K_n}=p_n(H)\ket{K_0},
   \qquad
   p_n=\frac{P_n}{\sqrt{h_n}}\,.
   \label{eq:orthonormal-polynomial-representative-main}
\end{equation}
Its leading coefficient is $(b_1b_2\cdots b_n)^{-1}$, since
$h_n=b_1^2b_2^2\cdots b_n^2$. In finite cyclic dimension $d$, this statement ends at $n=d-1$. The formal degree-$d$ continuation is the terminal annihilating polynomial $T_d$ introduced below, not an additional positive-norm Krylov vector. Before termination, the three-term recurrence takes the monic form~\cite{Chihara1978}
\begin{equation}
   E P_n(E)=P_{n+1}(E)+a_nP_n(E)+b_n^2P_{n-1}(E),
   \qquad
   P_{-1}=0,\qquad b_0=0\,,
   \label{eq:base-recurrence}
\end{equation}
where $a_n\in\RR$ and $b_n^2=h_n/h_{n-1}>0$ for $n\ge1$ as long as the chain has not terminated. In the orthonormal basis $p_n=P_n/\sqrt{h_n}$, the restriction of $H$ is represented by the symmetric tridiagonal Jacobi operator $\Jac$, acting according to
\begin{equation}
   H\ket{K_n}
   =
   b_{n+1}\ket{K_{n+1}}
   +a_n\ket{K_n}
   +b_n\ket{K_{n-1}},
   \qquad b_0=0\,.
   \label{eq:jacobi-action}
\end{equation}
The reference Fourier--orthogonal-polynomial (OP) moments and Krylov amplitudes are
\begin{equation}
   I_n(t)=\int \ee^{-\ii Et}P_n(E)\dd\mu(E),
   \qquad
   \phi_n^{(0)}(t)=\frac{I_n(t)}{\sqrt{h_n}}\,.
   \label{eq:base-moments}
\end{equation}
Under the standing determinacy assumptions, the measure $\mu$, the Jacobi data $\{a_n,b_n\}$, and the Fourier--OP moments are equivalent descriptions of the same cyclic spectral problem. The scalar measure is attached to the pair $(H,\ket{K_0})$, not to $H$ alone. Changing the seed changes the measure and, in general, the complete Jacobi data.

We shall keep the monic and orthonormal normalizations visible throughout. The polynomials $P_n$ are monic because Christoffel transformations and terminal quotients take their simplest form in this convention, whereas the Hilbert-space vectors are normalized by $h_n^{-1/2}$. Thus, if $p_n=P_n/\sqrt{h_n}$, then 
$\phi_n^{(0)}(t)=\int\ee^{-\ii Et}p_n(E)\dd\mu(E)$, while multiplication by $E$ in the $p_n$-basis is represented by the symmetric Jacobi matrix with off-diagonal entries $b_n$. We therefore write Christoffel coefficients in the monic convention and probabilities and reproducing kernels with orthonormal denominators.\footnote{For self-adjoint $H$, the Lanczos phases may be chosen so that $b_n>0$ until termination. This fixes a convenient phase convention for the Krylov basis. Relative phases in a finite superposition of Krylov states remain part of the seed data and enter through $|Q(E)|^2$.}

\paragraph{Finite terminal closure.}
Finite-dimensional cyclic subspaces require the following terminal convention. If the cyclic dimension is $d$, then the measure has $d$ distinct support points with positive weights, the positive-norm orthogonal family ends at $P_{d-1}$, and the monic terminal polynomial is the characteristic polynomial of the finite Jacobi matrix,
$
   T_d(E)=\prod_{\alpha=0}^{d-1}(E-E_\alpha).
$
Since the Jacobi representation is cyclic, $T_d$ is also its minimal polynomial and satisfies $T_d(H)\ket{K_0}=0$. Polynomial functions on the cyclic support are therefore represented by the quotient algebra
$
   \cA_d=\CC[E]/\langle T_d(E)\rangle.
$
This quotient is essential in polynomial calculations near the terminal edge. Products whose degree exceeds $d-1$ must not be set to zero. Their unique values on the support are obtained by reduction modulo $T_d$. Moreover, the symbols $P_{d+1},P_{d+2},\ldots$ are not determined by the finite Jacobi data as physical orthogonal polynomials. 

Although the symbols beyond $P_{d-1}$ carry no physical content, the ordinary-polynomial connector constructions of section~\ref{sec:finiteband}, both the root systems and the determinant formulas, can reach beyond the terminal degree, and it is only for such algebraic purposes that auxiliary terminal lifts are introduced. A proper lift is an ideal-compatible continuation whose beyond-terminal polynomials belong to the null ideal generated by $T_d$, for example
\begin{equation}
   \widehat P_m(E)=P_m(E),\quad 0\le m\le d-1,
   \qquad
   \widehat P_{d+k}(E)=T_d(E)\Lambda_k(E),\qquad k\ge0,
   \label{eq:proper-terminal-lift-main}
\end{equation}
where $\Lambda_k$ is any monic polynomial of degree $k$. Such a lift specifies polynomial representatives, not additional Krylov sites. Once a connector determinant or confluent-root construction has been evaluated, the physical finite-chain result is obtained by reducing modulo $T_d$. Finite-dimensional formulas are consequently identities of functions on the support, or equivalently identities in $\cA_d$, while terminal lifts provide a controlled representative-level implementation of the ordinary-polynomial connector formulas near the edge. Proposition~\ref{prop:terminal-lift-determinant} proves the lift independence of this prescription and its equivalence to the direct finite-support construction. Cumulative probabilities never require a lift. They are computed from the intrinsic quotient-image projector developed in section~\ref{sec:kernels}.

\paragraph{Polynomial filters.}
An admissible change of seed is specified by a nonzero polynomial $Q$ satisfying
\begin{equation}
   0<N_Q=\int |Q(E)|^2\dd\mu(E)<\infty\,,
   \qquad
   \ket{\psi_Q}=N_Q^{-1/2}Q(H)\ket{K_0}.
   \label{eq:seed-normalization}
\end{equation}
The spectral measure of $\ket{\psi_Q}$ is
\begin{equation}
   \dd\nu_Q(E)=\frac{|Q(E)|^2}{N_Q}\dd\mu(E).
   \label{eq:seed-measure}
\end{equation}
The condition in eq.~\eqref{eq:seed-normalization} places $\ket{K_0}$ in
the common domain of $Q(H)$ and $\Qsh(H)$, where $\Qsh(z)=\overline{Q(\bar z)}$ is the
polynomial with complex-conjugated coefficients, because
$|Q(E)|=|\Qsh(E)|$ on the real spectrum.  Equation~\eqref{eq:seed-measure}
then follows from the spectral theorem and
$Q(H)^\dagger=\Qsh(H)$. Multiplication of $Q$ by a nonzero scalar changes the normalized vector only by an overall phase, so the physical seed is determined by the projective class of $Q$. In finite cyclic dimension, it is determined by the projective class of a nonzero element $[Q]\in\cA_d$. Replacing $Q$ by $Q+T_dS$ changes neither the vector nor its scalar measure. Unless stated otherwise, the degree and roots of a finite-dimensional filter refer to its canonical representative of degree below $d$.

Let $R_n^{[Q]}$ denote the monic orthogonal polynomials for $\nu_Q$, with squared norms $\thh_n^{[Q]}$. The shifted Krylov basis is
$
   \ket{\widetilde K_n^{[Q]}}
   =
   R_n^{[Q]}(H)\ket{\psi_Q}/\sqrt{\thh_n^{[Q]}}
$
for every index before shifted termination. Apart from the embedding of the seed in the original Hilbert space, its scalar Jacobi problem, Krylov amplitudes and spread complexity are determined by $\nu_Q$. Although the transformation $\mu\mapsto\nu_Q$ reorganizes the full orthogonal family nonlinearly, section~\ref{sec:finiteband} derives an exact finite-band relation between the reference and shifted problems. For later use with complex coefficients, we collect the already defined involution $\Qsh$ together with the normalized filter $\Qhat$,
\begin{equation}
   \Qsh(z)=\overline{Q(\bar z)},
   \qquad
   \Qhat(z)=\frac{Q(z)}{\sqrt{N_Q}}\,,
   \label{eq:Qsharp-Qhat}
\end{equation}
so that for real $E$, $\Qsh(E)Q(E)=|Q(E)|^2$. For complex $z$, however, the polynomial continuation of $|Q(E)|^2$ is $\Qsh(z)Q(z)$, whereas $|Q(z)|^2$ generally depends on both $z$ and $\bar z$.

An interior seed must be distinguished from a truncation of the reference Jacobi chain. Starting from the reference state $\ket{K_r}$ at fixed $H$ does not delete the first $r$ sites. Associated polynomials are the orthogonal polynomials of the tail chain obtained after deleting its first $r$ sites; they describe this truncated problem  \cite{Chihara1978,SimonOPRL1}, whereas the interior seed retains both the backward and forward couplings of the original chain. If $e_r$ is the standard coordinate vector in the reference Jacobi representation, then
$
   \Jac e_r=b_re_{r-1}+a_re_r+b_{r+1}e_{r+1}.
$
The scalar measure associated with
$\ket{K_r}=P_r(H)\ket{K_0}/\sqrt{h_r}$ is
$
   \dd\nu_r(E)=\frac{P_r(E)^2}{h_r}\dd\mu(E),
$
not the spectral measure of the tail Jacobi matrix. Christoffel transformations, rather than associated-polynomial tails, are therefore the appropriate language for polynomial initial-state jumps at fixed $H$. A separate issue is whether a direct jump between two polynomial seeds factors through intermediate shifted measures. The following proposition answers this question and isolates the obstruction created by support loss.

\subsection{Relative filters, support loss and Jacobi reconstruction}
\label{subsec:relative-filters-support}

The preceding discussion compares an interior seed with a tail truncation.  We now compare
two polynomial seeds directly.  Their relative spectral multiplier is a quotient, so it defines
an ordinary isometric transport precisely when every component needed by the target survives in
the source measure.  The next proposition formulates this support-compatibility criterion and the resulting relative cocycle,
against the classical background of composing Christoffel steps~\cite{Zhedanov1997}.

\begin{proposition}[Relative spectral-filter cocycle]
\label{prop:relative-spectral-filter-cocycle}
Let $Q_i$ and $Q_j$ be nonzero polynomial seeds. Write
$N_i=N_{Q_i}$, $N_j=N_{Q_j}$ and
$\nu_i=\nu_{Q_i}$, $\nu_j=\nu_{Q_j}$, so that
\begin{equation}
   \dd\nu_i(E)=\frac{|Q_i(E)|^2}{N_i}\dd\mu(E),
   \qquad
   \dd\nu_j(E)=\frac{|Q_j(E)|^2}{N_j}\dd\mu(E).
   \label{eq:relative-cocycle-measures}
\end{equation}
Suppose that $\nu_j$ is absolutely continuous with respect to $\nu_i$. Equivalently,
$
   \mu\bigl(\{E:Q_i(E)=0,\ Q_j(E)\ne0\}\bigr)=0.
$
The relative multiplier
\begin{equation}
   F_{i\to j}(E)
   =
   \left(\frac{N_i}{N_j}\right)^{1/2}
   \frac{Q_j(E)}{Q_i(E)}
   \label{eq:relative-filter-main}
\end{equation}
is then defined $\nu_i$-almost everywhere, belongs to $L^2(\nu_i)$, and has norm one. More generally,
\begin{equation}
   (M_{i\leftarrow j}f)(E)
   =
   F_{i\to j}(E)f(E)
\end{equation}
defines an isometry
$M_{i\leftarrow j}:L^2(\nu_j)\to L^2(\nu_i)$. The reversed operator arrow indicates the direction of this isometric embedding between the two function spaces (pull back), while the measure reweighting itself runs from $i$ to $j$ (push forward) and obeys
$
   \dd\nu_j(E)=|F_{i\to j}(E)|^2\dd\nu_i(E).
$
For any compatible triple $i,j,k$, with
$\nu_k\ll\nu_j\ll\nu_i$, the multipliers may be represented on a common full-measure set so that
\begin{equation}
   F_{j\to k}(E)F_{i\to j}(E)=F_{i\to k}(E)
   \label{eq:relative-cocycle-main}
\end{equation}
holds $\nu_i$-almost everywhere. For the reference Krylov states,
$Q_j=p_j=P_j/\sqrt{h_j}$ and $N_j=1$, and hence
\begin{equation}
   \dd\nu_j(E)=p_j(E)^2\dd\mu(E),
   \qquad
   F_{j\to k}(E)=\frac{p_k(E)}{p_j(E)}\,.
   \label{eq:reference-state-relative-filter-main}
\end{equation}
Whenever $\nu_{\ell+1}\ll\nu_\ell$ for every $\ell=0,\ldots,r-1$, the reference-state filters telescope according to
\begin{equation}
   \frac{p_1}{p_0}
   \frac{p_2}{p_1}\cdots
   \frac{p_r}{p_{r-1}}
   =
   p_r,
   \qquad p_0=1\,.
   \label{eq:reference-state-telescope-main}
\end{equation}
Consequently, the sequential route
$0\to1\to\cdots\to r$ and the direct route $0\to r$ produce the same final scalar measure whenever the intermediate relative transformations are support compatible.
\end{proposition}

\begin{proof}
The relation
$\dd\nu_j=|F_{i\to j}|^2\dd\nu_i$ follows directly from
eqs.~\eqref{eq:relative-cocycle-measures} and
\eqref{eq:relative-filter-main}. The absolute-continuity assumption ensures that the part of the reference measure on which $Q_i$ vanishes but $Q_j$ does not is $\mu$-null. Therefore,
\begin{equation}
   \|M_{i\leftarrow j}f\|_{L^2(\nu_i)}^2
   =
   \int |f|^2|F_{i\to j}|^2\,\dd\nu_i
   =
   \int |f|^2\,\dd\nu_j
\end{equation}
for every $f\in L^2(\nu_j)$. The relation between the corresponding cyclic representations is particularly transparent in the common space $L^2(\mu)$. Define the canonical isometric embeddings
$
   (U_\ell f)(E)=f(E)Q_\ell(E)/\sqrt{N_\ell}.
$
Then
$
   U_iM_{i\leftarrow j}=U_j
$
under the stated support condition. In particular, the constant function representing the seed in $L^2(\nu_j)$ is carried to the spectral representative $Q_j/\sqrt{N_j}$ in $L^2(\mu)$. For a compatible triple, the normalization factors and polynomial quotients cancel in the product
$F_{j\to k}F_{i\to j}$ off $\{Q_j=0\}$, proving
eq.~\eqref{eq:relative-cocycle-main}. That set is $\nu_j$-null but may carry $\nu_i$-mass at an atom deleted by $Q_j$, where $\nu_k\ll\nu_j$ forces $Q_k$ to vanish as well, so $F_{i\to j}$ and $F_{i\to k}$ vanish $\nu_i$-almost everywhere on it and any finite value of $F_{j\to k}$ serves as the common representative. The specialization to
$Q_j=p_j$ follows from the orthonormality of the reference polynomials.
\end{proof}

At the level of densities, eq.~\eqref{eq:relative-cocycle-main} is the chain rule for Radon--Nikodym derivatives \cite{Folland1999} applied to $|F_{i\to j}|^2=\dd\nu_j/\dd\nu_i$. What the polynomial setting adds is the explicit rational multiplier lifting that identity from densities to the multipliers themselves, including their phases, together with the support-compatibility criterion deciding when the intermediate arrow exists at all.

The support condition in proposition~\ref{prop:relative-spectral-filter-cocycle} is necessary whenever the reference measure assigns positive mass to a zero of $Q_i$ which is not a zero of $Q_j$. Finite-dimensional chains make this obstruction explicit, and the same issue occurs at atomic components of an infinite scalar measure. Once such a component has been removed from $\nu_i$, the transition $i\to j$ requires information from the original reference measure $\mu$ that is absent from the $i$-cyclic problem. The direct transformation $\mu\to\nu_j$ remains well defined, but no multiplier on $L^2(\nu_i)$ can reconstruct the part of $\nu_j$ carried by $\{Q_i=0,\ Q_j\ne0\}$. Restriction to the common surviving support gives only the restricted measure, not the full $j$-cyclic problem. A nonzero finite-degree polynomial reweights a purely non-atomic measure without changing its measure class. Thus, in the present polynomial setting, support compatibility can fail only at atoms. A relative arrow is invertible precisely when
$\nu_i\sim\nu_j$. Equality of their topological supports is not sufficient. For a finitely supported reference measure with atoms $\{E_\alpha\}$, define
$
   S_i=\{E_\alpha:Q_i(E_\alpha)\ne0\}
$
and
$
   S_j=\{E_\alpha:Q_j(E_\alpha)\ne0\}.
$
Mutual absolute continuity is equivalent to $S_i=S_j$. Writing this common surviving set as $S_{ij}$ and defining
$
   T_{S_{ij}}(E)=\prod_{\xi\in S_{ij}}(E-\xi),
$
the relative multiplier has a polynomial representative which is a unit in
$\CC[E]/\langle T_{S_{ij}}\rangle$. Its values at atoms removed by both filters are not determined by the relative problem.

The relative cocycle should not be confused with restarting the Lanczos construction at each reference state. Starting Lanczos from the interior vector $\ket{K_j}$ gives the first residual
\begin{equation}
   H\ket{K_j}-a_j\ket{K_j}
   =
   b_{j+1}\ket{K_{j+1}}+b_j\ket{K_{j-1}},
   \label{eq:interior-first-residual-main}
\end{equation}
which is not proportional to $\ket{K_{j+1}}$ unless $j=0$ or the backward hopping $b_j$ vanishes. Moving between reference states uses instead the rational relative filters $p_{j+1}/p_j$. Their poles at zeros of the denominator are harmless only in the relative-measure sense specified above. Such filters need not be bounded multipliers on the original spectral support. Only the relative $L^2$ cancellation is guaranteed. By contrast, iterating the first Lanczos step in each newly generated Krylov chain generally produces a different sequence of cyclic vectors.

The support condition has a direct measure-theoretic characterization. For any positive Borel measure $\mu$ on $\mathbb R$ and any Borel set $A$, the shifted measure satisfies
\begin{equation}
   \nu_Q(A)
   =
   \frac{1}{N_Q}\int_A|Q(E)|^2\dd\mu(E),
   \qquad
   N_Q=\int|Q(E)|^2\dd\mu(E).
   \label{eq:absolute-continuity-seed-main}
\end{equation}
Equivalently, the Radon--Nikodym theorem \cite{Folland1999} gives 
$
   \dd\nu_Q/\dd\mu=|Q|^2/N_Q.
$
It follows immediately that $\nu_Q\ll\mu$. The converse holds precisely when
\begin{equation}
      \mu(\mathcal Z_Q)=0,
   \qquad
   \mathcal Z_Q=\{E\in\mathbb R:Q(E)=0\}.
   \label{eq:zero-set-support-loss-main}
\end{equation}
A nonzero finite-degree polynomial has only finitely many real zeros, so it cannot remove any part of a non-atomic measure in the measure-class sense. Multiplication by $|Q|^2$ may introduce isolated zeros into a continuous density without changing the measure class. A genuine measure-class change occurs only when $Q$ removes positive atomic mass. Even then, the topological support loses the corresponding point only when it is not accumulated by the remaining measure.

For finite cyclic dimension, let
\begin{equation}
   \dd\mu(E)
   =
   \sum_{\alpha=0}^{d-1}
   w_\alpha\delta(E-E_\alpha)\dd E,
   \qquad
   E_\alpha\ne E_\beta,\qquad
   w_\alpha>0\,.
\end{equation}
The shifted support consists exactly of the atoms for which
$Q(E_\alpha)\ne0$, with normalized weights
\begin{equation}
   w_\alpha^{[Q]}
   =
   \frac{|Q(E_\alpha)|^2w_\alpha}
   {\sum_\beta|Q(E_\beta)|^2w_\beta}\,.
   \label{eq:finite-shifted-weights-main}
\end{equation}
Equivalently, the class of
$W_Q=\Qsh Q/N_Q$ in $\cA_d$ is a unit precisely when
$Q(E_\alpha)\ne0$ at every visible support point. If $Q$ vanishes at one of these atoms, $W_Q$ is a zero-divisor and the shifted chain loses that spectral component. Its cyclic dimension is therefore reduced. More explicitly,
\begin{equation}
   r_{\rm zero}
   =
   \#\{\alpha:Q(E_\alpha)=0\},
   \qquad
   d_Q=d-r_{\rm zero}\,.
   \label{eq:support-loss-dimension}
\end{equation}
Degeneracies of the full Hamiltonian have already been compressed into the scalar weights $w_\alpha$, with one atom for each distinct energy visible to the seed. Hence $d_Q\le d$, with equality unless the filter annihilates a visible atom. If the reference scalar measure has infinite support and $Q$ is a nonzero finite-degree polynomial, multiplication by $|Q|^2$ may delete atoms at the finitely many real zeros of $Q$, but it cannot reduce the shifted support to a finite set.
Equation~\eqref{eq:support-loss-dimension} is the seed-dependent form of the general
statement that the Krylov dimension of a state equals the number of distinct energies to which its spectral measure assigns nonzero spectral weight, a count that is itself accessible to experiment \cite{CindrakEtAl2024}. What the polynomial
class adds is control of \emph{which} energies are lost and of the transport across the loss.

\paragraph{Resolvent and moment reconstruction.}
The surviving support determines which spectral components remain and, in finite dimension, fixes the shifted cyclic dimension. The full shifted measure, including its atomic, absolutely continuous and singular-continuous parts, then determines the local Jacobi data, which may be reconstructed without a new full-space Lanczos calculation, either from the resolvent or from the shifted moment sequence.
For
$z\in\CC\setminus\supp\mu$, define
\begin{equation}
\begin{aligned}
   m(z)
   &=
   \int\frac{\dd\mu(E)}{E-z},\\
   m_Q(z)
   &=
   \frac{1}{N_Q}
   \int\frac{|Q(E)|^2}{E-z}\dd\mu(E).
\end{aligned}
   \label{eq:shifted-weyl-main}
\end{equation}
Let
$L=\deg(\Qsh Q)=2\deg Q$ and write
\begin{equation}
   \Qsh(z)Q(z)
   =
   \sum_{k=0}^{L}c_k^{[Q]}z^k,
   \qquad
   \mu_j=\int E^j\dd\mu(E).
   \label{eq:Qmultiplier-moments-main}
\end{equation}
Polynomial division gives the finite identity, the Christoffel instance of the linear spectral transformations of the Stieltjes function classified in ref.~\cite{Zhedanov1997}
\begin{equation}
   m_Q(z)
   =
   \frac{\Qsh(z)Q(z)}{N_Q}m(z)
   +
   \frac{1}{N_Q}
   \sum_{k=1}^{L}c_k^{[Q]}
   \sum_{j=0}^{k-1}z^j\mu_{k-1-j}\,.
   \label{eq:shifted-weyl-closed-main}
\end{equation}
Indeed, subtracting and adding $\Qsh(z)Q(z)$ in the numerator makes the polynomial difference
$(\Qsh(E)Q(E)-\Qsh(z)Q(z))$ divisible by $E-z$, with quotient polynomial in $E$ and $z$. Once $N_Q$ is fixed, the additive correction in
eq.~\eqref{eq:shifted-weyl-closed-main} requires only
$\mu_0,\ldots,\mu_{L-1}$. Equivalently, $m_Q$ is determined by $m$, $Q$, and the reference moments through order $L$. Its large-$z$ expansion gives the shifted moments, while its Jacobi continued fraction determines
$\widetilde a_n^{[Q]}$ and $\widetilde b_n^{[Q]}$. The formula therefore provides an algebraic check of normalization and support loss and connects the shifted Jacobi problem directly to local Green functions.

The same expansion reconstructs the shifted Jacobi data from moments. This is the moment/Hankel realization of the seed change, with the shifted moments obtained by a finite Christoffel convolution of the reference sequence. If
$Q(E)=\sum_{p=0}^{r}q_pE^p$, define
\begin{equation}
   M_n^{[Q]}
   =
   \int E^n\Qsh(E)Q(E)\dd\mu(E),
   \qquad
   \widehat\mu_n^{[Q]}
   =
   \frac{M_n^{[Q]}}{M_0^{[Q]}}\,.
   \label{eq:shifted-raw-normalized-moments-main}
\end{equation}
Then $M_0^{[Q]}=N_Q$ and 
\begin{equation}
\begin{aligned}
   M_n^{[Q]}
   &=
   \sum_{p,q=0}^{r}\bar q_pq_q\,\mu_{n+p+q}
   =\sum_{k=0}^{L}c_k^{[Q]}\mu_{n+k},\\
   m_Q(z)
   &\sim
   -\sum_{n=0}^{\infty}
   \frac{\widehat\mu_n^{[Q]}}{z^{n+1}}\,,
   \qquad
   |z|\to\infty,\qquad
   |\operatorname{Im}z|\ge\varepsilon|z|,
   \qquad
   0<\varepsilon<1\ \text{fixed}\,.
\end{aligned}
   \label{eq:shifted-moment-compression-main}
\end{equation}

The sector condition keeps $z$ at a fixed angular distance from the real axis. The first shifted Jacobi coefficients obtained from these moments are 
\begin{equation}
\begin{gathered}
   \widetilde a_0^{[Q]}
   =
   \widehat\mu_1^{[Q]},
   \qquad
   \bigl(\widetilde b_1^{[Q]}\bigr)^2
   =
   \widehat\mu_2^{[Q]}
   -
   \bigl(\widehat\mu_1^{[Q]}\bigr)^2,\\
   \widetilde a_1^{[Q]}
   =
   \frac{
   \widehat\mu_3^{[Q]}
   -2\widehat\mu_1^{[Q]}\widehat\mu_2^{[Q]}
   +\bigl(\widehat\mu_1^{[Q]}\bigr)^3}
   {
   \widehat\mu_2^{[Q]}
   -\bigl(\widehat\mu_1^{[Q]}\bigr)^2}\,.
\end{gathered}
   \label{eq:first-shifted-coefficients-moments-main}
\end{equation}
Equivalently, let
$
   s_k^{[Q]}
   =
   \int(E-\widehat\mu_1^{[Q]})^k\dd\nu_Q(E)
$
denote the central moments of the normalized shifted measure. Then
\begin{equation}
   \widetilde a_1^{[Q]}
   =
   \widehat\mu_1^{[Q]}
   +
   \frac{s_3^{[Q]}}{s_2^{[Q]}},
   \qquad
   \bigl(\widetilde b_2^{[Q]}\bigr)^2
   =
   \frac{s_4^{[Q]}}{s_2^{[Q]}}
   -
   s_2^{[Q]}
   -
   \frac{\bigl(s_3^{[Q]}\bigr)^2}
   {\bigl(s_2^{[Q]}\bigr)^2}\,.
   \label{eq:central-moment-local-coefficients-main}
\end{equation}
These identities apply only up to shifted termination. The formula for
$\widetilde a_1^{[Q]}$ requires $s_2^{[Q]}>0$, equivalently that the shifted measure is not a point mass. In finite dimension this means $d_Q\ge2$. The expression for
$(\widetilde b_2^{[Q]})^2$ is strictly positive when the shifted measure has at least three support points, corresponding to $d_Q\ge3$ in finite dimension, and gives the terminal value zero for a two-point shifted measure. For compact support, the large-$z$ series converges for sufficiently large $|z|$. For general support, it is an order-by-order asymptotic expansion, valid whenever all power moments of $\mu$, and hence of $\nu_Q$, are finite.
The associated Hankel determinants reconstruct the shifted Jacobi data at all orders up to termination. Equations~\eqref{eq:shifted-moment-compression-main}--\eqref{eq:central-moment-local-coefficients-main} provide large-$z$ sum rules and exact low-level checks, while the finite-band connector developed below gives the all-index relation adapted to polynomial initial-state jumps. 

The first two shifted quantities already have a direct physical interpretation,
\begin{equation}
 \widetilde a_0^{[Q]}=\langle H\rangle_Q,
 \qquad
 \bigl(\widetilde b_1^{[Q]}\bigr)^2
 =\langle H^2\rangle_Q-\langle H\rangle_Q^2.
\end{equation}
They fix the prepared energy mean and variance. The variance is the local first-hop scale and equals the coefficient of the $t^2$ term in $K_Q(t)$ in finite support, or in an infinite chain under the graph-norm regularity hypothesis accompanying eq.~\eqref{eq:short-time-shifted-complexity-main}. Support
compatibility is equally physical. An atom annihilated by $Q(H)$ is absent from the prepared
cyclic subspace and no subsequent polynomial acting at the same fixed $H$ can restore that
spectral component.

\section{Finite-band Christoffel transfer}
\label{sec:finiteband}

Section~\ref{sec:setup} identified a polynomial seed change with the Christoffel reweighting $\dd\nu_Q=W_Q\dd\mu$ and obtained the first shifted Jacobi coefficients from its power moments. At every shifted index for which the relevant orthogonal polynomials exist, orthogonality confines $W_QR_n^{[Q]}$ to a finite window of reference polynomials, and integration against $\ee^{-\ii Et}$ turns this connection formula into a finite linear map from the reference Fourier--OP moments to the shifted Krylov amplitudes \cite{Uvarov1969,KautskyGolub1983,BuenoMarcellan2004,NarayanHesthaven2012,GutlebOlverSlevinsky2024}. The resulting connector also recovers the shifted Jacobi data, admits root, Gram and reduced-Jacobi realizations, and reproduces the degree-one and short-time consistency checks derived below.

\subsection{Finite-band connectors and shifted Jacobi data}
\label{subsec:finite-band-connectors}
Set
\begin{equation}
   W_Q(E)=\frac{\Qsh(E)Q(E)}{N_Q},
   \qquad
   L=\deg W_Q.
   \label{eq:WQ-definition-main}
\end{equation}
If $\deg Q=r$, then $L=2r$ and the leading coefficient of $W_Q$ is positive. For real coefficients, $W_Q=Q^2/N_Q$. For complex coefficients, $\Qsh(z)Q(z)/N_Q$ is the polynomial whose restriction to the real axis is the positive multiplier $|Q(E)|^2/N_Q$.

\begin{proposition}[Finite-band transfer]
\label{prop:finite-band}
Let $R_n^{[Q]}$ be the monic OPs for $W_Q\dd\mu$. In an infinite cyclic problem, suppose that the reference OPs and Fourier--OP moments exist through degree $n+L$ and that the shifted inner product is positive definite through degree $n$. Then there are unique coefficients $\Gamma_{n,m}^{[Q]}$ such that
\begin{equation}
\begin{aligned}
   W_Q(E)R_n^{[Q]}(E)
   &=
   \sum_{m=n}^{n+L}\Gamma_{n,m}^{[Q]}P_m(E),\\
   \thh_n^{[Q]}
   &=
   \Gamma_{n,n}^{[Q]}h_n,\\
   \tphi_n^{[Q]}(t)
   &=
   \frac{1}{\sqrt{\thh_n^{[Q]}}}
   \sum_{m=n}^{n+L}\Gamma_{n,m}^{[Q]}I_m(t).
\end{aligned}
   \label{eq:finite-band-main}
\end{equation}
If $w_L$ is the leading coefficient of $W_Q$, then
$\Gamma_{n,n+L}^{[Q]}=w_L$. In finite cyclic dimension $d$, the corresponding statement for $0\le n<d_Q$ is
\begin{equation}
\begin{aligned}
   W_Q(E)R_n^{[Q]}(E)
   &\equiv
   \sum_{m=n}^{d-1}\Gamma_{n,m}^{[Q]}P_m(E)
   \pmod{T_d(E)},\\
   \thh_n^{[Q]}
   &=
   \Gamma_{n,n}^{[Q]}h_n,\\
   \tphi_n^{[Q]}(t)
   &=
   \frac{1}{\sqrt{\thh_n^{[Q]}}}
   \sum_{m=n}^{d-1}\Gamma_{n,m}^{[Q]}I_m(t).
\end{aligned}
   \label{eq:finite-band-quotient}
\end{equation}
Here the coefficients in the first line are those of the canonical representative in the basis $P_0,\ldots,P_{d-1}$.
\end{proposition}

\begin{proof}
In the ordinary polynomial ring, $W_QR_n^{[Q]}$ has degree $n+L$. Since $R_n^{[Q]}$ and $P_{n+L}$ are monic, comparison of leading coefficients gives
$\Gamma_{n,n+L}^{[Q]}=w_L$. For $m<n$, the coefficient of $P_m$ in the reference expansion is
$h_m^{-1}$ times
\begin{equation}
   \int P_m(E)W_Q(E)R_n^{[Q]}(E)\dd\mu(E)
   =
   \int P_m(E)R_n^{[Q]}(E)\dd\nu_Q(E)
   =
   0\,,
\end{equation}
because $\deg P_m<n$. Hence the expansion begins at $P_n$. Further, since $R_n^{[Q]}=P_n+$ lower-degree terms, orthogonality with respect to $\mu$ gives
\begin{equation}
   \int P_m(E)R_n^{[Q]}(E)\dd\mu(E)
   =
   h_n\delta_{mn},
   \qquad m\ge n\,.
\end{equation}
Multiplying the first identity in eq.~\eqref{eq:finite-band-main} by $R_n^{[Q]}$ and integrating with respect to $\mu$ gives
$\thh_n^{[Q]}=\Gamma_{n,n}^{[Q]}h_n$. Integration against
$\ee^{-\ii Et}\dd\mu(E)$ then gives the amplitude formula.

In finite dimension, the quotient class of $W_QR_n^{[Q]}$ has a unique expansion in the orthogonal basis $P_0,\ldots,P_{d-1}$. Shifted orthogonality eliminates its coefficients below $P_n$, and the same two integrations give the remaining lines of eq.~\eqref{eq:finite-band-quotient}. If an ordinary-ring representative is needed near the terminal edge, the leading-coefficient statement is understood after choosing a proper terminal lift and is followed by reduction modulo $T_d$. The leading coefficient belongs to the lifted ordinary-polynomial representative. The intrinsic finite-chain object is its reduced quotient class.
\end{proof}

In monic normalization the first two lines of proposition~\ref{prop:finite-band} are the Christoffel connection and the Cholesky-diagonal norm relation of refs.~\cite{Uvarov1969,KautskyGolub1983}, in the banded operator form of ref.~\cite{GutlebOlverSlevinsky2024}; the third line transports them to the Fourier--OP amplitudes, and eq.~\eqref{eq:finite-band-quotient} carries the pair across the finite terminal edge in canonical quotient form.
The coefficients may also be extracted directly from
\begin{equation}
   \Gamma_{n,m}^{[Q]}
   =
   \frac{1}{h_m}
   \int P_m(E)W_Q(E)R_n^{[Q]}(E)\dd\mu(E),
   \qquad
   n\le m\le n+L\,.
   \label{eq:Gamma-inner-product-extraction-main}
\end{equation}
In finite dimension, this formula applies to the canonical quotient coefficients for $n\le m\le d-1$. It directly checks connector conventions and normalization, although the root, Gram, and reduced-Jacobi constructions below are usually more efficient.
The connector also determines the complete shifted Jacobi data. Define
$
   S_n^{[Q]}=W_QR_n^{[Q]}.
$
The following identities replace a new Lanczos construction in the full Hilbert space by finite algebra on each connector row of the solved reference Jacobi problem.

\begin{proposition}[Connector dictionary for shifted Jacobi data]
\label{prop:connector-dictionary}
Write
$\beta_n=b_n^2$ and
$\widetilde\beta_n^{[Q]}=(\widetilde b_n^{[Q]})^2$, with
$\beta_0=\widetilde\beta_0^{[Q]}=0$, and set
$\Gamma_{n,m}^{[Q]}=0$ outside the relevant connector row. 
Throughout, the hypotheses are those of proposition~\ref{prop:finite-band} at the level under consideration. In the infinite-support case the reference orthogonal polynomials exist through degree $n+L$, while the shifted inner product is positive definite through degree $n$; in finite cyclic dimension the quotient formulas are used for $0\le n<d_Q$. The identities below use only the connection coefficients, not the Fourier--OP moments. The norm relation $\thh_n^{[Q]}=\Gamma_{n,n}^{[Q]}h_n$ makes every displayed diagonal connector entry in a denominator strictly positive. Up to shifted termination,
\begin{align}
   \widetilde\beta_n^{[Q]}
   &=
   \beta_n
   \frac{\Gamma_{n,n}^{[Q]}}
   {\Gamma_{n-1,n-1}^{[Q]}},
   \qquad n\ge1\,,
   \label{eq:connector-beta-dictionary-main}\\
   \widetilde a_n^{[Q]}
   &=
   a_n
   +
   \beta_{n+1}
   \frac{\Gamma_{n,n+1}^{[Q]}}
   {\Gamma_{n,n}^{[Q]}}
   -
   \widetilde\beta_n^{[Q]}
   \frac{\Gamma_{n-1,n}^{[Q]}}
   {\Gamma_{n,n}^{[Q]}}\,.
   \label{eq:connector-a-dictionary-main}
\end{align}
The full connector satisfies the standard intertwining relation between the reference and modified Jacobi matrices~\cite{KautskyGolub1983,GutlebOlverSlevinsky2024},
\begin{equation}
\begin{aligned}
   \Gamma_{n,k-1}^{[Q]}
   +a_k\Gamma_{n,k}^{[Q]}
   +\beta_{k+1}\Gamma_{n,k+1}^{[Q]}
   &=
   \Gamma_{n+1,k}^{[Q]}
   +\widetilde a_n^{[Q]}\Gamma_{n,k}^{[Q]}
   +\widetilde\beta_n^{[Q]}\Gamma_{n-1,k}^{[Q]}\,.
\end{aligned}
   \label{eq:connector-recurrence-compatibility-main}
\end{equation}
In finite cyclic dimension, these formulas are interpreted after canonical reduction modulo $T_d$, with the reference terminal convention $\beta_d=b_d^2=0$, and only up to shifted termination. They give
$\widetilde\beta_n^{[Q]}$ for $1\le n<d_Q$, with
$\widetilde\beta_{d_Q}^{[Q]}=0$, and
$\widetilde a_n^{[Q]}$ for $0\le n<d_Q$. The final diagonal relation uses
\begin{equation}
   S_{d_Q}^{[Q]}
   =
   W_Q\widetilde T_{d_Q}
   \equiv0
   \pmod{T_d},
\end{equation}
where $\widetilde T_{d_Q}$ is the monic terminal polynomial of the shifted support.
\end{proposition}

\begin{proof}
Multiplication of the shifted monic recurrence
\begin{equation}
   ER_n^{[Q]}
   =
   R_{n+1}^{[Q]}
   +\widetilde a_n^{[Q]}R_n^{[Q]}
   +\widetilde\beta_n^{[Q]}R_{n-1}^{[Q]}
\end{equation}
by $W_Q$ gives
\begin{equation}
   ES_n^{[Q]}
   =
   S_{n+1}^{[Q]}
   +\widetilde a_n^{[Q]}S_n^{[Q]}
   +\widetilde\beta_n^{[Q]}S_{n-1}^{[Q]}\,.
\end{equation}
Expanding each $S_j^{[Q]}$ in the reference basis and using
$
   EP_k=P_{k+1}+a_kP_k+\beta_kP_{k-1}
$
gives eq.~\eqref{eq:connector-recurrence-compatibility-main} upon comparison of the coefficient of $P_k$. Setting $k=n-1$ gives
eq.~\eqref{eq:connector-beta-dictionary-main}; setting $k=n$ then gives
eq.~\eqref{eq:connector-a-dictionary-main}. In finite dimension, the same coefficient comparison is performed in the quotient algebra. At the last shifted level, the term $S_{d_Q}^{[Q]}$ is the zero quotient class, which fixes the terminal diagonal relation.
\end{proof}

For a degree-$r$ seed, the connector window contains at most $2r+1$ reference Fourier--OP moments,
$
   I_n(t),\ldots,I_{n+2r}(t),
$
before terminal reduction. Once the time-independent row
$\Gamma_{n,\bullet}^{[Q]}$ is known, it determines
$\tphi_n^{[Q]}(t)$ without requiring any of the preceding shifted amplitudes. The locality of this statement comes from the Fourier--OP representation of the amplitude. The shifted basis vector
$
   \ket{\widetilde K_n^{[Q]}}
   =
   R_n^{[Q]}(H)\Qhat(H)\ket{K_0}/\sqrt{\thh_n^{[Q]}}
$
before terminal reduction, is generated by a polynomial of degree $n+r$ and may have components throughout the reference levels $(0,1,2,\ldots,n+r)$. In finite dimension its representative is reduced in $\cA_d$. It is therefore not confined to the local connector window $n,\ldots,n+2r$. The lower reference components cancel only in the amplitude pairing, in which the adjoint seed factor contributed by the bra combines with the seed factor in the shifted vector to produce the numerator $W_QR_n^{[Q]}$, after which shifted orthogonality removes the components below $P_n$.

\subsection{Connector realizations, degree-one seeds and initial growth}
\label{subsec:connector-realizations}

The connector admits three complementary realizations. They differ in what they take as
given.  We name them once and use the following terminology throughout:
\begin{enumerate}[label=(\roman*),leftmargin=2.4em,itemsep=1pt]
\item the \emph{root realization}, which encodes divisibility by $W_Q$ as vanishing conditions at the zeros of $W_Q$ and
      solves the resulting confluent Christoffel system for the connector row directly;
\item the \emph{moment/Gram realization}, which is root free and builds the shifted Hankel matrix
      from the reference power moments, extracts $R_n^{[Q]}$, and forms
      $S_n^{[Q]}=W_QR_n^{[Q]}$;
\item the \emph{reduced-Jacobi realization}, which runs the Lanczos recursion for the filtered
      coordinate vector $v_Q=Q(\Jac)e_0/\sqrt{N_Q}$ \emph{inside the already constructed reference
      Jacobi representation}, and recovers the connector from the resulting basis.
\end{enumerate}
The three realizations produce the same connector. They are distinguished by the reference data they consume (roots, moments, or the Jacobi matrix itself) and by their numerical conditioning, which is
compared in appendix~\ref{app:numerics}.
The coefficient-space projector developed in section~\ref{sec:kernels} uses the related multiplication image $\Qhat\cP_\ell$, with $\cP_m$ the space of polynomials of degree at most $m$, but computes cumulative probabilities rather than connector rows. In finite dimension the distinction matters, because the connector realizations below produce ordinary-ring representatives that are subsequently reduced modulo $T_d$, whereas the cumulative probabilities of section~\ref{sec:kernels} come from a projector defined intrinsically on the quotient $\cA_d$. Appendix~\ref{app:christoffel} gives the determinant solution and proves its terminal-lift prescription; appendix~\ref{app:numerics} develops the corresponding reference-space implementations.

For the root construction, let $\cP_m$ denote the polynomials in $E$ of degree at most $m$, equipped with the $L^2(\mu)$ inner product, and factor
\begin{equation}
   W_Q(E)
   =
   w_L\prod_{a=1}^{s}(E-\zeta_a)^{m_a},
   \qquad
   \sum_{a=1}^{s}m_a=L\,.
   \label{eq:weight-factorization}
\end{equation}
Since $R_n^{[Q]}$ and $P_{n+L}$ are monic,
$\Gamma_{n,n+L}^{[Q]}=w_L$. The remaining $L$ coefficients are fixed by the divisibility condition $W_Q\mid S_n^{[Q]}$, equivalently
\begin{equation}
   \partial_E^\ell S_n^{[Q]}(\zeta_a)=0,
   \qquad
   a=1,\ldots,s,
   \qquad
   \ell=0,\ldots,m_a-1\,.
   \label{eq:root-conditions-main}
\end{equation}
This is an $L\times L$ system for
$\Gamma_{n,n}^{[Q]},\ldots,\Gamma_{n,n+L-1}^{[Q]}$. In the ordinary polynomial-ring realization, the system is nonsingular whenever the shifted inner product is positive definite through degree $n$. Indeed, a homogeneous solution defines
$
   \delta S_n\in\Span\{P_n,\ldots,P_{n+L-1}\}
$
which is divisible by $W_Q$, so that
$\delta S_n=W_QT$ with $\deg T<n$. Since
$\delta S_n\perp\cP_{n-1}$ in $L^2(\mu)$,
\begin{equation}
   0
   =
   \int \overline{T(E)}\,\delta S_n(E)\dd\mu(E)
   =
   \int W_Q(E)|T(E)|^2\dd\mu(E)
   =
   \|T\|_{\nu_Q}^2.
\end{equation}
Positive definiteness gives $T=0$, proving uniqueness. Near a finite terminal edge, the same argument is applied after choosing a proper terminal lift of the reference polynomials, and the resulting connector is then reduced modulo $T_d$.

The connector may also be constructed without finding the roots. Write
$
   \Qsh(E)Q(E)=\sum_{k=0}^{L}c_k^{[Q]}E^k.
$
The normalized shifted Hankel matrix is
\begin{equation}
   \mathsf H_{ij}^{[Q]}
   =
   \int E^{i+j}W_Q(E)\dd\mu(E)
   =
   \widehat\mu_{i+j}^{[Q]}
   =
   \frac{1}{N_Q}
   \sum_{k=0}^{L}c_k^{[Q]}\mu_{i+j+k}\,,
   \qquad i,j\ge0\,.
   \label{eq:shifted-Hankel-moments-main}
\end{equation}
Here the $\mu_j$ are the reference power moments defined in
eq.~\eqref{eq:Qmultiplier-moments-main}. Gram determinants or weighted Gram--Schmidt recover $R_n^{[Q]}$ from $\mathsf H^{[Q]}$, after which
$S_n^{[Q]}=W_QR_n^{[Q]}$ gives the connector. At higher degrees, orthogonal factorizations of the corresponding Gram or coefficient matrix are preferable to raw determinants. Complex seed coefficients require no modification of this moment construction, since $W_Q=\Qsh Q/N_Q$ has real coefficients and is nonnegative on the real spectrum. For a seed with complex coefficients, the cumulative probabilities are instead obtained from the Hermitian coefficient-space projector of section~\ref{sec:kernels}; as explained there, the kernel built from evaluations at real seed roots is not carried to nonreal roots by analytic continuation.

The reduced-Jacobi realization works directly with the reference Jacobi matrix. In the orthonormal reference basis, the filtered seed has coordinate vector
\begin{equation}
   v_Q
   =
   \frac{Q(\Jac)e_0}{\sqrt{N_Q}}\,,
   \qquad
   N_Q
   =
  e_0^\dagger\Qsh(\Jac)Q(\Jac)e_0\,.
   \label{eq:reduced-jacobi-seed-main}
\end{equation}
Lanczos applied to $\Jac$ with initial vector $v_Q$ gives the Jacobi matrix and basis of the shifted measure $W_Q\dd\mu$. This calculation takes place inside the already constructed reference Jacobi representation rather than in the full Hilbert space. For finite support one uses $\Jac_d$. If $Q$ vanishes at some support atoms, the associated components of $v_Q$ vanish and the reduced process terminates at the shifted cyclic dimension $d_Q$. The reduced-Jacobi construction therefore determines the shifted recurrence and basis data directly. The shifted norms and connector are then recovered as above, while section~\ref{sec:kernels} constructs the associated cumulative-probability projectors. Appendix~\ref{app:numerics} gives the exact scaled connector relation and the required finite-rank and truncation checks.

The root and moment constructions determine the connector directly, whereas the reduced-Jacobi realization determines equivalent shifted basis data from which the connector is recovered. The same transformation also admits an operator-factorization interpretation, relating the Christoffel modification to a Darboux interchange of Jacobi factors.
In this language a single Darboux step implements multiplication of the measure by one real linear factor $E-\xi$. The pure first jump $Q=P_1$ is therefore not a single Darboux step. Its physical multiplier $W_Q\propto P_1^2$ is quadratic and corresponds to two confluent linear factors. In the reference Krylov basis we can write
\begin{equation}
   \Jac
   =
   \mathsf A+\mathsf B_++\mathsf B_-\,,
   \qquad
   \mathsf A e_n=a_ne_n\,,
   \qquad
   \mathsf B_+e_n=b_{n+1}e_{n+1}\,,
   \qquad
   \mathsf B_-=\mathsf B_+^\dagger\,.
   \label{eq:Jacobi-ladder-main}
\end{equation}
This diagonal-plus-ladder split fixes the notation for the factorization that follows. Its weighted-shift structure is taken up again in section~\ref{sec:context} and appendix~\ref{app:christoffel}. For a finite Jacobi matrix, or for an infinite Jacobi operator for which the corresponding operator factorization is well defined, a simple real factor $E-\xi$ with $\xi$ below the spectral support gives
$
   \Jac-\xi\Id=\mathsf L_\xi\mathsf L_\xi^\dagger
$
and
$
   \widetilde\Jac=\mathsf L_\xi^\dagger\mathsf L_\xi+\xi\Id,
$
where $\mathsf L_\xi$ is the lower-bidiagonal Cholesky factor with positive diagonal. This Cholesky interchange preserves the symmetric orthonormal convention and gives the Jacobi matrix of $(E-\xi)\dd\mu/(a_0-\xi)$. The normalization does not affect the Jacobi coefficients.
The familiar $LU\leftrightarrow UL$ form applies instead to the monic, generally nonsymmetric Jacobi representation. General polynomial multipliers are obtained through algebraic Darboux or block factorizations, with confluent limits for repeated roots \cite{BuenoMarcellan2004,GutlebOlverSlevinsky2024}. For the seed problem, the complete positive multiplier $W_Q$ is the invariant object. It fixes the final transformed Jacobi matrix even when a particular ordering of its linear factors passes through intermediate functionals that are not positive and, for nonreal factors, not Hermitian. The transformed recurrence is encoded by $\widetilde a_n^{[Q]}$ and $\widetilde\beta_n^{[Q]}$. These coefficients need not retain the structural form of the reference chain. If, as in subsection~\ref{subsec:HW}, the reference data are those of a ladder algebra, so that $a_n$ is affine and $b_n^2$ linear in $n$, the shifted data need not be affine and linear, and in general they are not.

At a finite terminal edge, determinant formulas are evaluated with the proper lifts of section~\ref{sec:setup} and then reduced modulo $T_d$, as proved in proposition~\ref{prop:terminal-lift-determinant}. Equivalently, the reference polynomials may be evaluated on the spectral support, with $w_\alpha$ replaced by $W_Q(E_\alpha)w_\alpha$ and zero-weight atoms removed, followed by the degree-ordered finite Gram factorization. Both procedures give the same canonical quotient connector and introduce no post-terminal Krylov sites.

\paragraph{Degree-one polynomial seeds.} Assume that the reference chain has not terminated at its seed, so that $h_1=b_1^2>0$ and $\ket{K_1}$ exists. The first nontrivial seed already exhibits both conjugate-root and confluent constraints. Let
\begin{equation}
   \ket{\psi_c}
   =
   c_0\ket{K_0}+c_1\ket{K_1},
   \qquad
   |c_0|^2+|c_1|^2=1\,.
   \label{eq:degree-one-seed-main}
\end{equation}
Since $P_1(E)=E-a_0$ and $h_1=b_1^2$, the corresponding filter is
$
   Q(E)=c_0+c_1P_1(E)/\sqrt{h_1}.
$
For $c_1\ne0$, its root is
$
   z=a_0-b_1c_0/c_1,
$
while $\Qsh$ has root $\bar z$. If $z\notin\mathbb R$, the multiplier
$\Qsh Q$ has two distinct conjugate roots. If $z\in\mathbb R$, the two roots coalesce and the Christoffel constraints are confluent. The pure first jump $c_0=0$ is the special confluent case $z=a_0$. Away from a finite terminal edge, the connector has the ordinary-ring form
\begin{equation}
   W_Q(E)R_n^{[Q]}(E)
   =
   \Gamma_{n,n}^{[Q]}P_n(E)
   +
   \Gamma_{n,n+1}^{[Q]}P_{n+1}(E)
   +
   \Gamma_{n,n+2}^{[Q]}P_{n+2}(E)\,.
   \label{eq:degree-one-connector-main}
\end{equation}
When $n+2\ge d$, the same expression is understood in the lift-and-reduce sense of the preceding paragraph. 
The two remaining coefficients are fixed by the two, possibly confluent, root constraints. Thus, a degree-one seed produces a degree-two Christoffel multiplier and changes the full shifted Jacobi data. On finite support, the same connector follows from the reduced discrete measure without extracting the roots of $Q$.

In the ordinary polynomial range, all three constructions give the same connector polynomial $S_n^{[Q]}=W_QR_n^{[Q]}$. At a finite terminal edge they give the same canonical quotient class after reduction modulo $T_d$, although auxiliary lifted representatives may differ.  Equation~\eqref{eq:Gamma-inner-product-extraction-main} checks its normalization, while proposition~\ref{prop:connector-dictionary} recovers the shifted Jacobi data. Numerical implementations and stability tests are summarized in appendix~\ref{app:numerics} \cite{Gautschi2004,Parlett1998}.

\paragraph{Survival amplitude and initial growth.}
Each connector row determines
$\tphi_n^{[Q]}$ independently, without first constructing the lower shifted amplitudes. The zeroth connector row gives the survival amplitude of the shifted seed. Since
$R_0^{[Q]}=1$ and $\thh_0^{[Q]}=1$,
\begin{equation}
   \tphi_0^{[Q]}(t)
   =
   \int\ee^{-\ii Et}W_Q(E)\dd\mu(E)
   =
   \sum_m\Gamma_{0,m}^{[Q]}I_m(t),
   \label{eq:survival-connector-transfer-main}
\end{equation}
where the sum runs over the nonzero entries of the zeroth connector row. Equivalently, using
$(\ii\partial_t)\ee^{-\ii Et}=E\ee^{-\ii Et}$,
\begin{equation}
   \tphi_0^{[Q]}(t)
   =
   \frac{1}{N_Q}
   \Qsh(\ii\partial_t)Q(\ii\partial_t)I_0(t).
   \label{eq:survival-differential-transfer-main}
\end{equation}
The first representation is adapted to the reference Krylov basis, while the second provides a direct differential check.

Expanding either representation gives the energy variance which enters the Mandelstam--Tamm bound. Whenever the shifted measure is not supported at a single energy,
\begin{equation}
   \Delta_Q^2
   =
   \int E^2\dd\nu_Q
   -
   \left(\int E\dd\nu_Q\right)^2
   =
   \bigl(\widetilde b_1^{[Q]}\bigr)^2
   =
   \thh_1^{[Q]}.
   \label{eq:MT-speed-shifted-seed-main}
\end{equation}
In finite dimension the shifted measure fails to be a point mass precisely when at least two atoms survive the filter, that is, when $d_Q\ge2$. If $d_Q=1$, the shifted chain terminates at $n=0$, the complexity vanishes identically, and $\thh_1^{[Q]}$ is not part of the positive-norm shifted family. For a finite shifted chain with $d_Q\ge2$, the tridiagonal Schr\"odinger equation for the shifted amplitudes,
$
   \ii\,\partial_t\tphi_n^{[Q]}
   =
   \widetilde b_{n+1}^{[Q]}\tphi_{n+1}^{[Q]}
   +\widetilde a_n^{[Q]}\tphi_n^{[Q]}
   +\widetilde b_n^{[Q]}\tphi_{n-1}^{[Q]},
$
gives
$
   \tphi_1^{[Q]}(t)
   =
   -\ii\widetilde b_1^{[Q]}t+O(t^2).
$
Moreover,
$
   \tphi_n^{[Q]}(-t)=\overline{\tphi_n^{[Q]}(t)},
$
so the probabilities are even analytic functions of time. It follows that
\begin{equation}
   K_Q(t)=\Delta_Q^2t^2+O(t^4).
   \label{eq:short-time-shifted-complexity-main}
\end{equation}
For an infinite chain, the same quadratic coefficient holds under the corresponding graph-norm condition, although the general remainder is then only $o(t^2)$.\footnote{Let
$\mathsf N_{\rm K}^{[Q]}=\sum_{n\ge0}n\ket{\widetilde K_n^{[Q]}}\!\bra{\widetilde K_n^{[Q]}}$.
If
$t\mapsto\ee^{-\ii tH}\ket{\psi_Q}$
is differentiable at $t=0$ as a map into
$\operatorname{Dom}\bigl((\mathsf N_{\rm K}^{[Q]})^{1/2}\bigr)$
with graph norm
$\|v\|_{\rm gr}^2=\|v\|^2+\|(\mathsf N_{\rm K}^{[Q]})^{1/2}v\|^2$,
then
$K_Q(t)=\Delta_Q^2t^2+o(t^2)$.
An $O(t^4)$ remainder requires stronger graph-norm regularity or a uniform
$\mathsf N_{\rm K}^{[Q]}$-weighted tail estimate. It does not follow from the fourth energy moment alone.} For the reference-state jump $Q=P_r/\sqrt{h_r}$, the variance is read directly from the original Jacobi chain
\begin{equation}
   H\ket{K_r}
   =
   b_{r+1}\ket{K_{r+1}}
   +
   a_r\ket{K_r}
   +
   b_r\ket{K_{r-1}},
   \qquad
   \Delta_r^2=b_r^2+b_{r+1}^2\,.
   \label{eq:reference-state-speed-main}
\end{equation}
Here $b_0=0$, and the missing terminal hopping is set to zero in finite dimension. Thus, the initial complexity growth of the jumped seed is governed by the sum of the squared hopping amplitudes from level $r$ to its two neighboring reference levels.
This identity fixes the quadratic coefficient of the shifted-basis complexity; by itself it does not identify the full finite-time complexity $K_r(t)$ with a reference-basis distance from level $r$.

The connector therefore determines each shifted amplitude one row at a time and reconstructs the shifted Jacobi chain from the same time-independent coefficients, without a new ambient-space Lanczos calculation. Spread complexity, however, weights the entire shifted ladder; section~\ref{sec:kernels} obtains the required cumulative probabilities by finite-rank projection of the reference Christoffel--Darboux kernel.

\section{Projected kernels and complexity}
\label{sec:kernels}

Proposition~\ref{prop:finite-band} determines each shifted amplitude separately. Spread complexity, however, is naturally reconstructed from the cumulative probabilities of the shifted Krylov filtration. Before terminal reduction, multiplication by $\Qhat$ maps $(\cP_\ell,\langle\cdot,\cdot\rangle_{\nu_Q})$ isometrically onto its image in $\cP_{\ell+r}$; in finite support with $\ell+r\ge d$, the same image is understood intrinsically in the quotient $\cA_d$. The corresponding orthogonal projector gives $\Phi_\ell^{[Q]}(t)$ directly and hence, through the tail identity, determines $K_Q(t)$ \cite{SimonCDK}. Subsection~\ref{subsec:cumulative-currents} collects the tail, generating-function and current identities that reduce the complexity to cumulative data; subsection~\ref{subsec:projection-filtration} constructs those cumulative probabilities by a finite-rank projection and derives a general filtration bound; subsection~\ref{subsec:finite-time-first-jump} then establishes finite-time control of the complexity under a linear growth condition on the Jacobi coefficients and derives the first jump explicitly. For a real degree-$r$ filter whose roots are real and simple, proposition~\ref{prop:projected-kernel-main} uses $r$ evaluation constraints, with derivative jets replacing them at repeated roots. Filters with nonreal roots or complex coefficients, and finite-support rows at or beyond the terminal edge, are instead covered by the Hermitian coefficient-space and quotient-image realizations. The first jump is written out separately only because its rank-one correction is the case used in the Charlier analysis of section~\ref{sec:examples}. Appendix~\ref{app:projected-kernel} proves the jet and coefficient-space realizations and treats the finite terminal quotient in detail.

\subsection{Cumulative probabilities and Krylov currents}
\label{subsec:cumulative-currents}
We define the reference and shifted truncated kernels by\footnote{Here $M$ is the upper polynomial cutoff in the reference kernel; it is unrelated to the matrix-valued measure introduced later.}
\begin{equation}
   \mathbb K_M^\mu(E,F)
   =
   \sum_{m=0}^{M}\frac{P_m(E)P_m(F)}{h_m}\,,
   \qquad
   \tK_\ell^{[Q]}(E,F)
   =
   \sum_{n=0}^{\ell}
   \frac{R_n^{[Q]}(E)R_n^{[Q]}(F)}
   {\thh_n^{[Q]}}\,.
   \label{eq:kernels-def}
\end{equation}
In finite shifted dimension, $0\le\ell\le d_Q-1$ and every sum over shifted Krylov levels terminates at $d_Q-1$. We use $d=\infty$ and $d_Q=\infty$ for nonterminating reference and shifted chains. The cumulative probability is
\begin{equation}
   \Phi_\ell^{[Q]}(t)
   =
   \sum_{n=0}^{\ell}|\tphi_n^{[Q]}(t)|^2
   =
   \iint
   \ee^{-\ii t(E-F)}
   \tK_\ell^{[Q]}(E,F)
   \dd\nu_Q(E)\dd\nu_Q(F).
   \label{eq:cumulative-kernel}
\end{equation}
The quantity $\Phi_\ell^{[Q]}(t)$ is the probability that the evolved shifted state is found
at or below Krylov level $\ell$. It is the cumulative distribution function of the Krylov-level
random variable, so the complexity, which is the mean level, is the sum of the tail probabilities
$1-\Phi_\ell^{[Q]}$. For a finite shifted Krylov space,
\begin{equation}
   K_Q(t)
   =
   \sum_{n=0}^{d_Q-1}
   n|\tphi_n^{[Q]}(t)|^2
   =
   \sum_{\ell=0}^{d_Q-2}
   \bigl(1-\Phi_\ell^{[Q]}(t)\bigr).
   \label{eq:tail-complexity}
\end{equation}
For an infinite chain satisfying this shifted-density assumption, shifted-OP completeness in the shifted cyclic subspace and Parseval give
$\sum_{n\ge0}|\tphi_n^{[Q]}(t)|^2=1$. Tonelli's theorem \cite{Folland1999} then gives the extended nonnegative identity
\begin{equation}
\begin{aligned}
   \sum_{n\ge0}n|\tphi_n^{[Q]}(t)|^2
   &=
   \sum_{n\ge0}
   \sum_{\ell=0}^{n-1}
   |\tphi_n^{[Q]}(t)|^2
   =
   \sum_{\ell\ge0}
   \bigl(1-\Phi_\ell^{[Q]}(t)\bigr)
   \in[0,\infty].
\end{aligned}
   \label{eq:tail-complexity-infinite-main}
\end{equation}
Equation~\eqref{eq:tail-complexity-infinite-main} is the same mean-as-sum-of-tails identity, now valid in $[0,\infty]$. Consequently, $K_Q(t)$ is finite precisely when the shifted Krylov-level distribution has a finite first moment.

The same probabilities may be collected in
$
   G_Q(s;t)=\sum_{n=0}^{d_Q-1}
   s^n|\tphi_n^{[Q]}(t)|^2,
$
where the upper limit is understood as infinity when $d_Q=\infty$. Its kernel representation is 
\begin{equation}
   \mathscr G_Q(s;E,F)
   =
   \sum_{n=0}^{d_Q-1}
   s^n
   \frac{R_n^{[Q]}(E)R_n^{[Q]}(F)}
   {\thh_n^{[Q]}}\,.
   \label{eq:probability-kernel-generating-main}
\end{equation}
For an infinite chain and $|s|<1$, the series in
eq.~\eqref{eq:probability-kernel-generating-main} converges in
$L^2(\nu_Q\otimes\nu_Q)$, while in finite dimension it is a finite sum. In either case,
\begin{equation}
   G_Q(s;t)
   =
   \iint
   \ee^{-\ii t(E-F)}
   \mathscr G_Q(s;E,F)
   \dd\nu_Q(E)\dd\nu_Q(F).
\end{equation}
Independently of any pointwise boundary value of the kernel, Parseval gives $G_Q(1;t)=1$. For real $s\uparrow 1$, monotone convergence gives
$
   K_Q(t)=\left.s\partial_sG_Q(s;t)\right|_{s=1^-}
$
as an identity in $[0,\infty]$. Pointwise convergence of the infinite kernel series and its boundary limit as $s\to1^-$ require additional assumptions. None is needed for the finite-cutoff kernels.

\paragraph{Krylov-cut currents and complexity growth.}
Complexity grows exactly as fast as probability flows outward across the cuts of the Krylov
chain. Each tail probability $1-\Phi_\ell^{[Q]}$ changes only by transport across the bond
$(\ell,\ell+1)$, so the growth rate of $K_Q$ is a sum of bond currents, each of which the
finite-band transfer renders computable from two adjacent connector rows. In the shifted basis, the amplitudes satisfy the tridiagonal Schr\"odinger equation with hoppings $\widetilde b_n^{[Q]}$ and diagonals $\widetilde a_n^{[Q]}$. Define the signed current across the bond $(\ell,\ell+1)$ by
\begin{equation}
   J_{\ell+1}^{[Q]}(t)
   =
   2\widetilde b_{\ell+1}^{[Q]}
   \im\!\left(
      \overline{\tphi_\ell^{[Q]}(t)}
      \tphi_{\ell+1}^{[Q]}(t)
   \right).
   \label{eq:shifted-current-main}
\end{equation}
With $J_0^{[Q]}=0$, and $J_{d_Q}^{[Q]}=0$ at finite shifted termination, the continuity equation is
$
   \frac{\dd}{\dd t}|\tphi_n^{[Q]}|^2
   =
   J_{n+1}^{[Q]}-J_n^{[Q]}.
$
Consequently,
$
   \dot\Phi_\ell^{[Q]}(t)=J_{\ell+1}^{[Q]}(t).
$
Thus $J_{\ell+1}^{[Q]}$ is oriented toward the lower-level side of the cut, and the outward
current is $-J_{\ell+1}^{[Q]}$.
Substitution of the finite-band amplitudes, together with
$
   (\widetilde b_{\ell+1}^{[Q]})^2
   =
   \thh_{\ell+1}^{[Q]}/\thh_\ell^{[Q]},
$
gives
\begin{equation}
   \dot\Phi_\ell^{[Q]}(t)
   =
   \frac{2}{\thh_\ell^{[Q]}}
   \im\!\left[
      \overline{
      \sum_m\Gamma_{\ell,m}^{[Q]}I_m(t)}
      \left(
      \sum_k\Gamma_{\ell+1,k}^{[Q]}I_k(t)
      \right)
   \right].
   \label{eq:Phi-speed-transfer-main}
\end{equation}
Each sum is restricted to its finite connector window, with the terminal-quotient convention of proposition~\ref{prop:finite-band}. Thus, the probability current through the cut after level $\ell$ is expressed entirely through two adjacent connector rows and the reference Fourier--OP moments. In particular,
\begin{equation}
   \left|\dot\Phi_\ell^{[Q]}(t)\right|
   \le
   \frac{2}{\thh_\ell^{[Q]}}
   \left|
      \sum_m\Gamma_{\ell,m}^{[Q]}I_m(t)
   \right|
   \left|
      \sum_k\Gamma_{\ell+1,k}^{[Q]}I_k(t)
   \right|.
   \label{eq:Phi-speed-bound-main}
\end{equation}
Using the tail identity of eq.~\eqref{eq:tail-complexity},\footnote{Differentiating $K_Q=\sum_{\ell}(1-\Phi_\ell^{[Q]})$ term by term and inserting $\dot\Phi_\ell^{[Q]}=J_{\ell+1}^{[Q]}$ gives $\dot K_Q=-\sum_\ell J_{\ell+1}^{[Q]}$; in finite dimension the sum is finite and no further justification is needed.} these currents give
\begin{equation}
   \dot K_Q(t)
   =
   -2\sum_{\ell=0}^{d_Q-2}
   \frac{1}{\thh_\ell^{[Q]}}
   \im\!\left[
      \overline{
      \sum_m\Gamma_{\ell,m}^{[Q]}I_m(t)}
      \left(
      \sum_k\Gamma_{\ell+1,k}^{[Q]}I_k(t)
      \right)
   \right].
   \label{eq:K-speed-transfer-main}
\end{equation}

In infinite dimension, the same formula holds provided the tail sum may be differentiated term by term.  A sufficient condition on every compact time interval is uniform convergence of $\sum_{\ell\ge0}\widetilde b_{\ell+1}^{[Q]}|\tphi_\ell^{[Q]}(t)|\,|\tphi_{\ell+1}^{[Q]}(t)|$.  This makes the bond-current series uniformly absolutely convergent; because the tail sum converges at $t=0$, where $K_Q(0)=0$, the standard theorem on differentiating a series then justifies the above formula.
The shifted Krylov-level operator
$
   \mathsf N_{\rm K}^{[Q]}
   =
   \sum_{n=0}^{d_Q-1}
   n\ket{\widetilde K_n^{[Q]}}
   \bra{\widetilde K_n^{[Q]}},
$
with the upper limit understood as infinity when $d_Q=\infty$, has $K_Q(t)$ as its expectation
value. Whenever the evolved state has finite $\operatorname{Var}_t(\mathsf N_{\rm K}^{[Q]})$ and satisfies the operator-domain hypotheses needed to identify $\dot K_Q(t)=\ii\langle[H,\mathsf N_{\rm K}^{[Q]}]\rangle_t$, the Robertson inequality gives $|\dot K_Q(t)|\le 2\Delta_Q\sqrt{\operatorname{Var}_t(\mathsf N_{\rm K}^{[Q]})}$, where the subscript $t$ denotes expectation in $\ee^{-\ii tH}\ket{\psi_Q}$, as in the Krylov speed-limit literature \cite{DeffnerCampbell2017,Hornedal2022,GillSarkar2024}, and we do not use this bound below. What the
transfer calculus adds is instead eq.~\eqref{eq:K-speed-transfer-main}, the exact growth rate
written in reference data alone.

The preceding identities reduce the complexity problem to cumulative probabilities, which the reference filtration determines directly.

\subsection{Finite-rank projection and filtration bounds}
\label{subsec:projection-filtration}

To reconstruct the cumulative probability $\Phi_\ell^{[Q]}$, we need the orthogonal projector onto
the first $\ell+1$ shifted Krylov levels, rather than only the individual amplitudes. In the
spectral representation, multiplication by $\Qhat$ embeds this filtration into the reference
polynomial space, while divisibility by $Q$ is enforced by root constraints. The following
proposition therefore expresses the required projector as a finite-rank correction of the
reference Christoffel--Darboux kernel. The coefficient-space realization below covers the complex-root
and terminal cases for which point-evaluation constraints are not intrinsic.

\begin{proposition}[Projected Christoffel--Darboux kernel]
\label{prop:projected-kernel-main}
Let $0\le\ell<d_Q$. Assume first that $Q$ has real coefficients, degree $r$, and simple real roots
$\xi_1,\ldots,\xi_r$. Suppose either that the reference support is infinite or that it is finite with $\ell+r\le d-1$. Multiplication by
$\Qhat=Q/\sqrt{N_Q}$ is an isometry from
$(\cP_\ell,\langle\cdot,\cdot\rangle_{\nu_Q})$
onto the subspace of $\cP_{\ell+r}$ which vanishes at every $\xi_a$. Its reproducing kernel is 
\begin{equation}
\begin{gathered}
   \Qhat(E)\Qhat(F)
   \tK_\ell^{[Q]}(E,F)
   =
   \mathbb K_{\ell+r}^{\mu}(E,F)
   -
   \mathbf k_\ell(E)^T
   G_\ell^{-1}
   \mathbf k_\ell(F),\\
   (\mathbf k_\ell(E))_a
   =
   \mathbb K_{\ell+r}^{\mu}(E,\xi_a),
   \qquad
   (G_\ell)_{ab}
   =
   \mathbb K_{\ell+r}^{\mu}(\xi_a,\xi_b).
\end{gathered}
   \label{eq:projected-kernel-main}
\end{equation}
The simplicity assumption may be dropped. If $\xi_a$ has algebraic multiplicity $m_a$ as a root of $Q$, its evaluation constraint is replaced by the derivative jet of orders
$0,\ldots,m_a-1$. The corresponding jet Gram matrix is given in appendix~\ref{app:projected-kernel}. 
The proposition is restricted to the range in which the polynomial representation of $\cP_{\ell+r}$ is faithful. The following paragraph treats the complementary finite-support regime and explains why the quotient-image projector replaces the kernel formula.

For finite support with $\ell+r\ge d$, the exact object is instead the quotient-image projector constructed in eq.~\eqref{eq:projection-coeff-main} below, with $M_\ell=d-1$ and all products reduced modulo $T_d$. This formulation also covers support loss and arbitrary redundant spanning sets. Off-support evaluation is not an intrinsic functional on $\cA_d$, and terminal lifts cannot extend the reproducing-kernel sum past degree $d-1$.
\end{proposition}
Indeed, every beyond-terminal lift $\widehat P_m$ with $m\ge d$, in the sense of eq.~\eqref{eq:proper-terminal-lift-main}, lies in the ideal generated by $T_d$ and so has zero $L^2(\mu)$ norm, while off-support evaluation depends on the chosen polynomial representative rather than on its quotient class, and no reproducing-kernel sum can see past degree $d-1$.

\begin{proof}
Multiplication by $\Qhat$ is an isometry because
\begin{equation}
   \int
   \overline{\Qhat(E)f(E)}
   \Qhat(E)g(E)\dd\mu(E)
   =
   \int
   \overline{f(E)}g(E)\dd\nu_Q(E).
\end{equation}
For simple roots,
\begin{equation}
   Q\cP_\ell
   =
   \left\{
      S\in\cP_{\ell+r}:
      S(\xi_a)=0,\quad a=1,\ldots,r
   \right\}.
\end{equation}
The representer of evaluation at $\xi_a$ in $\cP_{\ell+r}$ is
$\mathbb K_{\ell+r}^\mu(\,\cdot\,,\xi_a)$. The reproducing kernel of the orthogonal complement of the span of these representers is the Schur-complement expression on the right-hand side of
eq.~\eqref{eq:projected-kernel-main}. The Gram matrix is positive definite because the evaluation constraints are independent. Independence holds because Lagrange interpolation at the $r$ distinct points $\xi_a$ realizes arbitrary prescribed values by a polynomial of degree at most $r-1$, and $\cP_{r-1}\subset\cP_{\ell+r}$, so no nontrivial combination of these functionals annihilates $\cP_{\ell+r}$. Under the hypothesis of the proposition the polynomial representation of $\cP_{\ell+r}$ is faithful, so the representers are linearly independent and $G_\ell$ is invertible. The confluent form of the same argument is given in appendix~\ref{app:projected-kernel}. For a root $\xi_a$ of multiplicity $m_a$ the evaluation functional is replaced by the jet $S\mapsto S^{(j)}(\xi_a)$, $0\le j<m_a$, whose reproducing-kernel representer is
$\partial_F^j\mathbb K^\mu_{\ell+r}(\cdot,F)|_{F=\xi_a}$. 

The Gram matrix of these representers is the jet Gram matrix of eq.~\eqref{eq:jet-gram-app}, and the same Schur-complement argument gives the confluent form of eq.~\eqref{eq:jet-kernel-app}. For finite support with $\ell+r\le d-1$, the polynomial representation remains faithful and the same proof applies. Once $\ell+r\ge d$, the correct finite-dimensional construction is projection onto the quotient image, which gives eq.~\eqref{eq:projection-coeff-main}.
\end{proof}

For nonreal root constraints or complex seed coefficients, the coordinate-free statement is the Hermitian projection onto
$\Qhat\cP_\ell$. Its kernel contains
$\Qhat(E)\overline{\Qhat(F)}$ rather than
$\Qhat(E)\Qhat(F)$. For complex evaluation points, the reference Hermitian kernel is
$
   \mathbb K_M^{\mu,\mathrm H}(z,w)
   =
   \sum_{m=0}^{M}p_m(z)\overline{p_m(w)}.
$
The bilinear kernel in eq.~\eqref{eq:kernels-def} is used for real spectral variables and real root constraints. It is not analytically continued unchanged to nonreal roots. The coefficient-space construction below implements the general Hermitian projector without choosing a root representation.

Let $p_m=P_m/\sqrt{h_m}$ and define the reference-basis amplitudes of the evolved shifted seed by
\begin{equation}
   g_m^{[Q]}(t)
   =
   \int
   \ee^{-\ii Et}
   \Qhat(E)p_m(E)\dd\mu(E).
   \label{eq:gq-def}
\end{equation}
Set
\begin{equation}
   M_\ell
   =
   \begin{cases}
      \ell+r\,,&d=\infty\,,\\
      \min\{\ell+r,d-1\},&d<\infty\,,
   \end{cases}
   \qquad
   g_{\le M_\ell}^{[Q]}
   =
   \bigl(
      g_0^{[Q]},\ldots,g_{M_\ell}^{[Q]}
   \bigr)^T.
   \label{eq:projection-cutoff-main}
\end{equation}
Choose any spanning set of the image $\Qhat\cP_\ell$, expand it in the orthonormal reference coordinates
$p_0,\ldots,p_{M_\ell}$, and collect the coefficient vectors as the columns of $C_{Q,\ell}$. In finite dimension, all products are first reduced modulo $T_d$. The corresponding orthogonal projector and cumulative probability are
\begin{equation}
\begin{aligned}
   \Pi_{Q,\ell}
   &=
   C_{Q,\ell}
   \bigl(C_{Q,\ell}^\dagger C_{Q,\ell}\bigr)^+
   C_{Q,\ell}^\dagger\,,\\
   \Phi_\ell^{[Q]}(t)
   &=
   \bigl(g_{\le M_\ell}^{[Q]}(t)\bigr)^\dagger
   \Pi_{Q,\ell}
   g_{\le M_\ell}^{[Q]}(t).
\end{aligned}
   \label{eq:projection-coeff-main}
\end{equation}
Here $+$ denotes the Moore--Penrose inverse \cite{BenIsraelGreville2003}. The formula is therefore independent of redundant choices in the spanning set and remains valid for real or complex seed coefficients. For the canonical columns, namely the coefficient vectors of $\Qhat p_0,\ldots,\Qhat p_\ell$ reduced modulo $T_d$ in finite dimension, and $0\le\ell<d_Q$, the image has dimension $\ell+1$. The pseudoinverse is retained because it also permits redundant spanning sets and rank-revealing numerical implementations.

Writing
$
   \phi^{(0)}(t)
   =
   (\phi_0^{(0)}(t),\phi_1^{(0)}(t),\ldots)^T,
$
the full reference-coordinate vector is
$
   g^{[Q]}(t)
   =
   \Qhat(\Jac)\phi^{(0)}(t).
$
Because $|\ee^{-\ii tE}|=1$ on the real spectrum, the spectral measure of $\ee^{-\ii t\Jac}e_0$ is again $\mu$. Since $N_Q=\int |Q(E)|^2\dd\mu(E)<\infty$, the spectral-theorem domain criterion gives $\ee^{-\ii t\Jac}e_0\in\Dom Q(\Jac)$. On this domain,
\begin{equation}
   \bigl\|Q(\Jac)\ee^{-\ii t\Jac}e_0\bigr\|^2
   =\int |Q(E)|^2|\ee^{-\ii tE}|^2\dd\mu(E)
   =N_Q<\infty\,.
   \label{eq:filtered-domain-norm-main}
\end{equation}
The spectral calculus gives
$Q(\Jac)\ee^{-\ii t\Jac}e_0=\ee^{-\ii t\Jac}Q(\Jac)e_0$. Hence
$g^{[Q]}(t)$ is the evolved shifted seed in reference Krylov coordinates, and
$\Phi_\ell^{[Q]}(t)$ is the squared norm of its projection onto
$\Qhat\cP_\ell$. The entries $g_m^{[Q]}$ are reference-basis amplitudes, not shifted Krylov amplitudes. The kernel formalism therefore separates the physical time evolution from the shifted Krylov filtration.

\paragraph{Filtration bounds.}
Multiplication by a degree-$r$ polynomial advances the polynomial filtration by at most $r$
levels, so the shifted space of degree $\ell$ embeds in the reference space of degree $\ell+r$.
Whatever probability the evolved seed carries beyond reference level $\ell+r$ must therefore sit
beyond shifted level $\ell$, and summing these tails bounds the complexity from below by
reference data alone. The hypotheses in force are those of section~\ref{sec:setup}, where the Hamiltonian is self-adjoint,
the seed satisfies $0<N_Q<\infty$ in eq.~\eqref{eq:seed-normalization}, and for an
infinite shifted chain the polynomials are dense in $L^2(\nu_Q)$. The last of these is imposed on
$\nu_Q$ directly rather than deduced from the corresponding statement for $\mu$.
\begin{proposition}[Filtration lower bound]
\label{prop:filtration-lower-bound}
Let $Q$ have degree $r$, and use the reference-basis amplitudes
$g_m^{[Q]}(t)$ defined in eq.~\eqref{eq:gq-def}. Let $d$ and $d_Q$ denote the reference and shifted cyclic dimensions, with infinity allowed, and let $M_\ell$ be given by
eq.~\eqref{eq:projection-cutoff-main}. Then
\begin{equation}
   K_Q(t)
   \ge
   \sum_{\ell=0}^{d_Q-2}
   \sum_{m=M_\ell+1}^{d-1}
   |g_m^{[Q]}(t)|^2,
   \label{eq:filtration-lower-bound}
\end{equation}
where an upper limit $d-1=\infty$ denotes an infinite sum. Equivalently, with $(x)_+=\max\{x,0\}$ and
$d_Q-1=\infty$ when $d_Q=\infty$,
\begin{equation}
   K_Q(t)
   \ge
   \sum_{m=0}^{d-1}
   \min\!\left\{
      d_Q-1,(m-r)_+
   \right\}
   |g_m^{[Q]}(t)|^2.
   \label{eq:filtration-lower-bound-expectation}
\end{equation}
For an infinite shifted support the cap is inactive. In finite dimension, a degree-$r$ polynomial can remove at most $r$ distinct support atoms, so
$d-d_Q\le r$ and
$(m-r)_+\le d_Q-1$ for $m\le d-1$. Hence the coefficient reduces to
$(m-r)_+$ for every polynomial seed considered here.
\end{proposition}

\begin{proof}
Multiplication by $\Qhat$ identifies the shifted degree-$\ell$ polynomial space with
$
   \Qhat\cP_\ell\subset\cP_{M_\ell},
$
with products interpreted in the terminal quotient when $d<\infty$. If
$\Pi_{\mathcal A}$ and $\Pi_{\mathcal B}$ are the orthogonal projectors onto closed subspaces
$\mathcal A\subset\mathcal B$, then
$
   \|\Pi_{\mathcal A}v\|^2
   \le
   \|\Pi_{\mathcal B}v\|^2.
$
Applying this to
$
   \mathcal A=\Qhat\cP_\ell,
   \mathcal B=\cP_{M_\ell},
$
and $v=g^{[Q]}(t)$ gives
\begin{equation}
   \Phi_\ell^{[Q]}(t)
   \le
   \sum_{m=0}^{M_\ell}
   |g_m^{[Q]}(t)|^2.
\end{equation}
Since $g^{[Q]}(t)$ is normalized,
\begin{equation}
   1-\Phi_\ell^{[Q]}(t)
   \ge
   \sum_{m=M_\ell+1}^{d-1}
   |g_m^{[Q]}(t)|^2.
\end{equation}
Summing this inequality and using
eq.~\eqref{eq:tail-complexity} in finite dimension or
eq.~\eqref{eq:tail-complexity-infinite-main} in infinite dimension proves
eq.~\eqref{eq:filtration-lower-bound}. Tonelli's theorem \cite{Folland1999} applies to the resulting nonnegative double sum. For fixed $m$, the contributing values of $\ell$ satisfy
$
   0\le\ell\le d_Q-2
$
and
$
   \ell<m-r.
$
Their number is
$
   \min\{d_Q-1,(m-r)_+\},
$
which proves eq.~\eqref{eq:filtration-lower-bound-expectation}. The final simplification follows from
$d-d_Q\le r$ in finite dimension and is immediate when $d_Q=\infty$.
\end{proof}

The filtration bound controls shifted complexity by the reference-level tail of the evolved shifted seed. In particular, since $(m-r)_+\ge m-r$ and the amplitudes are normalized, eq.~\eqref{eq:filtration-lower-bound-expectation} implies
$
   K_Q(t)\ge\sum_{m}m\,|g_m^{[Q]}(t)|^2-r,
$
so a degree-$r$ preparation can depress the mean Krylov level below the reference-basis first
moment of the evolved seed by at most $r$ levels. A comparison with the reference-vacuum complexity $K_0(t)$ requires additional dynamical information. For the Heisenberg--Weyl number-state jumps this information is available in subsection~\ref{subsec:HW}.

\subsection{Finite-time control and the first jump}
\label{subsec:finite-time-first-jump}
The finite-band relation is local in Krylov level and does not, by itself,
control the unbounded first moment defining spread complexity.  A linear
upper bound on the reference Jacobi coefficients gives a general global
criterion.

\begin{proposition}[Finite-time control for fixed-degree seed jumps]
\label{prop:finite-time-control-main}
Suppose that, for some $c\in\RR$ and $A,C\ge0$, the reference Jacobi
coefficients satisfy
\begin{equation}
   |a_n-c|+b_n+b_{n+1}\le An+C\,,
   \label{eq:reference-linear-bound-main}
\end{equation}
at every existing reference level, with the terminal hopping set to zero
in finite dimension.  If $Q$ has fixed degree $r$, then every existing
shifted hopping coefficient obeys
\begin{equation}
   \widetilde b_{n+1}^{[Q]}\le A(n+r)+C\,.
   \label{eq:shifted-linear-bound-main}
\end{equation}
More generally, a Jacobi chain satisfying $b_{n+1}\le An+C$ has finite
spread complexity at every finite time and
\begin{equation}
   K(t)\le
   \begin{cases}
   \left(\dfrac{C}{A}\right)^2\sinh^2(A|t|),&A>0\,,\\[3mm]
   \min\{C^2t^2,2C|t|\},&A=0\,.
   \end{cases}
   \label{eq:complexity-linear-growth-bound-main}
\end{equation}
Consequently, the shifted chain satisfies, for $A>0$,
\begin{equation}
   K_Q(t)\le
   \left(r+\frac{C}{A}\right)^2\sinh^2(A|t|),
   \label{eq:shifted-complexity-linear-growth-bound-main}
\end{equation}
while for $A=0$ the second line of
eq.~\eqref{eq:complexity-linear-growth-bound-main} applies with the same
$C$.
\end{proposition}

\begin{proof}
Let $\Pi_N^{\rm ref}$ project onto the first $N+1$ reference Krylov levels and put $N_*=n+r$, with $N_*$ saturated at $d-1$ when the
reference cyclic dimension $d$ is finite; projectors with a larger index
are understood to saturate at the same terminal level.  The shifted level-$n$ vector
is generated by a polynomial of degree at most $n+r$ and therefore lies
in $\Pi_{N_*}^{\rm ref}\cH$.  Its next hopping is bounded by the
rectangular compression
\begin{equation}
   \mathsf T_{N_*}
   =\Pi_{N_*+1}^{\rm ref}(H-c\Id)\Pi_{N_*}^{\rm ref},
   \qquad
   \widetilde b_{n+1}^{[Q]}
   \le\|\mathsf T_{N_*}\|.
   \label{eq:rectangular-compression-hopping-main}
\end{equation}
Indeed, the shifted Lanczos residual is the component of
$(H-c\Id)\ket{\widetilde K_n}$ orthogonal to the preceding shifted
levels.  Every row and column of the compressed reference Jacobi matrix
has absolute sum at most $AN_*+C$; before termination its last row
contains only $b_{N_*+1}$ and obeys the same estimate.  Hence
\begin{equation}
   \|\mathsf T_{N_*}\|_2
   \le
   \sqrt{\|\mathsf T_{N_*}\|_1
          \|\mathsf T_{N_*}\|_\infty}
   \le AN_*+C\,,
\end{equation}
which proves eq.~\eqref{eq:shifted-linear-bound-main}.  Only the degree
filtration entered this argument, so it also covers complex and
confluent polynomial seeds and rows meeting a finite terminal edge.

For the dynamical estimate, terminate the Jacobi chain after level $M$,
write $\phi_n^{[M]}(t)$ for the resulting amplitudes and set
$K_M(t)=\sum_{n=0}^{M}n|\phi_n^{[M]}(t)|^2$.  Then the exact bond current gives
\begin{equation}
   |\dot K_M|
   \le2\sum_{n=0}^{M-1}b_{n+1}
      |\phi_n^{[M]}||\phi_{n+1}^{[M]}|.
   \label{eq:finite-truncation-current-bound-main}
\end{equation}
For $A>0$, put $d_0=C/A$.  Since
$(n+d_0)^2\le(n+1)(n+d_0^2)$, Cauchy--Schwarz gives
\begin{equation}
   |\dot K_M|\le2A\sqrt{K_M(K_M+d_0^2)}\,.
\end{equation}

Assume first that $C>0$, so $d_0>0$. For $\varepsilon>0$, define
$y_\varepsilon(t)=\operatorname{arcsinh}\!\bigl(\sqrt{K_M(t)+\varepsilon}/d_0\bigr)$.
The differential inequality implies $|\dot y_\varepsilon(t)|\le A$ almost everywhere, and hence
$y_\varepsilon(t)\le \operatorname{arcsinh}(\sqrt{\varepsilon}/d_0)+A|t|$.
Letting $\varepsilon\downarrow0$ gives%

$K_M(t)\le d_0^2\sinh^2(A|t|)$.  If $A>0$ and $C=0$, the hypothesis
forces $b_1=0$ and the initial site is disconnected.  For $A=0$,
\begin{equation}
   \sum_{n=0}^{M-1}|\phi_n^{[M]}\phi_{n+1}^{[M]}|
   \le\sqrt{1-|\phi_0^{[M]}|^2}
   \le\sqrt{K_M}\,,
   \qquad
   \sum_{n=0}^{M-1}|\phi_n^{[M]}\phi_{n+1}^{[M]}|\le1\,,
\end{equation}
so eq.~\eqref{eq:finite-truncation-current-bound-main} gives both terms
in the second line of eq.~\eqref{eq:complexity-linear-growth-bound-main}.

These estimates are uniform in $M$.  The linear hopping bound implies Carleman's criterion of section~\ref{sec:setup} unless a zero hopping has already terminated the cyclic chain; the minimal Jacobi operator is then essentially self-adjoint, and its principal terminations converge to it in the strong-resolvent sense \cite{SimonMomentProblem1998}. Here each termination is extended by zero to a bounded self-adjoint operator on $\ell^2$ that agrees with the minimal Jacobi operator on any fixed finitely supported vector once $M$ is large enough; convergence on that core together with essential self-adjointness gives the resolvent statement, and the unitary groups then converge strongly~\cite{ReedSimonI}.  Each fixed
finite set of evolved amplitudes converges, and Fatou's lemma \cite{Folland1999} gives
\begin{equation}
   K(t)=\sum_{n\ge0}n|\phi_n(t)|^2
   \le\liminf_{M\to\infty}K_M(t).
\end{equation}
This proves finite-time finiteness.  Applying the same result to
eq.~\eqref{eq:shifted-linear-bound-main}, with $C$ replaced by $Ar+C$,
gives eq.~\eqref{eq:shifted-complexity-linear-growth-bound-main}.
\end{proof}

The proposition gives a model-independent sufficient condition, not the
sharp growth rate of a particular chain. 
The same $\sinh^2$ profile occurs in the exactly solvable linear-hopping chain of ref.~\cite{Parker2019}; eq.~\eqref{eq:complexity-linear-growth-bound-main} is the corresponding envelope under the linear hopping bound alone, with the prefactor fixed by the intercept.  In the Charlier problem the
large-index analysis in subsection~\ref{app:H1-large-index} determines
the shifted intercepts, treats atom-deletion resonances and provides the
stronger domain control used in the full-chain short-time expansion. 

\paragraph{First jump.}
Assume that the reference chain has at least two levels, so that
$h_1=b_1^2>0$, and take
\begin{equation}
   Q(E)
   =
   \frac{P_1(E)}{\sqrt{h_1}}
   =
   \frac{E-a_0}{\sqrt{h_1}}\,,
   \qquad
   \alpha=a_0\,,
   \qquad
   A_m=P_m(\alpha).
\end{equation}
In infinite support, or in finite support while $\ell+1\le d-1$, the rank-one projected kernel is
\begin{equation}
\begin{aligned}
   \frac{P_1(E)P_1(F)}{h_1}
   \tK_\ell^{(1)}(E,F)
   &=
   \mathbb K_{\ell+1}^{\mu}(E,F)
   -
   \frac{
      \mathbb K_{\ell+1}^{\mu}(E,\alpha)
      \mathbb K_{\ell+1}^{\mu}(\alpha,F)}
   {\mathbb K_{\ell+1}^{\mu}(\alpha,\alpha)}\,.
\end{aligned}
   \label{eq:first-jump-kernel}
\end{equation}
At a finite terminal edge beyond this range, the same cumulative probability is obtained from the quotient projector
eq.~\eqref{eq:projection-coeff-main}. 

The reference-basis amplitudes of the evolved first-jump state are
\begin{equation}
   g_m^{(1)}(t)
   =
   \frac{1}{\sqrt{h_1h_m}}
   \int
   \ee^{-\ii Et}
   P_1(E)P_m(E)\dd\mu(E).
   \label{eq:g1-def-main}
\end{equation}
Using the reference recurrence gives
\begin{equation}
   g_m^{(1)}(t)
   =
   \frac{
      I_{m+1}(t)
      +(a_m-a_0)I_m(t)
      +b_m^2I_{m-1}(t)}
   {\sqrt{h_1h_m}}\,,
   \qquad
   I_{-1}(t)=0\,.
   \label{eq:g1-reference-moments-main}
\end{equation}
In finite reference dimension, the terminal case $m=d-1$ is read with
$I_d(t)=0$, since $P_d=T_d$ vanishes on the support. Applying the rank-one projector to $g^{(1)}(t)$ gives
\begin{equation}
   \Phi_\ell^{(1)}(t)
   =
   \sum_{m=0}^{\ell+1}
   |g_m^{(1)}(t)|^2
   -
   \frac{
      \left|
         \sum_{m=0}^{\ell+1}
         A_mg_m^{(1)}(t)/\sqrt{h_m}
      \right|^2}
   {\displaystyle
      \sum_{m=0}^{\ell+1}A_m^2/h_m}\,.
   \label{eq:first-jump-cumulative}
\end{equation}
The first term is the reference-basis cumulative probability of the evolved first-jump state through level $\ell+1$. The second subtracts its squared component along the reproducing-kernel representer of evaluation at $\alpha$, that is, along $\mathbb K^\mu_{\ell+1}(\cdot,\alpha)$. For a real degree-$r$ seed with distinct real root locations $\xi_a$ with multiplicities $m_a$, this rank-one subtraction is replaced by the rank-$r$ jet-Gram projection built from derivative orders $0,\ldots,m_a-1$ at each $\xi_a$, where $\sum_a m_a=r$. For complex seed coefficients, the general Hermitian construction is eq.~\eqref{eq:projection-coeff-main}.

\paragraph{Heisenberg--Weyl specialization.}
The oscillator example of subsection~\ref{subsec:HW} shows explicitly how the projected kernel reconstructs a complete complexity series. We state the result here because it is the simplest closed realization of the rank-one correction; section~\ref{sec:examples} gives the Charlier reference data and connector amplitudes, while subsection~\ref{app:H1-first-jump} proves convergence. With $\omega>0$ the oscillator frequency and $\lambda>0$ its dimensionless displacement parameter, set $\tau=\omega t$ and define
\begin{equation}
   q(\tau)
   =
   \lambda(\ee^{-\ii\tau}-1),
   \qquad
   \kappa(\tau)
   =
   \frac{|q(\tau)|^2}{\lambda}
   =
   2\lambda(1-\cos\tau).
   \label{eq:kappa-def-main}
\end{equation}
We write $\kappa=\kappa(\tau)$ and $q=q(\tau)$. In the centered oscillator variable of subsection~\ref{subsec:HW} the reference chain has $a_0=0$, so the first-jump root sits at $\alpha=0$ and the constants of the preceding paragraph become
$
   A_j=P_j(0)
$
and
$
   D_j=h_j\mathbb K_j^\mu(0,0)>0,
$
with
$
   h_j=\lambda^j j!.
$
Then
$
   K_1(\tau)
   =
   \sum_{n\ge0}
   n|\tphi_n^{(1)}(\tau)|^2
$
is
\begin{equation}
\begin{aligned}
   K_1(\tau)
   &=
   \kappa(1+\ee^{-\kappa})
   +
   \sum_{j=1}^{\infty}
   \frac{
      \ee^{-\kappa}\kappa^j
      |A_jq-A_{j+1}|^2}
   {\lambda j!D_j}\\
   &=
   \kappa(1+\ee^{-\kappa})
   +
   \ee^{-\kappa}
   \sum_{j=1}^{\infty}
   \frac{
      \kappa^j
      \bigl(
         A_{j+1}^2
         +\kappa A_jA_{j+1}
         +\lambda\kappa A_j^2
      \bigr)}
   {\lambda j!D_j}\,.
\end{aligned}
   \label{eq:H1-first-jump-complexity}
\end{equation}
The first term is the reference-basis tail contribution
$
   \sum_{j\ge1}
   [1-\sum_{m=0}^{j}|g_m^{(1)}(\tau)|^2]
$
with the cutoff displaced by one reference level. The remaining series is the positive correction produced by the rank-one Christoffel projection. Its first representation is manifestly nonnegative term by term. The second is the algebraically equivalent real form obtained from
$
   |q|^2=\lambda\kappa
$
and
$
   q+\bar q=-\kappa.
$
The reference-basis first moment of the evolved first-jump state is $1+\kappa$, while the reference-vacuum complexity is $K_0(\tau)=\kappa$; the first term of eq.~\eqref{eq:H1-first-jump-complexity} alone already gives $K_1(\tau)\ge K_0(\tau)$, the comparison anticipated below proposition~\ref{prop:filtration-lower-bound} and sharpened in section~\ref{sec:examples}. Subsection~\ref{app:H1-first-jump} proves convergence of eq.~\eqref{eq:H1-first-jump-complexity} and derives the recurrences obeyed by $A_j$ and $D_j$; section~\ref{sec:examples} now provides the solved reference data on which all of these formulas operate.

The filtration bound compares two different measurements on the same evolved state.
In the oscillator, tight-binding and chord realizations, the reference probabilities resolve
Fock number, lattice site and chord number, respectively; signed or absolute displacements are
then model-specific moments of those distributions.  The shifted complexity instead measures
level in the Lanczos basis rebuilt from the prepared state.  Their inequalities constrain
re-Lanczos spreading by physical-basis transport without identifying the observables.

\section{Exact oscillator and tight-binding seed jumps}
\label{sec:examples}

Both families describe physical preparation questions at fixed Hamiltonian.  The Charlier
chain is the Lanczos representation of a linearly driven bosonic mode.  Preparing a number state
rather than the vacuum tests how an excited Fock seed reorganizes the Krylov basis even though the
evolution in the original Fock basis is known from displacement-operator methods
\cite{Perelomov1986}.  The Chebyshev chain is the one-particle sector of a uniform hopping model.
Changing its localized seed is a launch-site problem, relevant to propagation and state-transfer
questions, in which the couplings and energy levels remain fixed.  These preparations make the
distinction between motion in a physical basis and spread in the rebuilt Krylov basis directly
observable.

The transfer identities require only reference Jacobi data and a polynomial seed.  Exactly solvable Jacobi chains make this physical comparison
analytic at every Krylov level.  We choose the
Charlier and Chebyshev chains because their recurrence coefficients, spectral measures and
Fourier--orthogonal-polynomial moments are explicit,
so both the reference dynamics and the relative fixed-$H$ problem can be solved
at every level.  The two families expose
complementary analytic issues.  The Charlier oscillator is an infinite discrete chain with
atom-deletion resonances, root-free all-index connectors, large-index Jacobi asymptotics and a
convergent complexity series; its Charlier--Hermite scaling gives the continuous-support endpoint.
The Chebyshev chain turns the seed change into a localized-site jump and admits an exact half-line
connector together with a finite-volume fold through support loss and termination.

The main text gives the reference solutions and the principal shifted results. The two longer analytic appendices are needed for points that fixed-level formulas cannot settle. A finite connector row does not by itself control the first moment of an infinite Krylov distribution, while a preterminal polynomial identity does not determine all rows of a finite shifted chain after folding and support loss. Subsections~\ref{app:H1-first-jump} and~\ref{app:H1-general-r} develop the root-free Charlier recurrences and their low-rank reductions, while subsection~\ref{app:H1-large-index} establishes the fixed-$r$, fixed-$\lambda$ estimates required for finiteness and the domain estimates needed for short-time control. Appendix~\ref{app:finite} develops the finite product-Gram construction and uses it to obtain every Chebyshev connector through the terminal edge. Appendix~\ref{app:numerics}, in turn, gives rank-revealing reference-space implementations for generic Jacobi data, where closed generating functions are unavailable.

\subsection{The Heisenberg--Weyl chain and Charlier seed jumps}
\label{subsec:HW}

The Heisenberg--Weyl chain is an exactly solvable oscillator realization. The relevant algebra is $\mathfrak{osc}(1)=\mathfrak h_1\rtimes\mathfrak u(1)$; we nevertheless retain the conventional name Heisenberg--Weyl/Charlier oscillator.\footnote{For $\omega>0$, the number operator $N=a^\dagger a$ is essential, and $H$ does not belong to the bare Heisenberg--Weyl algebra $\mathfrak h_1=\Span\{\Id,a,a^\dagger\}$.} Consider the Hamiltonian
\begin{equation}
   H=\omega a^\dagger a+g a^\dagger+\bar g a+\delta\Id\,,
   \qquad g=\rho\ee^{\ii\chi},\quad \rho>0,\quad \delta\in\RR\,,
   \label{eq:H1-ham}
\end{equation}
with $\omega>0$, and define $\lambda=\rho^2/\omega^2$, $Y=(H-\delta\Id)/\omega$ and $\tau=\omega t$. Throughout this subsection, \(I_n(\tau)\) denotes the Fourier--OP moment for \(Y\). Evolution under \(H\) differs only by the common phase \(\ee^{-\ii\delta t}\), which cancels from all probabilities and complexities. In the phase-rotated Fock basis \(\ket{K_n}=\ee^{\ii n\chi}\ket{n}\), the Jacobi action is\footnote{This phase convention is a Lanczos gauge choice. In the unrotated Fock basis, the nearest-neighbor matrix elements carry the phase of \(g=\rho\ee^{\ii\chi}\). The rephasing \(\ket n\mapsto\ee^{\ii n\chi}\ket n\) removes this phase and makes the Jacobi hoppings real and positive, as required by the orthogonal-polynomial convention. A diagonal unitary implements it and therefore leaves the spectral measure, probabilities and Krylov complexity unchanged.}
\begin{equation}
Y\ket{K_n}=n\ket{K_n}+\sqrt{\lambda(n+1)}\ket{K_{n+1}}+\sqrt{\lambda n}\ket{K_{n-1}}.
   \label{eq:H1-jacobi}
\end{equation}
The reference monic polynomials satisfy
\begin{equation}
   yP_n(y)=P_{n+1}(y)+nP_n(y)+\lambda nP_{n-1}(y),
   \qquad h_n=\lambda^n n!\,.
   \label{eq:Charlier-recurrence-main}
\end{equation}
These are the centered monic Charlier polynomials \cite{Chihara1978}. Their spectral lattice is $y=x-\lambda$, with $x=0,1,\ldots$ and Poisson weights $\ee^{-\lambda}\lambda^x/x!$. The generating function
\begin{equation}
   \sum_{n=0}^{\infty}P_n(y)\frac{u^n}{n!}
   =\ee^{-\lambda u}(1+u)^{y+\lambda}
   \label{eq:Charlier-gen-main}
\end{equation}
may be integrated term by term against $\ee^{-\ii\tau y}\dd\mu(y)$, giving the reference moments
\begin{equation}
   I_n(\tau)=\cS(\tau)q(\tau)^n,
   \quad
   \cS(\tau)=\exp\{\lambda(\ee^{-\ii\tau}-1+\ii\tau)\},
   \quad q(\tau)=\lambda(\ee^{-\ii\tau}-1).
   \label{eq:H1-old-moments-main}
\end{equation}
The reference Krylov complexity is the Poisson mean $K_0(\tau)=2\lambda(1-\cos\tau)$. The simplification in eq.~\eqref{eq:H1-old-moments-main} is that every reference moment is a power of $q(\tau)$ multiplied by the same factor $\cS(\tau)$. Consequently, every fixed-degree polynomial seed shift reduces to finite algebra in $q(\tau)$, with the physical time dependence separated from the Christoffel coefficients. This separation makes the general construction explicit at every Charlier index.

\paragraph{Reference dynamics and connector amplitudes.}
The reference solution should be distinguished from the seed jump. The displacement/Bargmann method solves the reference dynamics directly \cite{Perelomov1986}. The related Fock-space generating-function treatment of stochastic birth--death systems is the Doi--Peliti construction, whose Charlier realization is discussed in refs.~\cite{Doi1976,Peliti1985,Ohkubo2012}. The common Charlier structure links the Markov and unitary settings, once the two are written in the same variable. The Lanczos data of eq.~\eqref{eq:H1-jacobi} belong to the centered operator $Y$, for which $a_n=n$ and $b_n^2=\lambda n$; the birth--death chain lives on the uncentered lattice $x=y+\lambda$, where the Karlin--McGregor recurrence of the linear immigration--death process, with birth rates $\lambda_n=\lambda$ and death rates $\mu_n=n$, has $a_n^{X}=\lambda_n+\mu_n=n+\lambda$ and the same $b_n^2=\lambda n$. The two differ by the constant shift $\lambda$, which leaves every hopping, probability and complexity unchanged.

The seed jump has a counterpart on the birth--death/Markov side. Multiplying the Charlier weight by $P_r^2/h_r$ is a stochastic Darboux transformation of the birth--death chain in the sense of ref.~\cite{GrunbaumDeLaIglesia2018}, whenever the transformed Jacobi data admit a birth--death factorization with nonnegative rates. For every fixed $r$, positivity of that algebraic factorization follows from the support of the transformed measure.  On the uncentered lattice, $\nu_r$ has infinitely many surviving nodes in $\mathbb Z_{\ge0}$, and therefore
$\int x|f(x)|^2\dd\nu_r(x)>0$ for every nonzero polynomial $f$.  Every finite leading principal block of the shifted Jacobi matrix $\widetilde{\Jac}_r^X$ is consequently positive definite.  Its consistent bidiagonal Cholesky factorization defines strictly positive rates through
$\widetilde\mu_0=0$, $\widetilde\lambda_0=\widetilde a_0^X$,
$\widetilde\mu_{n+1}=\widetilde\beta_{n+1}/\widetilde\lambda_n$ and
$\widetilde\lambda_{n+1}=\widetilde a_{n+1}^X-\widetilde\mu_{n+1}$.  This proves positivity at every level; conservativeness or nonexplosion of the associated continuous-time Markov process is a separate question.  For $r=1$, the first two factorization stages are explicit. Writing $\widetilde a_n^{X}=\widetilde\lambda_n+\widetilde\mu_n$ and $\widetilde\beta_{n+1}=\widetilde\lambda_n\widetilde\mu_{n+1}$, and using the $r=1$ entries of eq.~\eqref{eq:H1-local-sum-rule-check-main} on the uncentered lattice, which are $\widetilde a_0^{X}=1+\lambda$, $\widetilde\beta_1=3\lambda$ and $\widetilde a_1^{X}=\tfrac43+\lambda$, gives
\begin{equation}
   \widetilde\lambda_0=1+\lambda\,,
   \qquad
   \widetilde\mu_1=\frac{3\lambda}{1+\lambda}\,,
   \qquad
   \widetilde\lambda_1=\frac43+\lambda-\frac{3\lambda}{1+\lambda} \,,
   \label{eq:birth-death-first-jump}
\end{equation}
and $\widetilde\lambda_1$ attains its minimum $2\sqrt3-\tfrac83\approx0.7974$ at $\lambda=\sqrt3-1$, so all three explicit rates are strictly positive for every $\lambda>0$. The centering is not cosmetic here, since performed in the variable $Y$ the same algebra returns $\widetilde\lambda_1=\tfrac43-3\lambda$, which is negative for $\lambda>\tfrac49$, because a birth--death factorization presupposes a chain on the nonnegative lattice. The all-level Cholesky argument establishes positivity of the transformed rates, while the explicit first stages expose the Markov counterpart of the same Charlier seed transform.

In the present construction the group-theoretic solution enters through the reference moments \(I_n\). For example, completing the square gives
\begin{equation}
   Y=(a^\dagger+\sqrt\lambda\,\ee^{-\ii\chi})
     (a+\sqrt\lambda\,\ee^{\ii\chi})-\lambda\,,
\end{equation}
in the original Fock convention, with the corresponding phase-free form after rephasing, and eq.~\eqref{eq:Charlier-gen-main} then implies eq.~\eqref{eq:H1-old-moments-main}. The seed jump is a different operation. The standard route to the reference dynamics of eq.~\eqref{eq:H1-ham} is the displacement operator. Completing the square exhibits $Y$ as a displaced number operator, so the evolved vacuum is a coherent state and the reference amplitudes follow from the displacement algebra. A displaced vacuum is a coherent superposition over all Fock levels, whereas \(\ket{K_r}\) is one isolated Krylov/Fock level. Its shifted Krylov basis is determined instead by the Christoffel measure \(P_r^2\,\dd\mu/h_r\). Thus the group orbit of the vacuum and the reorthogonalization associated with an excited Krylov seed solve distinct problems.\footnote{The displacement operator determines the evolved number state in the original Fock basis, but not the Lanczos basis generated from that state at fixed $H$. The latter is the orthogonal-polynomial problem for $P_r^2\dd\mu/h_r$ developed in subsections~\ref{app:H1-first-jump} and~\ref{app:H1-general-r}.}

Finite superpositions require no new construction. If $\ket{\psi}=\sum_{r=0}^{R}c_r\ket{K_r}$ is normalized, then $\Qhat(y)=\sum_r c_rP_r(y)/\sqrt{h_r}$ and $|\Qhat(y)|^2\dd\mu(y)$ is its normalized spectral measure. The resulting Krylov basis cannot be obtained by linearly superposing the bases generated by the individual endpoint seeds. In the degree-one family \(c_0\ket{K_0}+c_1\ket{K_1}\), a generic seed with \(c_1\ne0\) gives a pair of conjugate Christoffel roots; the pair is confluent when \(c_0/c_1\) is real. The state \(\ket{K_1}\) is both the exact double-root specialization at \(y=0\) and the confluent limit \(c_0\to0\) of this degree-one family. Thus a linear state preparation already produces a nonlinear reorganization of the Krylov basis in the first nontrivial seed family.

The number-state jumps form a distinguished single-level subfamily. For the seed $\ket{K_r}=P_r(Y)\ket{K_0}/\sqrt{h_r}$, the shifted measure is
$
   \dd\nu_r(y)=\frac{P_r(y)^2}{h_r}\dd\mu(y).
$
The large-$z$ sum rules, equivalently the first rows of the connector dictionary, already determine the first shifted Jacobi data. Since $a_n=n$ and $b_n^2=\lambda n$ for the dimensionless operator $Y$, the first two shifted Lanczos steps for an arbitrary number-state seed are 
\begin{equation}
\begin{aligned}
   \widetilde a_0^{(r)}&=r\,,\qquad
   &\widetilde\beta_1^{(r)}&=\lambda(2r+1),\\
   \widetilde a_1^{(r)}&=r+\frac{1}{2r+1}\,,\qquad
   &\widetilde\beta_2^{(r)}&=
   \frac{2\lambda(r^2+r+1)}{2r+1}
   +\frac{4r(r+1)}{(2r+1)^2}\,.
\end{aligned}
   \label{eq:H1-local-sum-rule-check-main}
\end{equation}
Here $\widetilde\beta_n^{(r)}=(\widetilde b_n^{(r)})^2$. At $r=1$, the four quantities above reduce, in order, to $1$, $3\lambda$, $4/3$ and $2\lambda+8/9$. They agree both with a direct shifted recursion from $\ket{K_r}$ and with the determinant formula in subsection~\ref{app:H1-general-r}. If the corresponding connector polynomials are defined by
\begin{equation}
   \frac{P_r(y)^2}{h_r}R_n^{(r)}(y)=\sum_{k=0}^{2r}\Gamma_{n,n+k}^{(r)}P_{n+k}(y),
   \label{eq:H1-r-connector}
\end{equation}
then eq.~\eqref{eq:H1-old-moments-main} gives the shifted amplitudes
\begin{equation}
   \tphi_n^{(r)}(\tau)=\cS(\tau)
   \frac{q(\tau)^n\sum_{k=0}^{2r}\Gamma_{n,n+k}^{(r)}q(\tau)^k}{\sqrt{\Gamma_{n,n}^{(r)}\lambda^n n!}}\,.
   \label{eq:H1-r-amplitudes}
\end{equation}

Thus, all time dependence is carried by $\cS(\tau)$ and $q(\tau)$, while the seed dependence is isolated in the finite Christoffel coefficients $\Gamma_{n,n+k}^{(r)}$. The only remaining object in eq.~\eqref{eq:H1-r-amplitudes} is a finite, time-independent connector row. Appendix~\ref{app:christoffel} characterizes this row by the values and first derivatives of $P_n$ at the zeros of $P_r$. For Charlier polynomials, these data obey closed finite-dimensional recurrences, which subsection~\ref{app:H1-general-r} reformulates without extracting the zeros.\footnote{The measure \(P_r^2\dd\mu/h_r\) is the induced measure associated with the orthonormal polynomial \(P_r/\sqrt{h_r}\); see refs.~\cite{KautskyGolub1983,GautschiLi1993,DuranChristoffel2021} for related developments on polynomial modifications of orthogonality measures and Jacobi data.}

\paragraph{Root-free remainder coordinates.}
For $r\ge1$, the root-free construction is naturally formulated in the polynomial-remainder algebras
\begin{equation}
   \mathscr R_r=\CC[y]/\langle P_r\rangle,
   \qquad
   \mathscr R_r^{(2)}=\CC[y]/\langle P_r^2\rangle.
   \label{eq:H1-seed-residue-algebras-main}
\end{equation}
In the ordered monomial bases we write
\begin{equation}
\begin{aligned}
   [P_n]_{\mathscr R_r}
   &=\sum_{\ell=0}^{r-1}u_{n,\ell}^{(r)}y^\ell,
   &
   \mathbf u_n^{(r)}
   &=\bigl(u_{n,0}^{(r)},\ldots,u_{n,r-1}^{(r)}\bigr)^T,\\
   [P_n]_{\mathscr R_r^{(2)}}
   &=\sum_{\ell=0}^{2r-1}v_{n,\ell}^{(r)}y^\ell,
   &
   \mathbf v_n^{(r)}
   &=\bigl(v_{n,0}^{(r)},\ldots,v_{n,2r-1}^{(r)}\bigr)^T.
\end{aligned}
   \label{eq:H1-remainder-sequence-main}
\end{equation}
Let $\mathsf Y_r$ and $\mathsf Y_r^{(2)}$ represent multiplication by $y$ on $\mathscr R_r$ and $\mathscr R_r^{(2)}$, respectively. The Charlier recurrence descends to the two finite vector recurrences
\begin{equation}
\begin{aligned}
   \mathbf u_{n+1}^{(r)}
   &=(\mathsf Y_r-n\Id_r)\mathbf u_n^{(r)}
   -\lambda n\mathbf u_{n-1}^{(r)}\,,\\
   \mathbf v_{n+1}^{(r)}
   &=(\mathsf Y_r^{(2)}-n\Id_{2r})\mathbf v_n^{(r)}
   -\lambda n\mathbf v_{n-1}^{(r)}\,.
\end{aligned}
   \label{eq:H1-remainder-recurrence-main}
\end{equation}
The zeros $\xi_1,\ldots,\xi_r$ of $P_r$ are simple. The Chinese remainder theorem identifies the residue class of $P_n$ modulo $P_r$ with the evaluation vector
$
   \bigl(P_n(\xi_1),\ldots,P_n(\xi_r)\bigr).
$
Reduction modulo $P_r^2$ retains, in addition, the first derivative at each zero. Consequently, $\mathbf u_n^{(r)}$ is related to the evaluation vector by a fixed change of coordinates, while $\mathbf v_n^{(r)}$ encodes the corresponding values and first derivatives. The weighted cutoff matrices
\begin{equation}
   \mathsf H_L^{(r)}
   =\sum_{m=0}^{L}
   \frac{\mathbf u_m^{(r)}\mathbf u_m^{(r)\dagger}}{h_m}
\end{equation}
become the projected-kernel Gram matrices of appendix~\ref{app:projected-kernel} under this fixed change of coordinates. The sequence $\mathbf v_n^{(r)}$ similarly encodes the confluent connector of appendix~\ref{app:christoffel}. For fixed $r$, the two finite recurrences jointly determine the shifted Jacobi data, amplitudes and cumulative probabilities at every $n$. The quotient relations encode polynomial divisibility rather than annihilation of the oscillator operator; indeed, $P_r(Y)\ket{K_0}=\sqrt{h_r}\ket{K_r}\ne0$. Subsection~\ref{app:H1-general-r} gives the detailed construction, while subsection~\ref{app:H1-large-index} derives its large-index behavior.

\paragraph{Large-index behavior and the vacuum bound.}
The remainder recurrences determine every fixed connector row, but finiteness of spread complexity is a statement about the entire shifted chain. Large-index control at fixed $r$ and $\lambda$ is therefore required before the exact amplitude formulas may be summed with the Krylov-level weight. 
\begin{proposition}[Quantized intercepts for Charlier number-state jumps]
\label{prop:H1-intercepts-main}
If $s_r(\lambda)$ is the number of zeros of $P_r$ on the centered Poisson support, then, for fixed $r$ and $\lambda>0$, the shifted coefficients obey
\begin{equation}
   \widetilde a_n^{(r)}=n+s_r(\lambda)+O(n^{-1}),\qquad
   \widetilde\beta_n^{(r)}=
   \lambda\bigl[n+2r-s_r(\lambda)\bigr]+O(n^{-1}).
   \label{eq:H1-Jacobi-asymptotics-main}
\end{equation}
\end{proposition}
The proof is given in subsection~\ref{app:H1-large-index}, in particular in
lemmas~\ref{lem:H1-packed-columns-app} and~\ref{lem:H1-eta-packed-app}.
Thus, a fixed number-state jump preserves the linear slopes of both $\widetilde a_n^{(r)}$ and $\widetilde\beta_n^{(r)}$.
This is the Charlier instance of the asymptotic stability analyzed in
section~\ref{sec:rigidity}.  It implies invariance only for diagnostics that depend solely on
these leading slopes; it does not make the subleading chain or the finite-time dynamics
preparation independent.

Section~\ref{sec:rigidity} develops this point under explicit spectral hypotheses.  In the
Charlier chain, $s_r(\lambda)$ counts the atoms destroyed by the preparation and therefore the
resonance-dependent displacement from the generic intercepts.  Each deleted atom adds one to the
diagonal intercept and subtracts one from the intercept inside the hopping bracket. The bounded
changes distinguish generic parameters from the exceptional values at which the multiplier deletes
a Poisson atom. Equivalently, the two intercepts obey exact asymptotic sum and difference
relations 
\begin{equation}
\begin{aligned}
   \lim_{n\to\infty}
   \left[
      \widetilde a_n^{(r)}-n
      +\frac{\widetilde\beta_n^{(r)}}{\lambda}-n
   \right]
   &=2r\,,\\
   \lim_{n\to\infty}
   \left[
      \frac{\widetilde\beta_n^{(r)}}{\lambda}-n
      -\bigl(\widetilde a_n^{(r)}-n\bigr)
   \right]
   &=2\bigl[r-s_r(\lambda)\bigr].
\end{aligned}
   \label{eq:H1-intercept-budget-main}
\end{equation}
Thus the two intercepts have the fixed sum $2r$, while every deleted Poisson atom
transfers one unit from the hopping intercept to the diagonal intercept without changing that sum.
Within this family the tail recovers both the degree of the jump and the support-loss count.
The estimate is pointwise in $\lambda$ and is not uniform as a zero approaches the
Poisson lattice. The derivation in subsection~\ref{app:H1-large-index} also proves that
$K_r(\tau)$ is finite for every fixed $r$, every $\lambda>0$ and all $\tau$.

\begin{corollary}[Vacuum lower bound for Charlier number-state jumps]
\label{cor:H1-monotonicity}
For the Heisenberg--Weyl number-state jump \(\ket{K_0}\mapsto\ket{K_r}\), \(r\ge0\), write \(K_{r}(\tau)\) for complexity \(K_Q(\tau)\) with \(Q=P_r/\sqrt{h_r}\). Subsection~\ref{app:H1-large-index} proves that $K_r(\tau)$ is finite for every fixed $r$ and $\lambda>0$. Then
\begin{equation}
   K_{r}(\tau)\ge K_0(\tau)=2\lambda(1-\cos\tau).
   \label{eq:H1-monotonicity}
\end{equation}
For \(r\ge1\) the inequality is strict whenever \(K_0(\tau)>0\).
\end{corollary}

\begin{proof}
Apply Proposition~\ref{prop:filtration-lower-bound} to
\(Q=P_r/\sqrt{h_r}\).
Although \(P_r\) may vanish at \(s_r(\lambda)\) atoms of the
Poisson measure, it has finite degree and the surviving support remains
infinite. Hence \(d_Q=\infty\), including at the resonant values
\(s_r(\lambda)>0\), and the cap in
eq.~\eqref{eq:filtration-lower-bound-expectation} is inactive.
With
\(g_m^{(r)}(\tau)=
\bra{K_m}\ee^{-\ii\tau Y}\ket{K_r}\),
the filtration bound therefore gives
\begin{equation}
   K_r(\tau)\ge
   \sum_{m\ge0}(m-r)_+|g_m^{(r)}(\tau)|^2.
\end{equation}
The remaining identity follows from the fixed-basis oscillator dynamics. In the convention of eq.~\eqref{eq:H1-ham},
\begin{equation}
   \ee^{\ii\tau Y}a\,\ee^{-\ii\tau Y}
   =
   \ee^{-\ii\tau}a
   +\sqrt{\lambda}\,\ee^{\ii\chi}(\ee^{-\ii\tau}-1).
\end{equation}
The phase-rotated Krylov vector is \(\ket{K_r}=\ee^{\ii r\chi}\ket r\), so this overall phase does not affect diagonal number expectation values.  Since \(N=a^\dagger a=\sum_m m\ket{K_m}\bra{K_m}\), the reference-level mean of the evolved state is
\begin{equation}
   \sum_{m\ge0}m|g_m^{(r)}(\tau)|^2
   =
   \bra{K_r}\ee^{\ii\tau Y}N\ee^{-\ii\tau Y}\ket{K_r}
   =
   r+\lambda|\ee^{-\ii\tau}-1|^2
   =
   r+K_0(\tau),
   \label{eq:H1-old-number-mean}
\end{equation}
because the terms linear in \(a\) and \(a^\dagger\) have vanishing diagonal matrix elements in a number state. Both sides are defined in \([0,\infty]\) by the spectral theorem for \(N\), and both are finite. Since \(\ket{K_r}\) is a single Fock vector, \(a\,\ee^{-\ii\tau Y}\ket{K_r}=\ee^{-\ii\tau Y}\bigl(\ee^{-\ii\tau}a+\sqrt{\lambda}\,\ee^{\ii\chi}(\ee^{-\ii\tau}-1)\bigr)\ket{K_r}\) is a vector of squared norm \(r+K_0(\tau)\). The evolved seed therefore lies in the domain of \(a\), which is the domain of \(N^{1/2}\), and eq.~\eqref{eq:H1-old-number-mean} is an identity between finite quantities.  For each integer \(m\), \((m-r)_+-(r-m)_+=m-r\).  Therefore, 
\begin{equation}
   \sum_{m\ge0}(m-r)_+|g_m^{(r)}(\tau)|^2
   =
   K_0(\tau)+
   \sum_{m=0}^{r-1}(r-m)|g_m^{(r)}(\tau)|^2
   \ge K_0(\tau).
\end{equation}
This proves the vacuum lower bound. If \(r\ge1\) and \(K_0(\tau)>0\), the evolved state \(\ee^{-\ii\tau Y}\ket{K_r}\) has nonzero reference-basis overlap with \(\ket{K_0}\). Explicitly,
\begin{equation}
   |g_0^{(r)}(\tau)|^2
   =
   |\bra{K_0}\ee^{-\ii\tau Y}\ket{K_r}|^2
   =
   \ee^{-K_0(\tau)}\frac{K_0(\tau)^r}{r!}>0\,.
\end{equation}
The \(m=0\) term alone therefore makes the final nonnegative sum strictly positive.
\end{proof}
The vacuum lower bound is special to the Charlier oscillator dynamics. The general filtration bound remains valid for arbitrary polynomial seed jumps, but the final comparison with \(K_0\) used the identity \(\sum_m m|g_m^{(r)}(\tau)|^2=r+K_0(\tau)\). Finite Jacobi chains need not obey an analogous inequality and the failure can be exhibited
exactly. In the three-level chain of example~\ref{ex:spin-one-support-loss} the first jump deletes
the central atom, leaving two atoms of equal weight at $\pm1$, and both profiles are explicit,
\begin{equation}
   K_0(t)=\tfrac12\sin^2t+2\sin^4\tfrac t2\,,
   \qquad
   K_1(t)=\sin^2t\, .
   \label{eq:three-level-decrease}
\end{equation}
Hence $K_1(\pi)=0$ while $K_0(\pi)=2$, the maximal value permitted by this
three-level Krylov chain. A jump
to the first reference Krylov vector can strictly \emph{lower} the spread complexity, and by the
widest margin the dimension permits. The mechanism is support loss. Deleting the central atom
removes the unit gaps and halves the revival period from $2\pi$ to $\pi$, so at the re-timed
revival the jumped state has returned to its seed while the reference state sits at the far end of
its chain. The oscillator is protected against precisely this mechanism, since a fixed polynomial
seed deletes at most finitely many Poisson atoms and consecutive lattice points survive, so the
unit gap that sets the revival time is never removed.

The scalar and vector forms of the construction become explicit in two low-rank realizations.

\paragraph{The first jump.}
Section~\ref{sec:kernels} obtained the first-jump complexity from the rank-one projected kernel. Here we give the complementary connector data and the closed amplitude, while subsection~\ref{app:H1-first-jump} derives the underlying scalar remainder and jet recurrences. 
For the first jump $r=1$, let $A_n=P_n(0)$ and define $D_0=1$, $D_n=A_n^2+\lambda nD_{n-1}$. The three connector coefficients are
\begin{equation}
   \Gamma_{n,n}^{(1)}=\frac{D_{n+1}}{\lambda D_n}\,,\qquad
   \Gamma_{n,n+1}^{(1)}=\frac{(n+1)D_n-A_nA_{n+1}}{\lambda D_n}\,,\qquad
   \Gamma_{n,n+2}^{(1)}=\frac{1}{\lambda}\,.
   \label{eq:H1-first-Gamma-main}
\end{equation}
Iterating this recursion gives
\begin{equation}
   D_n=\lambda^n n!\sum_{m=0}^{n}\frac{A_m^2}{\lambda^m m!}
   =h_n\mathbb K_n^{\mu}(0,0)>0\,.
   \label{eq:Dn-main}
\end{equation}
The shifted polynomials for $\dd\nu_1=y^2\dd\mu/\lambda$ obey
\begin{equation}
   y^2R_n^{(1)}(y)=
   \frac{D_{n+1}P_n(y)+\bigl((n+1)D_n-A_nA_{n+1}\bigr)P_{n+1}(y)+D_nP_{n+2}(y)}{D_n}\,,
   \label{eq:H1-first-polys-main}
\end{equation}
and hence
\begin{equation}
   \tphi_n^{(1)}(\tau)=\cS(\tau)
   \frac{q(\tau)^n\left[D_{n+1}+\bigl((n+1)D_n-A_nA_{n+1}\bigr)q(\tau)+D_nq(\tau)^2\right]}{\lambda^{(n+1)/2}\sqrt{n!D_nD_{n+1}}}\,.
   \label{eq:H1-first-amplitude-main}
\end{equation}
The zeroth row gives the shifted survival amplitude directly from the reference one, 
\begin{equation}
   \tphi_0^{(1)}(\tau)
   =\frac{1}{\lambda}(\ii\partial_\tau)^2I_0(\tau)
   =\cS(\tau)\frac{\lambda+q(\tau)+q(\tau)^2}{\lambda}\,.
   \label{eq:H1-survival-speed-check-main}
\end{equation}
This is the \(n=0\) case of eq.~\eqref{eq:H1-first-amplitude-main}. Together with the local sum-rule data in eq.~\eqref{eq:H1-local-sum-rule-check-main}, it checks the first jump in two independent ways. The survival amplitude follows from the differential transfer rule, while the first shifted hopping follows from the reference local variance.

Three structural checks are contained in these formulas. First, the quantities $A_n$ and $D_n$ are reference Charlier data evaluated strictly at the Christoffel point; no shifted Lanczos recursion appears in eq.~\eqref{eq:H1-first-amplitude-main}. Subsection~\ref{app:H1-first-jump} identifies $A_n$ with the $r=1$ remainder sequence and obtains $D_n$ from its first-jet lift. Second, possible zeros of individual $A_n$, including their special behavior at integer $\lambda$, do not signal a singularity of the shifted problem because the cumulative sum
$D_n=h_n\mathbb K_n^\mu(0,0)$ remains strictly positive. Third, the complexity correction in eq.~\eqref{eq:H1-first-jump-complexity} comes from the kernel formulation. Its positivity is manifest in the modulus-square form, which gives $K_{1}(\tau)\ge K_0(\tau)$, whereas the equivalent real expansion is better adapted to asymptotic estimates.

The first jump also has a continuous-spectrum endpoint, obtained separately by taking $\omega\to0$ at fixed $\rho$ and physical time $t$. In this scaling, $\lambda=\rho^2/\omega^2\to\infty$ and $\tau_{\rm Ch}=\omega t\to0$, while $\tau_{\rm H}=\rho t=\sqrt\lambda\,\tau_{\rm Ch}$ remains fixed. It is not obtained by substituting $\omega=0$ into results expressed in terms of $Y=(H-\delta\Id)/\omega$. In the variable $y=(E-\delta)/\rho$, the vacuum measure is $\ee^{-y^2/2}\dd y/\sqrt{2\pi}$ and its monic OPs are the probabilists' Hermite polynomials. The first Fock-state jump multiplies this measure by $y^2$; subsection~\ref{app:H1-first-jump} derives the corresponding recurrence, amplitudes and complexity.

We give the first-level endpoint in full because in this case the complete
complexity resums to a short closed expression.
The shifted measure $y^2\dd\gamma(y)$, with
$\dd\gamma=\ee^{-y^2/2}\dd y/\sqrt{2\pi}$, is the generalized Hermite weight~\cite{Rosenblum1994}; its monic
orthogonal polynomials have $\widetilde a_n=0$ and
\begin{equation}
   \widetilde\beta_n=n+1-(-1)^n,
   \qquad\text{so that}\qquad
   \widetilde\beta_1=3,\quad\widetilde\beta_2=2,\quad\widetilde\beta_3=5,\ \ldots
   \label{eq:hermite-endpoint-recurrence-main}
\end{equation}
The parity alternation is the continuous-support signature of the seed, since multiplication by $y^2$
distinguishes even from odd degrees. With $c_n=n+1$ for even $n$ and $c_n=n+2$ for odd $n$, the
shifted amplitudes and complexity are
\begin{equation}
   \tphi_n^{(1)}(\tau_{\rm H})
   =\ee^{-\tau_{\rm H}^2/2}
    \frac{(-\ii\tau_{\rm H})^n\bigl(c_n-\tau_{\rm H}^2\bigr)}{\sqrt{\thh_n}}\,,
   \qquad
   \;K_1(\tau_{\rm H})=\tau_{\rm H}^2+1-\ee^{-2\tau_{\rm H}^2}\;
   \label{eq:hermite-endpoint-complexity-main}
\end{equation}
against the reference value $K_0(\tau_{\rm H})=\tau_{\rm H}^2$. The short-time behavior
$K_1=3\tau_{\rm H}^2-2\tau_{\rm H}^4+O(\tau_{\rm H}^6)$ reproduces
$\widetilde\beta_1=3$, as it must by
eq.~\eqref{eq:short-time-shifted-complexity-main}, and at late times
$K_1=\tau_{\rm H}^2+1+O(\ee^{-2\tau_{\rm H}^2})$.  The jump costs exactly one unit of complexity
asymptotically.  This is an exact statement about the first moment; it does not imply that
the full shifted Krylov distribution is a rigid one-level translate of the reference distribution.

Here the seed zero sits at $y=0$, which carries no Gaussian mass, so the measure class and the
cyclic dimension are unchanged. This is the continuous-support counterpart of the atom-deletion
discussion above, with the deletion channel switched off. Subsection~\ref{app:H1-first-jump}
derives eqs.~\eqref{eq:hermite-endpoint-recurrence-main}
and~\eqref{eq:hermite-endpoint-complexity-main}, including the Laguerre representation of the
shifted polynomials and the norms $\thh_n$. 

\paragraph{A three-component remainder example.}
Returning to the discrete Charlier chain, the case $r=3$ exhibits the vector character of the construction in a three-dimensional remainder space while keeping the local formulas short. Equation~\eqref{eq:H1-local-sum-rule-check-main} gives
\begin{equation}
   \widetilde a_0^{(3)}=3,\qquad
   \widetilde\beta_1^{(3)}=7\lambda,\qquad
   \widetilde a_1^{(3)}=\frac{22}{7},\qquad
   \widetilde\beta_2^{(3)}=\frac{26}{7}\lambda+\frac{48}{49}\,.
   \label{eq:H1-third-jump-low-data-main}
\end{equation}
These coefficients determine the short-time complexity through fourth order,
\begin{equation}
   K_3(\tau)=7\lambda\tau^2+
   \left(\frac{95\lambda}{84}-12\lambda^2\right)\tau^4+O(\tau^6).
   \label{eq:H1-third-jump-short-time-main}
\end{equation}
Subsection~\ref{app:H1-general-r} gives the all-index determinant and connector formulas and the three-component remainder sequence, while subsection~\ref{app:H1-large-index} derives the estimate controlling the infinite chain.

\paragraph{Arbitrary fixed Hermite level.}
Proposition~\ref{prop:Hermite-all-r-complexity-app} reconstructs every rebuilt
level for each fixed finite $r$.  With $x=\tau_{\rm H}^2$, it writes every amplitude as
\begin{equation}
   \tphi_n^{(r)}(\tau_{\rm H})
   =\ee^{-x/2}\frac{(-\ii\tau_{\rm H})^n}
      {\sqrt{\thh_n^{(r)}}}F_n^{(r)}(x),
   \qquad \deg F_n^{(r)}=r\,.
   \label{eq:Hermite-all-r-amplitude-main}
\end{equation}
The root-free moment/Gram construction of subsection~\ref{subsec:connector-realizations}
determines $\thh_n^{(r)}$ and $\widetilde\beta_n^{(r)}$, after which
eq.~\eqref{eq:Hermite-F-recurrence-app} generates the degree-$r$ polynomials.  The resulting
positive series gives the exact finite-time complexity.  At $r=1$, its two parity sectors sum to
eq.~\eqref{eq:hermite-endpoint-complexity-main}.
The same Hermite chain is also the fixed-level $q\uparrow1$ endpoint of the
continuous-$q$-Hermite chord problem in subsection~\ref{subsec:chord-context}.  It therefore links
the oscillator and DSSYK realizations without identifying their distinct physical bases.

The original Fock basis provides a second, directly physical coordinate.
Let
$D_r^{\rm H}(x)$ denote the mean absolute displacement from the prepared Fock level in the
original Fock basis.  We call it the physical Fock distance from the seed.  In what follows,
``radial'' always refers to this mean absolute displacement and not to a spatial radius.
\begin{equation}
\begin{gathered}
   D_r^{\rm H}(x)
   =x+2\ee^{-x}\sum_{m=0}^{r-1}(r-m)\frac{m!}{r!}
      x^{r-m}\bigl[L_m^{(r-m)}(x)\bigr]^2,\\
   K_r(\tau_{\rm H})\ge D_r^{\rm H}(x),\\
   D_r^{\rm H}(x)
   \ge x+2\ee^{-x}\frac{x^r}{(r-1)!}>x=K_0(\tau_{\rm H}),
   \qquad r\ge1,\quad x>0\,.
\end{gathered}
   \label{eq:Hermite-radial-bound-main}
\end{equation}
The first inequality is the continuous Hermite counterpart of the fixed-chord radial filtration in
proposition~\ref{prop:chord-radial-filtration}; the explicit factor of two strengthens the
one-sided Fock-tail estimate.

The short-time expansion is explicit through the first term that distinguishes these coordinates,
\begin{equation}
\begin{gathered}
   K_r(\tau_{\rm H})
   =(2r+1)\tau_{\rm H}^2-r(r+1)\tau_{\rm H}^4
     +\frac{2r(r+1)(r^2+r+1)}{3(2r+1)}\tau_{\rm H}^6
     +O_r(\tau_{\rm H}^8),\\
   K_r(\tau_{\rm H})-D_r^{\rm H}(\tau_{\rm H}^2)
   =\frac{r(r+1)}{2(2r+1)}\tau_{\rm H}^6
     +O_r(\tau_{\rm H}^8).
   \label{eq:Hermite-all-r-short-time-main}
\end{gathered}
\end{equation}
Physical  Fock motion and rebuilt Krylov spread therefore coincide through order $\tau_{\rm H}^4$.
For $r\ge2$, the exact finite-time series does not by itself control the late-time interchange of
the infinite first-moment sum; no such interchange is used here.

The oscillator illustrates the infinite-chain closure of the transfer
calculus. %
A finite-volume lattice model presents the complementary global issue.  It is not large-index
summability, but canonical folding at the terminal edge.

\subsection{The tight-binding chain and Chebyshev seed jumps}
\label{subsec:tight-binding}

The constant-coefficient tight-binding chain gives the polynomial seed transformation a direct lattice interpretation. For the reference seed \(\ket0\), the Krylov vectors coincide with the physical site states, whereas replacing \(\ket0\) by \(\ket k\) changes the initially occupied site without changing the Hamiltonian. This fixed-$H$ problem is complementary to the finite-size and localized-state analyses of ref.~\cite{Balasubramanian:2025variations}. We take the half-line to consist of sites \(n\ge0\), with the left Dirichlet condition \(\psi_{-1}=0\), and obtain the finite open chain by restricting to \(n=0,\ldots,N_{\rm lat}-1\) and imposing \(\psi_{N_{\rm lat}}=0\).\footnote{The Dirichlet ghost notation keeps the nearest-neighbor difference equation in its bulk form at the endpoints, with the auxiliary amplitudes outside the physical lattice set to zero. Thus the half-line has only \(\psi_{-1}=0\), while the finite open chain has \(\psi_{-1}=\psi_{N_{\rm lat}}=0\). With \(0\)-based site indexing, the finite-chain sine modes are proportional to \(\sin[(n+1)(j+1)\pi/(N_{\rm lat}+1)]\), where \(j=0,\ldots,N_{\rm lat}-1\).} Here \(n\) labels a physical site, \(j\) a finite-chain eigenmode and \(k\) the localized seed \(\ket k\). A negative hopping may be made positive by the site-dependent phase \(\ket n\mapsto(-1)^n\ket n\); this unitary gauge changes only site phases and leaves probabilities and spread complexity invariant. We therefore take \(b>0\) and write
\begin{equation}
   H_{\rm tb}=a\sum_n\ket n\bra n
   +b\sum_n\bigl(\ket{n+1}\bra n+\ket n\bra{n+1}\bigr),
   \label{eq:tight-binding-main}
\end{equation}
where the sums run over the physical sites and allowed bonds of the half-line or finite open chain.

\paragraph{Reference chain and localized seeds.}
Starting from \(\ket0\), the physical site basis is the Krylov basis until the terminal boundary is reached. On the half-line the Jacobi coefficients are \(a_n=a\), \(b_0=0\) and \(b_n=b\) for \(n\ge1\). On a finite chain the same coefficients hold before termination, with \(b_{N_{\rm lat}}=0\). The corresponding non-terminal monic polynomials and positive norms are
\begin{equation}
   P_n(E)=b^n U_n\left(\frac{E-a}{2b}\right),
   \qquad h_n=b^{2n},
   \label{eq:Chebyshev-polys-main}
\end{equation}
where \(U_n\) denotes the Chebyshev polynomial of the second kind. Equation~\eqref{eq:Chebyshev-polys-main} holds for every \(n\ge0\) on the half-line. On the finite chain, the norm formula holds for \(0\le n<N_{\rm lat}\), while the same polynomial expression at degree \(N_{\rm lat}\) gives the zero-norm terminal polynomial
\(
T_{N_{\rm lat}}(E)=P_{N_{\rm lat}}(E)
\).
The recurrence is encoded in
\begin{equation}
   \sum_{n=0}^{\infty}P_n(E)u^n
   =\frac{1}{1-(E-a)u+b^2u^2}\,,
   \qquad |u|\ \hbox{small}\,,
   \label{eq:Chebyshev-gen-main}
\end{equation}
which is the physical half-line generating function and, in finite volume, defines an auxiliary infinite-chain continuation of the preterminal Chebyshev polynomials. It follows that every localized site is a polynomial seed,
\begin{equation}
   \ket{k}=\frac{P_k(H_{\rm tb})}{\sqrt{h_k}}\ket0,
   \qquad
   \dd\nu_k(E)=\frac{P_k(E)^2}{h_k}\dd\mu_0(E).
   \label{eq:tight-binding-shift-main}
\end{equation}
On the half-line, this holds for every \(k\ge0\). On a finite open chain, it holds for \(0\le k<N_{\rm lat}\), with transformed-polynomial products interpreted as functions on the finite support and reduced modulo \(T_{N_{\rm lat}}\). Thus, the half-line connector is an identity in the ordinary polynomial ring, whereas the finite-chain connector is an identity in the ambient reference-chain quotient. Near the terminal edge, its coefficients may be evaluated directly by a quotient-space Gram construction, or through a proper terminal lift followed by reduction.

\paragraph{Local shifted Jacobi data and connector bandwidth.}
The first shifted Lanczos step has a direct site-space interpretation. The superscript \((k)\) denotes the shifted chain generated from \(\ket k\), while \(b_k,b_{k+1}\) on the right are the reference hoppings of the \(\ket0\)-based chain. Before the terminal edge, the local variance at a site \(k\ge1\) gives
\begin{equation}
   \widetilde a_0^{(k)}=a\,,
   \qquad
   \bigl(\widetilde b_1^{(k)}\bigr)^2=b_k^2+b_{k+1}^2\,.
   \label{eq:tight-binding-local-first-main}
\end{equation}
These Lanczos coefficients are also the reference-state specialization of the connector dictionary, so the independent site-space calculation gives a direct check of the general transfer formalism. On the half-line it gives \((\widetilde b_1^{(k)})^2=2b^2\) for every \(k\ge1\). The next hopping distinguishes the boundary from the bulk
\begin{equation}
   \bigl(\widetilde b_2^{(k)}\bigr)^2=
   \begin{cases}
      b^2/2, & k=1\,,\\
      b^2, & k\ge2\,,
   \end{cases}
   \label{eq:tight-binding-local-second-main}
\end{equation}
with the analogous reflected modification near the far endpoint of a finite open chain. The two branches emanating from a localized bulk seed are the site-space origin of the finite-rank correction in the projected Chebyshev kernel.

The local branching has an equivalent spectral description. For the finite open chain, the sine eigenvectors give the support points and boundary weights
\begin{equation}
   \theta_j=\frac{(j+1)\pi}{N_{\rm lat}+1}\,,\quad
   E_j=a+2b\cos\theta_j\,,\quad
   w_j=\frac{2}{N_{\rm lat}+1}\sin^2\theta_j\,,
   \quad j=0,\ldots,N_{\rm lat}-1\,.
   \label{eq:tight-binding-spectrum-main}
\end{equation}
The energies $E_j$ are precisely the zeros of $T_{N_{\rm lat}}$. As \(N_{\rm lat}\to\infty\), the boundary spectral measure converges to the half-line semicircle measure
\begin{equation}
   \dd\mu_0(E)=
   \frac{\sqrt{4b^2-(E-a)^2}}{2\pi b^2}
   \boldsymbol 1_{[a-2b,a+2b]}(E)\dd E.
   \label{eq:tight-binding-halfline-measure-main}
\end{equation}
Writing \(E=a+2b\cos\theta\), the half-line shift \(\ket0\to\ket k\) gives
\begin{equation}
   \dd\nu_k(\theta)=\frac{2}{\pi}\sin^2((k+1)\theta)\dd\theta,
   \qquad 0\le\theta\le\pi\,.
   \label{eq:tight-binding-shifted-theta-main}
\end{equation}
Equivalently, \(P_k\) has \(k\) simple zeros in the interior of the band,
and \(P_k^2\) gives double Christoffel zeros at those locations.  The
amplitude connector therefore has upper bandwidth \(2k\).  Reflection
about \(E=a\) restricts its half-line expansion to the \(k+1\)
parity-compatible terms \(P_n,P_{n+2},\ldots,P_{n+2k}\), whereas the
projected-kernel correction has rank \(k\).

\paragraph{Half-line dynamics for arbitrary localized seeds.}
Fourier transformation of eq.~\eqref{eq:tight-binding-halfline-measure-main}
gives the half-line reference amplitudes
\begin{equation}
   \phi_n^{(0)}(t)=
   \ee^{-\ii at}(-\ii)^n\frac{(n+1)J_{n+1}(2bt)}{bt}\,,
   \qquad I_n(t)=b^n\phi_n^{(0)}(t),
   \label{eq:tight-binding-reference-amplitudes-main}
\end{equation}
where \(J_\nu\) is the Bessel function of the first kind and the value at
\(t=0\) is understood by continuity. The Chebyshev structure makes the connector explicit for every localized
seed.  The case \(k=0\) is the reference chain.  For \(k\ge1\), let
\(R_n^{(k)}\) and \(\thh_n^{(k)}\) denote the monic OPs and their norms
for the shifted measure \(P_k^2\dd\mu_0/h_k\).  In the formulas below,
\(n\) labels the Krylov level in the chain generated from \(\ket k\),
rather than a physical site.  Write
\begin{equation}
   n=(\sigma-1)(k+1)+\varepsilon\,,
   \qquad \sigma=1,2,\ldots,
   \qquad 0\le\varepsilon\le k\,,
   \label{eq:tight-binding-block-decomposition-main}
\end{equation}
and, for \(0\le s\le k\), define
\begin{equation}
   \mathcal C_{\sigma,\varepsilon;s}^{(k)}=
   \begin{cases}
      \dfrac{\sigma+1}{\sigma}\,,
      &1\le\varepsilon\le k\,,
       \quad 0\le s\le k-\varepsilon,\\[6pt]
      1\,,
      &\varepsilon=0
       \quad\text{or}\quad k-\varepsilon<s\le k\,.
   \end{cases}
   \label{eq:tight-binding-all-k-coefficients-main}
\end{equation}
The connector, shifted norm and amplitude are
\begin{equation}
\begin{aligned}
   \frac{P_k(E)^2}{h_k}R_n^{(k)}(E)
   &=\sum_{s=0}^{k}
   \mathcal C_{\sigma,\varepsilon;s}^{(k)}
   b^{-2s}P_{n+2s}(E),\\
   \thh_n^{(k)}
   &=b^{2n}\mathcal C_{\sigma,\varepsilon;0}^{(k)}\,,\\
   \tphi_n^{(k)}(t)
   &=\frac{1}{\sqrt{\mathcal C_{\sigma,\varepsilon;0}^{(k)}}}
   \sum_{s=0}^{k}
   \mathcal C_{\sigma,\varepsilon;s}^{(k)}
   \phi_{n+2s}^{(0)}(t).
   \label{eq:tight-binding-all-k-connector-main}
\end{aligned}
\end{equation}
In the notation of section~\ref{sec:finiteband}, the nonzero connector
coefficients are
\(\Gamma_{n,n+2s}^{(k)}=b^{-2s}
\mathcal C_{\sigma,\varepsilon;s}^{(k)}\).  Thus, the upper coefficient
is \(b^{-2k}=h_k^{-1}\), while every odd-offset coefficient vanishes.
Appendix~\ref{app:finite-tight-binding-checks} proves
eq.~\eqref{eq:tight-binding-all-k-connector-main} by Chebyshev
divisibility and shows that its norm ratios reproduce the localized-site
hopping blocks. In finite volume, Gauss exactness makes the shifted polynomial, its norm and its unfolded connector agree with their half-line counterparts over three distinct index ranges. The same appendix identifies these ranges; the truncated Chebyshev product and its finite product-Gram matrix then implement the terminal folds and continue the connector through every remaining physical row. Equation~\eqref{eq:tight-binding-reference-amplitudes-main}
then expresses every half-line shifted amplitude as a finite Bessel sum. At \(n=0\), all the
coefficients in eq.~\eqref{eq:tight-binding-all-k-coefficients-main} are
unity, and the sum telescopes to\footnote{Our \(\ee^{-\ii Ht}\) convention is the complex conjugate of the
return-amplitude convention used in
ref.~\cite{Balasubramanian:2025variations}.}
\begin{equation}
   \tphi_0^{(k)}(t)=
   \ee^{-\ii at}\left[J_0(2bt)+(-1)^kJ_{2k+2}(2bt)\right].
   \label{eq:tight-binding-site-return-main}
\end{equation}

\paragraph{The first localized jump.}
For \(k=1\), set \(d_n=\lfloor n/2\rfloor+1\).  The values
\(P_{2m}(a)=(-1)^mb^{2m}\) and \(P_{2m+1}(a)=0\) give
\(h_n\mathbb K_n^\mu(a,a)=b^{2n}d_n\).  The \(k=1\) specialization of
eq.~\eqref{eq:tight-binding-all-k-connector-main} is
\begin{equation}
\begin{aligned}
   \frac{(E-a)^2}{b^2}R_n^{(1)}(E)
   &=\frac{d_{n+1}}{d_n}P_n(E)+\frac{1}{b^2}P_{n+2}(E),\\
   \thh_n^{(1)}&=b^{2n}\frac{d_{n+1}}{d_n}\,.
\end{aligned}
   \label{eq:tight-binding-first-connector-main}
\end{equation}
Its zeroth row can also be checked through the differential transfer of
the survival amplitude,
\begin{equation}
   \tphi_0^{(1)}(t)
   =\frac{1}{b^2}(\ii\partial_t-a)^2I_0(t)
   =I_0(t)+\frac{I_2(t)}{b^2}\,,
   \label{eq:tight-binding-survival-check-main}
\end{equation}
using
\(
   (E-a)^2/b^2=P_0(E)+P_2(E)/b^2
\).
Equation~\eqref{eq:tight-binding-local-first-main} independently fixes
the same short-time scale through the reference local variance.

Substitution in the finite-band transfer formula gives
\begin{align}
   \tphi_n^{(1)}(t)
   &=
   \sqrt{\frac{d_n}{d_{n+1}}}
   \left[
      \frac{d_{n+1}}{d_n}\phi_n^{(0)}(t)+\phi_{n+2}^{(0)}(t)
   \right]
   \label{eq:tight-binding-first-amplitude-transfer-main}\\
   &=
   \ee^{-\ii at}\frac{(-\ii)^n}{bt}
   \sqrt{\frac{d_n}{d_{n+1}}}
   \left[
      \frac{d_{n+1}}{d_n}(n+1)J_{n+1}(2bt)
      -(n+3)J_{n+3}(2bt)
   \right].
   \label{eq:tight-binding-first-amplitude-main}
\end{align}
\begin{samepage}
The parity dependence of \(d_n\) makes the two branches explicit.
Splitting eq.~\eqref{eq:tight-binding-first-amplitude-main} into even and
odd levels and using the three-term Bessel recurrence gives
\begin{align}
   \tphi_{2m}^{(1)}(t)
   &=
   (-1)^m\ee^{-\ii at}
   \left[J_{2m}(2bt)-J_{2m+4}(2bt)\right], \label{eq:tight-binding-first-amplitude-parity-main}
\\
   \tphi_{2m+1}^{(1)}(t)
   &=
   \frac{-\ii(-1)^m\ee^{-\ii at}}{\sqrt{(m+1)(m+2)}}
   \Bigl[
      (m+2)J_{2m+1}(2bt)+J_{2m+3}(2bt)
      -(m+1)J_{2m+5}(2bt)
   \Bigr]. \nonumber
\end{align}
\end{samepage}
Here \(m=0,1,2,\ldots\) and after conjugating the time evolution
convention, these expressions agree with
ref.~\cite{Balasubramanian:2025variations}.

The shifted complexity is therefore
\begin{equation}
   K_1(t)=
   \sum_{n\ge0}n\,\frac{d_n}{d_{n+1}}
   \left|
      \frac{d_{n+1}}{d_n}\phi_n^{(0)}(t)+\phi_{n+2}^{(0)}(t)
   \right|^2,
   \label{eq:tight-binding-first-complexity-main}
\end{equation}
or equivalently the sum over the parity-split Bessel amplitudes. The boundary seed gives the corresponding unshifted benchmark, whose complexity is the Chebyshev/Bessel sum\footnote{The resummation follows by setting \(x=2bt\) and using the weighted Bessel-square identities
\begin{equation}
\begin{aligned}
   \sum_{m\ge1}m^2J_m(x)^2
   &=\frac{x^2}{4}\,,\\
   \sum_{m\ge1}m^3J_m(x)^2
   &=
   \frac{x^2}{12}
   \Big[
      (4x^2+3)J_0(x)^2
      +(4x^2+1)J_1(x)^2  
      -4xJ_0(x)J_1(x)
   \Big]\,.
\end{aligned}
\label{eq:bessel-rayleigh-lommel-sums}
\end{equation}
These identities follow from the Bessel three-term recurrence and summation by parts \cite{DLMF}.}
\begin{align}
   K_0(t)
   &=\frac{1}{(bt)^2}\sum_{n\ge0}n(n+1)^2J_{n+1}(2bt)^2 \nonumber\\
   &=
   \frac{16b^2t^2+1}{3}J_1(2bt)^2
   -\frac{8bt}{3}J_0(2bt)J_1(2bt)
   +\frac{16b^2t^2+3}{3}J_0(2bt)^2
   -1\,.
   \label{eq:tight-binding-reference-complexity-main}
\end{align}

\paragraph{Finite-chain support loss.}
Finite volume retains the same seed transformation but adds an arithmetic mechanism for support loss. Let $d_{\rm cyc}^{(k,N_{\rm lat})}$ denote the cyclic dimension generated by the localized seed $\ket{k}$, equivalently the number of nonzero shifted weights. The weights and the number of deleted atoms are 
\begin{equation}
\begin{gathered}
   w_j^{(k)}
   =\frac{P_k(E_j)^2}{h_k}w_j
     =\frac{2}{N_{\rm lat}+1}\sin^2((k+1)\theta_j),\\
   r_{\rm zero}
   =\gcd(k+1,N_{\rm lat}+1)-1\,,
   \qquad d_{\rm cyc}^{(k,N_{\rm lat})}
   =N_{\rm lat}-r_{\rm zero}\,.
\end{gathered}
   \label{eq:tight-binding-support-loss-main}
\end{equation}
Support loss occurs exactly when \(\gcd(k+1,N_{\rm lat}+1)>1\). In particular, the first jump removes the band-center atom $E=a$ precisely when \(N_{\rm lat}\) is odd.

The zero weights in eq.~\eqref{eq:tight-binding-support-loss-main} do not
enter the shifted inner product.  The quotient by \(T_{N_{\rm lat}}\) is
the ambient finite-chain representation, while support loss reduces the
physical chain to the quotient associated with
\begin{equation}
   T_{\rm surv}^{(k,N_{\rm lat})}(E)
   =\prod_{\substack{0\le j<N_{\rm lat}\\ w_j^{(k)}>0}}(E-E_j).
   \label{eq:tight-binding-surviving-support-main}
\end{equation}
Equivalently, deleting the zero-weight terms from the spectral sum leaves
precisely the shifted inner product represented by this smaller quotient.
The finite amplitudes follow from the connector after canonical reduction
modulo \(T_{N_{\rm lat}}\), and the complexity is
\begin{equation}
   K_{k,N_{\rm lat}}(t)
   =\sum_{n=0}^{d_{\rm cyc}^{(k,N_{\rm lat})}-1}
   n\left|\tphi_n^{(k,N_{\rm lat})}(t)\right|^2.
   \label{eq:tight-binding-finite-complexity-main}
\end{equation}
Appendix~\ref{app:finite-tight-binding-checks} gives the finite-volume
amplitude and quotient-projector formulas and verifies the
$N_{\rm lat}=10$, $k=1$ recurrence against a direct site-space Lanczos 
computation.

\paragraph{Relation to spectral surgery of $XX$ chains.}
Christoffel transformations also underlie spectral-surgery constructions for quantum state
transfer.  There one selects or deletes spectral levels and reconstructs the Jacobi couplings of
an engineered $XX$ chain \cite{VinetZhedanov2012PST,VinetZhedanov2012Para}; recent applications
interpolate between Krawtchouk and homogeneous chains under coupling constraints and construct transfer chains exhibiting early state exclusion
\cite{VinetZhedanov2024Surgery,BaileyEtAl2024PST}.  The present fixed-H problem uses the same
spectral transformation for a complementary purpose.  The couplings and spectrum are given, the
launch site is varied, and the transformed Krylov wavefunction and its spread complexity are
computed.  Equation~\eqref{eq:tight-binding-support-loss-main} then gives a launch-site selection
rule in which $\left(\gcd(k+1,N_{\rm lat}+1)-1\right)$ counts the spectral levels invisible from site $k$.

\paragraph{Comparison across the solvable families.}
Taken together, these examples realize the construction on infinite discrete support, continuous support and compact finite support, and include both terminal closure and atom deletion.
The bounded continuous-$q$-Hermite chain adds the fixed-chord DSSYK realization.  Its
fixed-degree and deep-seed regimes are developed in sections~\ref{sec:rigidity}
and~\ref{sec:context}, respectively, and its $q\uparrow1$ and $q=0$ endpoints meet the Hermite and
constant-hopping chains studied here.
Beyond these solvable families, the finite-band and projected-kernel results apply to Meixner, Hahn and other Jacobi systems generated by birth--death processes \cite{KarlinMcGregor1957,DominguezBook}. They also apply to finite many-body systems once the reference Jacobi data are known. The exact statements hold for every finite polynomial seed degree. When an explicit connector becomes large, appendix~\ref{app:numerics} develops reduced-Jacobi and Gram realizations of the same transform, together with rank, residual and infinite-window checks.
Table~\ref{tab:universality_classes} places the four solvable regimes side by side.  Their
different spectral supports and physical seed coordinates make the common fixed-$H$ transfer
structure explicit.

\begin{table}[htbp]
\centering
\setlength{\tabcolsep}{3pt}
\begin{tabular}{@{}>{\raggedright\arraybackslash}p{0.275\textwidth}>{\raggedright\arraybackslash}p{0.27\textwidth}>{\raggedright\arraybackslash}p{0.16\textwidth}>{\raggedright\arraybackslash}p{0.25\textwidth}@{}}
\toprule
\textbf{Chain} & \textbf{Jacobi Data} & \textbf{OP Family} & \textbf{Natural Shifted Seeds} \\
\midrule
Constant hopping & $a_n=a,\ b_n^2=b^2\ (n\ge1)$ & Chebyshev $U_n$ & Localized site states \\
Heisenberg--Weyl, \(\omega>0\) & $a_n=\delta+\omega n$, $b_n^2=\rho^2n$ & Charlier & Number states \\
HW scaling endpoint, \(\omega\to0\) & $a_n=\delta$, $b_n^2=\rho^2n$ & Probabilists' Hermite & Fock-state seeds \\
DSSYK chord chain, $0\le q<1$ & $a_n=0,\ b_n^2=[n]_q$ & Continuous $q$-Hermite & Fixed-chord states \\
\bottomrule
\end{tabular}
\caption{Reference Jacobi data, orthogonal-polynomial (OP) families and natural shifted seeds for
the four solvable regimes.  The constant-hopping entry is understood before finite-chain
termination, and the DSSYK row uses the normalized chord Hamiltonian of
subsection~\ref{subsec:chord-context}.}
\label{tab:universality_classes}
\end{table}

The preceding examples each begin from a single scalar seed. The natural extension is to organize a finite family of possible seeds before selecting the direction whose scalar Krylov chain is to be constructed.

\section{Matrix-valued parent measures}
\label{sec:matrix}

We fix a finite-dimensional seed subspace and retain all its normalized pure seeds.  A direction
must be chosen before a scalar Krylov chain exists.  The scalar Lanczos bases generated by
different directions cannot be assembled linearly because orthogonalization depends nonlinearly
on the seed measure.  Their spectral measures can nevertheless be organized by a single
matrix-valued parent, whose quadratic compressions select the individual seed directions.
Compressing the projection-valued measure of $H$ to the seed subspace gives this positive
matrix-valued measure \cite{DamanikPushnitskiSimon2008,AlvarezFernandez2017}. The survey of Damanik, Pushnitski and Simon develops the matrix orthogonal polynomials and block Jacobi matrices carried by such a parent, and \'Alvarez-Fern\'andez and collaborators construct the Christoffel transformations acting on it. The proposition below fixes the isometry frame, the normalization to the identity and the quadratic compression in the form used immediately afterwards by the unitary covariance relation, the atom-deletion criterion and the matrix moment identity.

\begin{proposition}[Finite-seed parent measure]
\label{prop:matrix-parent-main}
Let
$V_0=(\ket{v_1},\ldots,\ket{v_p}):\CC^p\to\cH$
be an isometry onto the seed subspace and define
\begin{equation}
   \dd M(E)=V_0^\dagger\Pi(\dd E)V_0\,,
   \qquad
   \dd M_{ab}(E)=\bra{v_a}\Pi(\dd E)\ket{v_b}.
   \label{eq:matrix-measure}
\end{equation}
For every unit vector $c\in\CC^p$, the pure seed
$\ket{\psi_c}=V_0c$ has scalar spectral measure
\begin{equation}
   \dd\nu_c(E)=c^\dagger\dd M(E)c\,.
   \label{eq:scalar-compression}
\end{equation}
The measure $\dd M$ is positive semidefinite and normalized by
$\int_{\RR}\dd M(E)=I_p$.  Conversely, its scalar compressions determine
all matrix entries by polarization.
\end{proposition}

\begin{proof}
For every Borel set $B\subset\RR$ and $z\in\CC^p$,
\begin{equation}
   z^\dagger M(B)z
   =\bra{V_0z}\Pi(B)\ket{V_0z}
   =\|\Pi(B)V_0z\|^2\ge0\,,
\end{equation}
while $M(\RR)=V_0^\dagger V_0=I_p$.  The spectral measure of
$\ket{\psi_c}$ is
$\bra{V_0c}\Pi(\dd E)\ket{V_0c}=c^\dagger\dd M(E)c$.
Applying the complex polarization identity to the quadratic form
$z\mapsto z^\dagger M(B)z$ for every $B$ reconstructs the sesquilinear
matrix entries of $M$.
\end{proof}

The coordinate-free parent is the compression
$B\mapsto\mathsf P_{\mathcal S}\Pi(B)|_{\mathcal S}$
to the seed subspace
$\mathcal S=\operatorname{Ran}V_0=\Span\{\ket{v_1},\ldots,\ket{v_p}\}$, with
$\mathsf P_{\mathcal S}$ its orthogonal projector; $M(B)$ is the matrix of this operator in the
frame $V_0$. The compression need not remain projection-valued unless $\mathcal S$ reduces $H$; in
general it is a normalized positive operator-valued measure on the coefficient space. 
The construction depends only on the seed subspace.  Under a change of
orthonormal coordinates by $U\in U(p)$,
\begin{equation}
   V_0'=V_0U,\qquad
   \dd M'=U^\dagger\dd M U,\qquad
   c'=U^\dagger c\,,\qquad
   c'^\dagger\dd M'c'=c^\dagger\dd M c\,.
   \label{eq:matrix-parent-covariance-main}
\end{equation}
The normalized pure seeds are therefore parametrized by directions
$[c]\in\mathbb{CP}^{p-1}$.
In finite support, the direction $c$ deletes an atom $E_\alpha$ precisely when
$\Pi(\{E_\alpha\})V_0c=0$, equivalently when
$c^\dagger M(\{E_\alpha\})c=0$.  Thus a single parent measure carries the support-loss data of
the whole family.

If $\operatorname{Ran}V_0\subset\Dom(H^k)$, its $k$-th moment is
\begin{equation}
   \boldsymbol{\mu}_k
   =\int_{\RR}E^k\dd M(E)
   =V_0^\dagger H^kV_0\,,
   \qquad
   (\boldsymbol{\mu}_k)_{ab}=\bra{v_a}H^k\ket{v_b}.
   \label{eq:matrix-moments-main}
\end{equation}
These are single-index moments with matrix values.  The matrix structure encodes the seed
directions, not a second spectral variable.  The scalar moment sequence for a chosen
direction is obtained by the same quadratic compression,
\begin{equation}
   \mu_k^{(c)}
   =\int_{\RR}E^k\dd\nu_c(E)
   =c^\dagger\boldsymbol{\mu}_k c\,.
\end{equation}

\paragraph{Krylov-window specialization.}
Take $R\ge0$, with $R\le d-1$ in finite cyclic dimension, and define
\begin{equation}
   \mathcal S_R
   =\Span\{\ket{K_0},\ldots,\ket{K_R}\},
   \qquad
   V_R=(\ket{K_0},\ldots,\ket{K_R}),
\end{equation}
and set
\begin{equation}
   v_R(E)
   =\left(
      \frac{P_0(E)}{\sqrt{h_0}},\ldots,
      \frac{P_R(E)}{\sqrt{h_R}}
   \right)^{\mathsf T}.
\end{equation}
Since every vector in $\mathcal S_R$ lies in the same scalar cyclic
representation, the parent measure has the rank-one density
\begin{equation}
\begin{aligned}
   \dd M(E)&=v_R(E)v_R(E)^\dagger\dd\mu(E),\\
   \dd M_{rs}(E)
   &=\frac{P_r(E)P_s(E)}{\sqrt{h_rh_s}}\dd\mu(E),
   \qquad 0\le r,s\le R\,.
\end{aligned}
   \label{eq:krylov-block-measure}
\end{equation}
The Radon--Nikodym density is rank one $\mu$-almost everywhere, although
$\int\dd M=I_{R+1}$ by orthonormality.  For a general seed
subspace not contained in one scalar cyclic representation, choose a
scalar measure that dominates every entry of the compression. The
resulting matrix density may have rank greater than one and need not
factor through a single polynomial vector.

For a unit coefficient vector $c\in\CC^{R+1}$, define 
\begin{equation}
\begin{aligned}
   \Qhat_c(E)
   &=v_R(E)^{\mathsf T}c
   =\sum_{r=0}^{R}c_r\frac{P_r(E)}{\sqrt{h_r}},\\
   \dd\nu_c(E)
   &=|\Qhat_c(E)|^2\dd\mu(E).
\end{aligned}
   \label{eq:krylov-block-compression}
\end{equation}
Orthogonality of the reference polynomials gives
$\int|\Qhat_c|^2\dd\mu=c^\dagger c=1$, so this is precisely the
normalized polynomial-seed measure of the preceding sections.  The
off-diagonal entries of $\dd M$ retain the interference terms between
seed directions in the chosen Krylov frame.    The map $[c]\mapsto\nu_c$,
however, need not be injective.  Because $v_R(E)$ is real on the spectral support,
$c$ and $\bar c$ give the same scalar measure, and hence the same scalar
Krylov problem.  For the Krylov window, the entries of
eq.~\eqref{eq:matrix-moments-main} are finite linear combinations of the
reference scalar moments $\mu_j$, involving no order above $k+r+s$ in
the $(r,s)$ entry of $\boldsymbol{\mu}_k$.

\paragraph{Mixed preparations and scalar trace contractions.}
A density operator supported on $\mathcal S_R$ extends the pure-direction compressions above
from rank-one projectors $cc^\dagger$ to a general coefficient-space density matrix. If
$\sigma\ge0$ on $\CC^{R+1}$, $\operatorname{Tr}\sigma=1$, and
$\rho_R=V_R\sigma V_R^\dagger$, then 
\begin{equation}
\begin{aligned}
   \dd\nu_\sigma(E)
   =\dd\mu_{\rho_R}(E)
   =\operatorname{Tr}_{\cH}\!\left(\rho_R\Pi(\dd E)\right)
   =\operatorname{Tr}_{\CC^{R+1}}\!\left(\sigma\,\dd M(E)\right)
   =v_R(E)^\dagger\sigma v_R(E)\dd\mu(E)
\end{aligned}
   \label{eq:matrix-trace-compression-main}
\end{equation}
is a normalized scalar measure, with multiplier $W_\sigma(E)=v_R(E)^\dagger\sigma v_R(E)$, a polynomial of degree at most $2R$ that is nonnegative for every real $E$ because $\sigma\ge0$.  The rank-one choice
$\sigma=cc^\dagger$ reduces to eq.~\eqref{eq:scalar-compression}, whereas
a general $\sigma$ determines a single scalar Krylov chain through
$\dd\nu_\sigma$.  Its complexity is not an average of the pure-seed
complexities associated with a decomposition of $\sigma$. 

Because $W_\sigma$ is nonnegative on the whole real axis, its real roots have even multiplicity and its nonreal roots occur in conjugate pairs, so it admits a scalar spectral factorization $W_\sigma(E)=|Q_\sigma(E)|^2$ with $\deg Q_\sigma\le R$. Appendix~\ref{app:mixed-seeds}
proves this statement and develops the associated energy-distribution
chain. Consequently, every trace contraction from $\mathcal S_R$ is
measure-equivalent to an effective pure polynomial seed. This
equivalence concerns the scalar measure and its
Jacobi problem; it does not identify a genuinely mixed preparation with
a pure state. %
This scalar factorization is available because the consecutive Krylov window lies inside one
scalar cyclic representation, so eq.~\eqref{eq:krylov-block-measure} reduces every compression to
a single nonnegative polynomial density.  A general seed subspace with irreducible matrix-valued
spectral multiplicity need not admit such a scalar reduction. 
Appendix~\ref{app:operator-krylov} develops the distinct
Liouville-space construction for physical operator or density-matrix
evolution, whose spectral variable is an energy gap rather than an
energy.

\paragraph{Scalar compression versus block dynamics.}
Block Lanczos would instead build a common sequence of subspaces and assign complexity by
block level \cite{CrapsEvninPascuzzi2024}; $\dd M$ is the spectral input for that construction as
well, subject to deflation when the block moment problem becomes singular. Here it is used only
through its scalar compressions, and the consecutive Krylov window already shows why the block route degenerates there. If  
\begin{equation}
   \mathsf P_R=V_RV_R^\dagger\,,\qquad
   e_R=(0,\ldots,0,1)^{\mathsf T},
\end{equation}
and $b_{R+1}>0$, the reference three-term recursion gives
\begin{equation}
   (\Id-\mathsf P_R)HV_R
   =b_{R+1}\ket{K_{R+1}}e_R^\dagger\,.
   \label{eq:krylov-window-block-deflation-main}
\end{equation}
Thus, for $R\ge1$, the first block residual has rank one rather than
$R+1$, and a fixed square block recurrence deflates immediately.  

The present construction bypasses this block recursion by first choosing
a seed direction and then forming the corresponding scalar Jacobi
problem.  For the Krylov window this gives 
the projective family
\begin{equation}
   [c]\in\mathbb{CP}^{R}
   \longmapsto cc^\dagger
   \longmapsto
   \dd\nu_c=\operatorname{Tr}(cc^\dagger\dd M)
   \longmapsto
   \{R_n^{[c]},\tphi_n^{[c]},K_c(t)\}.
   \label{eq:matrix-parent-pipeline-main}
\end{equation}
Here, a unit representative of $[c]$ is understood.  
The final arrow uses the finite-band connector of
section~\ref{sec:finiteband} and the projected-kernel construction of
section~\ref{sec:kernels}, applied after the scalar seed direction has
been chosen. 
A common parent measure therefore organizes the family before the seed
direction is chosen, but it does not produce a linear family of scalar
Krylov bases.
In finite support, each member terminates at its own compressed cyclic
dimension, determined by the surviving atoms described above.

\paragraph{Where a seed subspace comes from.}
A common source of a finite seed family is a symmetry-resolved preparation subspace.  If a
self-adjoint charge $\mathcal Q$ strongly commutes with $H$, every charge sector reduces $H$.  A
finite-dimensional subspace $\mathcal S$ inside one sector need not itself be $H$-invariant or form
a degenerate multiplet.  It nevertheless defines the parent measure
$\dd M=V_0^\dagger\Pi(\dd E)V_0$, which retains the energy weights and interference data available
within $\mathcal S$.  A genuine symmetry multiplet requires an additional symmetry commuting with
$H$.  This construction is compatible with the charge-sector decompositions studied in
symmetry-resolved spread complexity and block-diagonal operator growth
\cite{CaputaDiGiulioLoc2025,CaputaDiGiulioLoc2025b}.  Matrix orthogonal polynomials provide the
coupled recurrence when the full matrix measure, rather than one scalar compression, is evolved
\cite{DuranVanAssche1995}.

A finite superposition of number, localized-site or chord states that lies in a fixed
reference Krylov window selects one scalar polynomial direction and is therefore covered by the
connector already constructed.  Matrix-valued dynamics is required only when the coupled
evolution of the whole seed subspace, rather than one chosen direction within it, is the physical
observable.

Whether the change is a single polynomial seed or a direction selected from a multiplet, the
next section asks which large-index data, and hence which candidate physical diagnostics, survive
the change of preparation.
 
\section{What polynomial preparations preserve at large Krylov depth}
\label{sec:rigidity}

The connector calculus shows that a polynomial seed can rearrange the Lanczos chain at every
finite level, while its action deep in the chain is more rigid.  At fixed degree, three distinct
statements answer which data forget the preparation and which retain it.  An asymptotically constant Jacobi chain  
returns to the same asymptotic Jacobi bulk, a linear Jacobi bound gives a degree-resolved
finite-time envelope on unbounded chains, and deleted Charlier atoms survive as a quantized
subleading memory.  We then take a separate controlled limit in which the seed itself moves deep
into such an asymptotically constant chain.  That growing-degree regime folds the two propagation directions about
the launch point and must not be conflated with fixed-degree rigidity.  The final conditioned-kernel
paragraphs give an auxiliary spectral-ensemble interpretation of the same finite-rank response.

\paragraph{The asymptotic Jacobi chain.}
\label{subsec:jacobi-limit-rigidity}

The cleanest statement is for bounded spectral support. The measure $\mu$ belongs to the Nevai class $M(a,b)$ \cite{Nevai1979,SimonOPRL1} if its Jacobi parameters converge, $a_n\to a$ and $b_n\to b>0$, which physically says that the Lanczos chain becomes asymptotically uniform far from the seed. Its essential spectrum is the band $[a-2b,a+2b]$, with at most countably many isolated spectral points outside it. 
 The half-line tight-binding chain of subsection~\ref{subsec:tight-binding} is exactly the 
constant representative of $M(a,b)$; the proposition below identifies which part of that
asymptotic bulk survives a more general polynomial preparation.
At every fixed $0\le q<1$, the normalized continuous-$q$-Hermite chain is the bounded
example $M\bigl(0,(1-q)^{-1/2}\bigr)$.  Its $q=0$ endpoint is precisely the constant half-line
chain, while the fixed-level $q\uparrow1$ Hermite limit leaves this compact Nevai regime.

\begin{proposition}[%
All-root rigidity of the Jacobi limit]
\label{prop:nevai-rigidity}
Let $\mu\in M(a,b)$ and let $Q\in\CC[E]$ be a nonzero polynomial of fixed degree.  Then the
normalized seed measure
\begin{equation}
   \dd\nu_Q(E)=\frac{|Q(E)|^2}{N_Q}\dd\mu(E),
   \qquad
   N_Q=\int |Q(E)|^2\dd\mu(E),
\end{equation}
also belongs to $M(a,b)$.  Equivalently,
\begin{equation}
   \widetilde a_n^{[Q]}\longrightarrow a,
   \qquad
   \widetilde b_n^{[Q]}\longrightarrow b \,.
   \label{eq:nevai-rigidity}
\end{equation}
No restriction is imposed on the real or complex roots of $Q$.  Real roots may lie in the
limiting band, at visible atoms, or outside the band; nonreal roots are also allowed. The multiplier is a modulus square, so every real root of $Q$ enters $|Q|^2$ to even order and
$\nu_Q$ is again a positive measure; the statement therefore concerns doubled real factors, and
not a single real linear Christoffel factor, which need not leave a positive measure at all.
\end{proposition}

\begin{proof}

It is enough to treat a real double-Christoffel factor and a nonreal conjugate pair.  For a
real root in the closed limiting band, the Nevai condition implies
$p_n(\xi)^2/\mathcal K_n(\xi)\to0$ and
$p_{n+1}(\xi)^2/\mathcal K_n(\xi)\to0$ \cite{BreuerLastSimon2009}.  The exact one-root formulas of
appendix~\ref{app:christoffel} then return the limits $(a,b)$.  At an atom,
$\mathcal K_n(\xi)\to\mu(\{\xi\})^{-1}$ and $p_n(\xi)\to0$; at an exterior non-atom, the ratio
asymptotics used above apply.  Nonreal roots are handled by the corresponding Hermitian kernel.
Iteration over the factors of $|Q|^2$ completes the proof.  The four cases and the kernel
estimates are given in appendix~\ref{app:all-root-nevai-proof}.\qedhere

\end{proof}

Proposition~\ref{prop:nevai-rigidity} is a fixed-degree statement and gives no universal
rate.  A Nevai tail can approach its constant limit arbitrarily slowly, and polynomial
preparation inherits that slow scale together with the relevant kernel asymptotics.  The
Denisov--Rakhmanov theorem remains a useful independent measure-side route when $\sigma_{\rm ess}(\Jac)=[a-2b,a+2b]$ and the absolutely
continuous density is positive almost everywhere on that band
\cite{Rakhmanov1982,Denisov2004}.  It reaches the same conclusion because multiplication by
$|Q|^2$ preserves positivity almost everywhere on the band, since the multiplier has only finitely many real zeros.  The kernel proof above is stronger in
spectral scope, since it requires only membership in $M(a,b)$ and therefore also covers singular
Nevai measures.  This conjunction is sufficient for Nevai asymptotics but is stronger than Nevai membership; positivity of the density alone cannot replace it without control of the essential spectrum \cite{Remling2011}. 
Bailey and Derevyagin prove Nevai invariance for a single Christoffel transform at a nonreal point and construct the two-step transform at points in opposite half-planes, including a conjugate pair~\cite{BaileyDerevyagin2021}. The proposition above instead gives a direct positive-measure proof for the full multiplier $|Q|^2$. It uses the Hermitian kernel without introducing a complex Jacobi matrix as an intermediate object, allows arbitrary fixed degree, and covers real roots in the limiting band and at atoms.   On unbounded
support there need not be a finite Jacobi limit.  The Charlier
analysis instead proves invariance of the leading linear growth.

\paragraph{A degree-resolved finite-time growth ceiling.}
\label{subsec:exponent-rigidity}

For an unbounded chain, a linear Jacobi bound gives a causal envelope rather than a finite
Jacobi limit.  Proposition~\ref{prop:finite-time-control-main} transports that envelope to every
fixed-degree preparation.
For $A>0$ and a seed of degree $r$, the shifted hopping obeys
$\widetilde b_{n+1}\le A(n+r)+C=An+(Ar+C)$, so the same proposition with shifted intercept
$Ar+C$ gives
\begin{equation}
   K_Q(t)\ \le\ \Bigl(r+\frac{C}{A}\Bigr)^{2}\sinh^{2}\bigl(A|t|\bigr).
   \label{eq:exponent-rigidity}
\end{equation}
Thus every degree-$r$ seed is controlled by the same admissible $\sinh^2(A|t|)$ envelope, with
seed dependence entering this estimate only through the prefactor $(r+C/A)^2$.  Neither $A$ nor
$2A$ is asserted to be an attained dynamical growth rate.  Equivalently,
$\limsup_{|t|\to\infty}|t|^{-1}\log[1+K_Q(t)]\le2A$ for every fixed $r$.
When $A=0$, the same proposition instead gives
\begin{equation}
   K_Q(t)\le \min\{C^2t^2,\,2C|t|\}.
   \label{eq:zero-slope-finite-time-ceiling}
\end{equation}

At fixed degree $r$, the bound is uniform over the coefficients and root positions of $Q$.
The degree enters only through the prefactor.  Whether a particular dynamics approaches the
envelope remains model dependent because the reference linear bound need not be sharp. Equality nevertheless occurs in the dynamical step itself. The exactly solvable operator-growth chain with $a_n\equiv0$ and $b_n=An$ \cite{Parker2019} meets $b_{n+1}\le An+C$ with $C=A$ and has $K(t)=\sinh^{2}(A|t|)$, so the $A>0$ line of eq.~\eqref{eq:complexity-linear-growth-bound-main} is attained there. The same chain satisfies eq.~\eqref{eq:reference-linear-bound-main} only with the doubled slope $2A$, so at $r=0$ the envelope in eq.~\eqref{eq:exponent-rigidity} exceeds the true growth by $\cosh^{2}(At)$, and that slack lies in the passage to the reference hypothesis rather than in the dynamical estimate.  In the
Charlier and half-line tight-binding families the shifted Jacobi data are known explicitly in
eqs.~\eqref{eq:H1-Jacobi-asymptotics-main} and
\eqref{eq:tight-binding-halfline-hopping-pattern-app}, so their preparation dependence can instead
be compared coefficient by coefficient. 

The fixed-$q$ continuous-$q$-Hermite chain is bounded and therefore obeys the $A=0$ form
with $C=2(1-q)^{-1/2}$ in the reference hypothesis
\eqref{eq:reference-linear-bound-main}, while its exact fixed-chord coefficients follow from
corollary~\ref{cor:qhermite-root-free-app}.  The Hermite endpoint has unbounded
$b_n=\sqrt n$ and lies outside the bounded case; proposition~\ref{prop:Hermite-all-r-complexity-app}
instead provides a direct convergent complexity series at every fixed Fock level.

Every rigidity statement in this section takes the
large-index limit at fixed filter degree.  If the degree grows with system size or Krylov depth,
the exact finite-level transfer remains valid but the estimates are not uniform.  That double-scaling regime is precisely where preparation may enter the leading asymptotics.

As a statement about speed, this completes a picture begun in
section~\ref{sec:finiteband}. 
A polynomial preparation sets the initial quadratic coefficient.  Under the differentiability
assumptions of eq.~\eqref{eq:short-time-shifted-complexity-main},
$K_Q(t)=\Delta_Q^2t^2+o(t^2)$ with $\Delta_Q^2=\thh_1^{[Q]}$; in finite support, or under the
stronger graph-norm regularity stated there, the remainder improves to $O(t^4)$.
Equation~\eqref{eq:MT-speed-shifted-seed-main} transports the same coefficient from the reference data.
At longer times the bound provides only a degree-dependent ceiling, whereas the exact
behavior remains dynamical and model dependent. This is also why
eq.~\eqref{eq:K-speed-transfer-main} is sharper here than a bound on $|\dot K_Q|$.  At fixed $H$,
the transfer identity determines the instantaneous response from reference data, while the speed
limit remains an inequality. 

\paragraph{A controlled deep-seed limit.}
Proposition~\ref{prop:nevai-rigidity} holds at fixed seed degree.  A different and equally
controlled limit is obtained by launching from the reference level
$\ket{K_r}=p_r(H)\ket{K_0}$ and then sending $r$ deep into a Jacobi chain in $M(a,b)$.  The induced measures converge weakly
to the arcsine measure \cite{VanAssche1995}.  Appendix
\ref{app:controlled-growing-degree} translates it into the shifted Lanczos problem.  For every
fixed shifted level $\ell$,
\begin{equation}
\begin{gathered}
   p_r(E)^2\dd\mu(E)\Longrightarrow
   \dd\rho_{\rm arc}^{a,b}(E)
   =\frac{\boldsymbol 1_{\{|E-a|<2b\}}\dd E}
          {\pi\sqrt{4b^2-(E-a)^2}}\,,\\[2pt]
   \widetilde a_\ell^{(r)}\longrightarrow a\,,\qquad
   \widetilde b_1^{(r)}\longrightarrow\sqrt2\,b\,,\qquad
   \widetilde b_\ell^{(r)}\longrightarrow b\,\quad(\ell\ge2)\,.
\end{gathered}
   \label{eq:deep-seed-folded-bulk-main}
\end{equation}
With $z=2bt$, the limiting shifted amplitudes and complexity are
\begin{equation}
\begin{gathered}
   \phi_0^{\rm arc}(t)=\ee^{-\ii at}J_0(z),\qquad
   \phi_n^{\rm arc}(t)=\sqrt2\,\ee^{-\ii at}(-\ii)^nJ_n(z),\\[2pt]
   \begin{aligned}
   K_{\rm arc}(t)
   =2\sum_{n\ge1}nJ_n(z)^2
   =z^2\bigl[J_0(z)^2+J_1(z)^2\bigr]-zJ_0(z)J_1(z).
   \end{aligned}
\end{gathered}
   \label{eq:deep-seed-folded-dynamics-main}
\end{equation}
The enhanced first hopping is the fold of the two propagation directions seen by a launch point
deep inside an asymptotically uniform half-line chain.  Re-Lanczos measures the distance traveled
away from that launch point.  The convergence is uniform on compact time intervals, but it is not
uniform when shifted depth grows with $r$.  The exact Chebyshev connectors exhibit moving
$O(1)$ defects in that joint scaling regime. 

The same folded limit applies to the fixed-$q$ continuous-$q$-Hermite chain and gives the
DSSYK realization developed in subsection~\ref{subsec:chord-context}; at $q=0$ it reduces to the
constant-chain fold.  The Charlier and Hermite chains have unbounded Jacobi coefficients and are
not covered by this compact-support proposition.  Their large-seed limits would require separate
scalings and are not inferred here.

\paragraph{Kernel saturation and atom deletion.}
\label{subsec:kernel-mass-dichotomy}

If the leading data are invariant, the seed dependence must live in the subleading data, and the
difference form \eqref{eq:kernel-difference-form-app} says exactly where it lives, in the growth of the
Christoffel--Darboux kernel at the seed roots. For a real
seed root $\xi\in\RR$, the relevant diagnostic is the total kernel mass
\begin{equation}
   \mathcal K_\infty(\xi)=\lim_{n\to\infty}\mathcal K_n(\xi)\in(0,\infty],
   \quad
      \mathcal K_\infty(\xi)=\frac{1}{\mu(\{\xi\})}\ \ \text{when }\mu\text{ is determinate and }\mu(\{\xi\})>0.
   \label{eq:kernel-mass-dichotomy}
\end{equation}

For the determinate measures considered here, finiteness of
$\mathcal K_\infty(\xi)$ is equivalent to $\xi$ being a mass point, and the second identity gives
its mass \cite{SimonCDK}. This leads to two cases:
\begin{itemize}[leftmargin=1.6em,itemsep=1pt]
\item \emph{$\mathcal K_\infty(\xi)=\infty$.} The evaluation direction does not saturate. The
multiplier reweights the chain; further asymptotic hypotheses are required to determine how the
subleading Jacobi data depend on the root position.
\item \emph{$\mathcal K_\infty(\xi)<\infty$.} The root sits on an atom of mass
$1/\mathcal K_\infty(\xi)$. The multiplier annihilates that atom and the corresponding direction
decouples. In the Charlier chain the resulting intercept offset is exactly one unit per deleted
Poisson atom, as counted by $s_r(\lambda)$ in
eq.~\eqref{eq:H1-Jacobi-asymptotics-main}; this quantization is not asserted for a general measure.
\end{itemize}
Complex seed roots remain covered by the Hermitian-kernel construction, but they have no
atom-deletion interpretation.

The second case is irreversible in a precise sense already established in
subsection~\ref{app:finite-multiplication-gram}.  Equation~\eqref{eq:finite-connector-pseudoinverse-app}
shows that the Moore--Penrose inverse of the transport reconstructs the surviving spectral sector
exactly and annihilates the deleted one.  A preparation that places a seed zero on an atom
therefore destroys spectral weight that no inverse built only from the shifted data can recover.
The Charlier tail nevertheless retains a coarse diagnostic of this loss.
Equation~\eqref{eq:H1-intercept-budget-main} shows that its two intercepts recover the preparation
degree $r$ and the number $s_r(\lambda)$ of deleted atoms.  This does not invert the deleted
measure; it recovers only these two discrete data.

\paragraph{Consequences for large-index diagnostics.}
\label{subsec:chaos-diagnostic}

Large-index Lanczos data enter several proposed diagnostics.  In operator Krylov space, the
growth of $b_n$ probes operator growth and the tail of the operator spectral function
\cite{Parker2019}.  In the state chains studied here, $b_n$ is instead recurrence data of the
cyclic pair $(H,\ket{K_0})$; an operator-growth interpretation requires a separate operator-Krylov
realization.  Recent
orthogonal-polynomial analyses in random-matrix settings likewise extract
spectral information from recursion-coefficient asymptotics \cite{Qu2025LanczosOP}.

The present results isolate a preparation-rigid part of those asymptotics.  Proposition~\ref{prop:nevai-rigidity} shows that
every fixed polynomial preparation preserves the limiting bulk of a Nevai chain, irrespective of
where its roots lie.  The convergence rate and the finite-$n$ corrections remain seed and measure
dependent.  A root on an atom can therefore leave a subleading imprint even though it cannot move
the Jacobi limit. The Charlier family lies outside every Nevai class, since $\widetilde b_n^{(r)}\sim\sqrt{\lambda n}$ diverges, so proposition~\ref{prop:nevai-rigidity} does not apply to it; the analogous rigidity there is of the two linear slopes, with the quantized intercept of proposition~\ref{prop:H1-intercepts-main} as the surviving imprint.  Preparation sensitivity is therefore organized by
scale.  Connector rows retain the full finite-level response, the limiting or leading data covered
by the stated hypotheses are rigid, and subleading coefficients remain model and seed dependent.
This sharpens the observation that spread complexity depends on the initial state
\cite{SreeramKannanModakAravinda2025,Balasubramanian:2025variations}.  The degree-resolved result
is a kinematic upper envelope; it neither determines an attained asymptotic
exponent nor defines a chaos diagnostic.  Saturation and any chaos interpretation remain
properties of the reference dynamics and of the Krylov construction being used.

\paragraph{Conditioned spectral ensembles.}
\label{subsec:palm}

There is a second way to see why the atomic case is special, which also explains the appearance of
the projected kernel of section~\ref{sec:kernels} and connects the construction to standard
random-matrix technology. Attach to the reference measure its \emph{auxiliary} $n$-point
orthogonal-polynomial ensemble, that is, the determinantal point process on $\RR$ with joint density
proportional to $\Delta(x)^2\prod_i\dd\mu(x_i)$, where $\Delta(x)=\prod_{i<j}(x_j-x_i)$ is the
Vandermonde determinant, and correlation kernel $\mathbb K_{n-1}^\mu$
\cite{Borodin2009DPP}. This ensemble is a device built from $\mu$; its particles are integration
variables and are not the spectrum of $H$, which may have any cardinality, finite or infinite.
If $\xi$ is an admissible point of the process with nonzero intensity, its reduced Palm
measure multiplies the remaining-particle weight by $(E-\xi)^2$ and is again determinantal, with
kernel \cite{Borodin2009DPP,Lazag2019Palm} \begin{equation}
   \mathbb K^{\rm Palm}(x,y)
   =\mathbb K(x,y)
    -\frac{\mathbb K(x,\xi)\mathbb K(\xi,y)}{\mathbb K(\xi,\xi)} \,.
   \label{eq:palm-kernel}
\end{equation}
For the cutoff $\ell$, eq.~\eqref{eq:first-jump-kernel} is therefore the reduced Palm
kernel of the auxiliary $(\ell+2)$-particle ensemble conditioned at one admissible point, after
the explicit multiplication by the Christoffel factor.  With $r$ distinct admissible points, the
rank-$r$ Schur complement has the analogous reduced-Palm interpretation. One starts from the
auxiliary $(\ell+r+1)$-particle ensemble and retains $\ell+1$ particles.  The algebraic projected-kernel identity itself is more general.  A generic zero of $P_r$ need not lie in the state space
of a discrete process --- off-lattice Charlier zeros are the simplest example --- and derivative jets
at repeated roots are confluent limits rather than literal conditioning of a simple point
process.  In those cases the result is a Christoffel deformation, not a probabilistic Palm event.

The same finite-rank identity passes directly to any controlled local kernel limit.  Suppose
that, under a scaling $x=x_0+u/\kappa_N$, the rescaled reference kernels converge locally uniformly
to $\mathbb K_*(u,v)$ and that the distinct conditioning nodes scale as
$\xi_{a,N}=x_0+s_a/\kappa_N$.  If the limiting Gram matrix
$G_*=[\mathbb K_*(s_a,s_b)]_{a,b=1}^{r}$ is nonsingular, continuity of finite-dimensional inversion
gives
\begin{equation}
   \mathbb K_*^{\rm Palm}(u,v)
   =\mathbb K_*(u,v)
    -\mathbf k_*(u)^\dagger G_*^{-1}\mathbf k_*(v),
   \qquad
   \bigl(\mathbf k_*(u)\bigr)_a=\mathbb K_*(s_a,u).
   \label{eq:palm-kernel-scaling-limit}
\end{equation}

Thus sine, Airy or other local limits, once proved for the reference cyclic measures, induce
their corresponding reduced-Palm seed-response limits without a second asymptotic calculation.
This is a universality statement for the auxiliary conditioned ensemble.  Its use as a physical
spectral diagnostic still requires a model-specific identification of the cyclic measure and its
scaling regime.

To translate this kernel limit into large-$n$ Jacobi corrections, one needs asymptotics strong
enough to control discrete differences. One sufficient hypothesis at a fixed admissible point $\xi$ is
\begin{equation}
   \log\mathcal K_n(\xi)=\log[c(\xi)n]+\frac{c_1(\xi)}{n}+r_n(\xi),
   \qquad
   \nabla^j r_n(\xi)=O(n^{-2-j}),\quad j=0,1,2,
   \label{eq:palm-difference-controlled-hypothesis}
\end{equation}
together with $p_n(\xi)=O(1)$ and $b_n=O(1)$.  It then follows that
$\nabla^2\log\mathcal K_n=O(n^{-2})$.  Defining the one-root endpoint contraction
\begin{equation}
   \eta_n(\xi)=
   b_n\frac{p_n(\xi)p_{n-1}(\xi)}{\mathcal K_{n-1}(\xi)},
   \label{eq:one-root-eta-main}
\end{equation}
one likewise obtains $\eta_n(\xi)=O(n^{-1})$ under these assumptions.
These stronger hypotheses carry the universal one-root kernel response into the shifted
Jacobi data.  They control the discrete hopping and diagonal corrections, not only the limiting
conditioned kernel.
The finite-rank Palm response therefore passes to a universal kernel limit.  Jacobi
corrections require the stronger discrete-difference data above, while a physical diagnostic also
requires a model-specific identification of the cyclic measure with that ensemble regime.

Fixed-degree preparations therefore preserve the Nevai bulk, deep reference seeds see its
folded two-sided limit, and subleading corrections retain local information about roots and
deleted atoms.  The next section applies this hierarchy to the continuous-$q$-Hermite chord chain
and to related spectral-response settings.

\section{%
Preparation transport in chord dynamics and spectral response}
\label{sec:context}

The oscillator, Hermite and lattice examples above already realize polynomial preparation
in physical systems at fixed Hamiltonian.  We now carry the same two-basis problem into chord
dynamics and then use the common spectral transformation to organize several response settings.
For a specified seed, the connector and projected kernel determine the prepared Lanczos data,
amplitudes and spread complexity.

In double-scaled SYK, chord number organizes the transfer Hilbert space and, in controlled
holographic regimes, is related to a bulk-length coordinate by a model-dependent normalization.
Empty-chord evolution is the solvable reference problem.  Taking a fixed nonzero chord level as
the seed instead asks for the Lanczos basis generated by that level and for the relation between
its rebuilt spread and motion in the original chord basis.  The subsequent realizations extract
response quantities from the same spectral transformation, subject to the hypotheses stated in
each case.

\subsection{Chord dynamics and Krylov-basis response}
\label{subsec:chord-context}
\label{subsec:information-context}

The double-scaled SYK transfer operator acts in the conventional chord-number basis as
\cite{Berkooz2019DSSYK}
\begin{equation}
   T\ket n=\ket{n+1}+[n]_q\ket{n-1},
   \qquad
   [n]_q=\frac{1-q^n}{1-q},
   \qquad 0\le q<1\,.
   \label{eq:chord-transfer}
\end{equation}
Here and throughout this subsection, $q$ is the DSSYK deformation parameter and is unrelated
to the earlier Charlier shorthand $q(\tau)$.
After the normalization $\ket n_{\Jac}=\ket n/\sqrt{[n]_q!}$, this is the continuous
$q$-Hermite Jacobi matrix, denoted by $\Jac_q$, with $a_n=0$ and $b_n^2=[n]_q$.
Berkooz et al. obtained its continuous-$q$-Hermite spectral resolution and arbitrary
fixed-chord matrix elements \cite{Berkooz2019DSSYK}.  For the zero-chord seed, the normalized
fixed chord-number states form the corresponding Krylov basis
\cite{RabinoviciEtAl2023BulkKrylov}.  We denote its spectral measure at
$\ket0_{\Jac}$ by $\mu_q$.  The mean level is therefore the mean chord
number, which becomes a bulk-length variable only after the model- and regime-dependent
normalization of the holographic dictionary \cite{Lin2022BulkHilbert}.

Zero-particle chord dynamics and, separately, matter-sector propagation have been developed
in ref.~\cite{Xu2025ChordDynamics}.  At finite $N$, the auxiliary chord chain agrees only with an
initial segment of the physical Krylov chain before finite-dimensional effects become important
\cite{BalasubramanianMaganNandiWu2026}.  We use the strict double-scaled, zero-particle chord
propagation as the reference dynamics.  The problem addressed here is instead to re-Lanczos a
fixed nonzero chord state and to determine its new Jacobi chain and amplitudes at every rebuilt
level.

A prepared state brings in two distinct bases.  Let $\boldsymbol\phi(t)$ be the empty-chord
amplitude vector.  A polynomial seed has the exact physical chord amplitudes
\begin{equation}
\begin{aligned}
   \boldsymbol g^{[Q]}(t)
   &=\frac{Q(\Jac_q)}{\sqrt{N_Q}}\boldsymbol\phi(t),\\
   \widetilde{\boldsymbol\phi}^{[Q]}(t)
   &=\mathsf U_Q^\dagger\boldsymbol g^{[Q]}(t),
   \qquad
   (\mathsf U_Q)_{mn}={}_{\Jac}\!\braket{m}{\widetilde K_n^{[Q]}}.
\end{aligned}
   \label{eq:chord-two-basis-map}
\end{equation}
The first line evolves the preparation in the fixed chord basis.  The second expresses the same
state in the Lanczos basis rebuilt from that preparation.  Accordingly,
\begin{equation}
   \overline n_Q(t)=\sum_{m\ge0}m|g_m^{[Q]}(t)|^2,
   \qquad
   K_Q(t)=\sum_{n\ge0}n|\widetilde\phi_n^{[Q]}(t)|^2
   \label{eq:prepared-chord-mean}
\end{equation}
are different observables.  Proposition~\ref{prop:filtration-lower-bound} gives
the general estimate
$K_Q(t)\ge\overline n_Q(t)-r$ for a seed of degree $r$.

For a general degree-$r$ filter, the reference mean need not begin at $r$, so this estimate is not
in general a displacement bound.  It becomes more informative for the fixed-chord preparation
$Q=p_r$, for which $p_r(\Jac_q)\ket0_{\Jac}=\ket r_{\Jac}$.
Propagation from this fixed chord is already determined by the reference transfer operator,
and the same squared continuous-$q$-Hermite weight has appeared in fixed-length entanglement
calculations \cite{Tang2024FixedChordEntropy}.  Our distinct object is the scalar Lanczos chain
for the cyclic pair $(\Jac_q,\ket r_{\Jac})$, equivalently the orthogonal-polynomial problem for
$p_r^2\dd\mu_q$, together with its comparison to the known chord-basis propagation.
Write 
\begin{equation}
\begin{gathered}
   \ket{\psi_r(t)}=\ee^{-\ii t\Jac_q}\ket r_{\Jac},
   \qquad
   g_{m|r}^{\rm ch}(t)\equiv g_m^{[p_r]}(t)
      ={}_{\Jac}\!\braket{m}{\psi_r(t)},\\
   D_r^{\rm ch}(t)=\sum_{m\ge0}|m-r|\,
      |g_{m|r}^{\rm ch}(t)|^2 .
\end{gathered}
   \label{eq:chord-radial-distance}
\end{equation}

Thus $D_r^{\rm ch}$ measures the mean absolute displacement in the normalized
chord-number basis, with the prepared chord level as its origin.  We call this displacement radial
because the two chord-number directions $r-d$ and $r+d$ are folded onto the same distance $d$.  It
is distinct both from the signed change $\overline n_r-r$ and from spread in the rebuilt Krylov
basis.

\begin{proposition}[Radial filtration for a fixed chord]
\label{prop:chord-radial-filtration}
For every $r\ge0$, $0\le q<1$ and finite $t$, let
\begin{equation}
   B_{r,\ell}=\sum_{\substack{m\ge0\\ |m-r|\le\ell}}
      \ket m_{\Jac}{}_{\Jac}\!\bra m,
   \qquad
   \widetilde\Pi_{\ell}^{(r)}=
      \sum_{n=0}^{\ell}\ket{\widetilde K_n^{(r)}}
      \bra{\widetilde K_n^{(r)}}.
   \label{eq:chord-radial-projectors}
\end{equation}
Then
\begin{equation}
\begin{aligned}
   K_r(t)-D_r^{\rm ch}(t)
   &=\sum_{\ell\ge0}\bra{\psi_r(t)}
     \bigl(B_{r,\ell}-\widetilde\Pi_{\ell}^{(r)}\bigr)
     \ket{\psi_r(t)}\ge0\,,\\
   K_r(t)&\ge D_r^{\rm ch}(t)\ge
      \bigl|\overline n_r(t)-r\bigr|.
   \label{eq:chord-radial-filtration}
\end{aligned}
\end{equation}
\end{proposition}

\begin{proof}
Tridiagonality implies that the first $\ell+1$ vectors generated from
$\ket r_{\Jac}$ have physical chord support only at distances at most $\ell$.
Consequently,
$\operatorname{Ran}\widetilde\Pi_{\ell}^{(r)}\subseteq
\operatorname{Ran}B_{r,\ell}$.  The two orthogonal projectors are nested, so
$B_{r,\ell}-\widetilde\Pi_{\ell}^{(r)}$ is itself an orthogonal projector.
For a nonnegative integer-valued random variable, its first moment is the sum of its
tail probabilities.  Applied in the two bases, this identity gives
\begin{equation}
\begin{aligned}
 K_r(t)&=\sum_{\ell\ge0}
 \left[1-\bra{\psi_r(t)}\widetilde\Pi_{\ell}^{(r)}\ket{\psi_r(t)}\right],\\
 D_r^{\rm ch}(t)&=\sum_{\ell\ge0}
 \left[1-\bra{\psi_r(t)}B_{r,\ell}\ket{\psi_r(t)}\right],
\end{aligned}
\end{equation}
where these sums converge.  Indeed,
$\|\Jac_q\|\le R_q$, where $R_q=2/\sqrt{1-q}$, and a matrix element between chord
levels separated by $d$ vanishes in every power $\Jac_q^k$ with $k<d$.
Hence, for $|t|\le T$,
\begin{equation}
 \left|{}_{\Jac}\!\bra m\ee^{-\ii t\Jac_q}\ket r_{\Jac}\right|
 \le \sum_{k=|m-r|}^{\infty}\frac{(R_qT)^k}{k!}\,.
\end{equation}
The same estimate holds in the rebuilt Jacobi representation, whose spectral support is
contained in that of $\mu_q$.  The squares of these bounds are summable with the respective
distance or level weights.  Subtracting the two tail sums proves the first line of
eq.~\eqref{eq:chord-radial-filtration}.  Finally,
$D_r^{\rm ch}(t)\ge|\overline n_r(t)-r|$ is the triangle inequality for the
physical chord-number distribution.
\end{proof}

The three observables separate already at short time.  Direct path expansion gives
\begin{equation}
\begin{aligned}
   \overline n_r(t)-r
   &=q^rt^2+O(t^4),\\
   D_r^{\rm ch}(t)
   &=K_r(t)+O(t^6)
    =\bigl([r]_q+[r+1]_q\bigr)t^2+O(t^4),\\
   K_r(t)-D_r^{\rm ch}(t)
   &=\frac{[r]_q[r+1]_q\bigl([r+2]_q-[r-1]_q\bigr)^2}
      {18\bigl([r]_q+[r+1]_q\bigr)}t^6+O(t^8),
   \qquad r\ge1\,.
   \label{eq:chord-two-basis-short-time}
\end{aligned}
\end{equation}
For $0<q<1$ the last coefficient is strictly positive and contains the depth factor
$q^{2r-2}$ through
$[r+2]_q-[r-1]_q=q^{r-1}(1+q+q^2)$.  At $q=0$ it remains nonzero for $r=1$ but vanishes at
this order for $r\ge2$.  Thus, for $0<q<1$, radial chord motion and rebuilt spread first separate
at order $t^6$; at $q=0$ and $r\ge2$ their separation
has no $t^6$ term and can first appear at higher order.

At $q=0$, we have $[n]_0=1$ for $n\ge1$, so $\Jac_0$ is exactly the
unit-hopping half-line chain of subsection~\ref{subsec:tight-binding}.  The boundary is then the
only source of a difference between physical mean absolute displacement and rebuilt spread.  In
fact, for every $r\ge1$,
\begin{equation}
   K_r(t)-D_r^{\rm ch}(t)
   =\frac{t^{2r+4}}{[(r+2)!]^2}+O(t^{2r+6})
   \qquad(q=0).
   \label{eq:qzero-boundary-defect}
\end{equation}
To obtain the leading coefficient, let
$\ket{A_r}=(\ket0_{\Jac}-\ket{2r}_{\Jac})/\sqrt2$.
No walk of length below $r+2$ distinguishes the two sides of the seed, whereas at length $r+2$
precisely one two-sided walk would cross the missing bond at the Dirichlet boundary.  Hence
${}_{\Jac}\!\bra{A_r}\Jac_0^{r+2}\ket r_{\Jac}=-1/\sqrt2$.
In the tail-projector identity of proposition~\ref{prop:chord-radial-filtration}, this leading
antisymmetric component occurs at the two cutoffs $\ell=r$ and $\ell=r+1$.  Each contributes
$t^{2r+4}/\bigl(2[(r+2)!]^2\bigr)$, while all other cutoffs begin at higher order.  This proves
eq.~\eqref{eq:qzero-boundary-defect}.  The $t^6$ defect therefore survives only for $r=1$; a seed
at depth $r\ge2$ postpones the first mismatch to the boundary-sensitive order $t^{2r+4}$.
The signed mean
chord number is already almost stationary for a deep seed even though both distributions spread.
As $q\uparrow1$, the coefficient of $t^6$ tends to
$r(r+1)/[2(2r+1)]$, exactly reproducing the Hermite radial-defect coefficient in
eq.~\eqref{eq:Hermite-all-r-short-time-app}. 

The hierarchy has a direct physical meaning.  At fixed $0<q<1$, the signed drift is
suppressed by $q^r$ as the seed moves into the chord bulk, whereas the common $O(t^2)$ coefficient
of $D_r^{\rm ch}$ and $K_r$ approaches $2/(1-q)$.  A deep seed can therefore spread appreciably in
both chord-number directions while its mean chord number barely changes.  The $t^6$ term is the
first local contribution that distinguishes rebuilt Krylov spread from mean absolute chord
displacement, and its $q^{2r-2}$ suppression anticipates the fixed-time radial fold proved below.
The $q\uparrow1$ value shows that this local basis defect contracts to the Hermite/Fock result at
fixed $r$; it is not a statement about the separate deep-seed limit.

The continuous-$q$-Hermite product formula expresses every
$g_{m|r}^{\rm ch}$ as a finite combination of
empty-chord amplitudes.  Rebuilding the basis is a separate problem.
Corollary~\ref{cor:qhermite-root-free-app} specializes the root-free remainder--Gram construction
to $p_r(E)^2\dd\mu_q(E)$ and determines every shifted hopping, Krylov vector, amplitude and
cumulative probability from fixed $r\times r$ data, for every fixed finite $r$ and every rebuilt
level.

The two endpoint limits of the coefficients are
\begin{equation}
   q\to0:\quad b_n^2\to1\quad(n\ge1),
   \qquad
   q\to1:\quad b_n^2\to n\quad\text{at fixed }n\,.
   \label{eq:chord-limits}
\end{equation}
The $q\to0$ endpoint is the infinite half-line constant chain of
subsection~\ref{subsec:tight-binding}.  As $q\uparrow1$, $b_n^2\to n$ at each fixed reference level
$n$.  Consequently, at fixed seed depth $r$ and fixed rebuilt level, the fixed-chord
reconstruction converges coefficientwise to the Hermite/Fock construction of
subsection~\ref{subsec:HW} and proposition~\ref{prop:Hermite-all-r-complexity-app}, with
$t=\tau_{\rm H}$.  This convergence is not uniform in reference or rebuilt level and does not by
itself permit the $q\uparrow1$ limit to pass through the infinite first-moment sum.  The deep-seed
limit below has a different order of limits.  It keeps $q<1$ and $t$ fixed before sending
$r\to\infty$.  Since $b_\infty=(1-q)^{-1/2}$ diverges as $q\uparrow1$, no interchange or joint
$q\uparrow1$, $r\to\infty$ limit is asserted here.

At every fixed $q<1$, one has $a_n=0$ and
$b_n\to(1-q)^{-1/2}$.  Hence $\mu_q$ belongs to the Nevai class
$M\bigl(0,(1-q)^{-1/2}\bigr)$ defined in
subsection~\ref{subsec:jacobi-limit-rigidity}.
  Proposition~\ref{prop:nevai-rigidity} therefore gives, for every fixed
polynomial chord preparation,
\begin{equation}
   \widetilde a_n\longrightarrow0,
   \qquad
   \widetilde b_n^2\longrightarrow\frac{1}{1-q}\,.
   \label{eq:chord-preparation-rigidity}
\end{equation}
The deep-seed limit of section~\ref{sec:rigidity} sharpens this statement.  Put
$b_\infty=(1-q)^{-1/2}$ and $z_q=2b_\infty t$.

The local path argument in the proof of
proposition~\ref{prop:deep-seed-folded-bulk-app} also applies before re-Lanczos reconstruction.
At fixed displacement $k$, every finite term in the exponential series sees only finitely many
Jacobi coefficients around the seed and therefore converges to the bilateral constant-chain
kernel.  The common factorial path bound is summable with weight $|k|$, uniformly on compact time
intervals, so the relative signed and absolute first moments are uniformly integrable.  This
proves the chord-basis limits stated below.
For fixed $q<1$, fixed time and each fixed $k\in\mathbb Z$,
\begin{equation}
\begin{gathered}
   \left|{}_{\Jac}\!\bra{r+k}\ee^{-\ii t\Jac_q}\ket r_{\Jac}\right|^2
   \longrightarrow J_k(z_q)^2,\\
   \overline n_r(t)-r\longrightarrow0\,,\\
   |\widetilde\phi_0^{(r)}(t)|^2\longrightarrow J_0(z_q)^2,\qquad
   |\widetilde\phi_n^{(r)}(t)|^2\longrightarrow2J_n(z_q)^2\quad(n\ge1),\\
   K_r(t)\longrightarrow2\sum_{n\ge1}nJ_n(z_q)^2=K_{\rm arc}(t).
\end{gathered}
   \label{eq:deep-chord-folding}
\end{equation}

The radial comparison becomes an equality in this limit, 
$D_r^{\rm ch}(t)\to2\sum_{n\ge1}nJ_n(z_q)^2=K_{\rm arc}(t)$ and therefore
$K_r(t)-D_r^{\rm ch}(t)\to0$.
Thus the physical chord distribution approaches a two-sided walk with zero mean signed
displacement, whereas the prepared Krylov chain folds the two directions and measures their mean
absolute displacement.  In length regimes where chord number is geometric, re-Lanczos spread is
the distance traveled from the prepared length, not the absolute mean length.  This is a
fixed-$q$, fixed-time bulk limit of the infinite chord chain.  It does not address the late-time
finite-$N$ saturation or the eventual departure of the auxiliary chord basis from the full
physical Krylov basis emphasized in ref.~\cite{BalasubramanianMaganNandiWu2026}.  Finite-temperature
and loop-corrected relations between Krylov growth and bulk length have been developed in
refs.~\cite{HellerPapaliniSchuhmann2025,AlfinitoBeccaria2026HigherLoop,AlfinitoBeccaria2026Onset}.
Matter and operator sectors instead use specially graded, two-sided or doubled chord spaces
\cite{Okuyama2024Doubled,AmbrosiniEtAl2025,AguilarGutierrez2025}.  Finite-$N$ null states and
state-dependent chord operators are studied in ref.~\cite{MiyajiMoriOkuyama2025}, while
ref.~\cite{FuEtAl2026} develops a broader Krylov-based holographic programme.  These constructions
motivate a block or bivariate extension of the present transport theory.  A generic matter
insertion, however, is not a scalar polynomial filter of the empty-chord chain.

\paragraph{Koherence of a polynomial preparation.}
The overlap matrix $\mathsf U_Q$ in eq.~\eqref{eq:chord-two-basis-map} is the exact interface with
Koherence, the Shannon coherence between perturbed and reference Krylov bases
\cite{Balasubramanian:2025variations}.  For the $n$th shifted vector,
\begin{equation}
   \mathfrak S_n(Q)
   =-\sum_m |(\mathsf U_Q)_{mn}|^2
       \log |(\mathsf U_Q)_{mn}|^2.
   \label{eq:polynomial-koherence}
\end{equation}
The amplitude connector is not this overlap matrix.  In reference coordinates it factorizes as
\begin{equation}
   \mathsf T_Q
   =\mathsf U_Q^\dagger\frac{Q(\Jac_q)}{\sqrt{N_Q}}\,.
   \label{eq:connector-koherence-factorization}
\end{equation}
Koherence measures the response of the ordered basis, while $\mathsf T_Q$ transports the prepared
dynamics.  The coefficient and root-Gram constructions determine both factors from reference
spectral data.  Support loss turns $\mathsf U_Q$ into the corresponding isometry onto the
surviving cyclic subspace.  For a degree-$r$ polynomial seed, its column at level $n$ has at most
$n+r+1$ nonzero entries, so the Koherence of every fixed column is finite.  A nonpolynomial
compact-filter limit would require separate entropy control.

For the fixed-chord seed, the first shifted vector for $r\ge1$ is
\begin{equation}
   \ket{\widetilde K_1^{(r)}}
   =\frac{\sqrt{[r]_q}\ket{r-1}_{\Jac}
          +\sqrt{[r+1]_q}\ket{r+1}_{\Jac}}
          {\sqrt{[r]_q+[r+1]_q}}\,.
   \label{eq:deep-chord-first-koherence}
\end{equation}
Its Koherence is the binary entropy of
$[r]_q/([r]_q+[r+1]_q)$ and tends to $\log2$ as $r\to\infty$.  The limiting bit is precisely the
left--right alternative that re-Lanczos folds into one radial Krylov direction.
For $r=0$, the preparation is unchanged, $\mathsf U_{p_0}=\Id$,
$\ket{\widetilde K_1^{(0)}}=\ket1_{\Jac}$ and $\mathfrak S_1(p_0)=0$.
The first column is the finite-$r$ instance that has the simple closed form above, but the
limit is not confined to it.  Proposition~\ref{prop:deep-seed-folded-bulk-app} and its proof imply
that, at fixed $q<1$, for every fixed $n\ge1$,
\begin{equation}
   \left\|
      \ket{\widetilde K_n^{(r)}}
      -\frac{\ket{r-n}_{\Jac}+\ket{r+n}_{\Jac}}{\sqrt2}
   \right\|\longrightarrow0,
   \qquad
   \mathfrak S_n(p_r)\longrightarrow\log2
   \quad(r\to\infty).
   \label{eq:deep-chord-all-column-koherence}
\end{equation}
For fixed $n$, the relevant column has support on at most the $2n+1$ physical levels
$r-n,\ldots,r+n$.  The norm limit therefore also implies the displayed entropy limit.
This is a fixed-$n$ statement and is not uniform when the shifted level grows with $r$. 

Physically, every fixed rebuilt level probes only a finite neighbourhood of a chord prepared
deep in the bulk.  In that neighbourhood the rebuilt coordinate retains the distance from the
seed but forgets whether the physical amplitude moved left or right; the residual binary choice
has entropy $\log2$ in natural-log units.  Levels that grow with $r$ probe distances comparable with the
seed depth and can resolve the half-line boundary, so the fixed-column conclusion need not survive.

	The first-column statement is not peculiar to chords.  For any reference-level seed
	$Q=p_r$ with $r\ge1$, tridiagonality gives
	\begin{equation}
		\ket{\widetilde K_1^{(r)}}=
		\frac{b_r\ket{r-1}_{\Jac}+b_{r+1}\ket{r+1}_{\Jac}}
		{\sqrt{b_r^2+b_{r+1}^2}}\,,
	\end{equation}
	so its Koherence is the binary entropy of $b_r^2/(b_r^2+b_{r+1}^2)$.
	It equals $\log2$ in the interior of the uniform chain and approaches $\log2$ as $r\to\infty$
	for the Charlier, Hermite and fixed-$q$ continuous-$q$-Hermite chains because neighbouring hoppings become locally
	equal in relative terms.  Near a finite terminal edge, or in a chain without this local symmetry,
	the two weights remain unequal and support loss can reduce the overlap to an isometry.  The
	all-fixed-column limit in eq.~\eqref{eq:deep-chord-all-column-koherence} is stronger than this
	general first-column observation and follows here from the proved Nevai deep-seed limit; we make
	no corresponding all-column claim for the unbounded Charlier or Hermite chains.

The chord problem therefore separates physical transport, re-Lanczos spreading and basis
response within one model.  We now turn to interfaces that apply to a general cyclic measure.

\subsection{Spectral deformations, filtering and transition tomography}
\label{subsec:spectral-response-context}
\label{subsec:toda-context}
\label{subsec:recursion-context}
\label{subsec:determinant-context}
\label{subsec:loschmidt-context}

A positive preparation filter changes the cyclic spectral weight without changing the
Hamiltonian.  The connector controls the resulting real-time amplitudes, derivatives of the
normalizer give static response, and determinant or return-amplitude formulas evaluate other
functionals of the same filtered measure.  Each interface below extracts a different observable
and therefore carries its own hypotheses and physical interpretation.

\paragraph{Toda response geometry.}
The Cholesky interchange following eq.~\eqref{eq:Jacobi-ladder-main} is a shifted $LR$ step and a
discrete B\"acklund move of the Toda hierarchy
\cite{SpiridonovZhedanov1995}. The discrete Darboux--Toda correspondence used here is the result of Spiridonov and Zhedanov; Symes's result instead concerns the QR algorithm and the finite nonperiodic Toda flow \cite{Symes1982}. On an open real parameter domain where the following
normalizer is finite and twice differentiable under the integral, the continuous and polynomial
directions combine into
\begin{equation}
\begin{aligned}
   \dd\mu_{Q,\boldsymbol s}(E)
   &=\frac{\exp(\sum_{j=1}^{m}s_jE^j)|Q(E)|^2}
           {Z_Q(\boldsymbol s)}\dd\mu(E),\\
   Z_Q(\boldsymbol s)
   &=\int\exp\!\left(\sum_{j=1}^{m}s_jE^j\right)|Q(E)|^2\dd\mu(E).
\end{aligned}
   \label{eq:commuting-deformation-lattice}
\end{equation}
The reference Toda deformation is obtained by setting $Q=1$ in eq.~\eqref{eq:commuting-deformation-lattice}.  
At the level of one algebraic Christoffel factor, $1-E/\xi$ represents the formal Miwa shift
$s_k\mapsto s_k-\xi^{-k}/k$.  A physical polynomial seed carries the positive multiplier
$|Q|^2$.  Thus a real seed factor $Q=1-E/\xi$ produces the doubled shift
$s_k\mapsto s_k-2\xi^{-k}/k$, while a nonreal root and its conjugate contribute
$s_k\mapsto s_k-(\xi^{-k}+\bar\xi^{-k})/k$.  These identities are understood as formal series, or
analytically where the corresponding expansions converge.
These shifts refer to the full hierarchy; in the finite-$m$ family displayed above, they are truncated to $1\le k\le m$.

The normalized pure state
\begin{equation}
   \ket{\psi_{Q,\boldsymbol s}}
   =Z_Q(\boldsymbol s)^{-1/2}
     \exp\!\left(\frac12\sum_{j=1}^{m}s_jH^j\right)Q(H)\ket{K_0}
   \label{eq:toda-times-deformation}
\end{equation}
has the exact response geometry
\begin{equation}
\begin{aligned}
   \partial_{s_j}\log Z_Q&=\langle H^j\rangle_{Q,\boldsymbol s},\\
   \partial_{s_j}\partial_{s_k}\log Z_Q
   &=\operatorname{Cov}_{Q,\boldsymbol s}(H^j,H^k)
    =F^{\rm Q}_{jk},
   \qquad
   g^{\rm FS}_{jk}=\frac14F^{\rm Q}_{jk}\,.
\end{aligned}
   \label{eq:toda-response-generator}
\end{equation}
Here $F^{\rm Q}$ is the pure-state quantum Fisher matrix.  A null vector is exactly a polynomial
combination of the visible charges that is
constant
$\mu_{Q,\boldsymbol s}$-almost everywhere.  For the first
Toda time, the same response gives
\begin{equation}
\begin{aligned}
   \partial_{s_1}a_n&=\beta_{n+1}-\beta_n,
      & n&\ge0\,,\\
   \partial_{s_1}\beta_n&=\beta_n(a_n-a_{n-1}),
      & n&\ge1,
      & \beta_0&=0\,.
\end{aligned}
   \label{eq:toda-first-flow-jacobi}
\end{equation}

The Toda and information-geometric identities are exact.  A GGE interpretation is conditional
because the powers $H^j$ need not form an independent, complete, local or quasilocal charge set
\cite{RigolDunjkoYurovskyOlshanii2007}.  Toda deformations of Krylov data and their
temperature-dependent extensions are developed in
refs.~\cite{DymarskyGorsky2020,TakahashiNandyDelCampo2026,AngelinosChakrabortyDymarsky2026}.
Physical time here remains generated by $H$; replacing it by $f(H)$ is a different operation.
The moment determinants and their Christoffel shifts also lie in the tau-function and Grassmannian
geometry of Jacobi transformations \cite{SegalWilson1985,AdlerVanMoerbeke1997,AdlerVanMoerbeke2000}.
This is an auxiliary integrable structure of the spectral problem and does not require the
physical Hamiltonian itself to be integrable.

\paragraph{Local spectral measures and determinant insertions.}
In the recursion method, a local orbital defines precisely the scalar local spectral measure and
Jacobi continued fraction of the cyclic pair \cite{HaydockHeineKelly1972}.
If a target orbital lies in a fixed finite Krylov window, the present transfer gives its local
density of states and continued fraction by finite algebra.  In a cyclic $d$-dimensional cluster,
spectral interpolation extends this statement to every target vector, with a polynomial degree
below $d$ modulo the terminal polynomial.  The qualifier matters when $d$ grows, since a generic
spatial orbital need not have a system-size-independent degree.

The same transform has an exact ensemble realization.  With
$\tau_n[\mu]=\det[\mu_{i+j}]_{i,j=0}^{n-1}$, Heine's formula gives the exact identity below for $\dd\widehat\mu=(E-\xi)\dd\mu $,
\begin{equation}
   \widehat\tau_n=(-1)^nP_n(\xi)\tau_n,
   \qquad n\ge0\,.
   \label{eq:tau-christoffel}
\end{equation}
Indeed, the factor $\prod_i(x_i-\xi)$ in the $n$-point Heine integral is the characteristic
polynomial average, with the displayed sign.  For a general seed,
\begin{equation}
   \frac{\tau_n[|Q|^2\mu]}{\tau_n[\mu]}
   =\left\langle\prod_{i=1}^{n}|Q(x_i)|^2\right\rangle_n
   =\bigl\langle|\det Q(M)|^2\bigr\rangle_n.
   \label{eq:multi-determinant-seed}
\end{equation}
The last form applies when the auxiliary ensemble has a matrix realization; normalization to
$\nu_Q$ adds $N_Q^{-n}$.  In minimal-string, JT-gravity or related double-scaled matrix integrals,
this is the corresponding paired multi-FZZT insertion at the roots of $Q$
\cite{FyodorovStrahov2003,OkuyamaSakai2021FZZT}.  The brane interpretation is
realization dependent.  The characteristic-polynomial identity itself holds for every cyclic
measure.

\paragraph{Filtered Loschmidt and thermofield-double amplitudes.}
The zeroth connector transfers the survival amplitude without reconstructing the prepared chain,
\begin{equation}
   \widetilde\phi_0^{[Q]}(t)
   =\frac{1}{N_Q}\Qsh(\ii\partial_t)Q(\ii\partial_t)I_0(t),
   \qquad
   I_0(t)=\bra{K_0}\ee^{-\ii tH}\ket{K_0}.
   \label{eq:filtered-work-statistics}
\end{equation}
The
spectral measure underlying eq.~\eqref{eq:filtered-work-statistics} is always the energy distribution
of the $Q$-prepared state with respect to $H$, and $\widetilde\phi_0^{[Q]}(t)$ is its Fourier transform.
That measure becomes a post-quench final-energy distribution only after the preparation is embedded
in a compatible quench protocol.  If the pre-quench energy measurement is sharp at $E_{\rm i}$,
the associated two-projective-measurement work distribution is its translate,
$P_W(W)=P_f(W+E_{\rm i})$; the two distributions coincide only after choosing the energy origin so
that $E_{\rm i}=0$ \cite{TalknerLutzHanggi2007,KrylovDQPT2023}.

Survival-amplitude zeros have been studied in
ref.~\cite{NovotnyEtAl2026}.  In any complex-time domain where the reference transform is
analytic, eq.~\eqref{eq:filtered-work-statistics} transports those zeros under polynomial
preparation.  For a
smooth seed family and an isolated simple zero $t_*(\theta)$,
\begin{equation}
   \frac{\dd t_*}{\dd\theta}
   =-\frac{\partial_\theta\widetilde\phi_0^{[Q_\theta]}(t_*)}
           {\partial_t\widetilde\phi_0^{[Q_\theta]}(t_*)}\,.
   \label{eq:filtered-zero-response}
\end{equation}

A dynamical quantum phase transition (DQPT) is signaled by such zero trajectories only if,
in a thermodynamic or other controlled scaling limit, they accumulate and pinch the real-time axis
and the intensive return rate becomes nonanalytic
\cite{HeylPolkovnikovKehrein2013,KrylovDQPT2024}.  Under a uniform finite,
nondegenerate saddle expansion, a filter whose logarithmic weight is subextensive at every
relevant saddle changes prefactors and finite-size zeros but not the leading rate branches or a
generic Stokes crossing, unless it annihilates a leading saddle.  A degree or filter strength that
scales extensively can alter that conclusion.  This is a conditional saddle argument for arbitrary many-body limits.

For a normalized thermofield double under one-sided evolution, define
its normalized return amplitude by
$\mathcal A_\beta(t)=\bra{\mathrm{TFD}_\beta}\ee^{-\ii tH_L}\ket{\mathrm{TFD}_\beta}$, where
\begin{equation}
   \ket{\mathrm{TFD}_\beta}
   =Z(\beta)^{-1/2}\sum_n\ee^{-\beta E_n/2}\ket n_L\ket n_R,
   \qquad
   \mathcal A_\beta(t)=\frac{Z(\beta+\ii t)}{Z(\beta)}\,.
   \label{eq:tfd-spectral-form-factor-context}
\end{equation}
Polynomial preparation by $Q(H_L)$ gives
\begin{equation}
\begin{aligned}
   Z_Q(s)&=\operatorname{Tr}\!\left[|Q(H)|^2\ee^{-sH}\right]
   =\Qsh(-\partial_s)Q(-\partial_s)Z(s),\\
   \mathcal A_{\beta,Q}(t)&=\frac{Z_Q(\beta+\ii t)}{Z_Q(\beta)}\,.
\end{aligned}
   \label{eq:filtered-tfd-partition-function}
\end{equation}
Its modulus squared exactly deforms the normalized unaveraged spectral form factor
\cite{DelCampoMolinaVilaplanaSantosSonner2018}.  Ramp and plateau statements require the relevant
averaging, smoothing or late-time limit and do not follow from the differential identity alone.

\paragraph{Compact filters and the quantum-signal-processing interface.}
Proposition~\ref{prop:compact-filter-closure-app} extends the scalar construction on bounded
support beyond a single polynomial.  If $Q_d\to F$ uniformly and
$A=\|F(H)\ket{K_0}\|>0$, then
\begin{equation}
   \epsilon_d=\|(F-Q_d)(H)\ket{K_0}\|
   =\|F-Q_d\|_{L^2(\mu)},
   \quad
   \left\|
      \frac{F(H)\ket{K_0}}{A}
      -\frac{Q_d(H)\ket{K_0}}{\|Q_d(H)\ket{K_0}\|}
   \right\|\le\frac{2\epsilon_d}{A},
   \label{eq:qsp-state-aware-approximation}
\end{equation}
provided $\epsilon_d<A$.  Unitary evolution and bounded observables inherit this estimate, while
the full spread complexity uses the compact-time tail control proved in the appendix.  Under a
block encoding or qubitized access model, QSP implements bounded parity-compatible polynomial
filters on a spectrum rescaled to $[-1,1]$
\cite{GilyenSuLowWiebe2019}.  Polynomial eigenstate filtering in this access model is developed in
ref.~\cite{LinTong2020Filtering}.  For an implementation of
$Q(H)/\alpha$, postselection succeeds with probability $N_Q/\alpha^2$ and amplitude amplification
multiplies the signal-processing query cost by $O(\alpha/\sqrt{N_Q})$.  The access model,
normalization, precision and preparation of the reference data remain separate costs.  State-aware
$L^2(\mu)$ approximation controls the filtered-state error entering the query analysis
 \cite{Adhikari2026Query}; the connector then predicts the conditioned Krylov dynamics after
that preparation.

\paragraph{Temperature tomography in Liouville space.}
The scalar gap measure at one temperature forgets how transitions with the same gap are
distributed over their source energies.  The bivariate measure of
appendix~\ref{app:operator-krylov} restores that coordinate.  After division by the known nonzero
thermal kernel and multiplication by the known partition function, its unnormalized temperature
family is the measure-valued bilateral Laplace transform in the source-energy variable,
\begin{equation}
   \dd\Lambda_\beta(\omega)
   =\int\ee^{-\beta F}\eta_O(\dd\omega,\dd F).
   \label{eq:temperature-tomography-main}
\end{equation}
Proposition~\ref{prop:temperature-tomography-app} proves that exact knowledge of this family on an
open interval determines the positive transition-resolved measure when the displayed exponential
moments are finite.  The statement requires unnormalized masses and known prefactors.  It
reconstructs aggregated transition intensities, not phases, eigenvectors, degenerate transition
labels or higher correlations.  Conditional derivatives give the source-energy mean and variance
at fixed gap.  For the normalized joint probability measure obtained after dividing out the known
thermal kernel, projection to the gap reduces the Fisher information for estimating
inverse temperature by exactly
\begin{equation}
   \mathcal I_{\rm full}(\beta)-\mathcal I_{\rm gap}(\beta)
   =\mathbb E_\beta\!\left[\operatorname{Var}_\beta(F\mid\omega)\right].
   \label{eq:temperature-fisher-deficit-main}
\end{equation}
Appendix~\ref{app:operator-krylov} proves these statements before imposing ETH.  Coarse-grained
off-diagonal ETH gives a physical interpretation of the recovered positive density as the
entropy-dressed transition intensity, but it remains an additional approximation.

These spectral operations must not be conflated.  At fixed $H$ and fixed inner product,
polynomial preparation gives the dynamical connector and the basis-overlap response.  Varying
temperature in Liouville space is a separate extension whose open unnormalized family restores
source-energy information that one projected gap measure loses in general.  Together with the
chord, Toda, determinant, TFD and QSP examples, it shows how the same spectral language organizes
dynamical transport, response geometry and filtered inference without identifying their
realization-dependent interpretations.

\section{Discussion and outlook}
\label{sec:discussion}

The central result is an exact relative transfer calculus at fixed Hamiltonian.  Once a reference
cyclic problem has been resolved, a fixed-degree polynomial preparation can be reconstructed
inside that solved Jacobi representation, without repeating Lanczos orthogonalization in the
ambient Hilbert space.  The connector determines individual amplitudes and shifted Jacobi data,
whereas the projected kernel determines cumulative probability and spread complexity.  Their
common spectral origin is
\begin{equation}
   \ket{\psi_Q}=N_Q^{-1/2}Q(H)\ket{K_0}
   \quad\Longrightarrow\quad
   W_Q\dd\mu
   \quad\Longrightarrow\quad
   \{\widetilde\phi_n^{[Q]}(t),\Phi_\ell^{[Q]}(t),K_Q(t)\}.
\end{equation}

Root, jet and coefficient-space data treat simple, repeated and complex seed zeros, while the
matrix-valued parent measure retains interference among several seed directions before scalar
compression.     Together, the connector and projected-kernel
constructions determine every fixed Krylov level.  The infinite sum defining
spread complexity 
requires additional tail control, and finite support requires quotient reduction at the terminal edge, with a smaller surviving
quotient when the seed deletes atoms.  The Charlier large-index
estimates control the oscillator first moment.  The fixed-$r$, all-level Hermite amplitudes and Fock-distance
bound close the Gaussian endpoint, while the continuous-$q$-Hermite remainder Gram reconstructs
every rebuilt level for any fixed finite chord seed.  In finite support, the product--Gram
construction provides the terminal rows absent from the infinite-support formulas.

\paragraph{Rigid bulk and preparation memory.}
A fixed-degree polynomial seed can change the Jacobi data at any prescribed finite depth while preserving the limiting coefficients of an asymptotically constant chain. The all-root proposition makes this statement exact for every measure in $M(a,b)$ and permits real roots in the band,
at visible atoms or outside the band, as well as nonreal roots.  It does not, however, determine a
universal convergence rate.  Finite
rows and subleading coefficients can therefore remain strongly seed dependent.  At an atomic root
the support loss is irreversible from the shifted scalar data alone.  In the Charlier family the
two subleading intercepts recover both the preparation degree and the number of deleted Poisson
atoms.  The Charlier
vacuum bound, the Hermite all-level amplitude closure and Fock-distance inequality, and the arithmetic support loss of the
constant-hopping chain are complementary model-dependent consequences.  The finite-time
ceiling controls only an admissible growth envelope and neither defines nor establishes a chaos exponent. Saturation and any such interpretation remain
properties of the reference dynamics and of the Krylov construction being used.

The deep-seed proposition probes a different scaling regime.  When $Q=p_r$ and $r$ tends to infinity,
the induced measure of a Jacobi chain in $M(a,b)$ approaches the arcsine measure. At every fixed rebuilt level, the shifted
Jacobi coefficients converge coefficientwise to the folded arcsine chain, whose first hopping is
$\sqrt2\,b$ and whose later hoppings are $b$.  This is the fold of the two directions
available around a launch point deep inside the reference chain.  Fixed-degree rigidity and this
controlled growing-degree limit are therefore compatible rather than competing statements. 

For the continuous-$q$-Hermite chord chain, the same proposition gives the fixed-$q$ folded
limit, while the $q=0$ endpoint exposes its exact boundary-sensitive correction.  At fixed seed
and rebuilt level, the separate $q\uparrow1$ limit reaches the Hermite construction for each fixed
finite $r$.
These extensions distinguish the relevant orders of limits rather than asserting a joint scaling
prescription.

\paragraph{Chord dynamics and the two-basis problem.}
Section~\ref{sec:context} makes the two-basis distinction explicit in the empty-chord
double-scaled SYK problem, where the reference Krylov level is the chord number.  For every fixed
finite prepared chord $r$, the continuous-$q$-Hermite product formula determines propagation in
the normalized chord-number basis, while the remainder-Gram construction determines the basis rebuilt
from that state at every Krylov level.  Their radial filtration gives
$K_r\ge D_r^{\rm ch}\ge|\overline n_r-r|$. 
Signed chord growth, physical distance from the prepared chord and rebuilt
spread are related, but are not the same observable.

The endpoint operations retain different information.  At fixed $r$, the coefficientwise
$q\uparrow1$ limit identifies the Hermite/Fock recursion; the Hermite endpoint separately admits
an exact amplitude closure and Fock-distance  bound.  At fixed $q<1$, the deep-seed limit instead produces
the two-sided profile $J_k(z_q)^2$ in the normalized chord-number
basis and folds its left--right motion into the rebuilt half-line chain.  At fixed $q<1$ and fixed time, as $r\to\infty$,
$D_r^{\rm ch}\to K_{\rm arc}$ and $K_r-D_r^{\rm ch}\to0$, whereas the signed mean displacement
vanishes.  The $q=0$ endpoint is the infinite half-line constant chain, not a finite-support
quotient.  Whenever chord number has a controlled bulk-length normalization, $D_r^{\rm ch}$
therefore measures distance from the prepared length and $K_r$ is its exact re-Lanczos upper
bound; the bulk interpretation remains tied to the model and regime.

The same two-basis structure gives a precise interface with Koherence.  The overlap matrix
$\mathsf U_Q$ measures the response of the ordered basis, while
$\mathsf T_Q=\mathsf U_Q^\dagger Q(\Jac)/\sqrt{N_Q}$ also includes the polynomial preparation and
transports its dynamics.  For the fixed-chord family in the deep-seed limit $r\to\infty$, and for every fixed $n\ge1$, the Koherence of the
$n$th shifted vector tends to $\log2$, the entropy of the left--right alternative that the new
half-line chain folds.  This fixed-$n$ limit is not uniform when the shifted level grows with the
preparation depth.
Characteristic-polynomial insertions, filtered Loschmidt amplitudes and filtered thermofield
doubles give further exact realizations.  Their matrix-model, work-statistics or spectral-form-factor interpretations require the realization, protocol and averaging assumptions stated in
section~\ref{sec:context}.

\paragraph{Response geometry, implementation and tomography.}
The normalizer of the commuting Toda family is also its response functional.  On a parameter
domain where it is finite and twice differentiable under the integral, the Hessian of $\log Z_Q$ is the covariance matrix of the visible powers of
$H$, the pure-state quantum Fisher matrix, and four times the Fubini--Study metric.  This geometry
is exact for the prepared state family.  Calling the deformation a GGE additionally requires a
physically appropriate set of independent local or quasilocal charges.

On compact spectral support, the polynomial calculus also controls bounded nonpolynomial filters
through approximation.  The relevant error is state aware because it is measured in $L^2(\mu)$,
and the compact-time tail proposition  carries it to spread complexity.  Quantum signal processing
provides an implementation only after specifying a block-encoding or qubitized access model,
rescaling the spectrum, enforcing the admissible parity, and accounting for polynomial degree,
target precision, normalization, success probability and any amplitude-amplification overhead. The connector predicts the Krylov dynamics
conditioned on a successful preparation.  It does not remove the cost of preparing the reference
data or the filtered state.

Temperature variation in Liouville space exposes the source-energy information discarded by a
single gap measure.  Under the known-prefactor and exponential-moment hypotheses,
proposition~\ref{prop:temperature-tomography-app} identifies the unnormalized open-temperature
family as a measure-valued bilateral Laplace transform and reconstructs the positive
transition-resolved measure.  The result is exact before ETH and does not recover phases,
eigenvectors or higher correlations.  Its conditional temperature derivatives give the
source-energy mean and variance at fixed gap; averaging the latter over the gap gives exactly the
Fisher-information deficit.

\paragraph{Open problems.}
For fixed $0\le q<1$, finite $r$ and
finite time, the fixed-chord DSSYK reconstruction is exact at every rebuilt level.
The first open regime is instead macroscopic.  The seed degree, observed level or time may scale
together with system size or Krylov depth.  The controlled limit $Q=p_r$, $r\to\infty$ resolves
the fixed-level, fixed-time folded regime but not the joint $n/r$ or $z_q/r$ crossover to the
zero-chord boundary.  At
the Hermite endpoint the finite-time amplitude closure is exact, while the fixed-$r$ late-time
first-moment asymptotics for $r\ge2$ require separate control.  DSSYK matter sectors and their
infinite $q$-products further suggest scalar, block and bivariate extensions whose tail estimates
are not yet available.  

Rational and resolvent-dressed preparations
give a complementary nonpolynomial direction when their poles, zeros and any compensating masses
are compatible with the support \cite{Zhedanov1997,AlishahihaVasli2026KrylovDistribution}.  A second problem is to classify subleading
preparation memory across compact, Freud-type, hard-edge and atomic measures, including the
nonuniform crossover as a root approaches an atom.  Reduced-Palm asymptotics organize the
auxiliary ensemble problem, but a physical spectral diagnostic still needs a model-specific
theorem connecting the cyclic measure to that scaling limit.  A third problem is stability.
Finite-depth and noisy Lanczos data, ill-conditioned root and jet Gram matrices,  postselection overhead and
inverse-Laplace conditioning must be propagated into quantitative errors on amplitudes and
complexity.  Time-dependent, moving-frame and Floquet generators add an evolving Krylov frame
\cite{ChowdhuryMahapatra2024,Grabarits2026}; determining which part of the fixed-generator transfer
survives is a separate problem.  These directions retain the central question of the paper:
which features of the dynamics generated from a new preparation can be transported from
a resolved reference problem while the physical generator remains fixed?

\section*{Acknowledgments}
A.C. acknowledges the School \& Workshop on Number Theory and Physics 2025 at ICTP, Trieste, and Refinements in Enumerative Geometry and Physics 2026 at ICTS, Bengaluru, for their hospitality and for creating stimulating environments where part of the work was completed. A.C. also acknowledges the ISM 2025, held at IIT Bhubaneswar, for supporting the string community in India and for useful discussions. 

\appendix

\section{Lanczos dictionary and finite terminal quotients}
\label{app:lanczos}

Section~\ref{sec:setup} uses three complementary descriptions of the cyclic problem, namely the Lanczos recursion, the orthogonal-polynomial representation of the seed spectral measure, and the Jacobi resolvent. We derive their precise relation here, including the closure of a finite chain and its quotient-algebra interpretation.

\paragraph{From the Lanczos recursion to the terminal quotient.} 
Starting from $\ket{K_0}=\ket{\psi_0}$, set $b_0=0$ and $\ket{K_{-1}}=0$. The Lanczos recursion \cite{Lanczos1950} defines
\begin{equation}
   a_n=\bra{K_n}H\ket{K_n},
   \qquad
   \ket{\mathcal R_n}=(H-a_n)\ket{K_n}-b_n\ket{K_{n-1}},
   \qquad
   b_{n+1}=\|\ket{\mathcal R_n}\|.
\end{equation}
Before termination, $b_{n+1}>0$ and $\ket{K_{n+1}}=b_{n+1}^{-1}\ket{\mathcal R_n}$. Induction, with $p_{-1}=0$ and $p_0=1$, gives
\begin{equation}
   \ket{K_n}=p_n(H)\ket{K_0},
   \qquad
   p_{n+1}(E)=
   \frac{(E-a_n)p_n(E)-b_np_{n-1}(E)}{b_{n+1}},
\end{equation}
where $p_n$ has degree $n$. The scalar spectral measure $\dd\mu(E)=\bra{K_0}\Pi(\dd E)\ket{K_0}$ then gives
\begin{equation}
   \delta_{mn}
   =\braket{K_m}{K_n}
   =\bra{K_0}p_m(H)p_n(H)\ket{K_0}
   =\int p_m(E)p_n(E)\dd\mu(E).
\end{equation}
Thus, the Lanczos polynomials are precisely the orthonormal polynomials of the spectral measure attached to the pair $(H,\ket{K_0})$ \cite{Chihara1978,SimonOPRL1}. Passing to the monic normalization $P_n=\sqrt{h_n}\,p_n$ gives eq.~\eqref{eq:base-recurrence}.

Suppose now that the cyclic dimension is $d<\infty$, so that $b_d=0$ and $b_n>0$ for $1\le n\le d-1$. The orthogonal family ends at $P_{d-1}$, while the final recurrence defines the monic terminal polynomial
\begin{equation}
   T_d(E)=(E-a_{d-1})P_{d-1}(E)-b_{d-1}^2P_{d-2}(E).
\end{equation}
Let $\Jac_n$ denote the $n\times n$ leading principal Jacobi truncation and let $\chi_n$ be its
characteristic polynomial,
\begin{equation}
      \chi_n(E)=\det(E\Id_n-\Jac_n),
   \qquad
   \chi_0=1,
   \qquad
   \chi_1=E-a_0\,.
\end{equation}
Expansion along the last row gives
\begin{equation}
      \chi_n(E)=(E-a_{n-1})\chi_{n-1}(E)
   -b_{n-1}^2\chi_{n-2}(E).
\end{equation}
Comparison with the monic recurrence shows that $\chi_n=P_n$ for $0\le n\le d-1$ and $\chi_d=T_d$. Consequently,
\begin{equation}
   T_d(E)=\det(E\Id_d-\Jac_d)
   =\prod_{\alpha=0}^{d-1}(E-E_\alpha),
   \qquad
   T_d(H)\ket{K_0}=0\,.
   \label{eq:terminal-characteristic-app}
\end{equation}
The finite Jacobi matrix is irreducible, hence $e_0$ is cyclic and its eigenvalues are simple. Its characteristic and minimal polynomials coincide. Polynomial functions on the visible spectrum are therefore represented by
$
   \cA_d=\CC[E]/\langle T_d\rangle,
$
which is the terminal quotient used throughout the main text.

\paragraph{Finite closure and support loss.}
The same closure admits several equivalent descriptions. For $d=\dim\cK(H,\ket{K_0})<\infty$,
\begin{enumerate}[label=\emph{(\roman*)},leftmargin=2.2em,itemsep=4pt,topsep=4pt,parsep=0pt]
\item $\ket{K_0},H\ket{K_0},\ldots,H^{d-1}\ket{K_0}$ form a basis of the cyclic subspace;
\item the Lanczos recursion satisfies $b_d=0$ and $b_n>0$ for $1\le n\le d-1$;
\item the least-degree monic annihilator of the cyclic vector is $T_d$;
\item the scalar spectral measure has exactly $d$ support points.
\end{enumerate}
To see the equivalence directly, write the spectral resolution of the cyclic vector as
$
   \ket{K_0}=\sum_{\alpha=0}^{s-1}u_\alpha\ket{E_\alpha},
$
where the visible energies are distinct and every $u_\alpha$ is nonzero. %
In this eigenbasis the vector $H^k\ket{K_0}$ has components $u_\alpha E_\alpha^k$, so the
coefficient matrix of $\{\ket{K_0},H\ket{K_0},\ldots\}$ factors as
$\operatorname{diag}(u_\alpha)$ times a Vandermonde matrix in the distinct $E_\alpha$. The
Vandermonde factor has rank $s$, so the cyclic dimension equals the number of visible
energies. The least-degree monic annihilator is $\prod_{\alpha=0}^{s-1}(E-E_\alpha)$, and the Lanczos recursion terminates at $b_s=0$.

For
$
   \dd\mu(E)=\sum_{\alpha=0}^{d-1}w_\alpha
   \delta(E-E_\alpha)\dd E,
$
the finite orthogonality relations take the form
\begin{equation}
   \sum_{\alpha=0}^{d-1}w_\alpha
   P_m(E_\alpha)P_n(E_\alpha)
   =h_n\delta_{mn},
   \qquad 0\le m,n\le d-1\,.
\end{equation}
The shifted weights are those of eq.~\eqref{eq:finite-shifted-weights-main}.
Deleting the zero-weight atoms and forming the degree-ordered
orthogonal-polynomial family of the reduced discrete measure gives the
finite shifted Jacobi chain directly. Ordinary polynomial expressions of
degree at least $d$ instead require reduction modulo $T_d$; both procedures
represent the same computation in $\cA_d$.

\begin{example}[Three-level support loss]
\label{ex:spin-one-support-loss}
Consider a three-level cyclic Jacobi chain with $a_0=a_1=a_2=0$ and
$b_1=b_2=1/\sqrt2$, generated from its endpoint $\ket{K_0}$.  Its
spectral measure and terminal polynomial are
\begin{equation}
   \dd\mu(E)=\frac14\delta(E+1)\dd E
   +\frac12\delta(E)\dd E
   +\frac14\delta(E-1)\dd E,
   \qquad
   T_3(E)=E(E^2-1).
\end{equation}
The first interior Krylov vector is $\ket{K_1}=\sqrt2H\ket{K_0}$. It is generated by $Q(E)=E$, for which $N_Q=h_1=1/2$, and its normalized measure is
\begin{equation}
   \dd\nu_1(E)=\frac{E^2}{h_1}\dd\mu(E)
   =\frac12\delta(E+1)\dd E
   +\frac12\delta(E-1)\dd E.
   \label{eq:spin-one-support-loss-app}
\end{equation}
The root of $Q$ deletes the atom at $E=0$, reducing the shifted cyclic dimension to $d_Q=2$. Equivalently, the least-degree monic polynomial satisfying $\widetilde T(H)Q(H)\ket{K_0}=0$ is
$
   \widetilde T_2(E)=E^2-1,
$
since $T_3=Q\widetilde T_2$. This is the quotient realization of the support loss in eq.~\eqref{eq:support-loss-dimension}. It also makes explicit why a fixed-$H$ interior seed is not described by the associated-polynomial tail discussed in section~\ref{sec:setup}. Section~\ref{sec:examples} returns to this chain, where the deleted atom is what allows the jump to \emph{lower} the spread complexity, in contrast with the oscillator bound of corollary~\ref{cor:H1-monotonicity}.
\end{example}

\paragraph{Weyl-function reconstruction.}
For $z\in\CC\setminus\RR$, the Weyl function is the Stieltjes transform
\begin{equation}
   m(z)=\bra{K_0}(H-z)^{-1}\ket{K_0}
   =\int\frac{\dd\mu(E)}{E-z}
   =\frac{1}{a_0-z-
   \dfrac{b_1^2}{a_1-z-
   \dfrac{b_2^2}{a_2-z-\cdots}}}\,.
\end{equation}
The spectral theorem gives the integral representation. Decomposing $\Jac-z\Id$ into the first coordinate and the Jacobi tail beginning at $e_1$ gives
$
   m(z)=[a_0-z-b_1^2m_1(z)]^{-1}
$
by the Schur complement, and iteration produces the continued fraction \cite{Chihara1978,SimonOPRL1,GesztesySimon1997}. In a finite chain it ends with $a_{d-1}-z$, because $b_d=0$; in an infinite chain it is understood under the determinacy and self-adjointness assumptions stated in section~\ref{sec:setup}. Applied to eq.~\eqref{eq:shifted-weyl-closed-main}, this recursion determines $\widetilde a_n^{[Q]}$ and $\widetilde b_n^{[Q]}$ at every level for which the required moments exist, terminating at $\widetilde b_{d_Q}^{[Q]}=0$ in finite support.

The Lanczos, finite-support and resolvent descriptions are thus mutually consistent. The next appendix turns from the underlying cyclic problem to explicit formulas for its Christoffel connector coefficients.

\section{Christoffel coefficients and confluent determinants}
\label{app:christoffel}

Section~\ref{sec:finiteband} places the connector polynomial $S_n=W_QR_n^{[Q]}$ in the finite reference window of eq.~\eqref{eq:finite-band-main}, while eq.~\eqref{eq:root-conditions-main} enforces its divisibility by $W_Q$. %
Here we solve the resulting constraint system explicitly, reduce it for real seeds to
diagonal-kernel and root-Gram data, the forms in which the rigidity analysis of
section~\ref{sec:rigidity} and the Charlier asymptotics of appendix~\ref{app:H1} consume the
transfer, prove that the determinant construction survives finite-chain termination, and
specialize it to degree-one seeds. %
We write $W$ for any positive multiple of the nonnegative polynomial multiplier $W_Q$ and
suppress the superscript $[Q]$ on $R_n$ and $\Gamma_{n,m}$ when no ambiguity can arise. Rescaling
$W$ by a positive constant leaves the monic orthogonal polynomials and their Jacobi coefficients
unchanged, while multiplying every connector coefficient by that constant; the physical
normalization is recovered by setting $W=W_Q$.

\paragraph{Confluent connector systems and Christoffel determinants.}
Factor
\begin{equation}
   W(E)=w_L\prod_{a=1}^{s}(E-\zeta_a)^{m_a},
   \qquad
   w_L>0\,,
   \qquad
   \sum_{a=1}^{s}m_a=L\,,
\end{equation}
where the $\zeta_a$ are pairwise distinct and $m_a\ge1$. In the absolute-index convention of proposition~\ref{prop:finite-band}, we write
\begin{equation}
   S_n(E)=\sum_{m=n}^{n+L}\Gamma_{n,m}P_m(E),
   \qquad
   \Gamma_{n,n+L}=w_L\,.
\end{equation}
The identity $S_n=WR_n$ is equivalent to the vanishing of $S_n$ at each root of $W$ with the prescribed algebraic multiplicity. The remaining connector coefficients therefore satisfy
\begin{equation}
   \sum_{m=n}^{n+L-1}\Gamma_{n,m}
   \partial_E^\ell P_m(\zeta_a)
   =-w_L\partial_E^\ell P_{n+L}(\zeta_a),
   \qquad
   \substack{a=1,\ldots,s\,,\\[1pt] \ell=0,\ldots,m_a-1\,.}
   \label{eq:confluent-linear-system-app}
\end{equation}
Conversely, these root conditions make $S_n/W$ a monic degree-$n$ polynomial. Since $S_n\in\Span\{P_n,\ldots,P_{n+L}\}$, the quotient is orthogonal with respect to $W\dd\mu$ to every polynomial of degree below $n$ and is therefore $R_n$. Positive definiteness of the shifted inner product through degree $n$ gives uniqueness.

For the number-state seed $Q=P_r/\sqrt{h_r}$, with $r<d$ in finite support, the multiplier $W=P_r^2/h_r$ has double roots at the simple zeros $\xi_1,\ldots,\xi_r$ of $P_r$. Setting $L=2r$ and $\Gamma_{n,n+2r}=1/h_r$ gives
\begin{equation}
\begin{pmatrix}
P_n(\xi_1)&\cdots&P_{n+2r-1}(\xi_1)\\
P_n'(\xi_1)&\cdots&P_{n+2r-1}'(\xi_1)\\
\vdots&&\vdots\\
P_n(\xi_r)&\cdots&P_{n+2r-1}(\xi_r)\\
P_n'(\xi_r)&\cdots&P_{n+2r-1}'(\xi_r)
\end{pmatrix}
\begin{pmatrix}
\Gamma_{n,n}\\
\vdots\\
\Gamma_{n,n+2r-1}
\end{pmatrix}
=-\frac{1}{h_r}
\begin{pmatrix}
P_{n+2r}(\xi_1)\\ P_{n+2r}'(\xi_1)\\ \vdots\\
P_{n+2r}(\xi_r)\\ P_{n+2r}'(\xi_r)
\end{pmatrix}.
\label{eq:general-root-system}
\end{equation}
In finite support, the system above is an identity among ordinary polynomials while $n+2r\le d-1$. The physical shifted rows exist for $0\le n<d_Q$; rows reaching the terminal edge are evaluated with the proper terminal lifts of eq.~\eqref{eq:proper-terminal-lift-main}, whose determinant form is established in proposition~\ref{prop:terminal-lift-determinant} below, and then reduced modulo $T_d$.

The same solution has a compact confluent determinant form \cite{Chihara1978,Szego1975,BuenoMarcellan2004}. Define
\begin{equation}
   \mathbf P_n(E)=
   \begin{pmatrix}
      P_n(E)&P_{n+1}(E)&\cdots&P_{n+L}(E)
   \end{pmatrix}.
\end{equation}
Let $\mathsf C_n$ be the $L\times L$ coefficient matrix of the linear system \eqref{eq:confluent-linear-system-app}, with rows labeled by the derivative conditions $(a,\ell)$ and columns by $m=n,\ldots,n+L-1$, and let $\Delta_n=\det\mathsf C_n$. The uniqueness established above gives $\Delta_n\ne0$. Then
\begin{equation}
   R_n(E)=\frac{(-1)^Lw_L}{\Delta_nW(E)}
   \det\begin{pmatrix}
      \mathbf P_n(E)\\
      \hline
      \mathbf P_n(\zeta_1)\\
      \partial_E\mathbf P_n(\zeta_1)\\
      \vdots\\
      \partial_E^{m_1-1}\mathbf P_n(\zeta_1)\\
      \hline
      \vdots\\
      \hline
      \mathbf P_n(\zeta_s)\\
      \partial_E\mathbf P_n(\zeta_s)\\
      \vdots\\
      \partial_E^{m_s-1}\mathbf P_n(\zeta_s)
   \end{pmatrix}.
\end{equation}
At $E=\zeta_a$, the top row and its first $m_a-1$ derivatives reproduce the corresponding constraint rows, so the determinant is divisible by $W(E)$. Expansion through the final entry of the top row gives leading coefficient $(-1)^L\Delta_n$. The prefactor consequently makes the quotient monic and fixes the overall sign, which a proportionality statement would leave undetermined; the sign matters because the connector coefficients enter the amplitudes linearly.

\paragraph{One-root kernel reduction.}
A real degree-one seed produces a normalized double-Christoffel modification whose Jacobi
coefficients can be written entirely in terms of the diagonal reference kernel, the object
whose growth at the seed root controls the preparation response in
section~\ref{sec:rigidity}. Conversely, the $r=1$ case of the root-Gram construction below is the single
node $\xi=a_0$, the zero of $P_1$, at which the two realizations agree; the
statement here holds for every real $\xi$.

\begin{proposition}[Jacobi data for a real double root]
\label{prop:double-Christoffel-kernel-Jacobi-app}
Let $\mu$ be a positive measure on $\RR$, either with infinite support
and finite moments of all orders or with finite support of cardinality
$d$.  Write $p_n=P_n/\sqrt{h_n}$ for its physical orthonormal
polynomials and fix $\xi\in\RR$.  %
On infinite support define the following kernels for every $n\ge0$; on support of cardinality
$d$, define them for $0\le n\le d-1$.
\begin{equation}
   \mathcal K_n(\xi)=\mathbb K_n^\mu(\xi,\xi)
   =\sum_{m=0}^{n}p_m(\xi)^2,
   \qquad \mathcal K_{-1}(\xi)=0\,.
   \label{eq:one-root-kernel-sequence-app}
\end{equation}
Assume also that
\begin{equation}
   N_\xi=\int_{\RR}(E-\xi)^2\dd\mu(E)>0\,,
   \qquad
   \dd\widehat\mu(E)=\frac{(E-\xi)^2}{N_\xi}\dd\mu(E).
\end{equation}
If the support is infinite, the Jacobi coefficients
$\widehat a_n$ and $\widehat\beta_n=\widehat b_n^2$ of
$\widehat\mu$ satisfy
\begin{equation}
   \widehat\beta_n
   =\beta_{n+1}
      \frac{\mathcal K_{n+1}(\xi)\mathcal K_{n-1}(\xi)}
           {\mathcal K_n(\xi)^2}\,,
   \qquad n\ge1\,,
   \label{eq:double-Christoffel-beta-kernel-app}
\end{equation}
and
\begin{equation}
   \widehat a_n
   =a_n-\Theta_n(\xi)
       \frac{\mathcal K_n(\xi)}{\mathcal K_{n+1}(\xi)}
      +\Theta_{n-1}(\xi)
       \frac{\mathcal K_{n-1}(\xi)}{\mathcal K_n(\xi)}\,,
   \label{eq:double-Christoffel-a-kernel-app}
\end{equation}
where the last term is absent for $n=0$ and
\begin{equation}
   \Theta_n(\xi)
   =\xi-a_{n+1}
      +b_{n+1}\frac{p_n(\xi)p_{n+1}(\xi)}{\mathcal K_n(\xi)}\,.
   \label{eq:double-Christoffel-Theta-app}
\end{equation}
If instead $\mu$ has finite support, let
$d_\xi=\#\{\alpha:E_\alpha\ne\xi\}$.
Equations~\eqref{eq:double-Christoffel-beta-kernel-app}--\eqref{eq:double-Christoffel-Theta-app}
remain valid for the physical rows $0\le n<d_\xi$ satisfying
$n+1\le d-1$.  Any remaining physical terminal data are determined in
$\cA_d$ by the quotient product-Gram construction of
subsection~\ref{app:finite-multiplication-gram}; the diagonal-kernel
formulas are not continued by introducing a post-terminal polynomial.
\end{proposition}

\begin{proof}
Multiplication by $E-\xi$~\cite{KautskyGolub1983} identifies the transformed degree-$n$ space
with the polynomials in $\cP_{n+1}$ which vanish at $\xi$; this subspace
has codimension one.  If $\widehat p_n$ is orthonormal for $\widehat\mu$, its
normalized representative in $L^2(\mu)$ is
\begin{equation}
   \frac{E-\xi}{\sqrt{N_\xi}}\widehat p_n(E)
   =\left[
      \frac{\mathcal K_n(\xi)}{\mathcal K_{n+1}(\xi)}
   \right]^{1/2}
   \left[
      p_{n+1}(E)
      -\frac{p_{n+1}(\xi)}{\mathcal K_n(\xi)}
       \mathbb K_n^\mu(E,\xi)
   \right].
   \label{eq:double-Christoffel-orthonormal-polynomial-app}
\end{equation}
The bracket vanishes at $\xi$ because $\mathbb K_n^\mu(\xi,\xi)=\mathcal K_n(\xi)$, and the reproducing property makes it orthogonal to every polynomial of degree at most $n$ which vanishes at $\xi$, hence to $(E-\xi)\cP_{n-1}$. Since $p_{n+1}$ is orthogonal to $\mathbb K_n^\mu(\,\cdot\,,\xi)$, the bracket has squared $\mu$-norm $\mathcal K_{n+1}(\xi)/\mathcal K_n(\xi)$, so the above prefactor normalizes it. Its leading coefficient is positive, which fixes the sign in the convention $\widehat p_n=\widehat P_n/\sqrt{\widehat h_n}$ with $\widehat P_n$ monic. Comparison of leading coefficients gives
eq.~\eqref{eq:double-Christoffel-beta-kernel-app}.  For the diagonal
coefficient, the reference three-term recurrence and the reproducing
property give
\begin{equation}
\begin{aligned}
   \left\langle p_{n+1},E \,\mathbb K_n^\mu(\,\cdot\,,\xi)\right\rangle_\mu
   &=b_{n+1}p_n(\xi),\\
   \left\langle \mathbb K_n^\mu(\,\cdot\,,\xi),
      E \,\mathbb K_n^\mu(\,\cdot\,,\xi)\right\rangle_\mu
   &=\xi\mathcal K_n(\xi)
     -b_{n+1}p_n(\xi)p_{n+1}(\xi).
\end{aligned}
   \label{eq:one-root-kernel-E-elements-app}
\end{equation}
Taking the expectation of multiplication by $E$ in
eq.~\eqref{eq:double-Christoffel-orthonormal-polynomial-app} and using
$\mathcal K_n=\mathcal K_{n-1}+p_n(\xi)^2$ gives
eq.~\eqref{eq:double-Christoffel-a-kernel-app}.
\end{proof}

\paragraph{All-root closure of the Nevai class.}
\label{app:all-root-nevai-proof}
We now prove the roots-anywhere statement used in
proposition~\ref{prop:nevai-rigidity}.  It is enough to treat one real squared factor and one
nonreal conjugate pair.  For a real root $\xi$, set
\begin{equation}
   d_n(\xi)=b_{n+1}
   \frac{p_{n+1}(\xi)p_n(\xi)}{\mathcal K_n(\xi)}\,.
   \label{eq:all-root-endpoint-contraction-app}
\end{equation}
Besides eq.~\eqref{eq:double-Christoffel-beta-kernel-app}, comparison of the two leading
coefficients in eq.~\eqref{eq:double-Christoffel-orthonormal-polynomial-app} gives
\begin{equation}
   \widehat a_n=a_{n+1}+d_{n+1}(\xi)-d_n(\xi).
   \label{eq:all-root-diagonal-difference-app}
\end{equation}

Suppose first that $\xi\in[a-2b,a+2b]$.  The Jacobi coefficients are bounded and the
off-diagonal ones are bounded away from zero, since $a_n\to a$ and $b_n\to b>0$.  Under these
conditions the Nevai condition in kernel form gives, at every point of this closed band
\cite{Nevai1979,BreuerLastSimon2009},
\begin{equation}
   \frac{p_n(\xi)^2}{\mathcal K_n(\xi)}\longrightarrow0\,,
   \qquad
   \frac{p_{n+1}(\xi)^2}{\mathcal K_n(\xi)}\longrightarrow0\,.
   \label{eq:nevai-condition-full-band-app}
\end{equation}
Consequently $\mathcal K_{n+1}/\mathcal K_n\to1$ and
$\mathcal K_{n-1}/\mathcal K_n\to1$, while Cauchy--Schwarz and boundedness of $b_{n+1}$ give
$d_n(\xi)\to0$.  Equations~\eqref{eq:double-Christoffel-beta-kernel-app}
and~\eqref{eq:all-root-diagonal-difference-app} therefore give
$\widehat b_n\to b$ and $\widehat a_n\to a$.
If $\xi$ is a spectral atom, kernel saturation gives
$\mathcal K_n(\xi)\to\mu(\{\xi\})^{-1}$ and $p_n(\xi)\to0$.  The same two formulas again return
$(a,b)$.  Every remaining real root outside the limiting band is an exterior non-atom. 
Every spectral point outside the essential band is a discrete eigenvalue and, because $e_0$ is cyclic, an atom of $\mu$. An exterior non-atom therefore lies outside $\supp\mu$. The exterior ratio-asymptotic theorem then applies and, in particular, gives eventual nonvanishing of $p_n(\xi)$.  There the
Nevai ratio asymptotics imply
$p_{n+1}(\xi)/p_n(\xi)\to\varphi(\xi)$ with $|\varphi(\xi)|>1$
\cite{Nevai1979,SimonOPRL1}.
Summing the resulting geometric tail gives
$\mathcal K_n(\xi)/p_n(\xi)^2\to(1-|\varphi(\xi)|^{-2})^{-1}$.  Hence the kernel factor in
eq.~\eqref{eq:double-Christoffel-beta-kernel-app} tends to one and
$d_n(\xi)\to b\varphi(\xi)(1-|\varphi(\xi)|^{-2})$.  In particular,
$d_{n+1}(\xi)-d_n(\xi)\to0$.

For a nonreal root $z$, define
\begin{equation}
\begin{aligned}
   \mathcal K_n(z)&=\mathbb K_n^{\mu,\mathrm H}(z,z)=\sum_{j=0}^n|p_j(z)|^2,\\
   q_n(E)&=p_{n+1}(E)
      -\frac{p_{n+1}(z)}{\mathcal K_n(z)}
       \mathbb K_n^{\mu,\mathrm H}(E,z),\\
   d_n(z)&=b_{n+1}
      \frac{p_{n+1}(z)\overline{p_n(z)}}{\mathcal K_n(z)}.
\end{aligned}
   \label{eq:all-root-complex-kernel-app}
\end{equation}

Then $q_n(z)=0$ and $q_n/(E-z)$, after normalization, is orthogonal for
$|E-z|^2\dd\mu$.  The same norm and leading-coefficient calculation as in the real case gives
\begin{equation}
\begin{aligned}
   \widehat\beta_n
   =\beta_{n+1}\frac{\mathcal K_{n+1}(z)\mathcal K_{n-1}(z)}
      {\mathcal K_n(z)^2}\,,
   &\qquad n\ge1\,,\\
   \widehat a_n=a_{n+1}+d_{n+1}(z)-d_n(z),
   &\qquad n\ge0\,.
\end{aligned}
   \label{eq:all-root-complex-Jacobi-app}
\end{equation}
The nonreal ratio asymptotics
$p_{n+1}(z)/p_n(z)\to\varphi(z)$, with $|\varphi(z)|>1$, imply
\begin{equation}
   \frac{\mathcal K_n(z)}{|p_n(z)|^2}
   \longrightarrow\frac{1}{1-|\varphi(z)|^{-2}}\,,
   \qquad
   d_n(z)\longrightarrow b\varphi(z)
      \bigl(1-|\varphi(z)|^{-2}\bigr).
   \label{eq:all-root-complex-ratio-limit-app}
\end{equation}
Therefore the kernel factor in the first formula tends to one and
$d_{n+1}(z)-d_n(z)\to0$.  Thus every real squared factor and every nonreal conjugate-pair factor
preserves $M(a,b)$.  Each factor of $|Q|^2$ is nonnegative on $\RR$, so every intermediate measure is again positive
with infinite support and finite moments of all orders, and lies in $M(a,b)$ by the step just
proved. Its limiting band is unchanged, and its atoms are those of the previous measure less the
transformed real root when that root was an atom.  The classification of each later root is
therefore read off the intermediate measure it acts on. Iteration over the factors of $|Q|^2$ proves
proposition~\ref{prop:nevai-rigidity}.  No step gives a measure-independent convergence rate.

Having established asymptotic closure for arbitrary root locations, we return to exact
finite-level Jacobi formulas for the special preparation $Q=P_r/\sqrt{h_r}$.

\paragraph{Exact root-Gram Jacobi data.}
For a jump to the reference level $r$, the double-Christoffel multiplier
$P_r^2/h_r$ admits a second finite reduction which is particularly well
adapted to Jacobi coefficients.  Let $\mu$ have either infinite support
or finite support of cardinality $d$; in the latter case assume
$0\le r<d$.  Let $\xi_1,\ldots,\xi_r$ be the distinct real zeros of
$P_r$.  With $p_m=P_m/\sqrt{h_m}$, define the following matrices for
$L\ge0$ in infinite support and for $0\le L\le d-1$ in finite support, 
\begin{equation}
\begin{aligned}
   \boldsymbol p_m
   &=\bigl(p_m(\xi_1),\ldots,p_m(\xi_r)\bigr)^{\mathsf T},\\
   \mathsf G_L^{(r)}
   &=\bigl[\mathbb K_L^\mu(\xi_a,\xi_b)\bigr]_{a,b=1}^{r}
   =\sum_{m=0}^{L}\boldsymbol p_m\boldsymbol p_m^{\mathsf T}\,,
   \qquad
   \mathcal D_L^{(r)}=\det\mathsf G_L^{(r)}.
\end{aligned}
   \label{eq:root-Gram-definition-app}
\end{equation}
For $r\ge1$, polynomial interpolation makes the evaluation map
$\cP_L\to\RR^r$ surjective whenever $L$ lies in these ranges and
$L\ge r-1$, so $\mathsf G_L^{(r)}$ is positive definite.
Equivalently, the Cauchy--Binet formula \cite{HornJohnson2013} gives
\begin{equation}
   \mathcal D_L^{(r)}
   =\sum_{0\le m_1<\cdots<m_r\le L}
      \left(
      \det\!\bigl[p_{m_j}(\xi_a)\bigr]_{a,j=1}^{r}
      \right)^2>0\,.
   \label{eq:root-Gram-Cauchy-Binet-app}
\end{equation}

\begin{proposition}[Root-Gram formulas for a reference-level jump]
\label{prop:root-Gram-Jacobi-app}
Suppose $\mu$ has infinite support, or finite support of cardinality
$d$ with $0\le r<d$, and use the root-Gram data above in their stated
ranges.  Let $\dd\nu_r=P_r^2\dd\mu/h_r$, the measure induced by $p_r$ \cite{GautschiLi1993}, and denote its monic orthogonal
polynomials, norms and Jacobi coefficients by $R_n^{(r)}$,
$\widetilde h_n^{(r)}$, $\widetilde a_n^{(r)}$ and
$\widetilde\beta_n^{(r)}=(\widetilde b_n^{(r)})^2$.  In finite
support let $d_r$ be the number of atoms not annihilated by $P_r$; in
infinite support put $d_r=\infty$.  For $L\ge r$ in infinite support,
and for $r\le L\le d-1$ in finite support, set
\begin{equation}
   \eta_L^{(r)}
   =b_L\boldsymbol p_L^{\mathsf T}
      \bigl(\mathsf G_{L-1}^{(r)}\bigr)^{-1}
      \boldsymbol p_{L-1}\,.
   \label{eq:root-Gram-eta-app}
\end{equation}
If the support is infinite, the following formulas hold for every
$n\ge0$.  On finite support, the norm formula holds for the physical rows
$0\le n<d_r$ satisfying $n+r\le d-1$; the hopping formula has the same
range, with $n\ge1$; and the diagonal formula holds while
$n+r\le d-2$,
\begin{align}
   \widetilde h_n^{(r)}
   &=\frac{h_{n+r}}{h_r}
      \frac{\mathcal D_{n+r}^{(r)}}
           {\mathcal D_{n+r-1}^{(r)}}\,,
   \label{eq:root-Gram-norm-app}\\
   \widetilde\beta_n^{(r)}
   &=\beta_{n+r}
      \frac{\mathcal D_{n+r}^{(r)}
            \mathcal D_{n+r-2}^{(r)}}
           {\bigl(\mathcal D_{n+r-1}^{(r)}\bigr)^2}\,,
   \qquad n\ge1\,,
   \label{eq:root-Gram-beta-app}\\
   \widetilde a_n^{(r)}
   &=a_{n+r}+\eta_{n+r+1}^{(r)}-\eta_{n+r}^{(r)}\,.
   \label{eq:root-Gram-a-app}
\end{align}
For $r=0$, every empty determinant, including $\mathcal D_{-1}^{(0)}$,
is one and $\eta_L^{(0)}=0$; the three formulas then reduce to the
reference identities at every physical level. On finite support, all
remaining physical terminal data, including the boundary diagonal at
$n+r=d-1$, are determined by the quotient product-Gram construction of
subsection~\ref{app:finite-multiplication-gram}.  The distinction between the rows covered by eqs.~\eqref{eq:root-Gram-norm-app}--\eqref{eq:root-Gram-a-app} and the terminal rows completed in the quotient matters when fewer than $r$ zeros of $P_r$ coincide with support points, since if $s$ atoms are deleted, then $d_r=d-s$, which may exceed $d-r$.
\end{proposition}

\begin{proof}
Put $L=n+r$ and
$\boldsymbol P_m=(P_m(\xi_1),\ldots,P_m(\xi_r))^{\mathsf T}$.
Multiplication by $P_r$ identifies $R_n^{(r)}$ with the monic
degree-$L$ polynomial that vanishes at the $r$ roots and is orthogonal
to the constrained subspace $P_r\cP_{n-1}$.  Orthogonal projection onto
the value constraints gives 
\begin{equation}
\begin{aligned}
   P_r(E)R_n^{(r)}(E)
   =P_L(E)-\sum_{m=0}^{L-1}\frac{P_m(E)}{h_m}
      \boldsymbol P_m^{\mathsf T}
      \bigl(\mathsf G_{L-1}^{(r)}\bigr)^{-1}
      \boldsymbol P_L\,.
\end{aligned}
   \label{eq:root-Gram-polynomial-app}
\end{equation}
Evaluation at any $\xi_a$ gives zero.  The reproducing property makes
the correction orthogonal to every polynomial of degree at most $L-1$
which vanishes at all $\xi_a$, and hence to $P_r\cP_{n-1}$.
Since $P_L$ is orthogonal to the correction, the matrix determinant
lemma \cite{HornJohnson2013} gives
\begin{equation}
   \bigl\|P_rR_n^{(r)}\bigr\|_\mu^2
   =h_L\frac{\mathcal D_L^{(r)}}{\mathcal D_{L-1}^{(r)}}\,.
\end{equation}
Division by $h_r$ proves eq.~\eqref{eq:root-Gram-norm-app}, and the
ratio of successive norms gives eq.~\eqref{eq:root-Gram-beta-app}.
The coefficient of $P_{L-1}$ in the correction term of
eq.~\eqref{eq:root-Gram-polynomial-app} is
\begin{equation}
   \frac{\boldsymbol P_{L-1}^{\mathsf T}
      (\mathsf G_{L-1}^{(r)})^{-1}\boldsymbol P_L}{h_{L-1}}
   =b_L\boldsymbol p_{L-1}^{\mathsf T}
      (\mathsf G_{L-1}^{(r)})^{-1}\boldsymbol p_L
   =\eta_L^{(r)}.
\end{equation}
Thus the two leading coefficients of $P_rR_n^{(r)}$ are those of $P_L-\eta_L^{(r)}P_{L-1}$. 
Comparison of the next-to-leading coefficients in two consecutive monic
recurrences proves eq.~\eqref{eq:root-Gram-a-app}.
For $r=1$, $\mathcal D_L^{(1)}=\mathcal K_L(\xi_1)$ and the result is
equivalent to proposition~\ref{prop:double-Christoffel-kernel-Jacobi-app}
for the zero $\xi_1$ of $P_1$, on rows where both realizations are defined.
\end{proof}

The zeros of $P_r$ are real and simple~\cite{Chihara1978}, and the proof uses no further property of the reference polynomial: monicity, degree $r$ and distinct real zeros are exactly what make multiplication by $P_r$ identify $R_n^{(r)}$ with the monic degree-$(n+r)$ polynomials vanishing at every node. The same proof therefore gives eqs.~\eqref{eq:root-Gram-norm-app}--\eqref{eq:root-Gram-a-app} for any monic degree-$r$ multiplier with $r$ distinct real zeros, with $h_r$ replaced by its squared $\mu$-norm and $d_r$ by the number of atoms it does not annihilate. This is the arbitrary-polynomial modification treated by Kautsky and Golub~\cite{KautskyGolub1983}; the specialization $Q=P_r$ is the induced-measure problem of Gautschi and Li~\cite{GautschiLi1993}. The normalized reference-level preparation $Q=P_r/\sqrt{h_r}$ is the case carried through this work. The $r=1$ instance of the arbitrary-node extension agrees with proposition~\ref{prop:double-Christoffel-kernel-Jacobi-app} for a general real $\xi$, and specializes to the reference-level jump at $\xi=a_0$.

\paragraph{Continuous-$q$-Hermite specialization.}
For the normalized DSSYK chord Jacobi matrix, let
\begin{equation}
   P_{-1}(E)=0\,,\quad P_0(E)=1\,,\quad
   EP_n(E)=P_{n+1}(E)+[n]_qP_{n-1}(E),\quad
   h_n=[n]_q!\,,
   \label{eq:qhermite-monic-app}
\end{equation}
with $\mu_q$ denoting the spectral measure defined in
subsection~\ref{subsec:chord-context}, we have
$p_n=P_n/\sqrt{h_n}$ and
$p_r(\Jac_q)\ket0_{\Jac}=\ket r_{\Jac}$.  Continuous-$q$-Hermite
linearization \cite{DLMF}, in the DSSYK realization of
ref.~\cite{Berkooz2019DSSYK}, gives the physical chord amplitudes  
\begin{equation}
\begin{aligned}
   g_{m|r}^{\rm ch}(t)
   \equiv g_m^{[p_r]}(t)
   = {}_{\Jac}\!\bra m\ee^{-\ii t\Jac_q}\ket r_{\Jac}
   =\sum_{j=0}^{\min(m,r)}
   \frac{\sqrt{h_mh_rh_{m+r-2j}}}
        {h_jh_{m-j}h_{r-j}}\,
   \phi_{m+r-2j}(t).
\end{aligned}
   \label{eq:qhermite-physical-propagation-app}
\end{equation}
where $\phi_n(t)$ is the empty-chord amplitude.  These amplitudes refer
to the fixed normalized chord-number basis.  The basis rebuilt from $\ket r_{\Jac}$
is instead governed by the induced measure
$\dd\nu_r=P_r^2\dd\mu_q/h_r$, as follows from the general seed-measure identity
\eqref{eq:seed-measure} and the connector construction of section~\ref{sec:finiteband}.

\begin{corollary}[Root-free fixed-chord reconstruction]
\label{cor:qhermite-root-free-app}
Fix $0\le q<1$ and a finite $r\ge1$.
Using the root-free remainder coordinates introduced in
eqs.~\eqref{eq:H1-seed-residue-algebras-main}--\eqref{eq:H1-remainder-recurrence-main}
in the monomial basis of $\RR[E]/\langle P_r\rangle$,
let $\mathsf X_r$ represent multiplication
by $E$, and let $\boldsymbol u_m^{(r)}$ be the coefficient vector of
$P_m\bmod P_r$.  Then
\begin{equation}
\begin{gathered}
   \boldsymbol u_{m+1}^{(r)}
   =\mathsf X_r\boldsymbol u_m^{(r)}
      -[m]_q\boldsymbol u_{m-1}^{(r)}\,,\qquad
   \boldsymbol u_{-1}^{(r)}=0\,,\qquad
   \boldsymbol u_0^{(r)}=(1,0,\ldots,0)^{\mathsf T},\\
   \mathsf H_{L;q}^{(r)}
   =\sum_{m=0}^{L}
      \frac{\boldsymbol u_m^{(r)}
      \boldsymbol u_m^{(r)\mathsf T}}{h_m}\,,
   \qquad
   \Delta_{L;q}^{(r)}=\det\mathsf H_{L;q}^{(r)} \,.
   \label{eq:qhermite-remainder-gram-app}
\end{gathered}
\end{equation}
For $L\ge r-1$, $\mathsf H_{L;q}^{(r)}$ is positive definite.  Moreover
$\Delta_{r-1;q}^{(r)}=\prod_{m=0}^{r-1}h_m^{-1}$ and
$\Delta_{r;q}^{(r)}=\Delta_{r-1;q}^{(r)}$.  If
$R_n^{(r)}$ and $\thh_n^{(r)}$ are the monic orthogonal polynomials
and squared norms for $\nu_r$, their Jacobi data are
\begin{equation}
\begin{aligned}
   \widetilde a_n^{(r)}&=0\,,\\
   \thh_n^{(r)}
   &=\frac{h_{n+r}}{h_r}
     \frac{\Delta_{n+r;q}^{(r)}}{\Delta_{n+r-1;q}^{(r)}}\,,\\
   \bigl(\widetilde b_n^{(r)}\bigr)^2
   &=[n+r]_q
     \frac{\Delta_{n+r;q}^{(r)}
           \Delta_{n+r-2;q}^{(r)}}
          {(\Delta_{n+r-1;q}^{(r)})^2}\,,
   \qquad n\ge1\,.
   \label{eq:qhermite-shifted-data-app}
\end{aligned}
\end{equation}
For $L=n+r$, the corresponding polynomial is determined without
extracting any zero of $P_r$ by
\begin{equation}
   P_r(E)R_n^{(r)}(E)=P_L(E)
   -\sum_{m=0}^{L-1}\frac{P_m(E)}{h_m}
    \boldsymbol u_m^{(r)\mathsf T}
    (\mathsf H_{L-1;q}^{(r)})^{-1}\boldsymbol u_L^{(r)}.
   \label{eq:qhermite-shifted-polynomial-app}
\end{equation}
Its expansion in the orthonormal $p_m$ basis gives every rebuilt
Krylov vector and hence every amplitude
$\tphi_n^{(r)}(t)$.  More directly, for $L\ge r$ define
\begin{equation}
   \boldsymbol z_L^{(r)}(t)=
   \sum_{m=0}^{L}
      \frac{\boldsymbol u_m^{(r)}}{\sqrt{h_m}}
      g_{m|r}^{\rm ch}(t).
\end{equation}
Then the cumulative probability through rebuilt level $L-r$ and the
spread complexity are
\begin{equation}
\begin{aligned}
   \Phi_{L-r}^{(r)}(t)
   &=\sum_{m=0}^{L}|g_{m|r}^{\rm ch}(t)|^2
     -\boldsymbol z_L^{(r)}(t)^\dagger
      (\mathsf H_{L;q}^{(r)})^{-1}\boldsymbol z_L^{(r)}(t),\\
   K_r(t)&=\sum_{\ell\ge0}\bigl[1-\Phi_\ell^{(r)}(t)\bigr].
   \label{eq:qhermite-cumulative-app}
\end{aligned}
\end{equation}
The final series converges uniformly on compact time intervals.
\end{corollary}

\begin{proof}
Reducing eq.~\eqref{eq:qhermite-monic-app} modulo $P_r$ gives the
remainder recurrence.  It is the continuous-$q$-Hermite specialization of
eq.~\eqref{eq:H1-remainder-recurrence-main}, with $[m]_q$ in place of $\lambda m$ and no
diagonal term.  Because $P_0,\ldots,P_{r-1}$ are monic of successive degrees, their residue classes
form a basis of $\RR[E]/\langle P_r\rangle$.  Hence $\mathsf H_{L;q}^{(r)}$ is positive definite
for $L\ge r-1$.  At $L=r-1$ its Gram determinant is
$\prod_{m=0}^{r-1}h_m^{-1}$, while $P_r\bmod P_r=0$ gives
$\mathsf H_{r;q}^{(r)}=\mathsf H_{r-1;q}^{(r)}$ and proves the two initial determinant identities.

Evaluation at the distinct roots of $P_r$
maps the quotient basis to the root-evaluation basis by a nonsingular
Vandermonde matrix.  Thus $\mathsf H_{L;q}^{(r)}$ is congruent to the
root-Gram matrix of proposition~\ref{prop:root-Gram-Jacobi-app};
the fixed congruence factor cancels from every determinant ratio. Componentwise this is $\mathsf T_r\boldsymbol u_m^{(r)}=\boldsymbol P_m$, where $\mathsf T_r$ is
the Vandermonde matrix of $\xi_1,\ldots,\xi_r$ and
$\boldsymbol P_m=\bigl(P_m(\xi_a)\bigr)_{a=1}^{r}$, so that
$\mathsf G_L^{(r)}=\mathsf T_r\mathsf H_{L;q}^{(r)}\mathsf T_r^{\mathsf T}$ and the bilinear forms
transport together with the determinants, since
$\boldsymbol u_m^{(r)\mathsf T}(\mathsf H_{L-1;q}^{(r)})^{-1}\boldsymbol u_L^{(r)}
=\boldsymbol P_m^{\mathsf T}(\mathsf G_{L-1}^{(r)})^{-1}\boldsymbol P_L$.
In the same coordinates $\mathsf T_r\boldsymbol z_L^{(r)}(t)$ is the vector of overlaps of the
evolved state with the reference kernels $\mathbb K_L^\mu(\,\cdot\,,\xi_a)$, so the term subtracted
in eq.~\eqref{eq:qhermite-cumulative-app} is the squared norm of the projection onto their span.
That proposition gives the norm, hopping and polynomial formulas,
while parity of $P_r^2\dd\mu_q$ gives
$\widetilde a_n^{(r)}=0$.  Equation~\eqref{eq:qhermite-cumulative-app}
is the same projection written in quotient coordinates, obtained by applying
eq.~\eqref{eq:coeff-projector-app} to the remainder vectors.  Boundedness
of $\Jac_q$ for $q<1$ provides the compact-time first-moment tail bound.

Indeed, the shifted measure has the same compact support as the continuous-$q$-Hermite
measure, so $\|\widetilde\Jac_r\|\le R_q=2/\sqrt{1-q}$.  Tridiagonality then gives
$|\tphi_n^{(r)}(t)|\le
\sum_{k=n}^{\infty}(R_q|t|)^k/k!$.  The square of this factorial tail is summable with weight $n$,
uniformly for $t$ in a compact interval.  This proves convergence of the first-moment series and its
asserted compact-time uniformity. The shifted density assumption of section~\ref{sec:setup} is automatic on this compact support, so
$\sum_{n\ge0}|\tphi_n^{(r)}(t)|^2=1$ and eq.~\eqref{eq:tail-complexity-infinite-main} identifies
this first moment with the tail series displayed in
eq.~\eqref{eq:qhermite-cumulative-app}.
\end{proof}

The first two rebuilt hoppings provide useful local checks,
\begin{equation}
   \bigl(\widetilde b_1^{(r)}\bigr)^2=[r]_q+[r+1]_q,
   \qquad
   \bigl(\widetilde b_2^{(r)}\bigr)^2=
   \frac{[r]_q[r-1]_q+[r+1]_q[r+2]_q}
        {[r]_q+[r+1]_q}.
   \label{eq:qhermite-first-hoppings-app}
\end{equation}
In the normalization of eq.~\eqref{eq:qhermite-monic-app}, fixed $r$
and any fixed reconstructed level pass directly to the Hermite/Fock
data as $q\uparrow1$, with $t=\tau_{\rm H}$; this coefficientwise limit
does not by itself interchange $q\uparrow1$ with the infinite
first-moment sum.  At $q=0$ the reference problem is the infinite
half-line constant chain, whereas fixed $q<1$ and $r\to\infty$ gives
the folded bulk of proposition~\ref{prop:deep-seed-folded-bulk-app}.
Neither limit replaces the quotient construction required by a
finite terminal chain.

This corollary is the constructive, model-specific complement of the preceding root-Gram
proposition.  The root-Gram formulas apply to a general reference measure but require the zeros of
$P_r$.  Here the continuous-$q$-Hermite recurrence replaces those roots by an explicit $r$-component
remainder sequence and, in the same coordinates, reconstructs every shifted amplitude and the full
finite-time complexity.  The statement keeps $r$ fixed and finite.  It is therefore distinct from
the growing-degree limit below, and it does not address rank loss or terminal folding, for which the
physical quotient by $T_d$ in appendix~\ref{app:finite} is essential.

\paragraph{Controlled growing-degree limits on compact support.}
\label{app:controlled-growing-degree}
The fixed-degree closure of proposition~\ref{prop:nevai-rigidity} does not control a seed whose
degree grows with the depth of the reference chain.  Two limits remain accessible without a
uniform Christoffel estimate.

\begin{proposition}[Deep-seed arcsine limit and folded bulk]
\label{prop:deep-seed-folded-bulk-app}
Let $\mu\in M(a,b)$, the Nevai class defined in section~\ref{sec:rigidity} by
$a_n\to a$ and $b_n\to b>0$,
and let $p_r$ be its orthonormal polynomials.  The reference-level seed
measures $\dd\nu_r=p_r^2\dd\mu$ converge weakly to
\begin{equation}
   \dd\rho_{\rm arc}^{a,b}(E)
   =\frac{\boldsymbol 1_{\{|E-a|<2b\}}\dd E}
          {\pi\sqrt{4b^2-(E-a)^2}}\,.
   \label{eq:deep-seed-arcsine-app}
\end{equation}
This weak limit is due to Nevai~\cite{Nevai1979}; an operator-theoretic proof in the present recurrence-coefficient setting is given in ref.~\cite{VanAssche1995}.
For every fixed shifted level $\ell$,
\begin{equation}
   \widetilde a_\ell^{(r)}\longrightarrow a\,,
   \qquad
   \widetilde b_1^{(r)}\longrightarrow\sqrt2\,b\,,
   \qquad
   \widetilde b_\ell^{(r)}\longrightarrow b\quad(\ell\ge2).
   \label{eq:deep-seed-folded-Jacobi-app}
\end{equation}
In the positive Lanczos gauge, the corresponding shifted basis vectors obey, for every fixed
$\ell\ge1$,
\begin{equation}
   \left\|
      \ket{\widetilde K_\ell^{(r)}}
      -\frac{\ket{K_{r-\ell}}+\ket{K_{r+\ell}}}{\sqrt2}
   \right\|\longrightarrow0\,.
   \label{eq:deep-seed-folded-basis-app}
\end{equation}
The shifted complexities converge uniformly on every compact time interval to the complexity of
this folded constant chain.
\end{proposition}

\begin{proof}
For every fixed integer $m\ge0$,
\begin{equation}
   \int E^m p_r(E)^2\dd\mu(E)=\langle e_r,\Jac^m e_r\rangle.
   \label{eq:deep-seed-reference-bulk-app}
\end{equation}
Only Jacobi coefficients within distance $m$ of the site $r$ enter this matrix element.  They converge locally to those of the bilateral constant operator
$\Jac_\infty=a+b(S+S^{-1})$ on $\ell^2(\mathbb Z)$.  Its spectral measure at the origin is
eq.~\eqref{eq:deep-seed-arcsine-app}.  Compact support makes the moment problem determinate, so
moment convergence is weak convergence \cite{VanAssche1995}.  Positivity of the limiting finite
Hankel determinants then gives eq.~\eqref{eq:deep-seed-folded-Jacobi-app}.

For eq.~\eqref{eq:deep-seed-folded-basis-app}, apply the Lanczos recurrence to the reference
site $e_r$.  At fixed $\ell$, only Jacobi coefficients within distance $\ell$ of $r$ enter, and
they converge to the bilateral constant operator
$\Jac_\infty=a+b(S+S^{-1})$.  Lanczos at its origin gives
$v_0=e_0$ and
$v_\ell=(e_{-\ell}+e_\ell)/\sqrt2$ for $\ell\ge1$, with first hopping
$\sqrt2\,b$ and all subsequent hoppings $b$; this follows directly by induction in the
three-term recurrence.  The same finite-step induction proves the norm convergence in the
reference Jacobi basis.

Writing $z=2bt$, the limiting amplitudes are
\begin{equation}
   \phi_0^{\rm arc}(t)=\ee^{-\ii at}J_0(z),
   \qquad
   \phi_n^{\rm arc}(t)=\sqrt2\,\ee^{-\ii at}(-\ii)^nJ_n(z),
   \quad n\ge1\,.
   \label{eq:deep-seed-folded-dynamics-app}
\end{equation}
The Bessel recurrence and the telescoping identity for
$\sum_{n\ge0}J_n(z)J_{n+1}(z)$ give
\begin{equation}
   K_{\rm arc}(t)=2\sum_{n\ge1}nJ_n(z)^2
   =z^2\bigl[J_0(z)^2+J_1(z)^2\bigr]-zJ_0(z)J_1(z).
   \label{eq:deep-seed-folded-complexity-app}
\end{equation}
All shifted Jacobi matrices have a common norm bound $R$. Indeed $a_n$ and $b_n$ converge, so $\Jac$ is bounded and $\supp\mu$ is compact; each $\dd\nu_r=p_r^2\dd\mu$ is supported in $\supp\mu$, and $\widetilde\Jac_r$ is multiplication by $E$ on the associated polynomial subspace, so $\|\widetilde\Jac_r\|\le R=\sup_{E\in\supp\mu}|E|$ for every $r$.  Since a path from level zero to level
$n$ uses at least $n$ powers of the Jacobi matrix,
\begin{equation}
   \bigl|\langle e_n,\ee^{-\ii t\widetilde\Jac_r}e_0\rangle\bigr|
   \le\sum_{k=n}^{\infty}\frac{(R|t|)^k}{k!}\,.
   \label{eq:compact-Jacobi-path-tail-app}
\end{equation}
The square of this bound is summable with weight $n$, uniformly on compact time intervals.
Together with fixed-row convergence it proves the complexity limit.
\end{proof}

\begin{proposition}[Closure under compact spectral filters]
\label{prop:compact-filter-closure-app}
Let $H$ be bounded and let its cyclic spectral measure be supported in a compact interval.  Let
$F$ be continuous on that support, with
$N_F=\|F(H)\ket{K_0}\|^2>0$, and suppose that $|F|^2\mu$ has infinite support.  If polynomials
$Q_d$ converge uniformly to $F$ on the support, then the normalized seed states converge in
Hilbert norm, the normalized seed measures converge in total variation, and every fixed shifted
Jacobi coefficient converges.  Moreover, for every $T<\infty$,
\begin{equation}
   \sup_{|t|\le T}|K_{Q_d}(t)-K_F(t)|\longrightarrow0\,.
   \label{eq:compact-filter-complexity-closure-app}
\end{equation}
\end{proposition}

\begin{proof}
Uniform convergence and
$\bigl||Q_d|^2-|F|^2\bigr|\le|Q_d-F|(|Q_d|+|F|)$ give convergence of the normalizers, the states
and the normalized measures in total variation.  Since $N_{Q_d}\to N_F>0$, the normalizer is positive for every $d$ beyond some index, and the normalized seed state and seed measure are defined from that index onward; the convergence statements are read along that tail.  There $|Q_d|^2\dd\mu/N_{Q_d}$ again has infinite support, because $\mu$ does and $Q_d$ has finitely many zeros, so every fixed shifted Jacobi coefficient is defined.  Compact support then gives convergence of every
moment.  The finite Hankel determinants of the limiting infinite-support measure are positive, so
the moment formulas give convergence of each fixed Jacobi row and hence of each fixed amplitude.
The common path bound~\eqref{eq:compact-Jacobi-path-tail-app} controls the complexity tails
uniformly on $|t|\le T$, which proves eq.~\eqref{eq:compact-filter-complexity-closure-app}.
\end{proof}

If $F$ extends analytically to a complex neighborhood of the ambient interval, Chebyshev
approximants converge geometrically.  The state, measure and every fixed Jacobi row then inherit a
row-dependent geometric rate.  No uniform geometric rate for the full complexity follows without
an additional tail estimate. 

The preceding closure results concern infinite-support limits on compact spectral support.
We now return to finite support, where continuation beyond the terminal degree requires an explicit
quotient representative.

\paragraph{Terminal lifts and finite-support determinants.}
The preceding determinant is literal whenever every polynomial in its top row lies below the terminal degree. Determinantal Christoffel formulas for classical discrete measures provide related background~\cite{DuranChristoffel2021}. 
The following statement gives the quotient-algebra continuation through the finite-chain edge.

\begin{proposition}[Terminal lifts and determinant formulas]
\label{prop:terminal-lift-determinant}
Let $\mu$ have finite support of size $d$, terminal polynomial $T_d$, and quotient algebra $\cA_d=\CC[E]/\langle T_d\rangle$. Let $W$ be a degree-$L$ polynomial with leading coefficient $w_L>0$, nonnegative on $\supp\mu$ and positive at $d_W\ge1$ support points. Fix $0\le n<d_W$, equivalently a level for which $W\dd\mu$ is positive definite through degree $n$. Choose a proper terminal lift in the sense of eq.~\eqref{eq:proper-terminal-lift-main}.

Form the lifted row
\begin{equation}
   \widehat{\mathbf P}_n(E)=
   \begin{pmatrix}
      \widehat P_n(E)&\widehat P_{n+1}(E)&\cdots&
      \widehat P_{n+L}(E)
   \end{pmatrix},
\end{equation}
and let $\widehat{\mathsf C}_n$ and $\widehat\Delta_n$ denote its constraint minor and determinant. Then $\widehat\Delta_n\ne0$, and the preceding determinant formula, with $(\mathbf P_n,\Delta_n)$ replaced by $(\widehat{\mathbf P}_n,\widehat\Delta_n)$, gives $R_n$. Equivalently, its normalized numerator represents the ordinary polynomial $WR_n$. Reduction modulo $T_d$ gives the physical finite-dimensional connector, independently of the auxiliary polynomials $\Lambda_k$.
\end{proposition}

\begin{proof}
Since the lifted polynomials are monic of their indicated degrees,
\begin{equation}
   \widehat{\mathcal V}_{n,L}
   =\Span\{\widehat P_n,\ldots,\widehat P_{n+L}\}
   =\Span\{P_n,\ldots,P_{\min(n+L,d-1)}\}
   \oplus T_d\cP_{n+L-d}\,,
   \label{eq:terminal-lift-span-app}
\end{equation}
where $\cP_k=\{0\}$ for $k<0$. Thus, the second summand is absent when $n+L<d$. Let $R_n$ be the monic degree-$n$ polynomial orthogonal with respect to $W\dd\mu$, and let $\rho_n$ be the canonical remainder of $WR_n$ modulo $T_d$. Shifted orthogonality gives
\begin{equation}
   \rho_n\in\Span\{P_n,\ldots,P_{d-1}\}.
\end{equation}
If $n+L<d$, no reduction occurs and $\rho_n=WR_n\in\widehat{\mathcal V}_{n,L}$. If $n+L\ge d$, then
\begin{equation}
   WR_n-\rho_n=T_dA,
   \qquad
   \deg A\le n+L-d\,.
\end{equation}
The monic family $\Lambda_0,\ldots,\Lambda_{n+L-d}$ is a basis of $\cP_{n+L-d}$, so eq.~\eqref{eq:terminal-lift-span-app} again gives $WR_n\in\widehat{\mathcal V}_{n,L}$.

It remains to show that the lifted constraint system is nonsingular. Suppose $S\in\Span\{\widehat P_n,\ldots,\widehat P_{n+L-1}\}$ satisfies the homogeneous root conditions. Then $S=WT$ with $\deg T<n$. For $n=0$, this degree bound already gives $T=0$. For $n\ge1$, the quotient class of $S$ lies in $\Span\{P_n,\ldots,P_{d-1}\}$ and is orthogonal to $\cP_{n-1}$ in $L^2(\mu)$. Hence
\begin{equation}
   0=\int S(E)\overline{T(E)}\dd\mu(E)
   =\int W(E)|T(E)|^2\dd\mu(E).
\end{equation}
The condition $n<d_W$ makes the shifted inner product positive definite on $\cP_{n-1}$, so $T=0$ also in this case. The lifted constraint system is therefore nonsingular, $\widehat\Delta_n\ne0$, and the determinant formula applies to the lifted row. Multiplying the lifted determinant by $(-1)^Lw_L/\widehat\Delta_n$ produces a polynomial with the same root data and leading coefficient as $WR_n$. Expanding the lifted determinant along its top row exhibits it as a combination of $\widehat P_n,\ldots,\widehat P_{n+L}$, so the rescaled determinant and $WR_n$ both lie in $\widehat{\mathcal V}_{n,L}$. Since $\widehat P_{n+L}$ is monic and is the only spanning element of degree $n+L$, the matched leading coefficients give the two polynomials the same component along it. Their difference therefore lies in $\Span\{\widehat P_n,\ldots,\widehat P_{n+L-1}\}$ and satisfies the homogeneous root conditions, so the argument just given identifies the two. The shifted measure fixes \(WR_n\), and hence its class in \(\cA_d\), independently of the terminal lift. The auxiliary constraint minor and the unreduced connector coefficients may nevertheless depend on the chosen representatives.
\end{proof}

\paragraph{Degree-one seeds.}
Consider the normalized seed of eq.~\eqref{eq:degree-one-seed-main}, with $|c_0|^2+|c_1|^2=1$ and $h_1>0$. For $c_1\ne0$,
\begin{equation}
   Q(E)=c(E-z),
   \qquad
   c=\frac{c_1}{\sqrt{h_1}},
   \qquad
   z=a_0-\frac{c_0\sqrt{h_1}}{c_1}\,.
\end{equation}
Since $N_Q=1$, the multiplier is
$
   W_Q(E)=|c|^2(E-z)(E-\bar z).
$
For $z\notin\RR$, the conjugate pair is essential, since a single complex Christoffel factor does not define a positive measure on the real spectrum, whereas $W_Q(E)=|Q(E)|^2\ge0$. Away from a finite terminal edge, eq.~\eqref{eq:degree-one-connector-main} becomes
\begin{equation}
   W_Q(E)R_n^{[Q]}(E)
   =\Gamma_{n,n}P_n(E)
   +\Gamma_{n,n+1}P_{n+1}(E)
   +|c|^2P_{n+2}(E).
\end{equation}
The constraints at $z$ and $\bar z$ give
\begin{equation}
\begin{aligned}
   \Gamma_{n,n}
   &=|c|^2
   \frac{P_{n+1}(z)P_{n+2}(\bar z)-P_{n+2}(z)P_{n+1}(\bar z)}
   {P_n(z)P_{n+1}(\bar z)-P_n(\bar z)P_{n+1}(z)}\,,\\
   \Gamma_{n,n+1}
   &=|c|^2
   \frac{P_{n+2}(z)P_n(\bar z)-P_n(z)P_{n+2}(\bar z)}
   {P_n(z)P_{n+1}(\bar z)-P_n(\bar z)P_{n+1}(z)}\,.
\end{aligned}
\label{eq:degree-one-complex-root-app}
\end{equation}
The denominator is the corresponding two-by-two constraint determinant and is nonzero. In the Hermitian-kernel convention of section~\ref{sec:kernels}, the Christoffel--Darboux identity writes it as
$
   \bigl(-h_n(z-\bar z)\mathbb K_n^{\mu,\mathrm H}(z,z)\bigr),
$
where $\mathbb K_n^{\mu,\mathrm H}(z,z)=\sum_{k=0}^{n}|P_k(z)|^2/h_k>0$.

For a real root the two constraints coalesce to $S_n(z)=S_n'(z)=0$. Setting $z=x+\ii y$ in eq.~\eqref{eq:degree-one-complex-root-app} and letting $y\to0$, with the limiting real root again denoted by $z$, gives the Wronskian limits
\begin{equation}
\begin{aligned}
   \Gamma_{n,n}
   &=|c|^2
   \frac{P_{n+1}(z)P'_{n+2}(z)-P'_{n+1}(z)P_{n+2}(z)}
   {P_n(z)P'_{n+1}(z)-P'_n(z)P_{n+1}(z)}\,,\\
   \Gamma_{n,n+1}
   &=|c|^2
   \frac{P'_n(z)P_{n+2}(z)-P_n(z)P'_{n+2}(z)}
   {P_n(z)P'_{n+1}(z)-P'_n(z)P_{n+1}(z)}\,.
\end{aligned}
\label{eq:degree-one-real-root-app}
\end{equation}
Here the common denominator is
$h_n\mathbb K_n^\mu(z,z)>0$. The pure first jump has $c_0=0$, hence $z=a_0$ and $|c|^2=1/h_1$. These formulas give the explicit connector data used in the Charlier analysis of subsection~\ref{app:H1-first-jump}, where the Wronskians reduce to the determinant sequences $D_n$ and $E_n$. In finite dimension they apply directly when $n+2\le d-1$; for every remaining physical row $0\le n<d_Q$, the same expressions are evaluated with proper lifts and reduced modulo $T_d$.

Beyond the row-by-row constraint solution, the connector has an operator meaning on the Jacobi chain itself, which is the form matched by the ladder-algebra chains of section~\ref{sec:examples}. The Jacobi decomposition in eq.~\eqref{eq:Jacobi-ladder-main} obeys
\begin{equation}
   [\mathsf B_-,\mathsf B_+]e_n
   =(b_{n+1}^2-b_n^2)e_n\,.
\end{equation}
The Heisenberg--Weyl chain of subsection~\ref{subsec:HW} closes this weighted-shift structure into its familiar ladder algebra, while the constant-coefficient half-line chain of subsection~\ref{subsec:tight-binding} retains a boundary projection. In the single-factor Darboux step described after eq.~\eqref{eq:Jacobi-ladder-main}, denote the lower-bidiagonal Cholesky factor with positive diagonal by $\mathsf L_\xi$. For $\xi<\inf\supp\mu$, and whenever the infinite-operator factorization exists on its natural domain, interchanging the two Cholesky factors gives the Jacobi matrix of the normalized measure $(E-\xi)\dd\mu/(a_0-\xi)$; the normalization does not affect the Jacobi coefficients. For the complete seed multiplier $W_Q$, zero-weight atoms are removed at the measure level before the reduced Jacobi chain is formed.

The determinant formulas above resolve the connector constraints imposed
by divisibility by $W_Q$.  The next appendix turns to the distinct
constraints defining the image $\Qhat\cP_\ell$ and organizes their
orthogonal complement as a finite-rank projector, the form naturally
adapted to cumulative probabilities and spread complexity.

\section{Projected kernels and finite-rank constraints}
\label{app:projected-kernel}

Appendix~\ref{app:christoffel} determines individual connector rows by
imposing divisibility by the full polynomial multiplier
$W_Q=\Qsh Q/N_Q$.  For a degree-$r$ seed this is a degree-$2r$ constraint.
Cumulative probabilities involve a related but
lower-order constraint.  Multiplication by $\Qhat$ embeds the shifted
degree-$\ell$ filtration as the image $\Qhat\cP_\ell$ and imposes only the $r$ constraints associated with the roots of $Q$. Its
orthogonal projector gives the projected-kernel formula of
section~\ref{sec:kernels}.  We derive this projector first through
derivative jets, where the excluded directions are explicit, and then
in orthonormal reference coordinates, which gives the intrinsic
formulation for complex seeds and finite terminal quotients.  Throughout,
$\ell$ precedes shifted termination; in finite support,
$0\le\ell\le d_Q-1$.

\paragraph{Jet realization before the terminal edge.}
Let $\cP_m$ denote the polynomials of degree at most $m$, equipped with
the inner product inherited from $\mu$.  In finite support we restrict
here to $m\le d-1$.  The reference kernel $\mathbb K_m^\mu$ reproduces this
space, so that
\begin{equation}
   S(E)=\int \mathbb K_m^\mu(E,F)S(F)\dd\mu(F),
   \qquad S\in\cP_m\,.
\end{equation}
If $Q$ has degree $r$, multiplication by
$\Qhat=Q/\sqrt{N_Q}$ is an isometry from $(\cP_\ell,\nu_Q)$ onto
$\Qhat\cP_\ell\subset\cP_{\ell+r}$.  For the jet realization, assume
that $Q$ has real coefficients and only real roots.  With pairwise
distinct $\xi_a$ and $m_a\ge1$, write
\begin{equation}
   Q(E)=c\prod_{a=1}^{s}(E-\xi_a)^{m_a},
   \qquad
   \sum_{a=1}^{s}m_a=r\,.
\end{equation}
The image consists precisely of the polynomials in $\cP_{\ell+r}$ whose
jets vanish to the corresponding orders,
\begin{equation}
   S^{(j)}(\xi_a)=0,
   \qquad
   a=1,\ldots,s,
   \qquad
   j=0,\ldots,m_a-1,
   \qquad
   S\in\cP_{\ell+r}\,.
   \label{eq:kernel-jet-constraints-app}
\end{equation}
We use multi-indices $\gamma=(a,j)$ and $\eta=(b,i)$ for these $r$
functionals, and define their kernel representers and Gram matrix by 
\begin{equation}
\begin{aligned}
   k_{(a,j)}(E)
   &=\left.
      \partial_F^j \mathbb K_{\ell+r}^{\mu}(E,F)
      \right|_{F=\xi_a},\\
   G_{(a,j),(b,i)}
   &=\langle k_{(a,j)},k_{(b,i)}\rangle_\mu
   =\left.
      \partial_E^j\partial_F^i
      \mathbb K_{\ell+r}^{\mu}(E,F)
      \right|_{E=\xi_a,F=\xi_b}.
\end{aligned}
   \label{eq:jet-gram-app}
\end{equation}
Indeed,
$\langle k_{(a,j)},S\rangle_\mu=S^{(j)}(\xi_a)$.
The constraints are independent in infinite support and, for finite
support, whenever $\ell+r\le d-1$. Since $\sum_{a=1}^{s}m_a=r$, the confluent interpolation theorem gives,
for arbitrary prescribed values of the derivatives
$S^{(j)}(\xi_a)$, $0\le j<m_a$, a unique polynomial $S$ of degree at
most $r-1$. Since $\cP_{r-1}\subset\cP_{\ell+r}$, the corresponding jet
functionals are linearly independent on $\cP_{\ell+r}$  and $G$ is positive definite. Within $\cP_{\ell+r}$, the orthogonal complement of their span is
$\Qhat\cP_\ell$.  Subtracting the corresponding finite-rank projector
from the identity gives
\begin{equation}
   \Qhat(E)\Qhat(F)\tK_\ell^{[Q]}(E,F)
   =\mathbb K_{\ell+r}^{\mu}(E,F)
   -\sum_{\gamma,\eta}
      k_\gamma(E)(G^{-1})_{\gamma\eta}k_\eta(F).
   \label{eq:jet-kernel-app}
\end{equation}
This is the confluent derivative-jet form of
eq.~\eqref{eq:projected-kernel-main}.

The degree condition is essential in finite support.  Once
$\ell+r\ge d$, the polynomial Hilbert space is the quotient
$\cA_d$, rather than an ordinary degree-$\ell+r$ polynomial space.
Evaluation descends to this quotient only at points of the spectral
support, while derivative jets do not descend in general.  The jet
description is therefore no longer intrinsic, and the exact object is
the quotient-image projector constructed below.\footnote{Proper
terminal lifts extend connector determinants at the level of polynomial
representatives, but they cannot extend the reproducing-kernel sum.
Every polynomial in the null ideal $(T_d)$ has zero $L^2(\mu)$ norm.}

\paragraph{Hermitian coefficient-space realization.}
The preceding formula uses a bilinear kernel and real derivative data.
For complex seed coefficients, or near a finite terminal edge, we
represent $\Qhat\cP_\ell$ directly in the orthonormal reference basis.
Set $p_m=P_m/\sqrt{h_m}$ and let
$M_\ell$ be the cutoff in eq.~\eqref{eq:projection-cutoff-main}.  Define
the $(M_\ell+1)\times(\ell+1)$ coefficient matrix $C_{Q,\ell}$ by
\begin{equation}
   E^j\Qhat(E)
   =\sum_{m=0}^{M_\ell}(C_{Q,\ell})_{mj}p_m(E),
   \qquad j=0,\ldots,\ell,
\end{equation}
where, in finite dimension, both sides are understood after reduction
modulo $T_d$.  Its column space is precisely the coordinate image of
$\Qhat\cP_\ell$, and the reference inner product becomes the Euclidean
inner product on these columns.  The Hermitian orthogonal projector onto
the image is consequently
\begin{equation}
   \Pi_{Q,\ell}
   =C_{Q,\ell}C_{Q,\ell}^{+}
   =C_{Q,\ell}
   (C_{Q,\ell}^\dagger C_{Q,\ell})^+
   C_{Q,\ell}^\dagger\,.
   \label{eq:coeff-projector-app}
\end{equation}
Here $+$ denotes the Moore--Penrose inverse.  For the canonical columns
above and $0\le\ell<d_Q$, multiplication by $Q$ is injective on
$\cP_\ell$.  In finite support, a column relation would give a
polynomial of degree at most $\ell$ vanishing at all $d_Q$ surviving
atoms, and hence the zero polynomial.  In infinite support, the same
conclusion follows because a nonzero polynomial cannot vanish on the
support of $\mu$.  Thus $C_{Q,\ell}$ has full column rank and the
ordinary inverse suffices.  The pseudoinverse makes the projector independent of
redundant spanning columns and keeps the same formula valid for
rank-revealing numerical representations.

With the reference-coordinate amplitudes of
eq.~\eqref{eq:gq-def}, the squared norm of this projection is
\begin{equation}
\begin{aligned}
   \Phi_\ell^{[Q]}(t)
   &=
   \bigl(g_{\le M_\ell}^{[Q]}(t)\bigr)^\dagger
   \Pi_{Q,\ell}\,
   g_{\le M_\ell}^{[Q]}(t),\\
   g^{[Q]}(t)
   &=\Qhat(\Jac)\phi^{(0)}(t).
\end{aligned}
   \label{eq:coeff-cumulative-app}
\end{equation}
This proves the coordinate formula
eq.~\eqref{eq:projection-coeff-main}.  It is root-free, Hermitian for
arbitrary complex $Q$, and remains exact after quotient reduction.  It
also avoids evaluating Christoffel--Darboux kernels near removable
singularities.  Numerically, the same range projector may be formed by a
rank-revealing QR factorization or singular-value decomposition, without
forming the normal matrix, as discussed in appendix~\ref{app:numerics}.

For the first jump, $Q=P_1/\sqrt{h_1}$ has the single root
$\alpha=a_0$.  In infinite support, or in finite support with
$\ell+1\le d-1$, $\Qhat\cP_\ell$ is the codimension-one subspace of
$\cP_{\ell+1}$ annihilated by evaluation at $\alpha$.  Its excluded
direction is $\mathbb K_{\ell+1}^{\mu}(\,\cdot\,,\alpha)$, whose squared norm is
$\mathbb K_{\ell+1}^{\mu}(\alpha,\alpha)$.  The rank-one subtraction in
eq.~\eqref{eq:first-jump-kernel} follows immediately. Thus $\Phi_\ell^{(1)}(t)$ is the cumulative probability through reference
level $\ell+1$, with the evaluation direction removed.

Beyond the faithful polynomial range, the quotient projector in
eq.~\eqref{eq:coeff-projector-app} gives the same cumulative quantity.
In the Charlier problem, subsection~\ref{app:H1-first-jump} combines this rank-one
evaluation projector with the first-jet connector data of
appendix~\ref{app:christoffel} and converts both into the recurrence
sequences used for the explicit infinite-chain analysis.

\section{Charlier number-state jumps and remainder sequences}
\label{app:H1}

In the Charlier chain of subsection~\ref{subsec:HW}, the projected-kernel constraints of appendix~\ref{app:projected-kernel} and the confluent connector of appendix~\ref{app:christoffel} admit finite-dimensional recurrence realizations. The factorization $I_n(\tau)=\cS(\tau)q(\tau)^n$ in eq.~\eqref{eq:H1-old-moments-main} isolates the time dependence, so the transformation associated with a fixed seed level reduces to finite algebra. For $r=1$, this algebra is carried by value and first-jet sequences at the origin. At general $r$, it becomes a pair of vector remainder sequences modulo $P_r$ and $P_r^2$. The rank-one and three-component reductions recover the explicit $r=1$ and
$r=3$ constructions. 

At the Hermite endpoint, the shifted weight is $\He_r(y)^2\dd\gamma(y)/r!$ for every fixed
$r$.  Proposition~\ref{prop:Hermite-all-r-complexity-app} gives every rebuilt amplitude through a
root-free scalar recurrence, together with an exact convergent complexity series and radial
comparison.  The $r=1$ hierarchy is the special case whose parity sums reduce to a finite
expression.  Finiteness and the short-time expansion are proved directly for the untruncated
Hermite chain; they are not inferred by interchanging the $q\uparrow1$ limit with an infinite
continuous-$q$-Hermite first-moment sum.
Coefficientwise convergence as $q\uparrow1$ controls every fixed rebuilt level, but the
complexity is an unbounded first moment over all levels.  Such convergence does not imply uniform
integrability of the Krylov-level distributions, so it does not justify exchanging the endpoint
limit with the infinite sum.  The direct Hermite argument below establishes the missing tail
control at every finite time.

\subsection{The first Charlier jump and the Hermite endpoint}
\label{app:H1-first-jump}

The double Christoffel point for the first jump is $y=0$, because the multiplier $P_1(y)^2/h_1=y^2/\lambda$ has a double zero there, and a double zero constrains the first derivative as well as the value, where a simple zero would constrain the value alone. The relevant reference data are therefore the value and first derivative of $P_n$ at the origin. We denote this jet by
\begin{equation}
   A_n=P_n(0),
   \qquad
   B_n=P_n'(0).
\end{equation}
Evaluating eq.~\eqref{eq:Charlier-gen-main} at $y=0$ and differentiating with respect to $y$ gives the exponential generating functions
\begin{equation}
   \sum_{n\ge0}A_n\frac{u^n}{n!}=\ee^{-\lambda u}(1+u)^\lambda,
   \qquad
   \sum_{n\ge0}B_n\frac{u^n}{n!}=\log(1+u)\ee^{-\lambda u}(1+u)^\lambda.
   \label{eq:AB-gen-app}
\end{equation}
The same two-component data close under the three-term recurrence. Evaluating that recurrence at the origin, before and after differentiation, gives, for $n\ge1$,
\begin{equation}
   A_{n+1}=-nA_n-\lambda nA_{n-1},
   \qquad
   B_{n+1}=A_n-nB_n-\lambda nB_{n-1}\,,
   \label{eq:AB-rec-app}
\end{equation}
with $A_0=1$, $A_1=0$, $B_0=0$ and $B_1=1$. Their parameter dependence is linked by $B_n=\partial_\lambda A_n+nA_{n-1}$ for $n\ge1$. Because the Christoffel point is double, the connector is confluent, and the jet enters it only through the two Wronskian-type combinations formed from adjacent degrees, 
\begin{equation}
   D_n=A_nB_{n+1}-B_nA_{n+1}\,,
   \qquad
   E_n=A_nB_{n+2}-B_nA_{n+2}\,.
\end{equation}
The quantity $D_n$ is the numerator of the confluent Christoffel--Darboux formula at the origin; $E_n$ is instead the corresponding gap-two Casoratian. The kernel identity below shows that $D_n$ is $h_n$ times a diagonal reproducing kernel and hence strictly positive, the property
on which the shifted norms and the whole correction series rest.
The Charlier recurrence reduces these determinants to
\begin{equation}
   D_0=1,
   \qquad
   D_n=A_n^2+\lambda nD_{n-1}\,,
   \qquad
   E_n=A_nA_{n+1}-(n+1)D_n\,.
   \label{eq:DE-app}
\end{equation}
Here the recurrence for $D_n$ holds for $n\ge1$, whereas the identity for $E_n$ holds for $n\ge0$. Coefficient extraction and iteration now give finite expressions at arbitrary index. With $\lambda^{\underline{k}}=\lambda(\lambda-1)\cdots(\lambda-k+1)$ and $\lambda^{\underline{0}}=1$, we have
\begin{equation}
\begin{aligned}
   A_n&=\sum_{k=0}^{n}\binom{n}{k}\lambda^{\underline{k}}(-\lambda)^{n-k},\\
   B_n&=\sum_{j=1}^{n}(-1)^{j+1}\frac{n!}{j(n-j)!}\,A_{n-j}\,,\\
   D_n&=\lambda^n n!\sum_{m=0}^{n}\frac{A_m^2}{\lambda^m m!}\,,\\
   E_n&=A_nA_{n+1}-(n+1)D_n\,.
\end{aligned}
   \label{eq:ABDE-closed-app}
\end{equation}
The first two identities are the coefficient extractions from $\ee^{-\lambda u}(1+u)^\lambda$ and its product with $\log(1+u)$; iterating the first-order recurrence for $D_n$ gives the third, while the last restates the determinant reduction of $E_n$. The connector coefficients are therefore explicit at every $n$ in reference Charlier data alone. The sequence $D_n$ also has an intrinsic kernel interpretation. Indeed, $D_n=h_n \mathbb K_n^{\mu}(0,0)$ follows from the confluent Christoffel--Darboux identity
\begin{equation}
   h_n \mathbb K_n^\mu(F,F)=P'_{n+1}(F)P_n(F)-P'_n(F)P_{n+1}(F).
   \label{eq:Dn-kernel-proof-app}
\end{equation}
At $F=0$, the right-hand side is $B_{n+1}A_n-B_nA_{n+1}=D_n$. Hence $D_n>0$ for every $\lambda>0$, independently of possible zeros of an individual $A_n$. These same determinants are the coefficients of the degree-one double-Christoffel transform.

For the measure $\dd\nu_1=y^2\dd\mu/\lambda$, the confluent identity is
\begin{equation}
   y^2R_n^{(1)}(y)=
   \frac{D_{n+1}P_n(y)-E_nP_{n+1}(y)+D_nP_{n+2}(y)}{D_n}\,.
   \label{eq:H1-app-poly}
\end{equation}
Its norm is $\thh_n^{(1)}=\lambda^{n-1}n!D_{n+1}/D_n$. Integrating this three-term connector against the reference Fourier factor gives the shifted amplitude
\begin{equation}
   \tphi_n^{(1)}(\tau)=\cS(\tau)
   \frac{q(\tau)^n\bigl[D_{n+1}-E_nq(\tau)+D_nq(\tau)^2\bigr]}{\lambda^{(n+1)/2}\sqrt{n!D_nD_{n+1}}}\,.
   \label{eq:H1-app-amplitude}
\end{equation}
Eliminating $E_n$ with eq.~\eqref{eq:DE-app} reproduces eq.~\eqref{eq:H1-first-amplitude-main}.

\paragraph{Reference-basis tails and global bounds.}
For the complexity it is more efficient to use cumulative probabilities than to sum the shifted amplitudes directly. The rank-one identity~\eqref{eq:first-jump-cumulative} expresses them through the evolved seed in the reference basis. Since \(P_1(y)=y\), \(h_1=\lambda\) and \(h_m=\lambda^m m!\), the required reference-basis amplitudes are
\begin{equation}
   g_m^{(1)}(\tau)
   =
   \frac{1}{\sqrt{\lambda h_m}}
   \int \ee^{-\ii \tau y}P_1(y)P_m(y)\dd\mu(y).
   \label{eq:H1-gm-first-jump-app}
\end{equation}
Applying the reference recurrence to $P_1P_m=yP_m$ gives
\begin{equation}
   g_m^{(1)}(\tau)
   =
   \frac{I_{m+1}(\tau)+mI_m(\tau)+\lambda m I_{m-1}(\tau)}
        {\sqrt{\lambda h_m}},
   \qquad I_{-1}(\tau)=0\,.
   \label{eq:H1-gm-reference-moments-app}
\end{equation}
The factorization \(I_m(\tau)=\cS(\tau)q(\tau)^m\) holds for every $m\ge0$ and reduces this expression, for $m\ge1$, to
\begin{equation}
   g_m^{(1)}(\tau)
   =
   \cS(\tau)
   \frac{q(\tau)^{m-1}
   \left(q(\tau)^2+mq(\tau)+\lambda m\right)}
   {\lambda^{(m+1)/2}\sqrt{m!}}\,.
   \label{eq:H1-gm-closed-app}
\end{equation}
The restriction is on the written form only. The prefactor $q^{m-1}$ has an apparent pole at $m=0$, but the bracket cancels it, since $q^{m-1}(q^2+mq+\lambda m)\big|_{m=0}=q$ identically. Removing that singularity extends the closed form to $m=0$ and returns 
$
   g_0^{(1)}(\tau)=\cS(\tau)\frac{q(\tau)}{\sqrt{\lambda}}.
$
The probabilities depend on time through the real combination
\begin{equation}
   \kappa(\tau)=4\lambda\sin^2(\tau/2)
   =2\lambda(1-\cos\tau)
   =\frac{|q(\tau)|^2}{\lambda}\,.
\end{equation}
The remaining polynomial factor obeys
\begin{equation}
   q(\tau)^2+mq(\tau)+\lambda m
   =
   \lambda\ee^{-\ii\tau}\bigl(m-\kappa(\tau)\bigr),
\end{equation}
and therefore, for $m\ge1$,
\begin{equation}
   |g_m^{(1)}(\tau)|^2
   =
   \ee^{-\kappa(\tau)}
   \frac{\kappa(\tau)^{m-1}\bigl(m-\kappa(\tau)\bigr)^2}{m!}\,.
   \label{eq:H1-reference-basis-probability-app}
\end{equation}
The same removable singularity at $m=0$ returns
$
   |g_0^{(1)}(\tau)|^2=\kappa(\tau)\ee^{-\kappa(\tau)}.
$
At shifted cutoff \(L=\ell+1\), the rank-one subtraction in eq.~\eqref{eq:first-jump-cumulative} contributes the nonnegative term
\begin{equation}
\begin{aligned}
   C_L(\tau)
   &=\frac{\ee^{-\kappa(\tau)}\kappa(\tau)^L}
           {\lambda L!D_L}
     \bigl|A_Lq(\tau)-A_{L+1}\bigr|^2 .
\end{aligned}
   \label{eq:H1-Christoffel-correction-app}
\end{equation}
We henceforth suppress the argument $\tau$ in $q(\tau)$ and $\kappa(\tau)$. Equation~\eqref{eq:H1-reference-basis-probability-app} for reference-basis amplitudes gives $\sum_{m\ge0}m|g_m^{(1)}|^2=1+\kappa$. The corresponding shifted-cutoff tail is
\begin{equation}
   \sum_{L\ge1}\left(1-\sum_{m=0}^{L}|g_m^{(1)}|^2\right)
   =
   \sum_{m\ge2}(m-1)|g_m^{(1)}|^2
   =
   \kappa(1+\ee^{-\kappa}).
   \label{eq:H1-reference-tail-app}
\end{equation}
Together with the positive correction $\sum_{L\ge1}C_L$, this is precisely eq.~\eqref{eq:H1-first-jump-complexity}.

The same representation gives global control of the correction series. Equations~\eqref{eq:AB-rec-app} and~\eqref{eq:DE-app}, followed by a weighted Cauchy--Schwarz estimate, give
\begin{equation}
\begin{aligned}
   |A_Lq-A_{L+1}|^2
   &=
   \left|(q+L)A_L+\lambda L A_{L-1}\right|^2\\
   &\le
   \left(|q+L|^2+\lambda L\right)
   \left(A_L^2+\lambda L A_{L-1}^2\right)\\
   &\le
   \left(|q+L|^2+\lambda L\right)D_L\,.
\end{aligned}
   \label{eq:H1-correction-bound-app}
\end{equation}
The last step uses the recurrence $D_L=A_L^2+\lambda LD_{L-1}$ together with
$A_{L-1}^2\le D_{L-1}$, which follows for every $L\ge1$ because $A_m^2/h_m$ is one term of $\mathbb K_m^\mu(0,0)=D_m/h_m$, with $m=L-1$.
All remaining time dependence enters through $q$. Since \(\operatorname{Re}q=-\kappa/2\) and \(|q|^2=\lambda\kappa\),
\begin{equation}
   |q+L|^2+\lambda L
   =
   L^2+(\lambda-\kappa)L+\lambda\kappa\,.
\end{equation}
The polynomial on the right has no manifest sign, since its linear coefficient $(\lambda-\kappa)$ turns negative once $\kappa>\lambda$, but it equals a modulus squared plus a nonnegative term, so it is nonnegative for every $L\ge1$ and every $\tau$.  Equation~\eqref{eq:H1-Christoffel-correction-app} consequently obeys
\begin{equation}
   0\le C_L(\tau)\le
   \frac{\ee^{-\kappa}\kappa^L}{\lambda L!}
   \left[
      L^2+(\lambda-\kappa)L+\lambda\kappa
   \right].
   \label{eq:H1-correction-majorant-app}
\end{equation}
Summing this Poisson majorant gives
\begin{equation}
   0\le\sum_{L\ge1}C_L(\tau)
   \le
   \kappa(2-\ee^{-\kappa})+\frac{\kappa}{\lambda}\,.
   \label{eq:H1-correction-summed-bound-app}
\end{equation}
The correction series therefore converges absolutely, and its truncation error is controlled by Poisson tails. Combining this estimate with eq.~\eqref{eq:H1-reference-tail-app} and using $C_L(\tau)\ge0$ gives the global two-sided bound
\begin{equation}
   \kappa(\tau)\bigl(1+\ee^{-\kappa(\tau)}\bigr)
   \le K_1(\tau)
   \le
   \left(3+\frac{1}{\lambda}\right)\kappa(\tau).
   \label{eq:H1-complexity-global-bound-app}
\end{equation}
The lower estimate sharpens corollary~\ref{cor:H1-monotonicity} at $r=1$. For fixed $\lambda>0$, the upper estimate holds for every $\tau$ and compares $K_1$ uniformly with the vacuum complexity $K_0(\tau)=\kappa(\tau)$.

\paragraph{Hermite scaling endpoint.}
The continuous-spectrum endpoint of the first-jump construction is not obtained by setting
$\omega=0$ at fixed centered Charlier coordinate. Instead, $\omega\to0$ at fixed $\rho$ and physical
time produces the correlated central-limit scaling
\begin{equation}
   \lambda=\frac{\rho^2}{\omega^2}\to\infty\,,
   \qquad
   y_{\rm H}=\frac{E-\delta}{\rho}=\frac{x-\lambda}{\sqrt{\lambda}}\,,
   \qquad
   \tau_{\rm Ch}=\omega t,\qquad
   \tau_{\rm H}=\rho t=\sqrt{\lambda}\,\tau_{\rm Ch}\,.
   \label{eq:Hermite-scaling-variables-app}
\end{equation}
We write $y\equiv y_{\rm H}$ from now on. The two limits are correlated by construction, taking $\lambda\to\infty$ with $\tau_{\rm Ch}=\tau_{\rm H}/\sqrt\lambda\to0$ at fixed $\tau_{\rm H}$, and it is the pairing, not either limit alone, that leaves the reference dynamics finite. With $q(\tau_{\rm Ch})=\lambda(\ee^{-\ii\tau_{\rm Ch}}-1)$, both time-dependent factors of eq.~\eqref{eq:H1-old-moments-main} converge at fixed $\tau_{\rm H}$, and with them the reference complexity,
\begin{equation}
   \frac{q}{\sqrt\lambda}
   =\sqrt\lambda\bigl(\ee^{-\ii\tau_{\rm H}/\sqrt\lambda}-1\bigr)
   \longrightarrow-\ii\tau_{\rm H}\,,
   \qquad
   \cS\longrightarrow\ee^{-\tau_{\rm H}^2/2},
   \qquad
   \kappa\longrightarrow\tau_{\rm H}^2\,,
   \label{eq:Hermite-scaling-moment-limits-app}
\end{equation}
so that at each fixed $n$ the reference amplitude
$\phi_n^{(0)}=\cS q^n/\sqrt{h_n}=\cS(q/\sqrt\lambda)^n/\sqrt{n!}$ passes directly to
eq.~\eqref{eq:Hermite-reference-amplitudes-app} below. The last of these carries the physical content
of the correlation. Here $\kappa$ is the reference complexity $K_0$, and only the pairing
$\tau_{\rm Ch}\sim\lambda^{-1/2}$ holds it fixed. Taking $\lambda\to\infty$ at fixed
$\tau_{\rm Ch}\ne0$ instead sends $\kappa\to\infty$, and no limiting chain survives. In this regime the Poisson measure converges to the standard Gaussian measure
$
   \dd\gamma(y)=\frac{\ee^{-y^2/2}}{\sqrt{2\pi}}\dd y,
$ 
while the rescaled monic Charlier polynomials converge to the probabilists' Hermite family
\begin{equation}
   y\He_n(y)=\He_{n+1}(y)+n\He_{n-1}(y),
   \qquad h_n=n!.
   \label{eq:Hermite-reference-recurrence-app}
\end{equation}
The limiting Jacobi data are $a_n=0$ and $b_n^2=n$. Although the spectrum has become continuous, the Krylov index remains discrete, with basis $p_n(y)=\He_n(y)/\sqrt{n!}$ in $L^2(\gamma)$. We suppress the common physical phase $\ee^{-\ii\delta t}$ and regard the amplitudes as functions of $\tau_{\rm H}$; restoring that phase gives the physical amplitudes. The reference solution is
\begin{equation}
   \phi_n^{(0)}(\tau_{\rm H})=\ee^{-\tau_{\rm H}^2/2}
   \frac{(-\ii\tau_{\rm H})^n}{\sqrt{n!}},
   \qquad
   K_0(\tau_{\rm H})=\tau_{\rm H}^2\,.
   \label{eq:Hermite-reference-amplitudes-app}
\end{equation}
The first number-state jump remains $Q=p_1=y$ and changes the measure to
$
   \dd\nu_1(y)=y^2\dd\gamma(y).
$ 
Since the Gaussian measure has no atom at the zero of $Q$, the density vanishes at $y=0$ without removing a positive-mass spectral component; the measure class and cyclic dimension are unchanged.

The modified weight $y^2\dd\gamma(y)$ gives the generalized Hermite family~\cite{Rosenblum1994}. Its monic recurrence is
\begin{equation}
   y R_n(y)=R_{n+1}(y)+\widetilde\beta_n R_{n-1}(y),
   \qquad
   \widetilde a_n=0,
   \qquad
   \widetilde\beta_n=n+1-(-1)^n.
   \label{eq:Hermite-shifted-recurrence-app}
\end{equation}
The parity dependence begins with \(\widetilde\beta_1=3\), \(\widetilde\beta_2=2\) and \(\widetilde\beta_3=5\), making its alternating form explicit. It is also manifest in the Laguerre representation
\begin{equation}
\begin{aligned}
   R_{2m}(y)&=(-1)^m2^m m!
      L_m^{(1/2)}\left(\frac{y^2}{2}\right),\\
   R_{2m+1}(y)&=(-1)^m2^m m!
      y L_m^{(3/2)}\left(\frac{y^2}{2}\right), 
\end{aligned}
   \label{eq:Hermite-shifted-polys-Laguerre-app}
\end{equation}
and in the corresponding norms
\begin{equation}
   \thh_{2m}=(2m+1)!,
   \qquad
   \thh_{2m+1}=(2m+3)(2m+1)!\,.
   \label{eq:Hermite-shifted-norms-app}
\end{equation}
The same parity dependence can be absorbed into the coefficient
\begin{equation}
   c_n=\begin{cases}
      n+1, & n\text{ even},\\
      n+2, & n\text{ odd}.
   \end{cases}
   \label{eq:Hermite-cn-app}
\end{equation}
With this notation, direct Gaussian integration, or equivalently the degree-two connector for $W_Q=y^2$, gives
\begin{equation}
   \tphi_n^{(1)}(\tau_{\rm H})=\ee^{-\tau_{\rm H}^2/2}
   \frac{(-\ii\tau_{\rm H})^n(c_n-\tau_{\rm H}^2)}{\sqrt{\thh_n}}\,, 
   \label{eq:Hermite-shifted-amplitudes-app}
\end{equation}
and the shifted complexity sums exactly to
$
   K_1(\tau_{\rm H})=\tau_{\rm H}^2+1-\ee^{-2\tau_{\rm H}^2}.
$ 
At small time, $K_1(\tau_{\rm H})=3\tau_{\rm H}^2-2\tau_{\rm H}^4+O(\tau_{\rm H}^6)$, in agreement with $\widetilde\beta_1=3$, while at large time $K_1(\tau_{\rm H})=\tau_{\rm H}^2+1+O(\ee^{-2\tau_{\rm H}^2})$. This provides a continuous-spectrum check of both the amplitude connector and the criterion governing support loss.%

\paragraph{Arbitrary fixed Hermite level.}
The Gaussian endpoint has an additional scalar
amplitude-level structure. 
At every fixed Fock
level, all rebuilt amplitudes close under a scalar differential recurrence.  This keeps the
general root-free construction while using the additional structure of Gaussian propagation.

\begin{proposition}[Fixed-seed, all-level Hermite amplitude closure and radial filtration]
\label{prop:Hermite-all-r-complexity-app}
Fix $r\in\mathbb Z_{\ge1}$, put $x=\tau_{\rm H}^2$ and let
$R_n^{(r)}$ be monic and orthogonal for
\begin{equation}
\begin{gathered}
   \dd\nu_r(y)=\frac{\He_r(y)^2}{r!}\dd\gamma(y),
   \\
   yR_n^{(r)}(y)=R_{n+1}^{(r)}(y)
      +\widetilde\beta_n^{(r)}R_{n-1}^{(r)}(y),
   \qquad n\ge0\,,
   \\
   \widetilde\beta_0^{(r)}=0\,,
   \qquad
   \widetilde\beta_n^{(r)}=
      \frac{\thh_n^{(r)}}{\thh_{n-1}^{(r)}}\,,
   \qquad n\ge1\,.
\end{gathered}
   \label{eq:Hermite-induced-recurrence-app}
\end{equation}
Here $R_{-1}^{(r)}=0$, $R_0^{(r)}=1$ and $\thh_0^{(r)}=1$; all diagonal coefficients vanish
because $\nu_r$ is even.  The root-free moment/Gram realization of
subsection~\ref{subsec:connector-realizations} determines
$\thh_n^{(r)}$ and $\widetilde\beta_n^{(r)}$ from the Gaussian moments; the scalar recurrence
below uses these coefficients rather than leaving them as additional input.
The physical Fock amplitudes are 
\begin{equation}
   g_{m|r}^{\rm H}(\tau_{\rm H})
   \equiv g_m^{[p_r]}(\tau_{\rm H})
   =\bra m\ee^{-\ii\tau_{\rm H}(a+a^\dagger)}\ket r
   =\ee^{-x/2}(-\ii\tau_{\rm H})^{|m-r|}
    \sqrt{\frac{\min(m,r)!}{\max(m,r)!}}
    L_{\min(m,r)}^{(|m-r|)}(x).
   \label{eq:Hermite-Fock-transition-app}
\end{equation}

The above expression is the number-basis matrix element of the
displacement evolution \cite{Perelomov1986}; it is the reference-basis input to the rebuilt
amplitude problem.
For every $n\ge0$ there is a real polynomial $F_n^{(r)}(x)$ of degree exactly $r$ such that
\begin{equation}
   \tphi_n^{(r)}(\tau_{\rm H})
   =\ee^{-x/2}\frac{(-\ii\tau_{\rm H})^n}{\sqrt{\thh_n^{(r)}}}
      F_n^{(r)}(x).
   \label{eq:Hermite-scalar-amplitude-app}
\end{equation}
The full sequence follows from
\begin{equation}
\begin{aligned}
   F_0^{(r)}(x)&=L_r(x),\qquad
   F_1^{(r)}(x)=F_0^{(r)}(x)-2\partial_xF_0^{(r)}(x),\\
   xF_{n+1}^{(r)}(x)
   &=\widetilde\beta_n^{(r)}F_{n-1}^{(r)}(x)
     -(n-x)F_n^{(r)}(x)\\
   &\hspace{2.2em}-2x\partial_xF_n^{(r)}(x),
   \qquad n\ge1\,,
   \label{eq:Hermite-F-recurrence-app}
\end{aligned}
\end{equation}
with $F_n^{(r)}(0)=\thh_n^{(r)}/n!$.  Consequently
\begin{equation}
\begin{aligned}
   1&=\ee^{-x}\sum_{n=0}^{\infty}
      \frac{x^n[F_n^{(r)}(x)]^2}{\thh_n^{(r)}}\,,\\
   K_r(\tau_{\rm H})
   &=\ee^{-x}\sum_{n=1}^{\infty}
      n\frac{x^n[F_n^{(r)}(x)]^2}{\thh_n^{(r)}}\,.
   \label{eq:Hermite-all-r-complexity-app}
\end{aligned}
\end{equation}
Both series converge for every finite $\tau_{\rm H}$.

The physical radial displacement from the prepared Fock level,
$D_r^{\rm H}(x)=\sum_{m\ge0}|m-r|\,
|g_{m|r}^{\rm H}(\sqrt x)|^2$, obeys
\begin{equation}
   D_r^{\rm H}(x)=x+2\ee^{-x}\sum_{m=0}^{r-1}
      (r-m)\frac{m!}{r!}x^{r-m}
      \bigl[L_m^{(r-m)}(x)\bigr]^2,
   \qquad
   K_r(\tau_{\rm H})\ge D_r^{\rm H}(x).
   \label{eq:Hermite-radial-exact-app}
\end{equation}
In particular,
\begin{equation}
   K_r(\tau_{\rm H})\ge D_r^{\rm H}(x)
   \ge x+2\ee^{-x}\frac{x^r}{(r-1)!}
   >x=K_0(\tau_{\rm H}),
   \qquad r\ge1\,,\quad x>0\,.
   \label{eq:Hermite-all-r-vacuum-bound-app}
\end{equation}
At fixed $r$ and $x\to0$,
\begin{equation}
\begin{aligned}
   K_r(\tau_{\rm H})
   &=(2r+1)x-r(r+1)x^2
     +\frac{2r(r+1)(r^2+r+1)}{3(2r+1)}x^3+O_r(x^4),\\
   K_r(\tau_{\rm H})-D_r^{\rm H}(x)
   &=\frac{r(r+1)}{2(2r+1)}x^3+O_r(x^4).
   \label{eq:Hermite-all-r-short-time-app}
\end{aligned}
\end{equation}
Thus physical radial motion and rebuilt spread agree through order $\tau_{\rm H}^4$; the first
strictly positive re-Lanczos contribution occurs at order $\tau_{\rm H}^6$.
\end{proposition}

\begin{proof}
The survival amplitude of the prepared Fock state is
$\bra r\ee^{-\ii\tau_{\rm H}(a+a^\dagger)}\ket r=\ee^{-x/2}L_r(x)$, and the spectral theorem gives
\begin{equation}
   \sqrt{\thh_n^{(r)}}\,\tphi_n^{(r)}(\tau_{\rm H})
   =R_n^{(r)}(\ii\partial_{\tau_{\rm H}})
      \bigl[\ee^{-x/2}L_r(x)\bigr].
   \label{eq:Hermite-differential-amplitude-app}
\end{equation}
Orthogonality removes every Taylor power below $n$, and parity leaves a function of $x$ after
$(-\ii\tau_{\rm H})^n$ is divided out.  Each derivative $(\ii\partial_{\tau_{\rm H}})^{n-2j}$ permitted by parity returns $\ee^{-x/2}$
times a polynomial in $\tau_{\rm H}$ of degree at most $n-2j+2r$, so the whole sum has degree at
most $n+2r$ and $F_n^{(r)}$ has degree at most $r$; the coefficient of $\tau_{\rm H}^{\,n+2r}$ comes
from the monic highest derivative alone, where differentiating the Gaussian $n$ times against the
leading Laguerre coefficient leaves the degree-$r$ coefficient of $F_n^{(r)}$ equal to $(-1)^r/r!$
at every $n$.
This proves eq.~\eqref{eq:Hermite-scalar-amplitude-app}; its first nonvanishing Taylor coefficient
uses $\int y^nR_n^{(r)}\dd\nu_r=\thh_n^{(r)}$ and gives the stated value at $x=0$.
Applying $\ii\partial_{\tau_{\rm H}}$ to the Fourier integral of the recurrence in
eq.~\eqref{eq:Hermite-induced-recurrence-app} gives
eq.~\eqref{eq:Hermite-F-recurrence-app}.  
The reference Hermite chain has $a_n=0$ and $b_n=\sqrt n$, so
$b_n+b_{n+1}\le n+3/2$ and eq.~\eqref{eq:reference-linear-bound-main} holds with
$c=0$, $A=1$ and $C=3/2$. Proposition~\ref{prop:finite-time-control-main} therefore gives
$\widetilde b_{n+1}^{(r)}\le n+r+3/2$. Unless the shifted chain has already terminated, the
resulting Carleman divergence establishes determinacy, polynomial density and essential
self-adjointness.  Unitarity gives
the normalization series in
eq.~\eqref{eq:Hermite-all-r-complexity-app}, while
proposition~\ref{prop:finite-time-control-main}
proves finiteness of the Krylov-index
first moment and hence convergence of the second line, the complexity series.

The signed Fock displacement obtained from eq.~\eqref{eq:Hermite-Fock-transition-app} has mean
$\sum_m(m-r)|g_{m|r}^{\rm H}|^2=x$.  Adding twice its downward part gives the first formula in
eq.~\eqref{eq:Hermite-radial-exact-app}.  The projector inclusion used in
proposition~\ref{prop:chord-radial-filtration} applies equally to the Hermite chain.  All moments
of the displaced-number distribution are finite, so the same tail argument gives
$K_r\ge D_r^{\rm H}$ without a bounded-generator assumption.  Finiteness of $D_r^{\rm H}$ replaces the norm bound on the physical side, while on the
rebuilt side the two ingredients are the finiteness of $K_r$ established above and the
completeness of $\left\{\ket{\widetilde K_n^{(r)}}\right\}$ just established through the shifted Carleman bound;
both tail sums then converge and may be subtracted.  Retaining the $m=0$ term proves
eq.~\eqref{eq:Hermite-all-r-vacuum-bound-app}.

It remains to justify the short-time expansion of the unbounded first moment.  Degree filtration
places $\ket{\widetilde K_n^{(r)}}$ in the span of the first $n+r+1$ reference Hermite levels.
On that subspace, the reference operator $a+a^\dagger$ has norm at most
$2\sqrt{n+r+1}$.  Therefore
\begin{equation}
   \widetilde b_{n+1}^{(r)}
   =\bigl|\langle\widetilde K_{n+1}^{(r)}|
      (a+a^\dagger)|\widetilde K_n^{(r)}\rangle\bigr|
   \le2\sqrt{n+r+1}\,.
   \label{eq:Hermite-shifted-hopping-bound-app}
\end{equation}
Let $\widetilde\Jac_r$ denote the shifted Hermite Jacobi operator, and let
$\mathsf N\ket n=n\ket n$ be the Krylov-level counting operator of its rebuilt chain, where
$\ket n$ denotes $\ket{\widetilde K_n^{(r)}}$ rather than the physical Fock state of level $n$.
The norm of $\widetilde\Jac_r$ on levels at
most $k$ is consequently bounded by $4\sqrt{k+r+1}$.  Iteration from $\ket0$ gives
\begin{equation}
   \|(\widetilde\Jac_r)^k\ket0\|
   \le4^k\sqrt{\frac{(r+k)!}{r!}}\,.
   \label{eq:Hermite-analytic-vector-bound-app}
\end{equation}
The exponential series with this bound converges uniformly on compact time intervals both in
norm and in the graph norm of $\mathsf N^{1/2}$; the additional factor $\sqrt{k}$ does not affect
convergence.  Hence the Taylor expansion may be inserted into
$K_r=\|\mathsf N^{1/2}\exp(-\ii\tau_{\rm H}\widetilde\Jac_r)\ket0\|^2$ with a controlled
remainder through the required order.

Finally, exact moment determinants, or direct finite-path Lanczos expansion, give
\begin{equation}
   \widetilde\beta_1^{(r)}=2r+1\,,
   \qquad
   \widetilde\beta_2^{(r)}=\frac{2(r^2+r+1)}{2r+1}\,,
   \qquad
   \widetilde\beta_3^{(r)}=
   \frac{2r^4+4r^3+19r^2+17r+3}
        {(2r+1)(r^2+r+1)}\,.
   \label{eq:Hermite-first-three-beta-app}
\end{equation}
Expanding the zero-diagonal Jacobi evolution through six steps and the finite Laguerre sum in
eq.~\eqref{eq:Hermite-radial-exact-app} gives
eq.~\eqref{eq:Hermite-all-r-short-time-app}.
\end{proof}

For $r=1$, the hierarchy reduces to
$F_n^{(1)}(x)=n+1-x$ at even $n$ and $F_n^{(1)}(x)=n+2-x$ at odd $n$, and its two parity sums
recover $K_1=x+1-\ee^{-2x}$.  For $r=2s+\delta\ge2$, $\delta\in\{0,1\}$, the even and odd sectors
instead reduce under $z=y^2/2$ to weights proportional to
\begin{equation}
   z^{\delta+\epsilon-1/2}\ee^{-z}
   \bigl[L_s^{(\delta-1/2)}(z)\bigr]^2\dd z\,,
   \qquad \epsilon=0,1\,.
   \label{eq:Hermite-parity-weights-app}
\end{equation}
The nonconstant squared Laguerre factor explains why the $r=1$ generalized-Hermite summation does
not extend by replacing its parity parameter with $r$.  Equations
\eqref{eq:Hermite-F-recurrence-app} and \eqref{eq:Hermite-all-r-complexity-app} nevertheless give
every amplitude and the exact finite-time complexity.  Their fixed-$r$ late-time analysis for
$r\ge2$ requires control beyond the finite-time convergence proved here.

\paragraph{Why the oscillator group does not determine the shifted chain.}
The distinction between group evolution and the shifted Lanczos problem is
already visible for the first jump.  Equations~\eqref{eq:H1-ham}--\eqref{eq:H1-jacobi} give
$Y\ket{K_0}=\sqrt\lambda\ket{K_1}$ and hence
$\ee^{-\ii\tau Y}\ket{K_1}
=\lambda^{-1/2}Y\ee^{-\ii\tau Y}\ket{K_0}$.  Equivalently, in the
original Fock convention, $[H,a^\dagger]=\omega a^\dagger+\bar g$ gives
\begin{equation}
   \ee^{-\ii Ht}\ket{1}
   =
   \left[
      \ee^{-\ii\omega t}a^\dagger
      +\frac{\bar g}{\omega}
       \bigl(\ee^{-\ii\omega t}-1\bigr)
   \right]\ee^{-\ii Ht}\ket{0}.
   \label{eq:H1-excited-Fock-evolution-app}
\end{equation}
This identity determines the evolved state in the original
Fock/reference-Krylov basis; it does not construct the Lanczos basis of
the new cyclic pair $(Y,\ket{K_1})$.  In particular, unitary
Heisenberg--Weyl transformations send the vacuum to coherent states,
whereas $a^\dagger\ket0=\ket1$ is generated by a nonunitary ladder
operator.

A nonlinear oscillator with $\ket1$ as its vacuum does not bypass this
reorthogonalization.  On $P_+\cH$, where
$P_+=\Id-\ket0\bra0$, the formal choice
$d=a\sqrt{(N-1)/N}$ gives $d\ket1=0$ and
$d\ket n=\sqrt{n-1}\ket{n-1}$ for $n\ge2$.  It either removes the
boundary state $\ket0$, which belongs to the orbit generated by
$Y\ket{K_1}$, or represents the projected hopping by singular
number-dependent coefficients.  For example, with $M_d=d^\dagger d$,
$P_+aP_+=d\sqrt{(M_d+1)/M_d}$ holds only on the positive-$M_d$ sector
unless a generalized inverse is prescribed.

Once the shifted basis has been solved, it does admit an a posteriori
oscillator representation.  Let
$\mathfrak a\ket{K_n}=\sqrt n\ket{K_{n-1}}$ and
$W=\sum_n\ket{\widetilde K_n}\bra{K_n}$, with $W$ understood as the
corresponding partial isometry when a spectral atom is deleted.  Setting
$c=W\mathfrak a W^\dagger$ and $\mathcal N=c^\dagger c$ gives, on the
shifted cyclic subspace,
\begin{equation}
   Y=A(\mathcal N)+c^\dagger F(\mathcal N)+F(\mathcal N)c\,,
   \qquad
   A(n)=\widetilde a_n\,,\qquad
   F(n)=\frac{\widetilde b_{n+1}}{\sqrt{n+1}}\,.
   \label{eq:H1-aposteriori-oscillator-app}
\end{equation}
For the first jump, eq.~\eqref{eq:H1-local-sum-rule-check-main} gives
$A(0)=1$, $A(1)=4/3$, $F(0)^2=3\lambda$ and
$F(1)^2=\lambda+4/9$.  This nonlinear-oscillator description encodes
the already determined shifted Jacobi chain; it is not a replacement
for the Christoffel problem generated by $P_r^2\dd\mu/h_r$. That problem is solved next at
arbitrary seed level, where the four scalar sequences $A_n$, $B_n$, $D_n$, $E_n$ of this
subsection reappear as the rank-one reduction of a vector remainder calculus.

\subsection{Root-free remainder calculus for arbitrary number-state jumps}
\label{app:H1-general-r}

We return to the discrete Charlier chain at fixed $\lambda>0$ and consider a number-state seed at an arbitrary level $r\ge1$. In the normalization of subsection~\ref{subsec:HW}, the seed and its induced measure are
\begin{equation}
   \ket{K_r}=\frac{P_r(Y)}{\sqrt{h_r}}\ket{K_0},
   \qquad h_r=\lambda^r r!\,,
   \qquad
   \dd\nu_r(y)=\frac{P_r(y)^2}{h_r}\dd\mu(y).
   \label{eq:H1-general-r-seed-app}
\end{equation}
\paragraph{Remainder coordinates and Jacobi data.}
Divisibility is the operative notion here for a direct reason. The shifted inner product
is $h_r^{-1}\langle P_r\,\cdot\,,P_r\,\cdot\,\rangle_\mu$, so a shifted polynomial $f$ enters
the reference problem through the expansion of the product $P_rf$; the resulting
constraints depend on $f$ only modulo $P_r$, and, once first derivatives enter, modulo $P_r^2$.
Consequently how much each reference level contributes is bounded by the seed level $r$ and not by
the Krylov index, which is what makes the construction finite at every $n$.

Two related divisibility problems enter the shifted construction. The projected-kernel constraints of appendix~\ref{app:projected-kernel} depend only on polynomial classes modulo $P_r$, whereas the first-jet constraints of appendix~\ref{app:christoffel} depend on classes modulo $P_r^2$. Both can therefore be implemented without finding the zeros of $P_r$. We work in the ordered monomial bases of the finite quotient algebras
\begin{equation}
   \mathscr R_r=\CC[y]/\langle P_r\rangle,
   \qquad
   \mathscr R_r^{(2)}=\CC[y]/\langle P_r^2\rangle.
   \label{eq:H1-residue-jet-algebras-app}
\end{equation}
These auxiliary coordinate algebras encode polynomial divisibility. They impose no relation $P_r(Y)=0$ on the oscillator operator and do not describe termination of the infinite chain, since $P_r(Y)\ket{K_0}=\sqrt{h_r}\ket{K_r}\ne0$. The contrast with the terminal quotient $\cA_d=\CC[E]/\langle T_d\rangle$ of appendices~\ref{app:lanczos} and~\ref{app:finite} is exact and worth stating, since the two constructions look alike and play opposite roles. There the quotient is the physical algebra of functions on a finite support, $T_d(H)\ket{K_0}=0$ genuinely holds, and passing to the quotient encodes the end of the chain. Here the Charlier support is infinite, $\langle P_r\rangle$ is not an ideal of relations satisfied by $Y$, and the quotient is a bookkeeping device for remainders alone. If $\xi_1,\ldots,\xi_r$ are the simple zeros of $P_r$, the Chinese remainder theorem identifies the two quotient descriptions with evaluation and first-jet data according to
\begin{equation}
\begin{aligned}
   \mathscr R_r
   &\simeq\bigoplus_{a=1}^{r}\CC\,,
   &[f]_{\mathscr R_r}
   &\longmapsto\bigl(f(\xi_a)\bigr)_{a=1}^{r}\,,\\
   \mathscr R_r^{(2)}
   &\simeq\bigoplus_{a=1}^{r}\CC[\varepsilon_a]/\langle\varepsilon_a^2\rangle,
   &[f]_{\mathscr R_r^{(2)}}
   &\longmapsto
   \bigl(f(\xi_a)+f'(\xi_a)\varepsilon_a\bigr)_{a=1}^{r}\,.
\end{aligned}
   \label{eq:H1-Chinese-remainder-app}
\end{equation}
Let $\mathbf u_m^{(r)}$ and $\mathbf v_m^{(r)}$ be the remainder vectors defined in eq.~\eqref{eq:H1-remainder-sequence-main}, and let $\mathsf Y_r$ and $\mathsf Y_r^{(2)}$ represent multiplication by $y$ on the two quotient spaces. The Vandermonde and confluent Vandermonde matrices $\mathsf T_r$ and $\mathsf T_r^{(2)}$ convert the monomial remainders into the corresponding root coordinates.\footnote{For $\lambda>0$, the Charlier polynomials, their zeros and all remainder data used here are real. Thus the transposes in the Gram formulas agree with the Hermitian adjoints used in subsection~\ref{subsec:HW}.} Explicitly,
\begin{gather}
   \mathsf T_r\mathbf u_m^{(r)}
    =\bigl(P_m(\xi_1),\ldots,P_m(\xi_r)\bigr)^T, \quad
   \mathsf T_r\mathsf Y_r\mathsf T_r^{-1}
   =\diag(\xi_1,\ldots,\xi_r), \label{eq:H1-remainder-root-coordinate-map-app}\\
   \mathsf T_r^{(2)}\mathbf v_m^{(r)} 
   =\bigl(P_m(\xi_1),P_m'(\xi_1),\ldots,
   P_m(\xi_r),P_m'(\xi_r)\bigr)^T, \quad
   \mathsf T_r^{(2)}\mathsf Y_r^{(2)}
   (\mathsf T_r^{(2)})^{-1}
   =\bigoplus_{a=1}^{r}
   \begin{pmatrix}\xi_a&0\\1&\xi_a\end{pmatrix}. \nonumber
\end{gather}
The Jordan block follows from $(yf)'(\xi_a)=f(\xi_a)+\xi_af'(\xi_a)$. Thus polynomial remainders and root evaluations are two coordinate descriptions of the same constraint spaces.%

The remainder recurrence can be implemented directly in monomial coordinates. Write
$ 
   P_r(y)=y^r+\sum_{\ell=0}^{r-1}\pi_\ell^{(r)}y^\ell.
$ 
In the ordered basis $(1,y,\ldots,y^{r-1})$, with coefficient vectors written as columns, multiplication by $y$ is represented by the companion matrix
\begin{equation}
   \mathsf Y_r=
   \begin{pmatrix}
      0&0&\cdots&0&-\pi_0^{(r)}\\
      1&0&\cdots&0&-\pi_1^{(r)}\\
      0&1&\cdots&0&-\pi_2^{(r)}\\
      \vdots&&\ddots&&\vdots\\
      0&0&\cdots&1&-\pi_{r-1}^{(r)}
   \end{pmatrix}.
   \label{eq:H1-companion-matrix-app}
\end{equation}
Reducing the Charlier recurrence modulo $P_r$ gives the closed $r$-component system
\begin{gather}
   \mathbf u_{n+1}^{(r)}
   =(\mathsf Y_r-n\Id_r)\mathbf u_n^{(r)}
   -\lambda n\mathbf u_{n-1}^{(r)},
   \qquad
   \mathbf u_{-1}^{(r)}=0,\qquad
   \mathbf u_0^{(r)}=e_0,\qquad
   e_0=(1,0,\ldots,0)^T,\nonumber \\
   \sum_{n=0}^{\infty}\mathbf u_n^{(r)}\frac{u^n}{n!}
   =\ee^{-\lambda u}(1+u)^\lambda
   \exp\!\bigl(\log(1+u)\mathsf Y_r\bigr)e_0\,. \label{eq:H1-remainder-recurrence-EGF-app}
\end{gather}
For fixed seed level $r$, the matrix size is independent of $n$. Thus the entire sequence of remainder constraints is generated by a finite recurrence without extracting any roots.

The remainder vectors enter the shifted Jacobi problem through the positive-semidefinite cutoff Gram matrix
\begin{equation}
   \mathsf H_L^{(r)}
   =\sum_{m=0}^{L}\frac{\mathbf u_m^{(r)}
      \mathbf u_m^{(r)T}}{h_m},
   \qquad
   \Delta_L^{(r)}=\det\mathsf H_L^{(r)}.
   \label{eq:H1-remainder-Gram-app}
\end{equation}
In root coordinates it obeys the congruence
\begin{equation}
   \bigl[\mathbb K_L^\mu(\xi_a,\xi_b)\bigr]_{a,b=1}^{r}
   =\mathsf T_r\mathsf H_L^{(r)}\mathsf T_r^T.
   \label{eq:H1-Gram-coordinate-congruence-app}
\end{equation}
Equation~\eqref{eq:H1-Gram-coordinate-congruence-app} identifies $\mathsf H_L^{(r)}$ with the projected-kernel Gram matrix of appendix~\ref{app:projected-kernel}, expressed in remainder coordinates. It is positive definite for $L\ge r-1$, since $\mathbf u_0^{(r)},\ldots,\mathbf u_{r-1}^{(r)}$ form a unit-triangular polynomial basis. In the chosen monomial coordinates, its initial determinant and the subsequent rank-one updates are 
\begin{equation}
\begin{gathered}
   \Delta_{r-1}^{(r)}
   =\frac{1}{\prod_{m=0}^{r-1}h_m}
   =\frac{\lambda^{-r(r-1)/2}}{\prod_{m=0}^{r-1}m!}\,,
   \qquad
   \Delta_r^{(r)}=\Delta_{r-1}^{(r)}\,,\\
   \frac{\Delta_L^{(r)}}{\Delta_{L-1}^{(r)}}
   =1+\frac{\mathbf u_L^{(r)T}
      (\mathsf H_{L-1}^{(r)})^{-1}\mathbf u_L^{(r)}}{h_L}\,,
   \qquad L\ge r \,.
\end{gathered}
   \label{eq:H1-Delta-update-app}
\end{equation}
Here $\Delta_r^{(r)}=\Delta_{r-1}^{(r)}$ because $\mathbf u_r^{(r)}=0$, while the last line follows from the matrix determinant lemma and holds for $L\ge r$, the range on which $\mathsf H_{L-1}^{(r)}$ is invertible; below $L=r-1$ the determinant vanishes identically. Although $\Delta_L^{(r)}$ depends on the chosen remainder basis, each consecutive ratio $\Delta_L^{(r)}/\Delta_{L-1}^{(r)}$ is invariant under a fixed change of coordinates.

The same Gram matrix constructs the shifted polynomial. For $L=n+r$, the polynomial $P_rR_n^{(r)}$ is obtained by correcting the monic reference polynomial $P_L$ so that its remainder modulo $P_r$ vanishes. The vector-valued kernel polynomial
$ 
   \mathbf k_{L-1}^{(r)}(y)=
   \sum_{m=0}^{L-1}\frac{P_m(y)}{h_m}\mathbf u_m^{(r)}
$
collects the corresponding remainder representers. The minimum-norm lower-degree correction with remainder $\mathbf u_L^{(r)}$ lies in their span and gives
\begin{equation}
   P_r(y)R_n^{(r)}(y)=P_L(y)
   -\mathbf k_{L-1}^{(r)}(y)^T
   (\mathsf H_{L-1}^{(r)})^{-1}\mathbf u_L^{(r)}.
   \label{eq:H1-remainder-shifted-polynomial-app}
\end{equation}
Reduction of the right-hand side modulo $P_r$ gives
\(
\mathbf u_L^{(r)}-\mathsf H_{L-1}^{(r)}
(\mathsf H_{L-1}^{(r)})^{-1}\mathbf u_L^{(r)}=0.
\)
The right-hand side is therefore divisible by $P_r$ and defines a monic polynomial $R_n^{(r)}$ of degree $n$. Its shifted orthogonality follows from the same representer construction. Let $\deg T<n$ and write $P_rT=\sum_{m=0}^{L-1}d_mP_m$. The vanishing remainder of $P_rT$ gives $\sum_m d_m\mathbf u_m^{(r)}=0$. The $P_L$ term in eq.~\eqref{eq:H1-remainder-shifted-polynomial-app} is orthogonal to $P_rT$, while the inner product with the correction is
\begin{equation}
   \left(\sum_{m=0}^{L-1}d_m\mathbf u_m^{(r)}\right)^T
   (\mathsf H_{L-1}^{(r)})^{-1}\mathbf u_L^{(r)}=0\,.
\end{equation}
Consequently,
\(
\langle R_n^{(r)},T\rangle_{\nu_r}
=h_r^{-1}\langle P_rR_n^{(r)},P_rT\rangle_\mu=0.
\)
The squared $\mu$-norm of the right-hand side of eq.~\eqref{eq:H1-remainder-shifted-polynomial-app} is
\(
h_L+\mathbf u_L^{(r)T}(\mathsf H_{L-1}^{(r)})^{-1}\mathbf u_L^{(r)}
=h_L\Delta_L^{(r)}/\Delta_{L-1}^{(r)}.
\)
The first identity fixes the shifted norm, while comparison at degree $L-1$ fixes the diagonal recurrence coefficient. Writing $\widetilde\beta_n^{(r)}=(\widetilde b_n^{(r)})^2$ and
\begin{equation}
   \eta_L^{(r)}=
   \frac{\mathbf u_{L-1}^{(r)T}
   (\mathsf H_{L-1}^{(r)})^{-1}\mathbf u_L^{(r)}}{h_{L-1}}\,,
   \qquad L\ge r\,,
   \label{eq:H1-eta-definition-app}
\end{equation}
we find
\begin{equation}
\begin{aligned}
   \widetilde h_n^{(r)}
   &=\frac{h_{n+r}}{h_r}
      \frac{\Delta_{n+r}^{(r)}}{\Delta_{n+r-1}^{(r)}}\,,\\
   \widetilde\beta_n^{(r)}
   &=\lambda(n+r)
      \frac{\Delta_{n+r}^{(r)}\Delta_{n+r-2}^{(r)}}
      {(\Delta_{n+r-1}^{(r)})^2}\,,\qquad n\ge1\,,\\
   \widetilde a_n^{(r)}
   &=n+r+\eta_{n+r+1}^{(r)}-\eta_{n+r}^{(r)}\,,
   \qquad n\ge0\,.
\end{aligned}
   \label{eq:H1-general-r-Jacobi-app}
\end{equation}
This is the root-free remainder-coordinate realization of
proposition~\ref{prop:root-Gram-Jacobi-app}.%

Equation~\eqref{eq:H1-general-r-Jacobi-app}
determines the shifted Lanczos coefficients at every index, even when the
determinant ratios do not simplify to rational functions of
$n$. At the first two levels it reproduces
eq.~\eqref{eq:H1-local-sum-rule-check-main}, obtained independently in
subsection~\ref{subsec:HW} by applying $Y$ twice to $\ket{K_r}$ in the
reference Jacobi chain. This provides a local check of the all-index
determinant construction. Its large-index behavior is derived in
subsection~\ref{app:H1-large-index}.

\paragraph{The first-jet connector and shifted amplitudes.}
The preceding construction uses ordinary remainders modulo $P_r$ to determine norms and Jacobi coefficients. Individual amplitudes instead require the finite reference-band expansion of $P_r^2R_n^{(r)}$. Writing that expansion as in eq.~\eqref{eq:H1-unnormalized-connector-app} below and fixing its monic normalization $G_{n,2r}^{(r)}=1$, which is legitimate because $P_r^2R_n^{(r)}$ is monic of degree $n+2r$, the  remaining coefficients are determined by requiring the expansion to have a double zero at every zero of $P_r$. Algebraically, this replaces the divisor by its first infinitesimal thickening; the residue class modulo $P_r^2$ retains both the value and first derivative at every simple zero. The connector is therefore encoded by the first-jet sequence $\mathbf v_n^{(r)}$. It obeys the recurrence in eq.~\eqref{eq:H1-remainder-recurrence-EGF-app}, with $\mathsf Y_r,\Id_r,\mathbf u_n^{(r)}$ replaced by $\mathsf Y_r^{(2)},\Id_{2r},\mathbf v_n^{(r)}$, and has initial data $\mathbf v_{-1}^{(r)}=0$ and $\mathbf v_0^{(r)}=e_0^{(2r)}$. We write the unnormalized connector as
\begin{equation}
   P_r(y)^2R_n^{(r)}(y)
   =\sum_{j=0}^{2r}G_{n,j}^{(r)}P_{n+j}(y),
   \qquad G_{n,2r}^{(r)}=1\,.
   \label{eq:H1-unnormalized-connector-app}
\end{equation}
Divisibility by $P_r^2$ follows from the single vector relation $\sum_{j=0}^{2r}G_{n,j}^{(r)}\mathbf v_{n+j}^{(r)}=0$. Hence, with
\begin{equation}
   \mathsf V_n^{(r)}
   =\begin{pmatrix}
      \mathbf v_n^{(r)}&\mathbf v_{n+1}^{(r)}&\cdots&
      \mathbf v_{n+2r-1}^{(r)}
   \end{pmatrix},
   \label{eq:H1-jet-window-matrix-app}
\end{equation}
$\mathsf T_r^{(2)}\mathsf V_n^{(r)}$ is exactly the confluent value-and-derivative matrix in eq.~\eqref{eq:general-root-system}. The complete connector row in the remainder coordinates is therefore
\begin{equation}
   \begin{pmatrix}
      G_{n,0}^{(r)}\\ \vdots\\ G_{n,2r-1}^{(r)}
   \end{pmatrix}
   =-(\mathsf V_n^{(r)})^{-1}\mathbf v_{n+2r}^{(r)}\,.
   \label{eq:H1-jet-connector-app}
\end{equation}
A singular $\mathsf V_n^{(r)}$ would give
$0\ne S=P_r^2T\in\Span\{P_n,\ldots,P_{n+2r-1}\}$ with $\deg T<n$, for which
orthogonality gives the contradiction
$0=\langle T,S\rangle_\mu=\int P_r^2|T|^2\dd\mu>0$.

Equation~\eqref{eq:H1-jet-connector-app} is consequently the root-free form of the $2r$ value-and-derivative equations in appendix~\ref{app:christoffel}. Starting from the reference expansion of $P_r^2$, proposition~\ref{prop:connector-dictionary} gives the compatibility recurrence \begin{equation}
\begin{aligned}
   G_{n+1,j}^{(r)}=
   G_{n,j}^{(r)}
   +(n+j+1-\widetilde a_n^{(r)})G_{n,j+1}^{(r)}
   +\lambda(n+j+2)G_{n,j+2}^{(r)}
   -\widetilde\beta_n^{(r)}G_{n-1,j+2}^{(r)}\,,
\end{aligned}
   \label{eq:H1-G-recurrence-app}
\end{equation}
for $n\ge0$ and $0\le j\le2r$, where coefficients outside $0\le j\le2r$ vanish, $G_{-1,j}^{(r)}=0$, and $\widetilde\beta_0^{(r)}=0$. The lowest-band coefficient is fixed by the shifted norm and is strictly positive
\begin{equation}
   G_{n,0}^{(r)}
   =\lambda^r\frac{(n+r)!}{n!}
   \frac{\Delta_{n+r}^{(r)}}{\Delta_{n+r-1}^{(r)}}>0\,.
   \label{eq:H1-G0-Delta-app}
\end{equation}
The normalization relating this root-free connector to the
connector used in the main text is

\begin{equation}
   \Gamma_{n,n+j}^{(r)}
   =\frac{G_{n,j}^{(r)}}{h_r}\,,
   \qquad
   \widetilde h_n^{(r)}
   =\frac{h_nG_{n,0}^{(r)}}{h_r}\,.
   \label{eq:H1-Gamma-G-bridge-app}
\end{equation}

Equations~\eqref{eq:H1-jet-connector-app} and
\eqref{eq:H1-G-recurrence-app} determine the time-independent
connector data entering the individual amplitudes in
eq.~\eqref{eq:H1-r-amplitudes}. Cumulative probabilities require no
first-jet data and return to the $r$-dimensional Gram matrix.

\paragraph{Cumulative probabilities and spread complexity.}
The reference-basis amplitudes of the evolved number-state seed, together with their remainder-coordinate projection data, are 
\begin{equation}
\begin{aligned}
   g_m^{(r)}(\tau)
   &=\bra{K_m}\ee^{-\ii\tau Y}\ket{K_r}
   =\frac{1}{\sqrt{h_rh_m}}\int
   \ee^{-\ii\tau y}P_r(y)P_m(y)\dd\mu(y),\\
   \mathbf z_L^{(r)}(\tau)
   &=\sum_{m=0}^{L}
   \frac{\mathbf u_m^{(r)}}{\sqrt{h_m}}g_m^{(r)}(\tau), \\
   \mathcal C_L^{(r)}(\tau)&=
   \mathbf z_L^{(r)\dagger}
   (\mathsf H_L^{(r)})^{-1}\mathbf z_L^{(r)}.
\end{aligned}
   \label{eq:H1-general-r-z-app}
\end{equation}

The remainder-coordinate version of the projected kernel gives, for $L\ge r$,
\begin{equation}
   \Phi_{L-r}^{(r)}(\tau)
   =\sum_{m=0}^{L}|g_m^{(r)}(\tau)|^2-\mathcal C_L^{(r)}(\tau).
   \label{eq:H1-general-r-cumulative-app}
\end{equation}
To pass from the cumulative probabilities to spread complexity, we separate the reference-basis contribution below the seed level. The matrix elements of the oscillator displacement operator $D(\alpha)=\exp(\alpha a^\dagger-\bar\alpha a)$ between number states, which are the transition amplitudes generated by the reference evolution, give 
\begin{equation}
   |g_m^{(r)}(\tau)|^2=
   \ee^{-\kappa}\frac{m!}{r!}\kappa^{r-m}
   \left[L_m^{(r-m)}(\kappa)\right]^2,
   \qquad 0\le m<r\,.
   \label{eq:H1-number-transition-probability-app}
\end{equation}
Here $\kappa=|q|^2/\lambda$, and $L_m^{(\alpha)}$ denotes an associated Laguerre polynomial. Equation~\eqref{eq:H1-number-transition-probability-app} is the standard number-basis transition probability for the oscillator displacement operator \cite{Perelomov1986}. We suppress the arguments of $q=q(\tau)$ and $\kappa=\kappa(\tau)$ when no ambiguity can arise. Combining these probabilities with $\sum_{m\ge0}m|g_m^{(r)}|^2=r+\kappa$ gives the tail identity 
\begin{equation}
\begin{aligned}
   K_r(\tau)=
   \kappa+\ee^{-\kappa}
   \sum_{m=0}^{r-1}(r-m)\frac{m!}{r!}\kappa^{r-m}
   \left[L_m^{(r-m)}(\kappa)\right]^2
   +\sum_{L=r}^{\infty}\mathcal C_L^{(r)}(\tau).
\end{aligned}
   \label{eq:H1-general-r-complexity-app}
\end{equation}
Equation~\eqref{eq:H1-general-r-cumulative-app} is exact at every finite cutoff. Moreover, $\mathcal C_L^{(r)}(\tau)\ge0$ for $L\ge r$. Passing to the infinite complexity sum therefore reduces to proving convergence of this nonnegative correction series. The required large-index estimate is established in subsection~\ref{app:H1-large-index} for every fixed $r$ and $\lambda>0$.

\paragraph{Rank-one reduction.}
For $r=1$, the two quotient spaces reduce to $\mathscr R_1=\CC[y]/\langle y\rangle\simeq\CC$ and $\mathscr R_1^{(2)}=\CC[y]/\langle y^2\rangle$. In the bases $(1)$ and $(1,y)$,
\begin{equation}
   \mathbf u_n^{(1)}=A_n,\qquad
   [P_n]_{\mathscr R_1^{(2)}}=A_n+B_ny,\qquad
   \mathsf H_L^{(1)}=\sum_{m=0}^{L}\frac{A_m^2}{h_m}
   =\frac{D_L}{h_L}.
   \label{eq:H1-r1-residue-reduction-app}
\end{equation}
The $2\times2$ determinants with columns $(\mathbf v_n^{(1)},\mathbf v_{n+1}^{(1)})$ and $(\mathbf v_n^{(1)},\mathbf v_{n+2}^{(1)})$ are exactly $D_n$ and $E_n$ in eq.~\eqref{eq:DE-app}. Equation~\eqref{eq:H1-jet-connector-app} gives
\begin{equation}
   \bigl(G_{n,0}^{(1)},G_{n,1}^{(1)},G_{n,2}^{(1)}\bigr)
   =\left(\frac{D_{n+1}}{D_n},-\frac{E_n}{D_n},1\right).
   \label{eq:H1-r1-G-reduction-app}
\end{equation}
Substitution into eq.~\eqref{eq:H1-unnormalized-connector-app} recovers eq.~\eqref{eq:H1-app-poly}. Likewise, eq.~\eqref{eq:H1-general-r-complexity-app} reduces to the baseline $\kappa(1+\ee^{-\kappa})$ and the rank-one correction in eq.~\eqref{eq:H1-first-jump-complexity}. Thus the four sequences $A_n,B_n,D_n,E_n$ developed in subsection~\ref{app:H1-first-jump} are precisely the $r=1$ remainder and first-jet data.

\paragraph{Three-component remainder example.}
At $r=3$, the single-component remainder coordinate is replaced by a three-component sequence, while all finite-index data given below are still obtained from fixed-dimensional recurrences. These data extend the local coefficients in eq.~\eqref{eq:H1-third-jump-low-data-main} beyond the first two shifted levels. The example is also where ``root-free'' can be seen rather than asserted. None of the three zeros of the cubic below is a rational function of $\lambda$, yet every $\Delta_L^{(3)}$, $\eta_L^{(3)}$, $\widetilde a_n^{(3)}$ and $\widetilde\beta_n^{(3)}$ given below is rational in $\lambda$. Two general features of the construction are visible in the same data, namely $\mathbf u_r^{(r)}=0$, which is why $\Delta_r^{(r)}=\Delta_{r-1}^{(r)}$, and the cancellation of the launch-site component $\ket{K_3}$ from the first Lanczos residual. Here, we have
\begin{equation}
\begin{aligned}
   P_3(y)=y^3-3y^2+(2-3\lambda)y+2\lambda,\qquad
   h_3=6\lambda^3\,, \qquad
   \mathsf Y_3=
   \begin{pmatrix}
      0&0&-2\lambda\\
      1&0&3\lambda-2\\
      0&1&3
   \end{pmatrix}.
\end{aligned}
   \label{eq:H1-r3-P-M-app}
\end{equation}
Suppressing the superscript $(3)$ on the vectors, the recurrence in eq.~\eqref{eq:H1-remainder-recurrence-main} begins with
\begin{equation}
\begin{gathered}
   \mathbf u_0=(1,0,0)^T,\qquad
   \mathbf u_1=(0,1,0)^T,\qquad
   \mathbf u_2=(-\lambda,-1,1)^T,\qquad
   \mathbf u_3=0\,,\\
   \mathbf u_4=(3\lambda^2,3\lambda,-3\lambda)^T,\qquad
   \mathbf u_5=(-6\lambda^2,-6\lambda(1+\lambda),6\lambda)^T.
\end{gathered}
   \label{eq:H1-r3-u-app}
\end{equation}
In the monomial remainder basis, the first nonsingular Gram matrix and the determinant data needed through the third shifted hopping are
\begin{equation}
   \mathsf H_2^{(3)}=
   \begin{pmatrix}
      \frac32&\frac{1}{2\lambda}&-\frac{1}{2\lambda}\\
      \frac{1}{2\lambda}&\frac1\lambda+\frac{1}{2\lambda^2}&-\frac{1}{2\lambda^2}\\
      -\frac{1}{2\lambda}&-\frac{1}{2\lambda^2}&\frac{1}{2\lambda^2}
   \end{pmatrix},
   \label{eq:H1-r3-H2-app}
\end{equation}
\begin{gather}
   \Delta_2^{(3)}=\Delta_3^{(3)}=\frac{1}{2\lambda^3}\,,
   \quad
   \Delta_4^{(3)}=\frac{7}{8\lambda^3}\,,
   \quad
   \Delta_5^{(3)}=\frac{24+91\lambda}{80\lambda^4}\,, \quad
   \Delta_6^{(3)}
   =\frac{3(312+840\lambda+715\lambda^2)}{1600\lambda^5}\,, \nonumber \\
   \eta_3^{(3)}=\eta_4^{(3)}=0\,, \quad
   \eta_5^{(3)}=-\frac67\,,\quad
   \eta_6^{(3)}=-\frac{18(4+5\lambda)}{24+91\lambda}\,. \label{eq:H1-r3-Delta-eta-app}
\end{gather}
Substituting in eq.~\eqref{eq:H1-general-r-Jacobi-app} gives
\begin{equation}
\begin{alignedat}{2}
   \widetilde a_0^{(3)}
   &=3\,,
   &\qquad
   \widetilde\beta_1^{(3)}
   &=7\lambda\,,\\
   \widetilde a_1^{(3)}
   &=\frac{22}{7}\,,
   &\qquad
   \widetilde\beta_2^{(3)}
   &=\frac{2(24+91\lambda)}{49}\,,\\
   \widetilde a_2^{(3)}
   &=\frac{480+3101\lambda}
   {7(24+91\lambda)}\,,
   &\qquad
   \widetilde\beta_3^{(3)}
   &=\frac{63\lambda(312+840\lambda+715\lambda^2)}
   {(24+91\lambda)^2}\,.
\end{alignedat}
\label{eq:H1-r3-Jacobi-app}
\end{equation}
The first few shifted Lanczos vectors provide a direct check in the Hilbert space. With $\Xi_3=24+91\lambda$,
\begin{equation}
\begin{aligned}
   \ket{\widetilde K_0^{(3)}}&=\ket{K_3},\\
   \ket{\widetilde K_1^{(3)}}&=
   \frac{\sqrt3\,\ket{K_2}+2\ket{K_4}}{\sqrt7}\,,\\
   \ket{\widetilde K_2^{(3)}}&=
   \sqrt{\frac{21\lambda}{\Xi_3}}\ket{K_1}
   -\frac{8\sqrt3}{\sqrt{14\Xi_3}}\ket{K_2}
   +\frac{12}{\sqrt{14\Xi_3}}\ket{K_4}
   +\sqrt{\frac{70\lambda}{\Xi_3}}\ket{K_5}.
\end{aligned}
   \label{eq:H1-r3-Krylov-vectors-app}
\end{equation}
The unnormalized connector starts from the exact reference expansion 
\begin{equation}
\begin{aligned}
   \bigl(G_{0,0}^{(3)},\ldots,G_{0,6}^{(3)}\bigr)=\bigl(
   6\lambda^3,\ 18\lambda^2,\ 18\lambda(1+\lambda),\ 6(1+6\lambda),\ 9(2+\lambda),\ 9,\ 1\bigr).
\end{aligned}
   \label{eq:H1-r3-initial-connector-app}
\end{equation}

Equation~\eqref{eq:H1-G-recurrence-app} determines every
seven-term connector row, and eq.~\eqref{eq:H1-r-amplitudes}, with
\(\Gamma_{n,n+j}^{(3)}=G_{n,j}^{(3)}/h_3\), gives every shifted amplitude
without solving for the three zeros of \(P_3\).

The reference-basis contribution to the complexity in eq.~\eqref{eq:H1-general-r-complexity-app} is explicitly
\begin{equation}
   \mathcal B_3(\kappa)=\kappa+\ee^{-\kappa}
   \left[
      \frac{\kappa^3}{2}
      +\frac{\kappa^2(3-\kappa)^2}{3}
      +\frac{\kappa(\kappa^2-6\kappa+6)^2}{12}
   \right],
   \label{eq:H1-r3-baseline-app}
\end{equation}
so that $K_3(\tau)=\mathcal B_3(\kappa)+\sum_{L=3}^{\infty}\mathcal C_L^{(3)}(\tau)$. The local shifted Jacobi data give the short-time expansion in eq.~\eqref{eq:H1-third-jump-short-time-main}; in physical units,
\begin{equation}
   K_3(t)=7\rho^2t^2+
   \frac{\rho^2(95\omega^2-1008\rho^2)}{84}t^4+O(t^6).
   \label{eq:H1-r3-short-time-physical-app}
\end{equation}
The preceding formulas solve the shifted problem at every finite index. %
The infinite chain is governed by the large-index behavior of exactly two sequences built from the same finite-dimensional Gram data, the determinant $\Delta_L^{(r)}$ of eq.~\eqref{eq:H1-remainder-Gram-app}, and the inverse-Gram contraction $\eta_L^{(r)}$ of eq.~\eqref{eq:H1-eta-definition-app}. The next subsection works in root rather than remainder coordinates and reconciles the two descriptions at the outset.%

\subsection{Large-index Jacobi data and analytic control}
\label{app:H1-large-index}

Within this subsection, $L$ denotes the root-Gram cutoff and not the
multiplier degree %
also denoted by $L=\deg W_Q$ in sections~\ref{sec:setup} and~\ref{sec:finiteband} and in
appendix~\ref{app:christoffel}.
The general finite-time finiteness of $K_r$ already follows from
proposition~\ref{prop:finite-time-control-main}.  The sharper analysis below
determines the resonance-dependent Jacobi intercepts and the
$\sqrt n$ hopping scale, and establishes the tail and graph-domain estimates
used for the periodic bound and the full-chain short-time expansion.

The root-Gram formulas of proposition~\ref{prop:root-Gram-Jacobi-app} express the shifted Jacobi
data through two objects, the consecutive ratios of the Gram determinants $\mathcal D_L^{(r)}$ and
the inverse-Gram contraction $\eta_L^{(r)}$. Writing the ratios as differences of a logarithm
turns both into statements about growth in $L$, which is what the rest of this subsection
computes.

\begin{proposition}[Difference form of the induced Jacobi data]
\label{prop:kernel-difference-form-app}
Write $\nabla f_L=f_L-f_{L-1}$ and $\nabla^2f_L=f_L-2f_{L-1}+f_{L-2}$ for the backward first and
second differences.\footnote{The backward difference is used, and denoted $\nabla$ rather than
$\Delta$, because $\Delta$ is already in use for the determinants of this paper, namely $\Delta_n$ in
appendix~\ref{app:christoffel}, $\Delta_L^{(r)}$ in eq.~\eqref{eq:H1-remainder-Gram-app},
$\Delta_n^{[Q]}$ in appendix~\ref{app:finite} and the Vandermonde determinant $\Delta(x)$ of
the conditioned-ensemble and characteristic-polynomial discussions in
sections~\ref{sec:rigidity} and~\ref{sec:context}, as well as for the energy
variance $\Delta_Q$ of eq.~\eqref{eq:MT-speed-shifted-seed-main}.} With $\mathcal D_L^{(r)}$ and $\eta_L^{(r)}$ as in
eqs.~\eqref{eq:root-Gram-definition-app} and~\eqref{eq:root-Gram-eta-app}, the identities of
proposition~\ref{prop:root-Gram-Jacobi-app} read
\begin{equation}
   \widetilde\beta_n^{(r)}
   =\beta_{n+r}\,
   \exp\!\Bigl[\nabla^2\log\mathcal D_{n+r}^{(r)}\Bigr],
   \qquad
   \widetilde a_n^{(r)}
   =a_{n+r}+\nabla\eta_{n+r+1}^{(r)} \,,
   \label{eq:kernel-difference-form-app}
\end{equation}
in the ranges of proposition~\ref{prop:root-Gram-Jacobi-app}. On infinite support the hopping
identity holds for $n\ge1$, the second difference at $n=0$ calling on $\mathcal D_{r-2}^{(r)}$,
which lies outside the positivity range, and the diagonal identity for every $n\ge0$; on finite
support, within the physical rows stated there. In the one-root case $r=1$ the Gram determinant is
the diagonal reproducing kernel, $\mathcal D_L^{(1)}=\mathcal K_L(\xi_1)$, so that
\begin{equation}
\begin{aligned}
   \widehat\beta_n
   &=\beta_{n+1}\exp\!\bigl[\nabla^2\log\mathcal K_{n+1}(\xi)\bigr],
   \qquad n\ge1\,,\\
   \widehat a_n
   &=a_n-\Theta_n(\xi)\,\ee^{-\nabla\log\mathcal K_{n+1}(\xi)}
        +\Theta_{n-1}(\xi)\,\ee^{-\nabla\log\mathcal K_{n}(\xi)},
\end{aligned}
   \label{eq:one-root-difference-form-app}
\end{equation}
with $\Theta_n$ as in eq.~\eqref{eq:double-Christoffel-Theta-app} and the last term of
$\widehat a_n$ absent at $n=0$, as in
eq.~\eqref{eq:double-Christoffel-a-kernel-app}.
\end{proposition}

\begin{proof}
The underlying Christoffel--Uvarov determinant ratios are in ref.~\cite{Uvarov1969} and give the recurrence coefficients of the induced measure $\nu_r$ studied by Gautschi and Li~\cite{GautschiLi1993}. Organizing those ratios as first and second differences of one logarithm is what lets the estimates in this subsection act directly on the shifted Jacobi data. For the hopping,
eq.~\eqref{eq:root-Gram-beta-app} gives
$\widetilde\beta_n^{(r)}/\beta_{n+r}
=\mathcal D_{n+r}^{(r)}\mathcal D_{n+r-2}^{(r)}/(\mathcal D_{n+r-1}^{(r)})^2$,
whose logarithm is
\begin{equation}
   \log\mathcal D_{n+r}^{(r)}-2\log\mathcal D_{n+r-1}^{(r)}+\log\mathcal D_{n+r-2}^{(r)}
   =\nabla^2\log\mathcal D_{n+r}^{(r)}\,.
\end{equation}
The determinants are strictly positive in the stated range
by eq.~\eqref{eq:root-Gram-Cauchy-Binet-app}, so the logarithm is defined. For the diagonal,
eq.~\eqref{eq:root-Gram-a-app} is
$\widetilde a_n^{(r)}=a_{n+r}+\eta_{n+r+1}^{(r)}-\eta_{n+r}^{(r)}$, which is $\nabla\eta_{n+r+1}^{(r)}$. The one-root specialization uses
$\mathcal D_L^{(1)}=\mathcal K_L(\xi_1)$ together with
eqs.~\eqref{eq:double-Christoffel-beta-kernel-app}--\eqref{eq:double-Christoffel-Theta-app}, and
$\mathcal K_n/\mathcal K_{n+1}=\ee^{-\nabla\log\mathcal K_{n+1}}$.
\end{proof}

The determinant $\mathcal D_L^{(r)}$ is the root-coordinate form of the remainder-coordinate
determinant $\Delta_L^{(r)}$ of eq.~\eqref{eq:H1-remainder-Gram-app}, which is what
subsection~\ref{app:H1-general-r} used.
Equation~\eqref{eq:H1-Gram-coordinate-congruence-app} gives
$\bigl[\mathbb K_L^\mu(\xi_a,\xi_b)\bigr]=\mathsf T_r\mathsf H_L^{(r)}\mathsf T_r^{T}$, hence
\begin{equation}
   \mathcal D_L^{(r)}=\bigl(\det\mathsf T_r\bigr)^2\,\Delta_L^{(r)},
   \qquad\text{so}\qquad
   d_L^{(r)}=\frac{\Delta_L^{(r)}}{\Delta_{L-1}^{(r)}}
            =\frac{\mathcal D_L^{(r)}}{\mathcal D_{L-1}^{(r)}}
            =\ee^{\nabla\log\mathcal D_L^{(r)}} .
   \label{eq:gram-hankel-reconciliation-app}
\end{equation}
The basis-dependent factor $(\det\mathsf T_r)^2$ cancels from every consecutive ratio, so the
sequence $d_L^{(r)}$ of the estimates below and the logarithmic differences of
eq.~\eqref{eq:kernel-difference-form-app} are the same object; in particular
$\nabla^2\log\mathcal D_L^{(r)}=\log\bigl(d_L^{(r)}/d_{L-1}^{(r)}\bigr)$. The same change of basis
identifies the two definitions of $\eta_L^{(r)}$. Substituting
$\boldsymbol p_m=\mathsf T_r\mathbf u_m^{(r)}/\sqrt{h_m}$ and
$\mathsf G_{L-1}^{(r)}=\mathsf T_r\mathsf H_{L-1}^{(r)}\mathsf T_r^{T}$ in
eq.~\eqref{eq:root-Gram-eta-app} cancels $\mathsf T_r$ and leaves
$\mathbf u_L^{(r)T}(\mathsf H_{L-1}^{(r)})^{-1}\mathbf u_{L-1}^{(r)}/\sqrt{h_Lh_{L-1}}$; the
prefactor $b_L=\sqrt{h_L/h_{L-1}}$ then returns
eq.~\eqref{eq:H1-eta-definition-app} exactly. The sequence estimated below is therefore the one
entering eq.~\eqref{eq:H1-general-r-Jacobi-app}.

Equation~\eqref{eq:kernel-difference-form-app} is the organizing statement of this subsection. It
separates the problem into a part that is common to every seed, the index shift $n\mapsto n+r$
inherited from the degree filtration, and a correction controlled entirely by how fast
the root-Gram data grow in $L$. Two features of that growth matter, one for each coefficient:
\begin{itemize}[itemsep=1pt]
\item %
      the diagonal correction is the difference $\nabla\eta_L^{(r)}$ of the full endpoint
      contraction in eq.~\eqref{eq:root-Gram-eta-app}.  A kernel ratio by itself does not force
      $\eta_L^{(r)}$ to converge; the Charlier endpoint estimate below is the additional input
      that determines its linear term and hence the intercept;
\item %
      the hopping correction is the discrete curvature
      $\nabla^2\log\mathcal D_L^{(r)}$.  In the Charlier regimes analyzed below, factorial
      $\Gamma$-growth produces its $L^{-1}$ term, while the all-lattice branch saturates
      superalgebraically.
\end{itemize}
For the Charlier problem the dichotomy is explicit and is the content of
eqs.~\eqref{eq:H1-fixed-point-Charlier-asymptotics-app} below. At an off-lattice root the
normalized evaluations grow factorially, while at a root on the Poisson lattice the evaluations
are square summable and $\mathcal K_L$ \emph{saturates}, its limit being the reciprocal of the
mass of the deleted atom,
\begin{equation}
   \lim_{L\to\infty}\mathcal K_L(\xi)=\frac{1}{\mu(\{\xi\})}
   \qquad\text{whenever }\mu(\{\xi\})>0 \,.
   \label{eq:christoffel-atom-limit-app}
\end{equation}
Saturation makes $\nabla^2\log\mathcal K_L$ superalgebraically small, of order $O(L^{-p})$ for every
fixed $p>0$ at the rate established in
eq.~\eqref{eq:H1-all-lattice-superalgebraic-app}, and $\nabla\log\mathcal K_L\to0$; factorial
growth makes $\nabla^2\log\mathcal K_L\sim L^{-1}$ and $\ee^{-\nabla\log\mathcal K_L}\to0$. The
two branches of eq.~\eqref{eq:H1-Jacobi-asymptotics-main}, with intercepts $s_r(\lambda)$ and
$2r-s_r(\lambda)$, are precisely these two behaviors, counted root by root.\footnote{Let $\xi_1,\ldots,\xi_r$ be the zeros of $P_r$ and define
\begin{equation}
   s_r(\lambda)=
   \#\bigl\{x\in\mathbb Z_{\ge0}:P_r(x-\lambda)=0\bigr\}.
   \label{eq:H1-support-zero-count-app}
\end{equation}
The integer $s_r(\lambda)$ counts the zeros of $P_r$ on the centered Poisson support and hence the atoms removed by the number-state multiplier.} 

For admissible finite deletion sets, Yermolayeva and Zhedanov obtain defected-lattice Charlier measures and recurrence coefficients rational in the index~\cite{YermolayevaZhedanov1999}. Related background includes Christoffel and Geronimus transformations of discrete classical families in the Pearson framework~\cite{Manas2022Pearson} and large-index Jacobi behavior under Darboux transformations~\cite{BaileyDerevyagin2021}. These works do not contain the reference-state choice $Q=P_r$ or the all-$r$ intercept results derived below. For fixed $r\ge1$ and $\lambda>0$, the asymptotic statement quoted in eq.~\eqref{eq:H1-Jacobi-asymptotics-main} is, as $n\to\infty$,
\begin{equation}
\begin{aligned}
   \widetilde a_n^{(r)}
   &=n+s_r(\lambda)+O(n^{-1}),\\
   \widetilde\beta_n^{(r)}
   &=\lambda\bigl(n+2r-s_r(\lambda)\bigr)+O(n^{-1}).
\end{aligned}
   \label{eq:H1-general-r-Jacobi-asymptotics-app}
\end{equation}
Generically, $s_r(\lambda)=0$. In particular, away from a support-zero resonance, the third jump satisfies $\widetilde a_n^{(3)}=n+O(n^{-1})$ and $\widetilde\beta_n^{(3)}=\lambda(n+6)+O(n^{-1})$.

The dependence of the intercepts on $s_r(\lambda)$ can be traced to the two distinct scales of the root Gram matrix. Set $\mathbf w_m=\mathsf T_r\mathbf u_m^{(r)}=(P_m(\xi_a))_{a=1}^{r}$. The Cauchy--Binet expansion of eq.~\eqref{eq:H1-Gram-coordinate-congruence-app} is
\begin{equation}
   \bigl(\det\mathsf T_r\bigr)^2\Delta_L^{(r)}
   =
   \sum_{0\le m_1<\cdots<m_r\le L}
   \frac{
      \left(
      \det\!\bigl[P_{m_j}(\xi_a)\bigr]_{a,j=1}^{r}
      \right)^2
   }{
      \prod_{j=1}^{r}h_{m_j}
   }\,.
   \label{eq:H1-Cauchy-Binet-Delta-app}
\end{equation}
The two asymptotic scales entering this determinant are already visible in the fixed-argument Charlier formulas. They follow directly from the generating function; see also ref.~\cite{HuangLinZhao2021}. With $x=\lambda+\xi$, the off-lattice and lattice forms are 
\begin{align}
   P_n(x-\lambda)
   &=n![u^n]\ee^{-\lambda u}(1+u)^x
   =\frac{(-1)^n\ee^\lambda}{\Gamma(-x)}
   \Gamma(n-x)\left(1+\frac{\lambda(x+1)}{n}
   +O(n^{-2})\right),
   \; x\notin\mathbb Z_{\ge0}\,, \nonumber \\
   P_n(k-\lambda)
   &=n!\sum_{j=0}^{\min(k,n)}
   \binom{k}{j}\frac{(-\lambda)^{n-j}}{(n-j)!},
   \quad k\in\mathbb Z_{\ge0}\,.  \label{eq:H1-fixed-point-Charlier-asymptotics-app}
\end{align}

For $x\notin\mathbb Z_{\ge0}$, the first line is the Darboux expansion about $u=-1$, obtained from $\ee^{-\lambda u}=\ee^\lambda\ee^{-\lambda(1+u)}$. The expansion may be continued to arbitrary finite order and thereby defines the full inverse-power asymptotic series. At $x=k\in\mathbb Z_{\ge0}$, the binomial expansion of $(1+u)^k$ terminates, the branch singularity at $u=-1$ is absent, and the second line is exact. In the determinant estimates below, the off-lattice expansion is retained to the finite order required at fixed $r$.

By proposition~\ref{prop:kernel-difference-form-app} and
eq.~\eqref{eq:gram-hankel-reconciliation-app}, the entire large-index problem is now the following
pair of estimates for the Charlier weight, which we state once and prove in the rest of this
subsection.

\begin{proposition}[Large-index behavior of the two kernel differences]
\label{prop:charlier-difference-estimates-app}
Fix $r\ge1$ and $\lambda>0$, and let $r_{\rm off}=r-s_r(\lambda)$ be the number of zeros of $P_r$
off the centered Poisson lattice. Then, as $L\to\infty$,
\begin{equation}
   \nabla^2\log\mathcal D_L^{(r)}=\frac{r_{\rm off}}{L}+O(L^{-2}),
   \qquad
   \nabla\eta_L^{(r)}=-r_{\rm off}+O(L^{-1}).
   \label{eq:charlier-difference-estimates-app}
\end{equation}
For $r_{\rm off}=0$ both differences are $O(L^{-p})$ for every fixed $p>0$; this case is settled by
the superalgebraic decay of the normalized lattice evaluations, established at
eq.~\eqref{eq:H1-all-lattice-superalgebraic-app} below, and is the saturation branch of the
dichotomy of eq.~\eqref{eq:kernel-mass-dichotomy}.
\end{proposition}

Substituted into eq.~\eqref{eq:kernel-difference-form-app}, these two lines give
eq.~\eqref{eq:H1-general-r-Jacobi-asymptotics-app} immediately, and they are the only place where
the Charlier weight enters. The slopes are fixed by the index shift $n\mapsto n+r$ and the
intercepts by the single integer $r_{\rm off}$. The proof occupies the remainder of the
subsection and has exactly two technical inputs, one for each line. They are
lemma~\ref{lem:H1-packed-columns-app} for the hopping difference, through the Casoratian
expansion and, when lattice roots are present, a Schur complement; and
lemma~\ref{lem:H1-eta-packed-app} for the diagonal difference, through an endpoint
interpolation. Both rest on the same fact, that the two root classes of
eq.~\eqref{eq:H1-fixed-point-Charlier-asymptotics-app} live on different scales. The factorially
growing off-lattice sector is resolved first; the bounded directions associated with deleted atoms
are then incorporated.

\paragraph{Off-lattice Gram asymptotics.}
When \(r_{\rm off}\ge1\), relabel the off-lattice roots as
\(\xi_a^{\rm off}\), and write
\(x_a=\lambda+\xi_a^{\rm off}\notin\mathbb Z_{\ge0}\),
\(a=1,\ldots,r_{\rm off}\). Factorial growth concentrates the off-lattice Gram matrix near its upper cutoff. Since the corresponding constraint space has dimension \(r_{\rm off}\), the last \(r_{\rm off}\) consecutive normalized evaluation vectors form the natural moving basis. Their leading determinant is the Casoratian, the discrete analogue of a Wronskian formed from consecutive shifts of a finite family of sequences,
\begin{equation}
\begin{aligned}
   \det\!\left[\Gamma(n+j-x_a)\right]_{
      \substack{1\le a\le r_{\rm off}\\
      0\le j\le r_{\rm off}-1}}
   &=\prod_{a=1}^{r_{\rm off}}\Gamma(n-x_a)
   \det\!\left[(n-x_a)_j\right]_{
      \substack{1\le a\le r_{\rm off}\\
      0\le j\le r_{\rm off}-1}}\\
   &=\prod_{a=1}^{r_{\rm off}}\Gamma(n-x_a)
   \prod_{1\le a<b\le r_{\rm off}}(x_a-x_b).
\end{aligned}
   \label{eq:H1-Charlier-Casoratian-app}
\end{equation}
Here \((z)_j=\Gamma(z+j)/\Gamma(z)\) denotes the rising factorial.
Since the rising factorials form a monic polynomial basis in $z=n-x$,
their evaluation determinant is the Vandermonde factor above. It is
nonzero because the Charlier zeros are distinct.\footnote{This notation is
unrelated to the falling factorial
\(\lambda^{\underline{k}}\) used in eq.~\eqref{eq:ABDE-closed-app}.} Taking
\(n=L-r_{\rm off}\) and using the first line of
eq.~\eqref{eq:H1-fixed-point-Charlier-asymptotics-app}, continued to
the finite order required at fixed $r$, shows that the determinant of
the actual consecutive evaluation matrix is asymptotic to a nonzero
multiple of the Casoratian in eq.~\eqref{eq:H1-Charlier-Casoratian-app}.
The matrix is therefore invertible for all sufficiently large \(L\).

In this moving high-index basis, write 
\begin{equation}
\begin{gathered}
   \mathbf w_m^{\rm off}
   =
   \bigl(P_m(\xi_1^{\rm off}),\ldots,
   P_m(\xi_{r_{\rm off}}^{\rm off})\bigr)^T,
   \qquad
   \mathbf z_m=\frac{\mathbf w_m^{\rm off}}{\sqrt{h_m}}\,,\\
   \mathsf Z_L
   =
   \bigl(\mathbf z_{L-r_{\rm off}},\ldots,\mathbf z_{L-1}\bigr),
   \qquad
   \mathbf t_{\ell,L}
   =\mathsf Z_L^{-1}\mathbf z_{L-r_{\rm off}-\ell}\,,\\
   \mathsf K_{L-1}^{\rm off}
   =\sum_{m=0}^{L-1}\mathbf z_m\mathbf z_m^T
   =\mathsf Z_L\mathsf A_L\mathsf Z_L^T\,,\\
   \mathsf A_L
   =\Id_{r_{\rm off}}
   +\sum_{\ell=1}^{L-r_{\rm off}}
   \mathbf t_{\ell,L}\mathbf t_{\ell,L}^T\,.
\end{gathered}
   \label{eq:H1-normalized-packed-Gram-app}
\end{equation}
The vector \(\mathbf t_{\ell,L}\) gives the coordinates of the
degree-\((L-r_{\rm off}-\ell)\) evaluation column in the moving
basis \(\mathsf Z_L\). The following estimate makes precise the factorial localization of these coordinates near the upper cutoff; it is the first of the two inputs to proposition~\ref{prop:charlier-difference-estimates-app}, and is used again in the endpoint contraction.

\begin{lemma}[Uniform packing of the off-lattice Charlier columns]
\label{lem:H1-packed-columns-app}
Fix \(r\), \(\lambda>0\), and a parameter value for which
\(s=r_{\rm off}\ge1\), and put \(n=L-s\) and
\(S=s(s-1)/2\). The abbreviation $s$ for $r_{\rm off}$ is used throughout this lemma,
lemma~\ref{lem:H1-eta-packed-app} and their proofs; the assembly returns to $r_{\rm off}$. There exist \(L_0,C_\star>0\), depending only on $\lambda$ and the
fixed root configuration, such that, for
\(L\ge L_0\),
\begin{equation}
\begin{gathered}
   |(\mathbf t_{\ell,L})_j|
   \le
   C_\star^\ell L^{-\ell/2+S},
   \qquad
   1\le\ell\le\frac{n}{2},\quad 1\le j\le s\,,\\
   \sum_{m=0}^{\lfloor n/2\rfloor}
   \left\|\mathsf Z_L^{-1}\mathbf z_m\right\|^2
   =O(L^{-p})
   \qquad\text{for every fixed }p>0\,.
\end{gathered}
   \label{eq:H1-terminal-column-bound-app}
\end{equation}
For each fixed \(\ell\) and \(j\), the coordinate has a complete
inverse-power expansion; in particular,
\begin{equation}
   (\mathbf t_{\ell,L})_j
   =L^{-(\ell+j-1)/2}
   \left(
      \tau_{\ell j}^{(0)}+
      \frac{\tau_{\ell j}^{(1)}}{L}+O(L^{-2})
   \right).
   \label{eq:H1-packed-column-fixed-expansion-app}
\end{equation}
Consequently,
\begin{equation}
   \bigl(\mathsf A_L-\Id_s\bigr)_{jk}
   =L^{-(j+k)/2}\bigl(e_{jk}^{(0)}+O(L^{-1})\bigr),\qquad
   \det\mathsf A_L
   =1+\frac{s^2\lambda}{L}+O(L^{-2}).
   \label{eq:H1-packed-Gram-expansion-app}
\end{equation}
Here $e_{jk}^{(0)}$ denotes the leading coefficient of the $(j,k)$ entry at its natural
order; only $e_{s1}^{(0)}$ is used below.
\end{lemma}

\begin{proof}
Set
\(\mathsf W_L=(\mathbf w_n^{\rm off},\ldots,
\mathbf w_{n+s-1}^{\rm off})\), so that its normalized counterpart satisfies
\(\mathsf Z_L=\mathsf W_L
\diag(h_n^{-1/2},\ldots,h_{n+s-1}^{-1/2})\).
Cramer's rule gives the exact identity
\begin{equation}
   (\mathbf t_{\ell,L})_j
   =\sqrt{\frac{h_{n+j-1}}{h_{n-\ell}}}\,
   \frac{
      \det\mathsf W_L^{(j\leftarrow\mathbf w_{n-\ell}^{\rm off})}
   }{\det\mathsf W_L}\,,
   \qquad 1\le j\le s\,,
   \label{eq:H1-packed-column-Cramer-app}
\end{equation}
where the superscript means that the \(j\)-th column is replaced, the denominator being
nonzero for all sufficiently large \(L\) by
eq.~\eqref{eq:H1-Charlier-Casoratian-app}.%

For any fixed integer \(M\ge1\), expansion about
\(u=-1\) gives
\begin{equation}
   P_m(x-\lambda)
   =(-1)^m\ee^\lambda
   \sum_{\nu=0}^{M-1}
   \frac{(-\lambda)^\nu}{\nu!\,\Gamma(-x-\nu)}
   \Gamma(m-x-\nu)
   +O\bigl(\Gamma(m-x-M)\bigr).
   \label{eq:H1-uniform-finite-Darboux-app}
\end{equation}
Here \(x\) ranges over the fixed off-lattice roots, so the remainder is
uniform over this finite set. Indeed, after writing \(v=1+u\), the
remainder in \(\ee^{-\lambda v}\) is \(v^M\) times an \emph{entire} function of $v$; coefficient extraction then gives the last term in
eq.~\eqref{eq:H1-uniform-finite-Darboux-app}. Entirety, rather than a local radius of
convergence, is what is needed, because the extraction pairs every Taylor coefficient of that function
against a rising-factorial ratio $\Gamma(m-x-M-\nu)/\Gamma(-x-M-\nu)$ over all $\nu\ge0$, and only
the $1/\nu!$ decay of the coefficients of $\ee^{-\lambda v}$ dominates that growth. 
A quantitative bound follows directly from the Taylor tail. If $T_k$ denotes its term with power $v^{M+k}$, then
\begin{equation}
   \frac{|T_k|}{|T_0|}
   =\frac{\lambda^kM!}{(M+k)!}
     \prod_{j=1}^{k}
     \left|\frac{x+M+j}{m-x-M-j}\right|.
\end{equation}
For all sufficiently large $m$ and $0\le k\le m/2$, this is at most $(C/m)^k$. For $k>m/2$, the polynomial form of the gamma quotient gives
$|T_k|\le C\lambda^k(k+m+C)^m/k!$. The ratio of successive quantities on the right is at most $C/m$ throughout that range; hence their sum is bounded by twice its first term for large $m$. Stirling's formula then gives
\begin{equation}
   \sum_{k>m/2}|T_k|
   \le C\,\Gamma(m-x-M)\exp[-c\,m\log m].
\end{equation}
The near and far ranges are therefore both $O(\Gamma(m-x-M))$. Because the off-lattice root set is finite, $C,c$ and the threshold $m_0$ can be chosen uniformly in $x$.%

For fixed \(\ell\), substitution in
eq.~\eqref{eq:H1-packed-column-Cramer-app} and continuation to
arbitrary finite \(M\) give the expansion in
eq.~\eqref{eq:H1-packed-column-fixed-expansion-app}. After
\(\Gamma(n-\ell-x_a)\) is removed from row \(a\), the leading numerator
has the fixed degree set
\(\{0,\ell,\ell+1,\ldots,\ell+s-1\}\) with
\(\ell+j-1\) omitted. Its excess over the ordinary Vandermonde degree
is \((s-1)\ell-j+1\); the corresponding Schur factor is homogeneous of that degree in the
shifted arguments $z_a=n-\ell-x_a$, and since $z_a=n\bigl(1+O(n^{-1})\bigr)$ uniformly over the
fixed root set, its leading value is that of the Schur polynomial at equal arguments, a fixed
nonzero constant.  Combining this degree with the gamma and norm ratios
gives the prefactor \(L^{-(\ell+j-1)/2}\) stated there. In particular,
factoring the leading gamma function from each row of the denominator
gives 
\begin{equation}
\begin{aligned}
   \det\mathsf W_L
   &={}
   (-1)^{sn+\binom{s}{2}}
   \frac{\ee^{s\lambda}}
   {\prod_{a=1}^{s}\Gamma(-x_a)}
   \prod_{a=1}^{s}\Gamma(n-x_a)
   \prod_{a<b}(x_a-x_b)
   \bigl(1+O(L^{-1})\bigr).
\end{aligned}
   \label{eq:H1-packed-denominator-app}
\end{equation}
Since the leading constant is nonzero, there is a \(c_0>0\) such that
\begin{equation}
   |\det\mathsf W_L|
   \ge
   c_0\prod_{a=1}^{s}\Gamma(n-x_a)
   \label{eq:H1-packed-denominator-lower-app}
\end{equation}
for all sufficiently large \(L\), a bound
independent of \(\ell\), so the same denominator estimate serves every term of the tail sum
below.

We now estimate the \(L\)-dependent tail without using a Darboux
expansion whose shift grows with \(L\). Coefficient extraction from the
generating function gives the exact finite sum
\begin{equation}
   P_m(x-\lambda)
   =
   \sum_{q=0}^{m}
   \binom{m}{q}(-\lambda)^{m-q}x^{\underline q}\,,
   \qquad
   x^{\underline q}
   =(-1)^q\frac{\Gamma(q-x)}{\Gamma(-x)}\,.
   \label{eq:H1-exact-fixed-point-finite-sum-app}
\end{equation}
The second identity holds for every off-lattice \(x\). Splitting the sum at \(m-q=m/2\), the gamma-product bound
\begin{equation}
   \frac{\Gamma(m-k-x)}{\Gamma(m-x)}
   \le \left(\frac{C}{m}\right)^k,
   \qquad 0\le k\le\frac m2\,,
\end{equation}
gives, after division by \(\Gamma(m-x)\), the two-range bound
\begin{equation}
   C\sum_{k=0}^{\lfloor m/2\rfloor}\frac{(C\lambda)^k}{k!}
   +
   Cm\bigl(2\max\{1,\lambda\}\bigr)^m
   \frac{\Gamma(m/2+C)}{\Gamma(m-x)}
   =O(1),
   \label{eq:H1-exact-sum-uniform-bound-app}
\end{equation}
because the second term is
\(O(\exp[-c_1m\log m])\) by Stirling's formula. Hence
\begin{equation}
   |P_m(x_a-\lambda)|
   \le C_0\Gamma(m-x_a),
   \qquad m\ge m_0,\quad 1\le a\le s\,.
   \label{eq:H1-uniform-fixed-point-bound-app}
\end{equation}
All constants are uniform over the finite root set.

Let \(1\le\ell\le n/2\), the cutoff that keeps the replaced column index $m=n-\ell$ above
$n/2\ge m_0$, so that eq.~\eqref{eq:H1-uniform-fixed-point-bound-app} is available and
$\Gamma(n-\ell-x_a)>0$. In the numerator of
eq.~\eqref{eq:H1-packed-column-Cramer-app}, factor
\(\Gamma(n-\ell-x_a)\) from row \(a\) after applying
eq.~\eqref{eq:H1-uniform-fixed-point-bound-app} entrywise.
For the unreplaced column indexed by \(k=0,\ldots,s-1\),
\begin{equation}
   \frac{\Gamma(n+k-x_a)}{\Gamma(n-\ell-x_a)}
   \le (C_1n)^{\ell+k},
\end{equation}
whereas
\begin{equation}
   \prod_{a=1}^{s}
   \frac{\Gamma(n-\ell-x_a)}{\Gamma(n-x_a)}
   \le \left(\frac{C_2}{n}\right)^{s\ell}.
\end{equation}
Expansion of the fixed-size determinant, followed by
eqs.~\eqref{eq:H1-packed-denominator-lower-app} and
\eqref{eq:H1-uniform-fixed-point-bound-app}, first gives
\begin{equation}
   \left|
      \det\mathsf W_L^{(j\leftarrow\mathbf w_{n-\ell}^{\rm off})}
   \right|
   \le
   C_3^\ell
   \prod_{a=1}^{s}\Gamma(n-\ell-x_a)\,
   n^{(s-1)\ell+S-j+1}.
   \label{eq:H1-packed-numerator-uniform-bound-app}
\end{equation}
Combining this with the product of gamma ratios above and the
denominator lower bound gives
\begin{equation}
   \left|
   \frac{
      \det\mathsf W_L^{(j\leftarrow\mathbf w_{n-\ell}^{\rm off})}
   }{\det\mathsf W_L}
   \right|
   \le
   C_4^\ell n^{-\ell+S-j+1},
   \qquad 1\le\ell\le\frac n2\,.
   \label{eq:H1-packed-unnormalized-ratio-app}
\end{equation}
Here \((s-1)\ell+S-j+1\) is the sum of the lengths
\(\ell+k\) for the \(s-1\) unreplaced columns. This explains the power
of \(n\) without invoking a generalized-Vandermonde estimate at growing
degree. Moreover, we have
\(
   \sqrt{\frac{h_{n+j-1}}{h_{n-\ell}}}
   \le C_5^\ell n^{(\ell+j-1)/2}.
\)
Since \(n=L-s\), eq.~\eqref{eq:H1-packed-column-Cramer-app} now proves
the first bound in eq.~\eqref{eq:H1-terminal-column-bound-app}.

The columns below the middle of the Gram sum are still farther from the
moving basis. Cofactor expansion of \(\mathsf Z_L^{-1}\), followed by
eqs.~\eqref{eq:H1-packed-denominator-lower-app} and
\eqref{eq:H1-uniform-fixed-point-bound-app}, gives constants
\(C_6,C_7>0\) such that, uniformly for \(0\le m\le n/2\),
\begin{equation}
   \left\|\mathsf Z_L^{-1}\mathbf z_m\right\|
   \le
   C_6^L L^{C_7}
   \frac{\Gamma(m+C_7)}{\sqrt{h_m}}
   \frac{\sqrt{h_n}}{\Gamma(n-C_7)}
   \le \exp\bigl[-cL\log L\bigr],
   \label{eq:H1-old-column-Stirling-app}
\end{equation}
for some \(c>0\) and all sufficiently large \(L\).
Indeed, each cofactor contains \(s-1\) consecutive high-index columns.
After division by eq.~\eqref{eq:H1-packed-denominator-lower-app}, their
gamma ratios contribute only \(L^{C_7}\). The replaced column contributes
at most \(C_6^m\Gamma(m+C_7)/\Gamma(n-C_7)\), with the finitely many
\(m<m_0\) absorbed by increasing \(C_6\).
The last inequality is Stirling's formula, with \(m\le n/2\) and
\(n=L-s\). It proves the
second line of eq.~\eqref{eq:H1-terminal-column-bound-app} after summing
over \(m\).

It remains to combine the fixed-\(\ell\) expansion with these uniform
bounds. Given any prescribed \(N>0\), choose a fixed integer
\(K>N+2S\). For \(L\) sufficiently large,
\begin{equation}
   \sum_{\ell=K}^{\lfloor n/2\rfloor}
   \|\mathbf t_{\ell,L}\|^2
   \le
   sL^{2S}\sum_{\ell=K}^{\infty}
   \left(\frac{C_\star^2}{L}\right)^\ell
   =O(L^{-N}).
   \label{eq:H1-packed-tail-split-app}
\end{equation}
Together with eq.~\eqref{eq:H1-old-column-Stirling-app}, this gives
\begin{equation}
   \mathsf A_L
   =
   \Id_s+\sum_{\ell=1}^{K-1}
   \mathbf t_{\ell,L}\mathbf t_{\ell,L}^T+O(L^{-N})
   \label{eq:H1-packed-fixed-K-reduction-app}
\end{equation}
entrywise. The indices not already included in these two sums are the
\(s\) basis columns \(m=n,\ldots,n+s-1\), which give the explicit
identity term in \(\mathsf A_L\). Thus no fixed-\(\ell\) expansion is
summed over a range that
grows with \(L\). Since \(N\) is arbitrary, the fixed-\(\ell\)
expansions give the complete asymptotic expansion of \(\mathsf A_L\).
The term \(\ell=1\) gives
\((\mathsf A_L-\Id_s)_{jk}=O(L^{-(j+k)/2})\), while every
\(\ell\ge2\) begins one inverse power later. At leading order,
the numerator degree set for \(\ell=j=1\) is \(\{0\}\) when \(s=1\)
and \(\{0,2,\ldots,s\}\) when \(s\ge2\). After the ordinary
Vandermonde is removed, its
leading Schur factor is \(s_{(1^{s-1})}(1^s)=s\), while the alternating
Charlier signs give the leading column ratio, 
\(\det\mathsf W_L^{(1\leftarrow\mathbf w_{n-1}^{\rm off})}/
\det\mathsf W_L=-sL^{-1}+O(L^{-2})\). The norm ratio in
eq.~\eqref{eq:H1-packed-column-Cramer-app} therefore gives
\begin{equation}
   (\mathbf t_{1,L})_1
   =-s\sqrt{\lambda}\,L^{-1/2}+O(L^{-3/2}).
   \label{eq:H1-packed-first-coordinate-app}
\end{equation}
Thus only the \((1,1)\) entry contributes to
\(\det\mathsf A_L\) at order \(L^{-1}\), and its contribution is
\(s^2\lambda/L\). This proves the last assertion.
\end{proof}
The remaining determinant scale is carried by the moving basis itself.
Combining eq.~\eqref{eq:H1-Charlier-Casoratian-app} with the fixed-point
expansion and \(h_m=\lambda^m m!\), we obtain
\begin{equation}
   (\det\mathsf Z_L)^2
   =C_{r,\lambda}
   \frac{\displaystyle
      \prod_{a=1}^{r_{\rm off}}
      \Gamma(L-r_{\rm off}-x_a)^2}
   {\displaystyle
      \prod_{j=0}^{r_{\rm off}-1}
      h_{L-r_{\rm off}+j}}
   \left(
      1+\frac{\alpha'_{r,\lambda}}{L}+O(L^{-2})
   \right),
   \qquad C_{r,\lambda}>0\,.
   \label{eq:H1-packed-Casoratian-expansion-app}
\end{equation}
Lemma~\ref{lem:H1-packed-columns-app} permits the fixed-index Darboux
expansions entering eq.~\eqref{eq:H1-packed-Casoratian-expansion-app}
to be passed through the complete off-lattice Gram sum, so
eq.~\eqref{eq:H1-packed-Casoratian-expansion-app} is an asymptotic expansion of the full
determinant and not a termwise expansion of finitely many near-edge columns.

\paragraph{Support-zero resonances.}
When \(s_r(\lambda)=0\), the root Gram matrix consists entirely of
the off-lattice sector, and
\(\det\mathsf K_{L-1}^{\rm off}
=(\det\mathsf Z_L)^2\det\mathsf A_L\).
When \(0<s_r(\lambda)<r\), the factorial sector is accompanied by a
bounded lattice sector. The latter must be separated by a Schur complement before the
large-$L$ determinant ratio can be read from the former. The remaining case
$s_r(\lambda)=r$, in which there is no off-lattice sector at all and $\mathsf Z_L$ is not defined,
is settled separately at eq.~\eqref{eq:H1-all-lattice-superalgebraic-app}. Relabel the lattice roots as
\(\xi_\alpha^{\rm lat}=k_\alpha-\lambda\), where
\(k_\alpha\in\mathbb Z_{\ge0}\) and
\(\alpha=1,\ldots,s_r(\lambda)\), and write 
\begin{equation}
\begin{gathered}
   \mathbf w_m^{\rm lat}
   =
   \bigl(P_m(\xi_1^{\rm lat}),\ldots,
   P_m(\xi_{s_r(\lambda)}^{\rm lat})\bigr)^T,
   \qquad
   \mathbf z_m^{\rm lat}
   =\frac{\mathbf w_m^{\rm lat}}{\sqrt{h_m}}\,,\\
   \mathsf K_{L-1}^{\rm lat}
   =
   \sum_{m=0}^{L-1}
   \mathbf z_m^{\rm lat}\mathbf z_m^{{\rm lat}\,T},
   \qquad
   \mathsf B_{L-1}
   =\sum_{m=0}^{L-1}
   \mathbf z_m^{\rm lat}\mathbf z_m^T\,,\\
   \mathsf K_{L-1}^{\rm root}
   =
   \begin{pmatrix}
      \mathsf K_{L-1}^{\rm lat}&\mathsf B_{L-1}\\
      \mathsf B_{L-1}^T&\mathsf K_{L-1}^{\rm off}
   \end{pmatrix},\\
   \widehat{\mathsf K}_{L-1}^{\rm off}
   =
   \mathsf K_{L-1}^{\rm off}
   -\mathsf B_{L-1}^T
   (\mathsf K_{L-1}^{\rm lat})^{-1}
   \mathsf B_{L-1}\,.
\end{gathered}
   \label{eq:H1-mixed-root-Schur-app}
\end{equation}
The Schur complement
\(\widehat{\mathsf K}_{L-1}^{\rm off}\) is the effective Gram matrix
on the off-lattice constraint space after the lattice directions have
been removed. Completeness of the Charlier polynomials on the Poisson
support fixes the lattice limit exactly,
\begin{equation}
   \mathsf K_{L-1}^{\rm lat}
   \longrightarrow
   \diag\!\left(
      \frac{\ee^\lambda k_1!}{\lambda^{k_1}},\ldots,
      \frac{\ee^\lambda k_{s_r(\lambda)}!}
      {\lambda^{k_{s_r(\lambda)}}}
   \right).
   \label{eq:H1-lattice-Gram-limit-app}
\end{equation}
For $s_r(\lambda)>0$, this limit is positive definite, and the exact lattice expression in
eq.~\eqref{eq:H1-fixed-point-Charlier-asymptotics-app} shows
that the remainder is \(O(L^{-p})\) for every fixed \(p>0\). In particular
$\mathsf K_{L-1}^{\rm lat}$ is invertible with bounded inverse for all sufficiently large $L$, so
the Schur complement of eq.~\eqref{eq:H1-mixed-root-Schur-app} is defined on that range, which is
where every statement below is made. Thus the lattice
block saturates, while the mixed block grows at most polynomially, since in the summand
$\mathbf z_m^{\rm lat}\mathbf z_m^{T}$ the superalgebraically small lattice factor and the
factorially growing off-lattice factor carry reciprocal $\sqrt{h_m}$ scales, which cancel and
leave a polynomially bounded contribution. Its
contribution to the unnormalized Schur complement need not vanish. It
becomes negligible only after the factorial scale of the off-lattice
block has been divided out.

Accordingly, write
$
    \widehat{\mathsf B}_L
    =
   \mathsf B_{L-1}(\mathsf Z_L^{-1})^T.
$
The sum defining the mixed block gives
\begin{equation}
   \widehat{\mathsf B}_L
   =
   \sum_{m=0}^{L-1}
   \mathbf z_m^{\rm lat}
   \bigl(\mathsf Z_L^{-1}\mathbf z_m\bigr)^T
   =O(L^{-p})
   \qquad\text{for every fixed }p>0\,.
   \label{eq:H1-normalized-mixed-block-decay-app}
\end{equation}
Indeed, for \(m\le n/2\), Cauchy--Schwarz combines the bounded lattice
Gram sum with the second estimate in
eq.~\eqref{eq:H1-terminal-column-bound-app}. For \(n/2<m<n\), write
\(\ell=n-m\) and use the first estimate there, whose $\ell$-independent constant does not
accumulate over the range because $C_\star^{\ell}L^{-\ell/2}\le1$ for every $\ell$ once
$L\ge C_\star^{2}$. The remaining
\(n\le m<L\) coordinates are columns of the identity. Throughout the
range \(m>n/2\), the exact
lattice evaluation in
eq.~\eqref{eq:H1-fixed-point-Charlier-asymptotics-app} is
superalgebraically small after division by \(\sqrt{h_m}\), whereas the
off-lattice coordinates have at most polynomial growth. This proves the superalgebraic decay asserted in eq.~\eqref{eq:H1-normalized-mixed-block-decay-app}.

In the normalized basis the full root Gram matrix becomes
\begin{equation}
\begin{aligned}
   \begin{pmatrix}
      \Id_{s_r(\lambda)}&0\\
      0&\mathsf Z_L^{-1}
   \end{pmatrix}
   \mathsf K_{L-1}^{\rm root}
   \begin{pmatrix}
      \Id_{s_r(\lambda)}&0\\
      0&(\mathsf Z_L^{-1})^T
   \end{pmatrix}
   &=
   \begin{pmatrix}
      \mathsf K_{L-1}^{\rm lat}&\widehat{\mathsf B}_L\\
      \widehat{\mathsf B}_L^T&\mathsf A_L
   \end{pmatrix}.
\end{aligned}
   \label{eq:H1-mixed-normalized-Gram-app}
\end{equation}
Its off-lattice Schur complement is therefore
\begin{equation}
\begin{aligned}
   \widehat{\mathsf S}_L
   =
   \mathsf Z_L^{-1}
   \widehat{\mathsf K}_{L-1}^{\rm off}
   (\mathsf Z_L^{-1})^T 
   =\mathsf A_L
   -\widehat{\mathsf B}_L^T
   (\mathsf K_{L-1}^{\rm lat})^{-1}
   \widehat{\mathsf B}_L
   =\mathsf A_L+O(L^{-p}),
\end{aligned}
   \label{eq:H1-normalized-Schur-deficit-app}
\end{equation}
entrywise for every fixed \(p>0\). Indeed,
eq.~\eqref{eq:H1-normalized-mixed-block-decay-app} and the positive
limit in eq.~\eqref{eq:H1-lattice-Gram-limit-app} make the Schur
deficit superalgebraically small. Hence
\begin{equation}
   \det\mathsf K_{L-1}^{\rm root}
   =
   \det\mathsf K_{L-1}^{\rm lat}\,
   (\det\mathsf Z_L)^2
   \det\widehat{\mathsf S}_L.
\end{equation}
The lattice determinant approaches a finite nonzero constant, and the
algebraic behavior of the consecutive ratio is carried, up to
superalgebraically small terms, by the off-lattice factor, so the uniform packing estimate
passes the fixed-index expansion through the mixed root Gram determinant as well. Since
\((\det\mathsf T_r)^2\Delta_{L-1}^{(r)}
=\det\mathsf K_{L-1}^{\rm root}\),
eqs.~\eqref{eq:H1-packed-Gram-expansion-app} and
\eqref{eq:H1-packed-Casoratian-expansion-app} give, for
\(r_{\rm off}\ge1\), with
\(\mathfrak c_{r,\lambda}^{(\Delta)}\) denoting the first subleading coefficient of the determinant ratio, which is never needed below,
\begin{equation}
   d_L^{(r)}
   =\lambda^{-r_{\rm off}}L^{r_{\rm off}}
   \left(
      1+\frac{\mathfrak c_{r,\lambda}^{(\Delta)}}{L}+O(L^{-2})
   \right),
   \qquad
   \frac{d_L^{(r)}}{d_{L-1}^{(r)}}
   =1+\frac{r_{\rm off}}{L}+O(L^{-2}).
   \label{eq:H1-Delta-ratio-asymptotics-app}
\end{equation}
By eq.~\eqref{eq:gram-hankel-reconciliation-app} the second line is exactly the hopping
input of proposition~\ref{prop:kernel-difference-form-app},
\begin{equation}
   \nabla^2\log\mathcal D_L^{(r)}
   =\log\frac{d_L^{(r)}}{d_{L-1}^{(r)}}
   =\frac{r_{\rm off}}{L}+O(L^{-2}),
   \label{eq:kernel-second-difference-charlier-app}
\end{equation}
which is the first line of proposition~\ref{prop:charlier-difference-estimates-app}.\footnote{A caution for anyone reproducing the resonant branch numerically.
The distinction between $r_{\rm off}=r$ and $r_{\rm off}<r$ turns on whether a root sits
\emph{exactly} on the Poisson lattice, %
whereas floating-point root finding generally returns a nearby off-lattice point unless the
resonance is imposed exactly. Two
independent things then go wrong, and both drive the computation toward the generic branch.
First, the condition $P_r(k-\lambda)=0$ is met only to rounding accuracy, so the atom at $k$
returns with weight of order the squared machine epsilon instead of zero, and the measure being
diagonalized is no longer the resonant one. Second, even at an exactly represented lattice root
the normalized evaluations $P_m(k-\lambda)/\sqrt{h_m}$ form the square-summable, and hence recessive, solution of the recurrence at that argument. This is the mass-point branch of eq.~\eqref{eq:kernel-mass-dichotomy}.  The second line of eq.~\eqref{eq:H1-fixed-point-Charlier-asymptotics-app} should be contrasted with the first. Forward recursion therefore contaminates them with the factorially growing solution at the level of the rounding
error, and the computed values track the dominant branch as soon as that contamination exceeds the
true one. %
The crossover index at which either effect becomes visible depends on the working precision
and on the distance from resonance; sufficiently high controlled precision can resolve the
resonant branch over any prescribed finite range. The resonance must therefore be imposed at the level of the measure, by deleting
the zero-weight atom before running Lanczos, as the construction of
appendix~\ref{app:numerics} does. Otherwise the recursion must be carried at precision high enough that
the contamination stays below the true lattice evaluation out to the index of interest.} %

This conclusion has a direct asymptotic interpretation. 
 A root off the Poisson lattice contributes $\log\Gamma$-type
growth to $\log\mathcal D_L^{(r)}$, and $\nabla^2\log\Gamma(L)=1/L+O(L^{-2})$. Each such root
contributes one unit, and a root on the lattice contributes none. This is the factorial-growth
branch of the dichotomy announced after eq.~\eqref{eq:kernel-difference-form-app}.

\paragraph{The diagonal difference.}
This is the second line of
proposition~\ref{prop:charlier-difference-estimates-app}. The contraction $\eta_L^{(r)}$ couples
the two endpoint evaluation vectors, and its leading behavior follows from interpolating the next
off-lattice vector in the consecutive basis.
The interpolation coefficients $\mathbf c_L$ are defined through
\begin{equation}
   \mathsf W_L
   =
   \bigl(
      \mathbf w_{L-r_{\rm off}}^{\rm off},
      \ldots,
      \mathbf w_{L-1}^{\rm off}
   \bigr),
   \qquad
   \mathbf w_L^{\rm off}
   =\mathsf W_L\mathbf c_L\,.
\end{equation}
With $s=r_{\rm off}$ and $n=L-s$ this is the matrix of the proof of
lemma~\ref{lem:H1-packed-columns-app} written out at its endpoints, so its Casoratian
invertibility and the normalization $\mathsf Z_L=\mathsf W_L\diag(\cdot)$ carry over unchanged.
\begin{lemma}[Endpoint interpolation and the inverse-Gram contraction]
\label{lem:H1-eta-packed-app}
Let \(s=r_{\rm off}\ge1\), and define \(\mathbf c_L\) as above. Then 
\begin{equation}
\begin{aligned}
   (\mathbf c_L)_j
   &=O\bigl(L^{s-j+1}\bigr),
   \qquad 1\le j\le s\,,\\
   (\mathbf c_L)_s
   &=-sL+\mathfrak c_{r,\lambda}^{({\rm int})}+O(L^{-1}),\\
   \widehat{\mathbf c}_L
   &=
   \frac{\mathsf Z_L^{-1}\mathbf w_L^{\rm off}}
   {\sqrt{h_{L-1}}}\\   (\widehat{\mathbf c}_L)_j
   &=
   \sqrt{\frac{h_{L-s+j-1}}{h_{L-1}}}\,
   (\mathbf c_L)_j
   =
   O\bigl(L^{1+(s-j)/2}\bigr)\,.
\end{aligned}
   \label{eq:H1-eta-normalized-packed-app}
\end{equation}
Each component admits a complete inverse-power expansion after the
leading power indicated has been removed. In particular,
\begin{equation}
   (\widehat{\mathbf c}_L)_s
   =-sL+\mathfrak c_{r,\lambda}^{({\rm int})}+O(L^{-1}).
   \label{eq:H1-eta-endpoint-refined-app}
\end{equation}
For the purely off-lattice root set,
\begin{equation}
   \eta_L^{(r)}
   =-sL+\mathfrak c_{r,\lambda}^{(\eta)}+O(L^{-1}),
   \quad\text{hence}\quad
   \nabla\eta_L^{(r)}=-s+O(L^{-1})=-r_{\rm off}+O(L^{-1}).
   \label{eq:H1-eta-packed-asymptotics-app}
\end{equation}
The second form is the diagonal input of
proposition~\ref{prop:kernel-difference-form-app}. Note that it is the \emph{first} difference of
$\eta$ that enters, so the linear term $(-sL)$, which is the large object here, contributes
only its slope. This is why the diagonal intercept is quantized by a root count while the
coefficient $\mathfrak c_{r,\lambda}^{(\eta)}$ never appears in the answer.
If lattice and off-lattice roots coexist, their contribution to this
formula is superalgebraically small, so
eq.~\eqref{eq:H1-eta-packed-asymptotics-app} remains valid with the same
off-lattice linear term.
\end{lemma}

\begin{proof}
The Casoratian estimate makes \(\mathsf W_L\) invertible for all
sufficiently large \(L\). Cramer's rule therefore gives
\begin{equation}
   (\mathbf c_L)_j
   =
   \frac{
      \det\mathsf W_L^{(j\leftarrow\mathbf w_L^{\rm off})}
   }{\det\mathsf W_L}.
   \label{eq:H1-forward-column-Cramer-app}
\end{equation}
After the common factors \(\Gamma(L-s-x_a)\) have been removed from the
rows, the leading numerator determinant has the degree set
\(\{0,1,\ldots,s\}\setminus\{j-1\}\). Its excess degree over the
denominator Vandermonde is \(s-j+1\). The finite Darboux expansion in
eq.~\eqref{eq:H1-uniform-finite-Darboux-app} then proves the first line
of eq.~\eqref{eq:H1-eta-normalized-packed-app} and gives a complete
inverse-power expansion of every component.

The coefficient of the highest basis element can be read off without
evaluating a determinant. Put \(z=L-s-x\). Up to its common row factor
and alternating column signs, the forward column is \((z)_s\), while
the consecutive basis consists of
\(1,(z)_1,\ldots,(z)_{s-1}\). Reduction modulo
\(\prod_{a=1}^{s}(z-L+s+x_a)\) shows that the coefficient of
\((z)_{s-1}\) in the remainder of \((z)_s\) is
\(sL+O(1)\). Restoring the alternating signs gives
\((\mathbf c_L)_s=-sL+O(1)\). Retaining two further inverse powers in
eq.~\eqref{eq:H1-uniform-finite-Darboux-app} gives the constant and the
\(O(L^{-1})\) remainder stated in
eq.~\eqref{eq:H1-eta-normalized-packed-app}. 
The normalization in eq.~\eqref{eq:H1-eta-normalized-packed-app} follows from
\(\mathsf Z_L=\mathsf W_L
\diag(h_{L-s}^{-1/2},\ldots,h_{L-1}^{-1/2})\). Since
\begin{equation}
   \sqrt{\frac{h_{L-s+j-1}}{h_{L-1}}}
   =
   \lambda^{-(s-j)/2}L^{-(s-j)/2}
   \bigl(1+O(L^{-1})\bigr),
   \label{eq:H1-endpoint-norm-ratio-app}
\end{equation}
it gives the stated orders and
eq.~\eqref{eq:H1-eta-endpoint-refined-app}.

Write \(\mathsf E_L=\mathsf A_L-\Id_s\). In this notation the first entry of
eq.~\eqref{eq:H1-packed-Gram-expansion-app} reads
\begin{equation}
   (\mathsf E_L)_{jk}
   =
   L^{-(j+k)/2}
   \left(e_{jk}^{(0)}+O(L^{-1})\right).
   \label{eq:H1-packed-error-entry-app}
\end{equation}
When \(s_r(\lambda)=0\), the root Gram matrix is $\mathsf G_{L-1}^{(r)}=\mathsf K_{L-1}^{\rm off}
=\mathsf Z_L\mathsf A_L\mathsf Z_L^{T}$, and the definition
eq.~\eqref{eq:root-Gram-eta-app} packs into the moving basis in three steps. The prefactor absorbs
the norm mismatch, since $b_L^2=\beta_L=h_L/h_{L-1}$ gives
$b_L\boldsymbol p_L=\mathbf w_L^{\rm off}/\sqrt{h_{L-1}}$, so
$\mathsf Z_L^{-1}b_L\boldsymbol p_L=\widehat{\mathbf c}_L$ exactly as defined in
eq.~\eqref{eq:H1-eta-normalized-packed-app}. Next, the last normalized endpoint vector is the last
column of \(\mathsf Z_L\), so $\mathsf Z_L^{-1}\boldsymbol p_{L-1}=\mathbf e_s$. 
With
\(\mathbf e_s=(0,\ldots,0,1)^T\in\RR^s\), and using
$(\mathsf Z_L\mathsf A_L\mathsf Z_L^{T})^{-1}
=\mathsf Z_L^{-T}\mathsf A_L^{-1}\mathsf Z_L^{-1}$ together with the symmetry of
$\mathsf A_L$, we therefore have
\begin{equation}
   \eta_L^{(r)}
   =\mathbf e_s^T(\Id_s+\mathsf E_L)^{-1}
      \widehat{\mathbf c}_L\,.
   \label{eq:H1-off-lattice-eta-contraction-app}
\end{equation}
The Neumann identity
\begin{equation}
   (\Id_s+\mathsf E_L)^{-1}
   =\Id_s-\mathsf E_L+
   \mathsf E_L^2(\Id_s+\mathsf E_L)^{-1}
   \label{eq:H1-packed-Neumann-app}
\end{equation}
is valid for all sufficiently large \(L\). It is an exact two-term truncation, not a series.
Writing $\mathsf E_L=\mathsf D\mathsf M_L\mathsf D$ with
$\mathsf D=\diag(L^{-1/2},\ldots,L^{-s/2})$ and $\mathsf M_L=O(1)$ entrywise, which is what
eq.~\eqref{eq:H1-packed-error-entry-app} says, gives
$(\Id_s+\mathsf E_L)^{-1}=\Id_s-\mathsf D\widetilde{\mathsf M}_L\mathsf D$ with
$\widetilde{\mathsf M}_L=\mathsf M_L+O(L^{-1})=O(1)$. The inverse appearing on the right of
eq.~\eqref{eq:H1-packed-Neumann-app} is therefore bounded and carries the same entrywise grading,
$[(\Id_s+\mathsf E_L)^{-1}]_{jk}=\delta_{jk}+O(L^{-(j+k)/2})$. In the first correction,
\begin{equation}
   (\mathsf E_L)_{sk}(\widehat{\mathbf c}_L)_k
   =O(L^{1-k}),
   \label{eq:H1-first-inverse-correction-app}
\end{equation}
so only \(k=1\) can alter the constant term. The remainder carries two factors of
$\mathsf E_L$ followed by that graded bounded inverse, so every contribution to it  is \(O(L^{-1})\), since a path
\(s\to i_1\to\cdots\to i_{q-1}\to k\) contributes
\(O(L^{1-k-i_1-\cdots-i_{q-1}})\) with $q\ge2$. Thus
\begin{equation}
   \mathfrak c_{r,\lambda}^{(\eta)}
   =
   \mathfrak c_{r,\lambda}^{({\rm int})}
   -e_{s1}^{(0)}\widehat c_1^{(0)},
   \label{eq:H1-eta-constant-app}
\end{equation}
for the leading coefficient
\((\widehat{\mathbf c}_L)_1
=L^{(s+1)/2}(\widehat c_1^{(0)}+O(L^{-1}))\), and
eq.~\eqref{eq:H1-eta-packed-asymptotics-app} follows. %

It remains to check that deleted Poisson atoms do not modify this
algebraic expansion. Define the normalized lattice endpoints
\begin{equation}
   \boldsymbol\ell_{L-1}
   =\frac{\mathbf w_{L-1}^{\rm lat}}{\sqrt{h_{L-1}}},
   \qquad
   \boldsymbol\ell_L^\sharp
   =\frac{\mathbf w_L^{\rm lat}}{\sqrt{h_{L-1}}}\,.
\end{equation}
The block inverse identity applied to
eq.~\eqref{eq:H1-mixed-normalized-Gram-app} gives
\begin{equation}
\begin{aligned}
   \eta_L^{(r)}
   =
   \boldsymbol\ell_{L-1}^T
   (\mathsf K_{L-1}^{\rm lat})^{-1}
   \boldsymbol\ell_L^\sharp
   +
   \left(
      \mathbf e_s
      -\widehat{\mathsf B}_L^T
      (\mathsf K_{L-1}^{\rm lat})^{-1}
      \boldsymbol\ell_{L-1}
   \right)^T
   \widehat{\mathsf S}_L^{-1}
   \left(
      \widehat{\mathbf c}_L
      -\widehat{\mathsf B}_L^T
      (\mathsf K_{L-1}^{\rm lat})^{-1}
      \boldsymbol\ell_L^\sharp
   \right).
\end{aligned}
   \label{eq:H1-mixed-eta-block-app}
\end{equation}

The exact lattice formula in
eq.~\eqref{eq:H1-fixed-point-Charlier-asymptotics-app} also shows that
\(\boldsymbol\ell_{L-1}\) and \(\boldsymbol\ell_L^\sharp\) are
\(O(L^{-p})\) for every fixed \(p>0\). Together with
eq.~\eqref{eq:H1-normalized-mixed-block-decay-app}, equations
\eqref{eq:H1-normalized-Schur-deficit-app} and
\eqref{eq:H1-mixed-eta-block-app} now show that the mixed contraction
differs from
\(\mathbf e_s^T\mathsf A_L^{-1}\widehat{\mathbf c}_L\) by a
superalgebraically small term.
The discarded pieces are not individually small. Each is an $O(L^{-p})$ factor multiplying
$\widehat{\mathsf S}_L^{-1}$ and $\widehat{\mathbf c}_L$, and $\widehat{\mathbf c}_L$ grows like
$L^{(s+1)/2}$ by the third line of eq.~\eqref{eq:H1-eta-normalized-packed-app}. What makes the
products superalgebraically small is that $p$ is arbitrary while the competing growth is a fixed
power, and that $\widehat{\mathsf S}_L^{-1}=\mathsf A_L^{-1}+O(L^{-p})$ stays bounded, again
because $\mathsf E_L\to0$.
\end{proof}

If \(r_{\rm off}=0\), every root lies on the Poisson lattice and
no high-index factorization is required. The full root Gram matrix
approaches the positive diagonal limit in
eq.~\eqref{eq:H1-lattice-Gram-limit-app}, with a remainder
\(O(L^{-p})\) for every fixed \(p>0\). Consequently,
\begin{equation}
   d_L^{(r)}=1+O(L^{-p}),\qquad
   \eta_L^{(r)}=O(L^{-p})
   \qquad\text{for every fixed }p>0\,.
   \label{eq:H1-all-lattice-superalgebraic-app}
\end{equation}
Equations~\eqref{eq:kernel-second-difference-charlier-app}
and~\eqref{eq:H1-eta-packed-asymptotics-app} are the two lines of
proposition~\ref{prop:charlier-difference-estimates-app}, which is thereby proved for
$r_{\rm off}\ge1$. The $r_{\rm off}=0$ branch follows from
eq.~\eqref{eq:H1-all-lattice-superalgebraic-app} in one step, since by
eq.~\eqref{eq:gram-hankel-reconciliation-app},
$\nabla^2\log\mathcal D_L^{(r)}=\log\bigl(d_L^{(r)}/d_{L-1}^{(r)}\bigr)=O(L^{-p})$ once
$d_L^{(r)}=1+O(L^{-p})$, and $\nabla\eta_L^{(r)}=\eta_L^{(r)}-\eta_{L-1}^{(r)}=O(L^{-p})$
directly. Inserting them into the
difference form \eqref{eq:kernel-difference-form-app}, or equivalently inserting
eqs.~\eqref{eq:H1-Delta-ratio-asymptotics-app} and~\eqref{eq:H1-eta-packed-asymptotics-app} into
eq.~\eqref{eq:H1-general-r-Jacobi-app}, gives, for $r_{\rm off}\ge1$,
\begin{equation}
   \widetilde\beta_n^{(r)}
   =\lambda\bigl(n+r+r_{\rm off}\bigr)+O(n^{-1}),
   \qquad
   \widetilde a_n^{(r)}
   =n+r-r_{\rm off}+O(n^{-1}).
\end{equation}
Since \(r_{\rm off}=r-s_r(\lambda)\), these formulas prove
eq.~\eqref{eq:H1-general-r-Jacobi-asymptotics-app}. Restoring physical
units gives
\begin{equation}
\begin{aligned}
   \widetilde a_n^{(r)}(H)
   &=\delta+\omega\bigl[n+s_r(\lambda)+O(n^{-1})\bigr],\\
   \bigl(\widetilde b_n^{(r)}(H)\bigr)^2
   &=\rho^2\bigl[n+2r-s_r(\lambda)\bigr]
   +O(\omega^2n^{-1}),
\end{aligned}
   \label{eq:H1-general-r-Jacobi-asymptotics-physical-app}
\end{equation}
for the physical Hamiltonian in eq.~\eqref{eq:H1-ham}. The number-state jump therefore preserves the slopes of the
Heisenberg--Weyl row in table~\ref{tab:universality_classes}. The
effect of the seed appears in the asymptotic intercepts.

This conclusion is pointwise in \(\lambda\) and is not uniform as a
zero approaches the Poisson lattice. As \(\lambda\) approaches a
support-zero resonance, an off-lattice root reaches
\(x_a\in\mathbb Z_{\ge0}\). The factor \(1/\Gamma(-x_a)\) in the first line of
eq.~\eqref{eq:H1-fixed-point-Charlier-asymptotics-app} then vanishes,
and the exact lattice expression in the second line replaces the
factorial asymptotic. The multiplier simultaneously deletes the
corresponding Poisson atom. Thus every lattice root shifts the two Jacobi intercepts accordingly. For \(r=3\) and
\(\lambda=3\), for example, 
\begin{equation}
\begin{gathered}
   P_3(y)
   =(y+2)(y^2-5y+3),\qquad s_3(3)=1,\\
   \widetilde a_n^{(3)}
   =n+1+O(n^{-1}),\qquad
   \widetilde\beta_n^{(3)}
   =3(n+5)+O(n^{-1}).
\end{gathered}
   \label{eq:H1-r3-resonant-asymptotics-app}
\end{equation}

\paragraph{Finiteness and short-time regularity.}
The Jacobi asymptotics have two analytic consequences for the
infinite chain. The domain facts needed below follow from the second
consequence.  Equation~\eqref{eq:H1-number-relative-bound-app} shows that
$\widetilde{\Jac}^{(r)}$ is self-adjoint on $\Dom\mathsf N$ and that $\delta_0$ belongs to the
domain of all its powers.  These facts make the unitary orbit and the continuity equation below
well defined.

The estimate \(\widetilde b_n^{(r)}=O(\sqrt n)\) controls
the first moment of the Krylov distribution uniformly under
finite-index truncation.
\begin{samepage}
\noindent The corresponding hopping constant and truncated complexity are
\begin{equation}
   C_{r,\lambda}^{\rm hop}
   =
   \sup_{n\ge0}
   \frac{\widetilde\beta_{n+1}^{(r)}}{n+1}<\infty,\qquad
   K_{r,M}(\tau)
   =
   \sum_{n\ge0}\min(n,M)
   |\tphi_n^{(r)}(\tau)|^2.
   \label{eq:H1-truncated-complexity-app}
\end{equation}
\end{samepage}
Equation~\eqref{eq:H1-general-r-Jacobi-asymptotics-app}, together with
the finitely many coefficients at small $n$, makes the supremum finite.
For $\mathsf N_Me_n=\min(n,M)e_n$, the commutator with the shifted
Jacobi operator has finite rank. Its continuity equation and the
Cauchy--Schwarz inequality give
\begin{equation}
\begin{aligned}
   \left|\frac{\dd K_{r,M}}{\dd\tau}\right|
   &\le
   2\sum_{n=0}^{M-1}
   \widetilde b_{n+1}^{(r)}
   |\tphi_n^{(r)}||\tphi_{n+1}^{(r)}|
   \le
   2\sqrt{C_{r,\lambda}^{\rm hop}K_{r,M}}\,\,.
\end{aligned}
   \label{eq:H1-truncated-complexity-current-bound-app}
\end{equation}
In the final step we used
\(\sum_{m=1}^{M}m|\tphi_m^{(r)}|^2\le K_{r,M}\) and
\(\sum_{n=0}^{M-1}|\tphi_n^{(r)}|^2\le1\), together with
\((\widetilde b_{n+1}^{(r)})^2=\widetilde\beta_{n+1}^{(r)}
\le C_{r,\lambda}^{\rm hop}(n+1)\), which is how the constant enters.
Since
\(K_{r,M}(0)=0\), regularize the square root as
\(\sqrt{K_{r,M}+\varepsilon'}\) with $\varepsilon'>0$, integrate, and then let
$\varepsilon'\downarrow0$.  This auxiliary regularizer is unrelated to the Kato--Rellich
relative-bound parameter $\varepsilon$ below and to the numerical truncation errors
$\varepsilon_0,\varepsilon_1$ of appendix~\ref{app:numerics}.  Monotone convergence as
$M\to\infty$ then gives
\(
   K_r(\tau)\le C_{r,\lambda}^{\rm hop}\tau^2.
\)
The probabilities are $2\pi$-periodic on the centered Poisson spectrum,
and the complexity vanishes at every $2\pi k$. Applying the same estimate
about the nearest recurrence time gives, with $\operatorname{dist}(\tau,2\pi\mathbb Z)=\min_{k\in\mathbb Z}|\tau-2\pi k|$ the distance to the nearest revival,
\begin{equation}
   K_r(\tau)
   \le
   C_{r,\lambda}^{\rm hop}
   \operatorname{dist}(\tau,2\pi\mathbb Z)^2
   <\infty\,.
   \label{eq:H1-general-r-complexity-bound-app}
\end{equation}

Because $\operatorname{dist}(\tau,2\pi\mathbb Z)\le\pi$, the bound is uniform in time and
$K_r(\tau)\le\pi^2C_{r,\lambda}^{\rm hop}$ for every $\tau$.  The complexity returns exactly to
zero at each $\tau\in2\pi\mathbb Z$. This is
the physical content of the $\sqrt n$ hopping scale on an integer spectrum, namely that the spread
complexity of a number-state jump stays bounded and recurrent rather than growing, and it is
what the estimate buys beyond convergence. Equation~\eqref{eq:H1-general-r-complexity-bound-app} proves
convergence of the nonnegative correction series in
eq.~\eqref{eq:H1-general-r-complexity-app}, including the \(r=3\)
specialization.

Second, finiteness of the first moment does not by itself justify the $O(\tau^6)$ remainder in eq.~\eqref{eq:H1-third-jump-short-time-main} on the untruncated chain. For
that purpose, let
\(\widetilde{\Jac}^{(r)}\) denote the Jacobi operator of the shifted
measure and let \(\mathsf Ne_n=ne_n\). The same large-index estimate
makes $\widetilde{\Jac}^{(r)}-\mathsf N$ infinitesimally
$\mathsf N$-bounded. On finitely supported sequences,
\begin{equation}
   \|(\widetilde{\Jac}^{(r)}-\mathsf N)\psi\|
   \le
   \varepsilon\|\mathsf N\psi\|
   +C_\varepsilon\|\psi\|,
   \qquad \varepsilon>0\,.
   \label{eq:H1-number-relative-bound-app}
\end{equation}
The diagonal part is bounded by
eq.~\eqref{eq:H1-general-r-Jacobi-asymptotics-app}.  For the
off-diagonal part,
\(\widetilde b_n^{(r)}=O(\sqrt n)\), and the inequality
\(n\le\varepsilon^2n^2+C_\varepsilon\) gives the stated
infinitesimal bound. Two standing hypotheses of Kato--Rellich follow from the same
tridiagonal structure. The perturbation is symmetric on the finitely supported sequences $c_{00}$,
since $\widetilde{\Jac}^{(r)}$ and $\mathsf N$ are both real symmetric there; and $c_{00}$ is a
core for $\mathsf N$, so eq.~\eqref{eq:H1-number-relative-bound-app} extends from $c_{00}$ to all
of $\Dom\mathsf N$ by continuity in the graph norm of $\mathsf N$, which is what identifies the
Kato--Rellich operator with the closure of the formal matrix. 
By the Kato--Rellich theorem \cite{ReedSimonII}, that closure is self-adjoint on \(\Dom\mathsf N\), and essentially
self-adjoint on any core of $\mathsf N$, in particular on $c_{00}$. The Jacobi matrix therefore has
a unique self-adjoint realization. Carleman's criterion,
\(\sum_{n\ge1}1/\widetilde b_n^{(r)}=\infty\), which holds here because
$\widetilde b_n^{(r)}=O(\sqrt n)$, gives that uniqueness independently
\cite{SimonMomentProblem1998}.

Hence
\(\Dom\widetilde{\Jac}^{(r)}=\Dom\mathsf N\), with equivalent graph
norms. Since every
finite power of a tridiagonal matrix maps $\delta_0$ to a finitely
supported vector, $\delta_0=(1,0,\ldots)^T$ lies in \(\bigcap_{m\ge1}\Dom((\widetilde{\Jac}^{(r)})^m)\). The orbit
\(\psi_r(\tau)=\ee^{-\ii\tau\widetilde{\Jac}^{(r)}}\delta_0\) is consequently
smooth to every order in the graph norm of \(\mathsf N\). The first moment is the quadratic form
$K_r(\tau)=\langle\psi_r(\tau),\mathsf N\psi_r(\tau)\rangle$, so smoothness transfers to it by the
product rule. Hence $K_r$ is $C^\infty$ in $\tau$, and its derivative of
any fixed order at $\tau=0$ is a finite combination of the matrix elements
$\langle\delta_0,(\widetilde{\Jac}^{(r)})^{j}\mathsf N(\widetilde{\Jac}^{(r)})^{k}\delta_0\rangle$,
each of which involves only finitely many Jacobi coefficients and so agrees with the truncated
computation. Finally,
the Jacobi matrix and \(\delta_0\) are real, so
\(\tphi_n^{(r)}(-\tau)
=\overline{\tphi_n^{(r)}(\tau)}\) and \(K_r(\tau)\) is even, which removes the odd orders.
Taylor's theorem, applied to this $C^\infty$ even function, now gives the \(O(\tau^6)\) remainder in
eq.~\eqref{eq:H1-third-jump-short-time-main} for the full infinite
chain.

This completes the Charlier construction. Finite-dimensional remainder
data determine every fixed Krylov level, while the root-Gram asymptotics
control the infinite tail. Both rest on the chain being infinite, so the tail estimates of
this subsection do not apply once the spectrum terminates; that regime is the subject of
appendix~\ref{app:finite}.%

\section{Finite-support multiplication maps and tight-binding quotients}
\label{app:finite}

Finite support changes the algebraic problem at the terminal edge.
Multiplication by a seed polynomial acts in the physical terminal algebra,
and its rank detects any spectral atoms deleted by the preparation.  The
resulting multiplication matrix determines the shifted inner product. A
degree-ordered factorization of its Gram matrix then gives the connector
through termination.  We formulate this finite-support construction first
and then specialize it to the truncated Chebyshev product of the open path.
Here the quotient is physical because the terminal polynomial vanishes on
the reference cyclic subspace, so $T_d(\Jac)=0$ is an operator identity on that space.
By contrast, the remainders modulo
$P_r$ and $P_r^2$ in subsection~\ref{app:H1-general-r}, and the
continuous-$q$-Hermite remainders in corollary~\ref{cor:qhermite-root-free-app},
are auxiliary coordinates for
infinite-support divisibility constraints.  In particular,
$P_r$ is not a terminal polynomial and does not annihilate the reference cyclic space, since
$P_r(\Jac)\ket{K_0}=\sqrt{h_r}\ket{K_r}\ne0$.

The remainder corollary therefore solves root constraints at fixed $r$ but cannot detect the
rank of multiplication on finite spectral support, remove annihilated atoms, or continue connector
rows through the last physical degree.  Those three terminal operations require reduction modulo
$T_d$ and the product-Gram factorization developed below.

\subsection{Multiplication maps and finite quotient connectors}
\label{app:finite-multiplication-gram}

The construction separates the common finite-dimensional linear algebra
from the product formula specific to each Jacobi family.  Seed multiplication
first gives a set of product vectors and their Gram matrix.
Degree-ordered residuals of these vectors then give the complete
connector rows.

Let the reference cyclic dimension be $d$, with spectral measure
\begin{equation}
   \dd\mu(E)=\sum_{\alpha=0}^{d-1}
   w_\alpha\delta(E-E_\alpha)\dd E,
   \qquad w_\alpha>0\,,
\end{equation}
terminal polynomial
$T_d(E)=\prod_{\alpha=0}^{d-1}(E-E_\alpha)$, and orthonormal
polynomials $p_m=P_m/\sqrt{h_m}$, $0\le m<d$, chosen real on the real
support.  Set $\Qhat=Q/\sqrt{N_Q}$ and
$W_Q=\Qsh Q/N_Q$.  Multiplication by the seed defines
\begin{equation}
   \cF_Q:\cA_d\longrightarrow\cA_d,
   \qquad
   \cF_Q[f]=[\Qhat f]_{T_d}\,,
   \qquad
   \cA_d=\CC[E]/\langle T_d\rangle.
   \label{eq:finite-product-map-app}
\end{equation}
The rank of this map is the shifted cyclic dimension,
\begin{equation}
\begin{aligned}
   d_Q&=\#\{\alpha:Q(E_\alpha)\ne0\},\\
   T_{\rm surv}^{[Q]}(E)
   &=\prod_{Q(E_\alpha)\ne0}(E-E_\alpha)
   =\frac{T_d(E)}{\gcd(T_d,Q)}\,,
\end{aligned}
   \label{eq:finite-product-surviving-polynomial-app}
\end{equation}
where the greatest common divisor is chosen monic.  Since $T_d$ has
simple zeros, its common factor with $Q$ removes precisely the atoms
annihilated by the seed.

Regard the quotient products
\begin{equation}
   \mathsf V_m^{[Q]}=\cF_Q[p_m],
   \qquad 0\le m<d\,,
   \label{eq:finite-product-vectors-app}
\end{equation}
as vectors in the reference polynomial Hilbert space.  This is
independent of the chosen representatives, since every multiple of
$T_d$ vanishes on the reference support and therefore has zero
$L^2(\mu)$ norm.  Their Gram matrix is the shifted inner product written
in the degree-ordered reference basis, 
\begin{equation}
\begin{aligned}
   \mathsf G_{mn}^{[Q]}
   &=\left\langle
      \mathsf V_m^{[Q]},\mathsf V_n^{[Q]}
      \right\rangle_\mu
   =\int p_m(E)p_n(E)\dd\nu_Q(E),\\
   \dd\nu_Q(E)&=W_Q(E)\dd\mu(E)
   =|\Qhat(E)|^2\dd\mu(E).
\end{aligned}
   \label{eq:finite-product-Gram-app}
\end{equation}
This matrix admits an exact spectral factorization.  We introduce the
weighted evaluation matrix
\begin{equation}
   (\mathsf E_\mu)_{\alpha m}
   =\sqrt{w_\alpha}\,p_m(E_\alpha),
   \qquad 0\le\alpha,m<d\,,
\end{equation}
such that finite orthogonality makes $\mathsf E_\mu$ real orthogonal.  If
$\mathsf M_Q$ denotes the matrix of $\cF_Q$ in the orthonormal
polynomial basis, then 
\begin{equation}
\begin{aligned}
   \mathsf M_Q
   &=\mathsf E_\mu^{\mathsf T}
   \diag\!\bigl(
      \Qhat(E_0),\ldots,\Qhat(E_{d-1})
   \bigr)\mathsf E_\mu\,,\\
   \mathsf G^{[Q]}
   &=\mathsf M_Q^\dagger\mathsf M_Q
   =\mathsf E_\mu^{\mathsf T}
   \diag\!\bigl(
      |\Qhat(E_0)|^2,\ldots,|\Qhat(E_{d-1})|^2
   \bigr)\mathsf E_\mu\,.
\end{aligned}
   \label{eq:finite-product-spectral-factorization-app}
\end{equation}
For a complex filter, $\mathsf M_Q$ is complex symmetric, while
$\mathsf G^{[Q]}$ is real symmetric and positive semidefinite.  Thus the
real Schur-complement and $LDL^{\mathsf T}$ constructions below depend
only on the positive multiplier $W_Q$.
Support loss is therefore the rank defect of seed multiplication in the
terminal algebra.

For $0\le n<d_Q$, define the degree-ordered residual product vectors and
their overlaps by
\begin{equation}
\begin{aligned}
   \mathsf V_{0,\perp}^{[Q]}&=\mathsf V_0^{[Q]},\\
   \mathsf V_{n,\perp}^{[Q]}
   &=\mathsf V_n^{[Q]}
   -\sum_{r=0}^{n-1}
   \frac{\widehat\Gamma_{r,n}^{[Q]}}
        {\widehat\Gamma_{r,r}^{[Q]}}
   \mathsf V_{r,\perp}^{[Q]}\,,\\
   \widehat\Gamma_{n,m}^{[Q]}
   &=\left\langle
      \mathsf V_{n,\perp}^{[Q]},\mathsf V_m^{[Q]}
     \right\rangle_\mu,
   \qquad 0\le m<d\,.
\end{aligned}
   \label{eq:finite-product-residuals-app}
\end{equation}
Equivalently, the complete residual row follows from the scalar
Schur-complement recurrence
\begin{equation}
   \widehat\Gamma_{n,m}^{[Q]}
   =\mathsf G_{nm}^{[Q]}
   -\sum_{r=0}^{n-1}
   \frac{
      \widehat\Gamma_{r,n}^{[Q]}
      \widehat\Gamma_{r,m}^{[Q]}}
      {\widehat\Gamma_{r,r}^{[Q]}}\,,
   \qquad
   0\le n<d_Q\,,\quad n\le m<d\,,
   \label{eq:finite-product-Schur-app}
\end{equation}
and the empty sum gives
$\widehat\Gamma_{0,m}^{[Q]}=\mathsf G_{0m}^{[Q]}$.  In matrix
language, the physical leading block has the unique factorization
\begin{equation}
   \mathsf G_{[d_Q]}^{[Q]}
   =\left[\mathsf G_{mn}^{[Q]}\right]_{m,n=0}^{d_Q-1}
   =\mathsf L_Q\mathsf D_Q\mathsf L_Q^{\mathsf T}\,,
   \quad
   (\mathsf L_Q)_{nr}
   =\frac{\widehat\Gamma_{r,n}^{[Q]}}
          {\widehat\Gamma_{r,r}^{[Q]}}\,,
   \quad
   \mathsf D_Q
   =\diag\!\bigl(
      \widehat\Gamma_{0,0}^{[Q]},\ldots,
      \widehat\Gamma_{d_Q-1,d_Q-1}^{[Q]}
   \bigr)\,,
   \label{eq:finite-product-LDL-app}
\end{equation}
where $\mathsf L_Q$ is unit lower triangular.  Thus the diagonal entries
of $\mathsf D_Q$ are the squared norms of the residual vectors. In the nondegenerate case this triangular factorization of the Gram matrix recovers the connection coefficients as in ref.~\cite{GutlebOlverSlevinsky2024}; what follows continues it through the rank defect $d_Q<d$.  This is
a coefficient-space factorization of an explicit matrix in the solved
reference quotient, not another Lanczos construction in the original
Hilbert space.

\begin{proposition}[Finite-support product-Gram transfer]
\label{prop:finite-product-Gram}
Let $R_n^{[Q]}$ be the monic orthogonal polynomials for $\dd\nu_Q$.
For every $0\le n<d_Q$,
\begin{equation}
   \mathsf V_{n,\perp}^{[Q]}
   =\left[
      \Qhat\frac{R_n^{[Q]}}{\sqrt{h_n}}
   \right]_{T_d}.
   \label{eq:finite-product-residual-polynomial-app}
\end{equation}
The complete quotient connector is consequently
\begin{equation}
\begin{aligned}
   \left[
      W_Q(E)R_n^{[Q]}(E)
   \right]_{T_d}
   &=\sum_{m=n}^{d-1}
   \Gamma_{n,m}^{[Q]}P_m(E),\\
   \Gamma_{n,m}^{[Q]}
   &=\sqrt{\frac{h_n}{h_m}}\,
   \widehat\Gamma_{n,m}^{[Q]}\,.
\end{aligned}
   \label{eq:finite-product-connector-app}
\end{equation}
Define the leading Gram determinants by
\begin{equation}
   \Delta_n^{[Q]}
   =\det\!\left[
      \mathsf G_{rs}^{[Q]}
   \right]_{r,s=0}^{n-1},
   \qquad \Delta_0^{[Q]}=1\,.
\end{equation}
Regarding $\mathsf V_{n,\perp}^{[Q]}$ as its coefficient column in the
orthonormal reference basis, we then have
\begin{equation}
\begin{aligned}
   \widehat\Gamma_{n,n}^{[Q]}
   &=\frac{\Delta_{n+1}^{[Q]}}{\Delta_n^{[Q]}}\,,\\
   \thh_n^{[Q]}
   &=h_n\frac{\Delta_{n+1}^{[Q]}}{\Delta_n^{[Q]}}\,,\\
   \bigl(\widetilde b_n^{[Q]}\bigr)^2
   &=b_n^2
   \frac{\Delta_{n+1}^{[Q]}\Delta_{n-1}^{[Q]}}
        {\bigl(\Delta_n^{[Q]}\bigr)^2}\,,
   \qquad 1\le n<d_Q\,,\\
   \widetilde a_n^{[Q]}
   &=\frac{
      \bigl(\mathsf V_{n,\perp}^{[Q]}\bigr)^\dagger
      \mathsf J_d\mathsf V_{n,\perp}^{[Q]}}
      {\widehat\Gamma_{n,n}^{[Q]}}\,,
   \qquad 0\le n<d_Q\,,
\end{aligned}
   \label{eq:finite-product-determinants-app}
\end{equation}
where $\mathsf J_d$ is the reference Jacobi matrix. Because the reference chain terminates, $b_d=0$ and the monic degree-$d$ continuation of the reference family is $T_d$ itself, so $\mathsf J_d$ represents multiplication by $E$ exactly on $\cA_d$, with $\mathsf J_d=\mathsf E_\mu^{\mathsf T}\diag\!\bigl(E_0,\ldots,E_{d-1}\bigr)\mathsf E_\mu$; the quadratic form above is therefore an exact shifted expectation and not the truncation of one.  Finally,
\begin{equation}
   \tphi_n^{[Q]}(t)
   =\frac{1}{\sqrt{\widehat\Gamma_{n,n}^{[Q]}}}
   \sum_{m=n}^{d-1}
   \widehat\Gamma_{n,m}^{[Q]}\phi_m^{(0)}(t),
   \qquad 0\le n<d_Q.
   \label{eq:finite-product-amplitude-app}
\end{equation}
Every pivot through $n=d_Q-1$ is positive and
$\widetilde b_{d_Q}^{[Q]}=0$.  If $d_Q<d$, then
$\Delta_{d_Q+1}^{[Q]}=0$. When $d_Q=d$, termination occurs at the
ambient dimension.
\end{proposition}

\begin{proof}
Equation~\eqref{eq:finite-product-spectral-factorization-app} gives
$\rank\mathsf G^{[Q]}=d_Q$.  A polynomial of degree below $d_Q$ with
zero shifted norm vanishes at all $d_Q$ surviving atoms and is therefore
the zero polynomial.  The leading $d_Q\times d_Q$ Gram block is thus
positive definite.

For $n<d_Q$, the restriction of $\cF_Q$ to
$\Span\{p_0,\ldots,p_n\}$ is injective, since a polynomial in this
space cannot vanish at all $d_Q$ surviving atoms unless it is zero.
The residual in eq.~\eqref{eq:finite-product-residuals-app} is therefore
the image of a unique polynomial in this space with unit coefficient
along $p_n$.  Orthogonality to the preceding product vectors means that
this polynomial is orthogonal to every polynomial of degree below $n$
in $L^2(\nu_Q)$.
Uniqueness of the shifted monic polynomial proves
eq.~\eqref{eq:finite-product-residual-polynomial-app}, and its squared
reference norm is
$
   \widehat\Gamma_{n,n}^{[Q]}
   =\frac{\thh_n^{[Q]}}{h_n}.
$
Taking its inner product with $\mathsf V_m^{[Q]}$ gives
eq.~\eqref{eq:finite-product-connector-app} while integration against
$\ee^{-\ii Et}\dd\mu(E)$ gives
eq.~\eqref{eq:finite-product-amplitude-app}.  The pivot identity is the
determinant formula for eq.~\eqref{eq:finite-product-LDL-app}, and
consecutive shifted norm ratios give the hopping coefficient.  The
expectation of multiplication by $E$ gives the diagonal coefficient.
If $d_Q<d$, the polynomial $T_{\rm surv}^{[Q]}$ has degree $d_Q$ and
zero shifted norm, which proves the final determinant statement. The shifted measure $\dd\nu_Q$ carries exactly $d_Q$ atoms, so the finite-closure equivalence of appendix~\ref{app:lanczos} gives $\widetilde b_{d_Q}^{[Q]}=0$ whether or not $d_Q<d$.  For $d_Q<d$ the vanishing of $\Delta_{d_Q+1}^{[Q]}$ results in the same termination in determinant form, while for $d_Q=d$ no further leading determinant exists and termination is that of the ambient chain.
\end{proof}

\paragraph{Inverse transport, conditioning and finite interpolation.}
The same spectral realization identifies exactly which information is
lost when the seed removes atoms.  Let
$\widetilde p_n^{[Q]}=R_n^{[Q]}/\sqrt{\thh_n^{[Q]}}$ and form the
$d_Q\times d_Q$ shifted evaluation matrix on the surviving support,
\begin{equation}
   (\mathsf E_{\nu_Q})_{\alpha n}
   =\sqrt{w_\alpha}\,|\Qhat(E_\alpha)|
      \widetilde p_n^{[Q]}(E_\alpha),
   \qquad Q(E_\alpha)\ne0\,.
\end{equation}
Finite orthogonality makes $\mathsf E_{\nu_Q}$ orthogonal.  Denote by
$\mathsf D_{|Q|}$ the $d_Q\times d$ row-selection matrix whose row for a
surviving atom $E_\alpha$ has the single nonzero entry
$|\Qhat(E_\alpha)|$ in column $\alpha$.  The orthonormal connection
defined on the surviving support by
\begin{equation}
   \bigl(p_0(E),\ldots,p_{d-1}(E)\bigr)
   =\bigl(\widetilde p_0^{[Q]}(E),\ldots,
      \widetilde p_{d_Q-1}^{[Q]}(E)\bigr)\mathsf R_Q
\end{equation}
has the exact spectral form
\begin{equation}
   \mathsf R_Q
   =\mathsf E_{\nu_Q}^{\mathsf T}
      \mathsf D_{|Q|}\mathsf E_\mu\,.
   \label{eq:finite-orthonormal-connection-spectral-app}
\end{equation}
If $\mathsf D_{|Q|}^{+}$ is obtained by replacing every nonzero entry
$|\Qhat(E_\alpha)|$ by its reciprocal and transposing the
row-selection pattern, the Moore--Penrose inverse is
\begin{equation}
\begin{aligned}
   \mathsf R_Q^{+}
   &=\mathsf E_\mu^{\mathsf T}
      \mathsf D_{|Q|}^{+}\mathsf E_{\nu_Q}\,,\\
   \mathsf R_Q\mathsf R_Q^{+}&=\Id_{d_Q}\,,\\
   \mathsf R_Q^{+}\mathsf R_Q
   &=\boldsymbol 1_{\{Q\ne0\}}(\mathsf J_d).
\end{aligned}
   \label{eq:finite-connector-pseudoinverse-app}
\end{equation}
Thus inverse transport reconstructs precisely the source component on
the surviving atoms and annihilates the deleted spectral sector.  The
nonzero singular values of $\mathsf R_Q$ are
$|\Qhat(E_\alpha)|$, and therefore
\begin{equation}
   \|\mathsf R_Q\|
   =\max_{Q(E_\alpha)\ne0}|\Qhat(E_\alpha)|,
   \qquad
   \|\mathsf R_Q^{+}\|
   =\frac{1}{
      \min_{Q(E_\alpha)\ne0}|\Qhat(E_\alpha)|}\,.
   \label{eq:finite-connector-conditioning-app}
\end{equation}
When no atom is deleted, $\mathsf R_Q^{+}$ is the ordinary inverse.
After support loss, no inverse acting on the surviving cyclic data can
recover the missing spectral component.

Finite support also makes the converse seed representation exact.  Let
$W(E_\alpha)\ge0$ satisfy
$\sum_{\alpha=0}^{d-1}W(E_\alpha)w_\alpha=1$.  Lagrange interpolation
produces a polynomial $Q$ of degree at most $d-1$ for which
\begin{equation}
   Q(E_\alpha)
   =\sqrt{W(E_\alpha)}\,\ee^{\ii\theta_\alpha},
   \qquad
   |Q(E_\alpha)|^2=W(E_\alpha),
   \label{eq:finite-pure-seed-interpolation-app}
\end{equation}
with arbitrary phases $\theta_\alpha$. Choosing phases $0$ or $\pi$
gives a real interpolant.  Every normalized finite-support reweighting
is therefore measure-equivalent, in the terminal quotient, to a pure
polynomial seed, including reweightings which delete atoms.  This
spectral equivalence does not make the interpolant a local fixed-depth initial-state jump, since its degree may be of order the terminal dimension.

\subsection{Localized-site connectors and finite tight-binding quotients}
\label{app:finite-tight-binding-checks}

The tight-binding chain of subsection~\ref{subsec:tight-binding} makes both the finite-band connector and its
terminal reduction unusually explicit.  On the half-line, Chebyshev
linearization proves the closed formula in
eq.~\eqref{eq:tight-binding-all-k-connector-main} for every localized
seed $\ket k$.  Finite volume retains this polynomial through a definite
range and then folds its reference expansion at the terminal edge.
Deeper in the shifted chain the polynomial itself changes, while zeros
of the seed on the sine spectrum may remove physical levels.  We first
separate these effects and then determine the complete connector from
the Chebyshev multiplication formula and the product-Gram transfer.

\paragraph{Half-line connector and shifted recurrence.}
Set $x=(E-a)/(2b)=\cos\theta$, so that
$P_n(E)=b^nU_n(x)$ and $h_n=b^{2n}$, as in
eq.~\eqref{eq:Chebyshev-polys-main}. The product and composition identities used below are standard Chebyshev formulas; see ref.~\cite{DLMF}. The product identity
\begin{equation}
   U_p(x)U_{n+p}(x)
   =\sum_{s=0}^{p}U_{n+2s}(x)
   \label{eq:tight-binding-Chebyshev-product-app}
\end{equation}
turns the connector formula into a polynomial divisibility statement.
With $n=(\sigma-1)(k+1)+\varepsilon$ as in
eq.~\eqref{eq:tight-binding-block-decomposition-main}, the right-hand
side of eq.~\eqref{eq:tight-binding-all-k-connector-main}, after removal
of its common factor $b^n$, is
\begin{equation}
   U_k(x)U_{n+k}(x)
   +\frac{1}{\sigma}U_{k-\varepsilon}(x)
   U_{n+k-\varepsilon}(x),
   \qquad 1\le\varepsilon\le k\,,
   \label{eq:tight-binding-connector-Chebyshev-app}
\end{equation}
whereas only the first term remains when $\varepsilon=0$.  Here $T_m$ denotes the Chebyshev polynomial of the first kind, $T_m(\cos\theta)=\cos(m\theta)$, and is unrelated to the terminal polynomial $T_d$ used elsewhere. The
composition identity
\begin{equation}
   U_{\sigma(k+1)-1}(x)
   =
   U_k(x)U_{\sigma-1}\!\left(T_{k+1}(x)\right)
   \label{eq:tight-binding-Chebyshev-composition-app}
\end{equation}
factors the $\varepsilon=0$ expression by $U_k^2$.  For
$\varepsilon>0$, the relation
$n+k-\varepsilon=\sigma(k+1)-1$ rewrites
eq.~\eqref{eq:tight-binding-connector-Chebyshev-app} as
\begin{equation}
   U_k(x)\left[
      U_{n+k}(x)
      +\frac{1}{\sigma}U_{k-\varepsilon}(x)
      U_{\sigma-1}\!\left(T_{k+1}(x)\right)
   \right].
   \label{eq:tight-binding-first-factor-app}
\end{equation}
At the simple zeros
\begin{equation}
   x_\alpha=\cos\frac{\alpha\pi}{k+1},
   \qquad \alpha=1,\ldots,k\,,
\end{equation}
we have $T_{k+1}(x_\alpha)=(-1)^\alpha$.  The sine representation of
the Chebyshev polynomials then gives
\begin{equation}
\begin{gathered}
   U_{\sigma-1}\!\left((-1)^\alpha\right)
   =\sigma(-1)^{\alpha(\sigma-1)}\,,\\
   U_{n+k}(x_\alpha)
   +\frac{1}{\sigma}U_{k-\varepsilon}(x_\alpha)
   U_{\sigma-1}\!\left((-1)^\alpha\right)
   =0\,.
\end{gathered}
   \label{eq:tight-binding-root-cancellation-app}
\end{equation}
The polynomial in square brackets in
eq.~\eqref{eq:tight-binding-first-factor-app} consequently contains a
second factor of $U_k$.  The connector numerator is divisible by
$U_k^2=P_k^2/h_k$.  Its highest reference component is
$b^{-2k}P_{n+2k}=h_k^{-1}P_{n+2k}$, so both the numerator and
$P_k^2/h_k$ have leading coefficient $h_k^{-1}$.  Their quotient is
therefore monic.  Since
the reference expansion starts at $P_n$, it is orthogonal to every
polynomial of degree below $n$ with respect to
$P_k^2\dd\mu_0/h_k$.  Uniqueness of the monic shifted polynomial proves
eq.~\eqref{eq:tight-binding-all-k-connector-main}.

The coefficient of $P_n$ gives the shifted norm in the same equation.
The measure in eq.~\eqref{eq:tight-binding-shifted-theta-main} is
symmetric about $E=a$, and hence
$\widetilde a_n^{(k)}=a$.  Consecutive norm ratios give the hopping
blocks
\begin{equation}
   b\sqrt{\frac{\sigma+1}{\sigma}},
   \qquad
   \underbrace{b,\ldots,b}_{k-1\ \mathrm{times}}\,,
   \qquad
   b\sqrt{\frac{\sigma}{\sigma+1}}\,,
   \qquad \sigma=1,2,\ldots.
   \label{eq:tight-binding-halfline-hopping-pattern-app}
\end{equation}
The middle string is absent for $k=1$.  For $k=0$, the multiplier is
unity and the reference hopping remains $b$.  Equation
\eqref{eq:tight-binding-halfline-hopping-pattern-app} agrees with
ref.~\cite{Balasubramanian:2025variations}.

\paragraph{Finite-volume ranges and terminal folding.}
For the remainder of this subsection, write $N=N_{\rm lat}$.  The
finite sine measure in eq.~\eqref{eq:tight-binding-spectrum-main} is the
$N$-point Gauss quadrature rule for the half-line Chebyshev-$U$ measure~\cite{Gautschi2004}
and is exact through degree $2N-1$.  We restrict throughout to physical
levels $0\le n<d_{\rm cyc}^{(k,N)}$.  Testing the degree-$n$ half-line
polynomial against a polynomial of degree at most $n-1$, with the factor
$P_k^2$ included, produces degree at most $2(n+k)-1$.  The finite and
half-line shifted polynomials therefore agree while
\begin{equation}
   n+k\le N.
   \label{eq:tight-binding-finite-polynomial-range-app}
\end{equation}
The norm integrand has degree $2(n+k)$, so the two norms agree only for
$n+k\le N-1$.  The shifted hopping
$\widetilde b_n^{(k,N)}$ retains its half-line value in the same range,
with its first possible modification at $n=N-k$.  Finally, the largest
reference index in the half-line connector is $n+2k$.  Its coefficient
list is already the canonical finite connector, without terminal
folding, precisely when
\begin{equation}
   n+2k\le N-1\,.
   \label{eq:tight-binding-unfolded-connector-range-app}
\end{equation}
The polynomial, norm and unfolded-connector ranges are therefore
distinct.  Each must also be intersected with
$0\le n<d_{\rm cyc}^{(k,N)}$, and support loss may make two or more of
their physical endpoints coincide.

The ambient reference-chain algebra is
\begin{equation}
   \cA_N=\CC[E]/\langle T_N\rangle,
   \qquad
   T_N(E)=\prod_{j=0}^{N-1}(E-E_j)=P_N(E).
   \label{eq:tight-binding-ambient-quotient-app}
\end{equation}
The Chebyshev recurrence gives, in this quotient,
\begin{equation}
\begin{aligned}
   P_N&\equiv0\,,\\
   P_m&\equiv-b^{2(m-N)}P_{2N-m}\,,
   \qquad N<m<2N.
\end{aligned}
   \label{eq:tight-binding-terminal-fold-app}
\end{equation}
Within the polynomial range
eq.~\eqref{eq:tight-binding-finite-polynomial-range-app}, this identity
evaluates every folded row of the half-line connector.  In particular,
an offset $s>(N-n)/2$ obeys
\begin{equation}
   b^{-2s}P_{n+2s}
   \equiv
   -b^{-2(N-n-s)}P_{2N-n-2s}
   \pmod{T_N},
   \label{eq:tight-binding-offset-fold-app}
\end{equation}
whereas the term with $2s=N-n$, when present, is proportional to $P_N$
and vanishes.  Every high-offset term in
eq.~\eqref{eq:tight-binding-all-k-coefficients-main} thus folds onto its
low-offset partner, with coincident contributions combined.

For $k\ge1$, the last level in the polynomial range has $n+k=N$.
Its block decomposition gives
$N+1=\sigma(k+1)+\varepsilon$.  If $\varepsilon=0$, then
$\gcd(k+1,N+1)=k+1$ and
$d_{\rm cyc}^{(k,N)}=N-k=n$, so this level lies beyond shifted
termination.  If $\varepsilon>0$, then
\begin{equation}
   \gcd(k+1,N+1)=\gcd(k+1,\varepsilon)\le k,
   \qquad
   d_{\rm cyc}^{(k,N)}
   =N-\gcd(k+1,N+1)+1>N-k=n\,.
\end{equation}
The level is therefore physical, and its upper connector term folds to the negative of $P_n$.  The resulting norm is 
\begin{equation}
   \thh_n^{(k,N)}
   =
   b^{2n}\left(
      \mathcal C_{\sigma,\varepsilon;0}^{(k)}-1
   \right)
   =\frac{b^{2n}}{\sigma}\,.
   \label{eq:tight-binding-first-terminal-norm-app}
\end{equation}
Dividing by the preceding half-line norm gives the first right-boundary
modification
\begin{equation}
   \bigl(\widetilde b_{N-k}^{(k,N)}\bigr)^2
   =
   \begin{cases}
      b^2/\sigma,&\varepsilon=1\,,\\[2pt]
      b^2/(\sigma+1),&2\le\varepsilon\le k\,.
   \end{cases}
   \label{eq:tight-binding-first-terminal-hopping-app}
\end{equation}
For $k=0$, the excluded level $n=N$ is the ordinary termination of the
reference chain.  When $n+k>N$, the half-line polynomial no longer
represents the finite shifted problem, and folding its connector is not
sufficient.

\paragraph{Fusion algebra and the exact finite connector.}
The support-loss formula in
eq.~\eqref{eq:tight-binding-support-loss-main} becomes especially
concrete in the Chebyshev quotient.  Set
\begin{equation}
   g=\gcd(k+1,N+1),
   \qquad
   d=d_{\rm cyc}^{(k,N)}=N-g+1\,.
\end{equation}
The deleted angles are $m\pi/g$, $m=1,\ldots,g-1$, and their energies
are the zeros of $P_{g-1}$.  The terminal polynomial of the surviving
support therefore factors as
\begin{equation}
   T_{\rm surv}^{(k,N)}(E)
   =\frac{P_N(E)}{P_{g-1}(E)}\,,
   \qquad
   \deg T_{\rm surv}^{(k,N)}=d\,.
   \label{eq:tight-binding-surviving-factor-app}
\end{equation}
The quotient by $T_N$ remains the ambient finite-chain representation.
The null ideal of the shifted inner product consists of the classes that
vanish on every surviving mode.  Quotienting the ambient algebra by this
ideal, equivalently by the radical of the shifted form, gives the
physical algebra
\begin{equation}
   \cA_{\rm surv}^{(k,N)}
   =\CC[E]/\left\langle T_{\rm surv}^{(k,N)}\right\rangle,
\end{equation}
in agreement with
eq.~\eqref{eq:tight-binding-surviving-support-main}.

The ambient quotient has a distinguished multiplication formula which makes
all remaining connector rows explicit.  In the affine coordinate
$x=(E-a)/(2b)$,
\begin{equation}
   \cA_N
   =\CC[E]/\langle T_N(E)\rangle
   \simeq\CC[x]/\langle U_N(x)\rangle.
   \label{eq:tight-binding-Chebyshev-quotient-app}
\end{equation}
The identity
$U_m(\cos\theta)=\sin[(m+1)\theta]/\sin\theta$
identifies $U_m$ with the character of the spin-$m/2$ representation of
$SU(2)$.  Ordinary Chebyshev linearization is the Clebsch--Gordan
product, while the terminal relation $U_N=0$ truncates it at level
$N-1$.  With its distinguished basis
$[U_0],\ldots,[U_{N-1}]$, the algebra in
eq.~\eqref{eq:tight-binding-Chebyshev-quotient-app} is the complexified
$\widehat{\mathfrak{su}}(2)_{N-1}$ fusion, or Verlinde, algebra
\cite{Verlinde1988}.
Equivalently, multiplication by $[U_1]$ has the open path $A_N$ as its
fusion graph.

The same identification is visible in the finite sine transform.  If
$x_j=\cos\theta_j$, then
\begin{equation}
   \sqrt{w_j}\,U_m(x_j)
   =
   \sqrt{\frac{2}{N+1}}
   \sin\!\left(\frac{(m+1)(j+1)\pi}{N+1}\right)
   =\mathsf S_{mj}^{(N-1)},
   \label{eq:tight-binding-Verlinde-S-app}
\end{equation}
where $\mathsf S^{(N-1)}$ is the modular $S$-matrix of
$\widehat{\mathfrak{su}}(2)_{N-1}$.  Its orthogonality is the finite
Chebyshev orthogonality relation, and multiplication by $[U_k]$ has
eigenvalues
$U_k(x_j)=\mathsf S_{kj}^{(N-1)}/\mathsf S_{0j}^{(N-1)}$.
Thus the discrete sine transform which diagonalizes the open chain also
diagonalizes quotient multiplication.  This identifies the based
terminal polynomial algebra; it does not assert that the tight-binding
Hamiltonian carries a physical $SU(2)$ or affine Kac--Moody symmetry.

For the localized seed $\ket k$, the specialization of
eq.~\eqref{eq:finite-product-map-app} is
\begin{equation}
   \cF_k^{(N)}:\cA_N\longrightarrow\cA_N,
   \qquad
   \cF_k^{(N)}[f]=[U_kf]_{U_N}\,.
   \label{eq:tight-binding-seed-multiplication-app}
\end{equation}
Its columns in the orthonormal Chebyshev basis are 
\begin{equation}
\begin{gathered}
   \mathsf V_m^{(k,N)}
   ={\cF}_k^{(N)}[U_m]
   =[U_kU_m]_{U_N}
   =\sum_{\substack{\ell_m\le j\le u_m\\
                    j\equiv k+m\ ({\rm mod}\ 2)}}U_j\,,\\
   \ell_m=|k-m|,
   \qquad
   u_m=\min\{k+m,\,2(N-1)-k-m\},
   \qquad 0\le m<N\,.
\end{gathered}
   \label{eq:tight-binding-fusion-product-app}
\end{equation}
This is the $\widehat{\mathfrak{su}}(2)_{N-1}$ fusion rule with labels
$k,m,j$.  It also follows directly by combining ordinary Chebyshev
linearization with the terminal fold in
eq.~\eqref{eq:tight-binding-terminal-fold-app}.

The corresponding product Gram matrix counts overlaps of
parity-restricted intervals,
\begin{equation}
\begin{aligned}
   \mathsf G_{rs}^{(k,N)}
   &=
   \left\langle
      \mathsf V_r^{(k,N)},\mathsf V_s^{(k,N)}
   \right\rangle_{\mu_0}\\
   &=
   \begin{cases}
      \displaystyle
      \max\left\{0,
      1+\frac{\min(u_r,u_s)-\max(\ell_r,\ell_s)}{2}
      \right\},
      &r\equiv s\pmod 2,\\[9pt]
      0,&r\not\equiv s\pmod 2.
   \end{cases}
\end{aligned}
   \label{eq:tight-binding-fusion-Gram-app}
\end{equation}
This explicit matrix is the model-dependent input to
proposition~\ref{prop:finite-product-Gram}.  Let
$\widehat\Gamma_{n,m}^{(k,N)}$ be its residual rows from
eq.~\eqref{eq:finite-product-Schur-app}.  Since
$\mathsf G^{(k,N)}$ is block diagonal in degree parity, the Schur
recurrence gives
$\widehat\Gamma_{n,m}^{(k,N)}=0$ whenever $m-n$ is odd.  Since
$P_m=b^mU_m$, the complete quotient connector is
\begin{equation}
\begin{aligned}
   \left[
      \frac{P_k(E)^2}{h_k}R_n^{(k,N)}(E)
   \right]_{T_N}
   &=
   \sum_{m=n}^{N-1}
   \Gamma_{n,m}^{(k,N)}P_m(E),\\
   \Gamma_{n,m}^{(k,N)}
   &=b^{n-m}\widehat\Gamma_{n,m}^{(k,N)},
   \qquad
   0\le n<d\,,\quad n\le m<N.
\end{aligned}
   \label{eq:finite-tight-binding-connector-app}
\end{equation}
In the unfolded range $n+2k\le N-1$,
\begin{equation}
   \widehat\Gamma_{n,n+2s}^{(k,N)}
   =\mathcal C_{\sigma,\varepsilon;s}^{(k)}\,,
   \qquad 0\le s\le k\,,
\end{equation}
so eq.~\eqref{eq:finite-tight-binding-connector-app} reproduces the
half-line connector in
eq.~\eqref{eq:tight-binding-all-k-connector-main}.  Within the larger
polynomial range it implements the terminal fold, and the same
Schur-complement recurrence continues through every later physical
level.

We define
\begin{equation}
   \Delta_n^{(k,N)}
   =
   \det\!\left[
      \mathsf G_{rs}^{(k,N)}
   \right]_{r,s=0}^{n-1},
   \qquad
   \Delta_0^{(k,N)}=1\,,
\end{equation}
such that the complete finite Jacobi data follow from
\begin{equation}
\begin{aligned}
   \widehat\Gamma_{n,n}^{(k,N)}
   &=\frac{\Delta_{n+1}^{(k,N)}}{\Delta_n^{(k,N)}}\,,\\
   \thh_n^{(k,N)}
   &=b^{2n}\frac{\Delta_{n+1}^{(k,N)}}{\Delta_n^{(k,N)}}\,,\\
   \bigl(\widetilde b_n^{(k,N)}\bigr)^2
   &=
   b^2
   \frac{\Delta_{n+1}^{(k,N)}\Delta_{n-1}^{(k,N)}}
        {\bigl(\Delta_n^{(k,N)}\bigr)^2}\,,
   \qquad 1\le n<d\,.
\end{aligned}
   \label{eq:tight-binding-finite-determinants-app}
\end{equation}
Symmetry of the shifted measure about $a$ gives
$\widetilde a_n^{(k,N)}=a$.  Thus $d$ is exactly the number of positive
degree-ordered pivots.  When $d<N$, the next determinant vanishes and
detects support loss; otherwise termination is fixed by the ambient
dimension $N$.

Equations~\eqref{eq:tight-binding-fusion-product-app}--
\eqref{eq:tight-binding-finite-determinants-app} reproduce the unfolded
coefficients of
eq.~\eqref{eq:tight-binding-all-k-coefficients-main}, implement the
folding in eq.~\eqref{eq:tight-binding-offset-fold-app}, and continue
through all remaining physical levels.  A proper terminal lift as in
proposition~\ref{prop:terminal-lift-determinant} gives the same quotient
class, but is unnecessary for evaluating these coefficients.

\paragraph{Finite amplitudes, projectors and checks.}
Define
\begin{equation}
   I_m^{(N)}(t)
   =\sum_{j=0}^{N-1}w_jP_m(E_j)\ee^{-\ii E_jt},
   \qquad
   \phi_m^{(0,N)}(t)=\frac{I_m^{(N)}(t)}{b^m}\,.
\end{equation}
The connector in
eq.~\eqref{eq:finite-tight-binding-connector-app} gives
\begin{equation}
\begin{aligned}
   \tphi_n^{(k,N)}(t)
   &=
   \frac{1}{\sqrt{\thh_n^{(k,N)}}}
   \sum_{m=n}^{N-1}
   \Gamma_{n,m}^{(k,N)}I_m^{(N)}(t)\\
   &=
   \frac{1}{\sqrt{\widehat\Gamma_{n,n}^{(k,N)}}}
   \sum_{m=n}^{N-1}
   \widehat\Gamma_{n,m}^{(k,N)}
   \phi_m^{(0,N)}(t),
   \qquad 0\le n<d\,.
\end{aligned}
   \label{eq:finite-tight-binding-amplitude-complexity-app}
\end{equation}
The finite complexity then follows from
eq.~\eqref{eq:tight-binding-finite-complexity-main}.

The quotient result has a direct site-space check.  After subtracting
$a\Id$, the hopping Hamiltonian exchanges the two site parities, which
again gives $\widetilde a_n^{(k,N)}=a$.  For $N=10$ and $k=1$, the
positive hoppings are
\begin{equation}
   \sqrt2,\ \frac1{\sqrt2}\,,\ \sqrt{\frac32}\,,\
   \sqrt{\frac23}\,,\ \frac2{\sqrt3}\,,\ \frac{\sqrt3}{2}\,,\
   \frac{\sqrt5}{2}\,,\ \frac2{\sqrt5}\,,\ \frac1{\sqrt5}\,,
\end{equation}
times $b$.  The first eight entries follow the half-line blocks in
eq.~\eqref{eq:tight-binding-halfline-hopping-pattern-app}; the last is
the terminal value in
eq.~\eqref{eq:tight-binding-first-terminal-hopping-app} with
$\sigma=5$.  The complete sequence agrees with
ref.~\cite{Balasubramanian:2025variations}.

For cumulative quantities, the root-space projector applies while
$\ell+k\le N-1$.  Beyond that range, the exact object is the quotient
projector onto the reduced image
$[P_k]\cP_\ell\subset\cA_N$. The Moore--Penrose form in
eq.~\eqref{eq:coeff-projector-app} makes the quotient projector
independent of redundant choices of spanning columns; for the canonical
columns and $0\le\ell<d$, the reduced image has dimension $\ell+1$.
Atom deletion is encoded by the vanishing of $[P_k]$ on the deleted
modes. The connector and
projector constructions therefore reproduce the amplitudes, cumulative
probabilities and spread complexity obtained from the finite sine
transform.

The open path completes the analytic check of terminal closure, since the
Chebyshev fusion rule resolves every localized-site connector and its
terminal fold.  Appendix~\ref{app:numerics} turns the same
multiplication, connector and projector constructions into an
implementation scheme for %
general reference Jacobi data, including finite-terminal and infinite-window checks.

\section{Implementation of the relative seed transform}
\label{app:numerics}

The constructions above are implemented entirely within a solved
reference Krylov problem. For each model, its reference data are
computed once and reused for every polynomially related seed. This
appendix collects the corresponding reference-space realizations, together with
the rank and truncation checks required at a finite terminal edge and
on an infinite chain.

\paragraph{Reference data and reduced realizations.}
For a normalized reference seed, the reusable data consist of the
Jacobi matrix $\Jac$, the monic norms and the Fourier--OP moments
\begin{equation}
\begin{gathered}
   h_0=1,\qquad h_n=b_1^2\cdots b_n^2\,,\\
   I_n(t)=\sqrt{h_n}\,
   \bra{K_n}\ee^{-\ii Ht}\ket{K_0},
   \qquad
   [\ee^{-\ii t\Jac}e_0]_n=\frac{I_n(t)}{\sqrt{h_n}}\,.
\end{gathered}
\end{equation}
  For a polynomial seed $Q$, the relative calculation
determines
\begin{equation}
   \Gamma^{[Q]},\qquad
   \thh_n^{[Q]},\qquad
   \tphi_n^{[Q]}(t),\qquad
   \Phi_\ell^{[Q]}(t),\qquad
   K_Q(t).
   \label{eq:numerical-relative-outputs-app}
\end{equation}
These are the connector, shifted norms, shifted amplitudes, cumulative
probabilities and spread complexity.

The three realizations named in subsection~\ref{subsec:connector-realizations}, the root,
moment/Gram and reduced-Jacobi constructions, are 
complementary, and table~\ref{tab:christoffel-hierarchy} indicates which is preferable in which
regime.  At low degree, the confluent-root system or Christoffel
determinant gives the connector directly, with a proper terminal lift
and quotient reduction near a finite edge.  At moderate degree, the
moment or Gram construction avoids root finding, while
eqs.~\eqref{eq:projection-coeff-main}
and~\eqref{eq:coeff-projector-app} give the coefficient-space
projectors.  Long finite chains may instead be treated directly in the
already constructed reference Jacobi representation.

\begin{table}[htbp]
\centering
\renewcommand{\arraystretch}{1.15}
\begin{tabular}{@{}p{0.30\linewidth} p{0.67\linewidth}@{}}
\toprule
\textbf{Regime} & \textbf{Preferred realization} \\
\midrule
Connector at low degree & Root or confluent Christoffel system \\
Connector and Jacobi data at moderate or high degree & Suitably conditioned Gram factorization or reduced-Jacobi realization; moment/Hankel methods at moderate degree only \\
Cumulative probabilities & Hermitian coefficient-space range projector \\
Finite terminal edge & Proper lift followed by quotient reduction for root-system or determinant connector formulas; quotient-image or finite weighted Gram factorization for intrinsic finite-support data \\
\bottomrule
\end{tabular}
\caption{Practical realizations of the relative seed transform.}
\label{tab:christoffel-hierarchy}
\end{table}

Let $r=\deg Q$, $L=\deg W_Q=2r$, and assume $N_Q>0$.  Before a terminal
edge, the common output of the connector constructions is
\begin{equation}
   S_n^{[Q]}(E)
   =W_Q(E)R_n^{[Q]}(E)
   =\sum_{m=n}^{n+L}\Gamma_{n,m}^{[Q]}P_m(E).
\end{equation}
In finite support this identity is understood modulo $T_d$, and the
physical expansion has $n\le m\le\min\{n+L,d-1\}$.  For floating-point
evaluation, it is convenient to scale the connector in orthonormal
coordinates, 
\begin{equation}
\begin{aligned}
   \mathsf T_{n,m}^{[Q]}
   &=
   \Gamma_{n,m}^{[Q]}
   \sqrt{\frac{h_m}{\thh_n^{[Q]}}}\,, \\
   \phi_m^{(0)}(t)&=\frac{I_m(t)}{\sqrt{h_m}},\\
   \tphi_n^{[Q]}(t)
   &=
   \sum_m\mathsf T_{n,m}^{[Q]}\phi_m^{(0)}(t)
   =
   \frac{1}{\sqrt{\thh_n^{[Q]}}}
   \sum_m\Gamma_{n,m}^{[Q]}I_m(t),\\
   \Phi_\ell^{[Q]}(t)
   &=\sum_{n=0}^{\ell}|\tphi_n^{[Q]}(t)|^2,\\
   K_Q(t)
   &=\sum_{n\ge0}n|\tphi_n^{[Q]}(t)|^2
   =\sum_{\ell\ge0}\bigl(1-\Phi_\ell^{[Q]}(t)\bigr).
\end{aligned}
   \label{eq:numerical-amplitude-complexity-app}
\end{equation}
The transfer row vanishes for $m<n$ and, before terminal reduction, for
$m>n+L$.  All time dependence thus resides in the reference amplitudes.  For finite shifted support the sums
terminate at $n=d_Q-1$ and $\ell=d_Q-2$; on an infinite chain the last
equality is understood in $[0,\infty]$.

For the reduced-Jacobi realization, the shifted seed is represented by
\begin{equation}
   v_Q=\frac{Q(\Jac)e_0}{\sqrt{N_Q}}\,,
   \qquad
   N_Q=e_0^\dagger\Qsh(\Jac)Q(\Jac)e_0
   =\|Q(\Jac)e_0\|^2>0\,.
   \label{eq:numerical-seed-vector-app}
\end{equation}
If $Q$ is given in the orthonormal reference basis, then
$\Qhat=\sum_{j=0}^{r}c_jP_j/\sqrt{h_j}$, with
$\sum_{j=0}^{r}|c_j|^2=1$, implies
$v_Q=(c_0,\ldots,c_r,0,\ldots)^T$.  More generally,
$Q(\Jac)e_0$ should be evaluated through the three-term recurrence or a
Clenshaw-type scheme, rather than by forming powers of $\Jac$. The spectral measure of $(\Jac,v_Q)$ is precisely $W_Q\dd\mu$.  For
finite $\Jac$, Hermitian Lanczos on this reference matrix therefore gives
the shifted Jacobi data exactly; on an infinite chain it is applied only
after the controlled truncation described below.  Let
$\mathsf U_Q=(u_0,\ldots,u_{d_Q-1})$ denote the resulting basis in a
finite realization.  With the positive-subdiagonal convention, we have
\begin{equation}
   u_n
   =\frac{R_n^{[Q]}(\Jac)v_Q}{\sqrt{\thh_n^{[Q]}}},
   \qquad
   \thh_0^{[Q]}=1\,,\qquad
   \thh_n^{[Q]}
   =\widetilde b_1^{\,2}\cdots\widetilde b_n^{\,2}\,.
\end{equation}
The same basis gives the scaled connector and the shifted amplitudes
directly, 
\begin{equation}
\begin{aligned}
   \mathsf T^{[Q]}
   &=\mathsf U_Q^\dagger
      \frac{Q(\Jac)}{\sqrt{N_Q}},\\
   \tphi^{[Q]}(t)
   &=\mathsf U_Q^\dagger\ee^{-\ii t\Jac}v_Q
   =\mathsf T^{[Q]}\phi^{(0)}(t).
\end{aligned}
   \label{eq:numerical-reduced-connector-app}
\end{equation}
The monic connector is recovered from
$\Gamma_{n,m}^{[Q]}
=\sqrt{\thh_n^{[Q]}/h_m}\,\mathsf T_{n,m}^{[Q]}$.
For complex $Q$, the matrices in
eq.~\eqref{eq:numerical-reduced-connector-app} may be complex during
the calculation, but $\mathsf T^{[Q]}$ is real in exact arithmetic
because it depends only on the positive multiplier $W_Q$.  Its residual
imaginary part is therefore another numerical diagnostic.
Thus the connector transfer and direct propagation in the shifted
Jacobi basis provide an exact finite-dimensional cross-check.  This Lanczos factorization acts only on the solved reference Jacobi matrix, not on the original Hilbert-space Hamiltonian, with loss of orthogonality monitored in the standard way \cite{ParlettScott1979}.

\paragraph{Finite support and numerical stability.}
Suppose that
$\dd\mu(E)=\sum_{\alpha=0}^{d-1}w_\alpha
\delta(E-E_\alpha)\dd E$.  We store the reference orthonormal polynomial
values and the shifted weights as 
\begin{equation}
\begin{gathered}
   p_m(E_\alpha)
   =\frac{P_m(E_\alpha)}{\sqrt{h_m}}\,,
   \qquad 0\le m<d\,,\\
   w_\alpha^{[Q]}
   =\frac{|Q(E_\alpha)|^2w_\alpha}{N_Q}\,,
   \qquad
   \mathcal I_Q=\{\alpha:w_\alpha^{[Q]}>0\},
   \qquad d_Q=|\mathcal I_Q|.
\end{gathered}
\end{equation}
Exact zeros of $Q(E_\alpha)$ are deleted.  Any further weight truncation
is numerical; $d_Q^{\rm(num)}$ is accepted only if it is stable under
threshold and precision changes.  Below, $d_Q$ and $\mathcal I_Q$
denote the retained support.

On the surviving atoms, form the weighted evaluation matrix 
\begin{equation}
\begin{aligned}
   (\mathsf E_Q)_{\alpha m}
   &=\sqrt{w_\alpha^{[Q]}}\,p_m(E_\alpha),
   \qquad
   \alpha\in\mathcal I_Q,\quad 0\le m<d\,,\\
   \mathsf G_{mn}^{[Q]}
   &=(\mathsf E_Q^\dagger\mathsf E_Q)_{mn}
   =\int p_m(E)p_n(E)\dd\nu_Q(E).
\end{aligned}
\end{equation}
This is the degree-ordered shifted Gram matrix, in agreement with
eq.~\eqref{eq:finite-product-spectral-factorization-app}.  A
rank-revealing QR factorization or SVD diagnoses the effective support
rank.  Since column pivoting changes the polynomial order, the shifted
orthogonal family is then formed by applying an order-preserving QR
factorization to the first $d_Q$ degree columns.  Extending the resulting
orthonormal basis across the remaining columns gives
\begin{equation}
   \mathsf E_Q=\mathsf Z_Q\mathsf T^{[Q]},
   \qquad
   \mathsf Z_Q^\dagger\mathsf Z_Q=I_{d_Q},
   \qquad
   \mathsf T_{n,n}^{[Q]}>0\,.
\end{equation}
The factors have the direct spectral interpretation
\begin{equation}
\begin{aligned}
   (\mathsf Z_Q)_{\alpha n}
   &=
   \sqrt{w_\alpha^{[Q]}}\,
   \frac{R_n^{[Q]}(E_\alpha)}
        {\sqrt{\thh_n^{[Q]}}}\,,\\
   \thh_n^{[Q]}
   &=h_n\bigl(\mathsf T_{n,n}^{[Q]}\bigr)^2,\\
   \Gamma_{n,m}^{[Q]}
   &=\sqrt{\frac{\thh_n^{[Q]}}{h_m}}\,
      \mathsf T_{n,m}^{[Q]}\,.
\end{aligned}
   \label{eq:numerical-finite-support-data-app}
\end{equation}
If $\mathsf D_E=\diag(E_\alpha)_{\alpha\in\mathcal I_Q}$, the shifted
Jacobi matrix is
\begin{equation}
   \widetilde\Jac^{[Q]}
   =\mathsf Z_Q^\dagger\mathsf D_E\mathsf Z_Q\,.
   \label{eq:numerical-finite-Jacobi-compression-app}
\end{equation}
It is tridiagonal in exact arithmetic.  The orthogonality residual of
$\mathsf Z_Q$ and the norm of the off-tridiagonal part of
$\widetilde\Jac^{[Q]}$ therefore test the inferred support rank and
working precision.  The degree-ordered factorization is the numerical
counterpart of the quotient Gram and Schur-complement constructions in
eqs.~\eqref{eq:finite-product-Schur-app}
and~\eqref{eq:finite-product-LDL-app}; it performs atom deletion before
terminal reduction, as required by the finite-support construction.

At high degree, determinant, Hankel and confluent-root realizations can
be severely ill-conditioned, particularly for clustered roots
\cite{Gautschi2004}; stabilized Christoffel algorithms remain useful at
lower degree \cite{BuenoDopico2007}.  For projected kernels, factor the
spanning matrix $C_{Q,\ell}$ directly.  If the numerically retained singular-value
decomposition is
\begin{equation}
   C_{Q,\ell}
   =\mathsf U_{\rm num}\mathsf\Sigma_{\rm num}
      \mathsf V_{\rm num}^\dagger\,,
   \qquad
   \Pi_{Q,\ell}
   =\mathsf U_{\rm num}\mathsf U_{\rm num}^\dagger\,,
\end{equation}
then neither an explicit inverse nor the normal matrix
$C_{Q,\ell}^\dagger C_{Q,\ell}$ need be formed as the latter would square
the condition number.  A thin QR factorization gives the same projector
when the columns are numerically independent.    These factorizations, applied equally to the root-constraint representers, expose nearly dependent constraints before they appear as cancellations among large determinants.  At moderate or high
degree, the most reliable route should be selected by comparing the
orthogonality, recurrence and connector residuals rather than by degree
alone.

\paragraph{Infinite-chain truncation.}
An infinite-chain computation contains two distinct truncations.  The
shifted amplitudes are retained through $M_{\rm cut}$, while the
reference evolution is evaluated in a finite Jacobi window.  Since a
degree-$r$ seed has connector width $L=2r$, amplitudes through
$M_{\rm cut}$ require reference moments at least through
$M_{\rm cut}+L$.  If a $D\times D$ truncation of $\Jac$ is used, it
should therefore satisfy
\begin{equation}
   D-1\ge M_{\rm cut}+L+B_{\rm ref}\,,
\end{equation}
where the additional buffer $B_{\rm ref}$ is increased until the result
is insensitive to the artificial terminal edge over the chosen time
interval.  A finite truncation of an infinite Jacobi matrix is not a
terminal quotient of the original problem; reflection from its imposed
boundary is a numerical error.  This propagation error has a direct
a posteriori estimate.  Let $P_D$ project onto reference levels
$0,\ldots,D-1$, set $\Jac_D=P_D\Jac P_D$, and embed
$\psi_D(t)=\ee^{-\ii t\Jac_D}v_Q$ in the infinite chain.  For $t\geq 0$, Duhamel's formula gives
\begin{equation}
   \left\|
      \ee^{-\ii t\Jac}v_Q-\psi_D(t)
   \right\|
   \le
   b_D\int_0^{|t|}
      \left|[\psi_D(s)]_{D-1}\right|\dd s\,.
   \label{eq:numerical-boundary-residual-app}
\end{equation}
For $t<0$, the same estimate holds with
$[\psi_D(s)]_{D-1}$ replaced by $[\psi_D(-s)]_{D-1}$ under the
integral.
The bound measures the amplitude reaching the artificial boundary.  A
finite truncation evolves unitarily and can therefore preserve
normalization even after a reflected component has returned; norm
conservation alone does not test $B_{\rm ref}$.

If the exact shifted amplitudes are retained through $M_{\rm cut}$, the
omitted normalization and complexity tails are
\begin{equation}
\begin{aligned}
   \varepsilon_0(M_{\rm cut},t)
   &=\sum_{n>M_{\rm cut}}|\tphi_n^{[Q]}(t)|^2,\\
   \varepsilon_1(M_{\rm cut},t)
   &=\sum_{n>M_{\rm cut}}n|\tphi_n^{[Q]}(t)|^2.
\end{aligned}
   \label{eq:numerical-tail-errors-app}
\end{equation}
They are related to the cumulative tails by
\begin{equation}
   \varepsilon_1(M_{\rm cut},t)
   =M_{\rm cut}\varepsilon_0(M_{\rm cut},t)
   +\sum_{\ell\ge M_{\rm cut}}
      \bigl(1-\Phi_\ell^{[Q]}(t)\bigr).
   \label{eq:numerical-tail-decomposition-app}
\end{equation}
In particular,
$\varepsilon_1(M_{\rm cut},t)
\ge(M_{\rm cut}+1)\varepsilon_0(M_{\rm cut},t)$, so a small
normalization tail is not an upper bound on the complexity error.
An analytic majorant gives a certified cutoff once both terms on the
right are below the prescribed tolerance, either at fixed time or
uniformly on a stated time interval.  For the Charlier first jump, the
cumulative-complexity remainder separates into the explicit reference
tail and the positive correction controlled by
eq.~\eqref{eq:H1-correction-majorant-app}; standard Poisson-tail
recurrences then bound the remaining sums.  By contrast,
eq.~\eqref{eq:H1-general-r-complexity-bound-app} proves finiteness for
every fixed number-state seed but is not a coefficientwise truncation
majorant.

Without an analytic majorant, $M_{\rm cut}$, $B_{\rm ref}$ and the
working precision should be increased independently.  Normalization is
monitored together with agreement between the amplitude transfer and
the projected-kernel evaluation of $\Phi_\ell^{[Q]}$.  Recurrence
residuals include
$\Gamma_{n,n}^{[Q]}h_n-\thh_n^{[Q]}$ and, before a terminal edge,
$\Gamma_{n,n+L}^{[Q]}-w_L$, where $w_L$ is the leading coefficient of
$W_Q$.  Positivity and monotonicity of the cumulative probabilities,
$\Phi_{d_Q-1}^{[Q]}=1$ in finite support, and the initial conditions
$\tphi_0^{[Q]}(0)=1$, $\tphi_{n>0}^{[Q]}(0)=0$ provide further checks.
These tests are diagnostic rather than rigorous.  The first omitted
term need not bound $\varepsilon_1$, and the amplitude and projector
calculations can share the same truncated reference input.

The same implementation applies when the normalized positive
multiplier is obtained by tracing the parent measure against a
coefficient-space density matrix. For a finite reference Krylov window,
appendix~\ref{app:mixed-seeds} gives a scalar factorization
$W_\sigma=Q_\sigma^\sharp Q_\sigma$.  The Gram and connector routes may
use $W_\sigma$ directly, while the projector and reduced-Jacobi
realizations may use any scalar factor $Q_\sigma$; no new reference data
are required.  With a fixed operator inner product, the same numerical
core applies in appendix~\ref{app:operator-krylov} after replacing the
state Jacobi data and Fourier--OP moments by their Liouvillian
counterparts.

\section{Scalar mixed-state compressions}
\label{app:mixed-seeds}

A density matrix does not determine a unique Krylov problem until the
cyclic representation and its inner product have been specified. This appendix develops the choice most directly connected with the scalar spectral construction of the main text by restricting the state to the commutative algebra generated by $H$, and delimits what that choice discards.  For density matrices supported on a finite reference Krylov window,
the contraction gives a single energy measure and hence a single scalar
Jacobi chain. Its polynomial density admits a scalar spectral
factorization, which places the resulting chain within the
polynomial-seed calculus without identifying the underlying mixed
preparation with a pure state.

\paragraph{The scalar energy-distribution functional.}
Let $\rho$ be a density matrix.  Its restriction to functions of $H$ is
the positive normalized functional
\begin{equation}
   \omega_\rho(f)=\operatorname{Tr}\!\left(\rho f(H)\right),
   \qquad
   \dd\mu_\rho(E)=\operatorname{Tr}\!\left(\rho\,\Pi(\dd E)\right).
   \label{eq:mixed-functional-app}
\end{equation}
For a discrete spectral resolution
$H=\sum_{\alpha}E_\alpha\Pi_\alpha$, this reduces to
\begin{equation}
   \omega_\rho(f)
   =\sum_{\alpha}f(E_\alpha)p_\alpha\,,
   \qquad
   p_\alpha=\operatorname{Tr}(\rho\Pi_\alpha).
\end{equation}
Thus $\mu_\rho$ retains only the total population of each energy
eigenspace.  Inter-energy coherences, together with traceless data within
a degenerate eigenspace, are invisible to this commutative restriction.

Let $P_n^{(\rho)}$ be the monic orthogonal polynomials of $\mu_\rho$ and
$h_n^{(\rho)}$ their norms.  The corresponding scalar Krylov amplitudes
are
\begin{equation}
   \phi_n^{(\rho)}(t)=
   \frac{1}{\sqrt{h_n^{(\rho)}}}
   \int_{\RR}\ee^{-\ii Et}P_n^{(\rho)}(E)\dd\mu_\rho(E).
   \label{eq:mixed-scalar-amplitudes-app}
\end{equation}
They are the orthogonal-polynomial coefficients of the energy
characteristic function
$\omega_\rho(\ee^{-\ii tE})=\operatorname{Tr}(\rho\ee^{-\ii Ht})$.
They should not be confused with amplitudes for the physical conjugation
$\rho(t)=\ee^{-\ii Ht}\rho\ee^{\ii Ht}$.  A thermal state makes the
distinction immediate; its density matrix is stationary, whereas its
energy characteristic function is generally time dependent.
Purification-based constructions make another choice by embedding the
mixed state in a larger Hilbert space before defining a state or operator
Krylov problem \cite{DasMori2024}.  We therefore call the chain defined
by eq.~\eqref{eq:mixed-functional-app} the scalar energy-distribution
chain.

\paragraph{Finite Krylov-window contractions and spectral factorization.}
Consider the reference window and its canonical isometry
\begin{equation}
   \mathcal S_R=\Span\{\ket{K_0},\ldots,\ket{K_R}\},
   \qquad
   V_R=(\ket{K_0},\ldots,\ket{K_R}).
\end{equation}
We distinguish a density operator $\rho_R$ on $\mathcal S_R$ from its
coefficient-space matrix $\sigma$ by writing
\begin{equation}
   \rho_R=V_R\sigma V_R^\dagger
   =\sum_{r,s=0}^{R}\sigma_{rs}\ket{K_r}\bra{K_s}\,,
   \qquad
   \sigma\ge0\,,\qquad \operatorname{Tr}\sigma=1\,.
   \label{eq:mixed-window-density-app}
\end{equation}
The parent measure of section~\ref{sec:matrix} is given by
\begin{equation}
\begin{aligned}
   \dd M(E)&=V_R^\dagger\Pi(\dd E)V_R\,,\\
   \dd M_{rs}(E)
   &=\frac{P_r(E)P_s(E)}{\sqrt{h_rh_s}}\dd\mu(E).
\end{aligned}
   \label{eq:mixed-parent-measure-app}
\end{equation}
Its trace contraction is consequently 
\begin{equation}
\begin{aligned}
   \dd\mu_{\rho_R}(E)
   =\operatorname{Tr}_{\cH}\!\left(\rho_R\Pi(\dd E)\right)
   =\operatorname{Tr}_{\CC^{R+1}}\!\left(\sigma\,\dd M(E)\right)
   =W_\sigma(E)\dd\mu(E).
\end{aligned}
   \label{eq:mixed-parent-compression-app}
\end{equation}
With $(v_R)_r(E)=P_r(E)/\sqrt{h_r}$, the multiplier takes the form
\begin{equation}
\begin{aligned}
   W_\sigma(E)
   =\sum_{r,s=0}^{R}\sigma_{sr}
     \frac{P_r(E)P_s(E)}{\sqrt{h_rh_s}}
   =v_R(E)^\dagger\sigma v_R(E)
   =\|\sigma^{1/2}v_R(E)\|^2\ge0\,.
\end{aligned}
   \label{eq:mixed-window-weight-app}
\end{equation}
Normalization follows  from eq.~\eqref{eq:mixed-functional-app}, or from $\int\dd M=I_{R+1}$, $\int W_\sigma\dd\mu=\operatorname{Tr}\sigma=1$. As $v_R(E)$ is real for real $E$, only the real symmetric part of $\sigma$ contributes; real off-diagonal entries retain interference between distinct Krylov directions, while the imaginary skew part is invisible to the scalar contraction.

For the rank-one state $\sigma=cc^\dagger$,
\begin{equation}
   W_\sigma(E)=|\Qhat_c(E)|^2,
   \qquad
   \Qhat_c(E)=\sum_{r=0}^{R}c_r
      \frac{P_r(E)}{\sqrt{h_r}}\,,
   \label{eq:mixed-rank-one-app}
\end{equation}
which recovers the pure polynomial seed
$\ket{\psi_c}=V_Rc$.  A diagonal coefficient state gives
\begin{equation}
   \sigma=\diag(p_0,\ldots,p_R),
   \qquad
   W_\sigma(E)=\sum_{r=0}^{R}p_r\frac{P_r(E)^2}{h_r}\,.
   \label{eq:mixed-diagonal-app}
\end{equation}
The uniform case $p_r=1/(R+1)$ is distinguished twice over. It is the maximally mixed state
on the window, and it is also the Haar average over pure window seeds, since
$\int cc^\dagger\dd c=I_{R+1}/(R+1)$ on the unit sphere. Its multiplier is therefore the normalized
kernel diagonal,
\begin{equation}
   W_\sigma(E)=\frac{\mathbb K_R^\mu(E,E)}{R+1}\,,
   \label{eq:mixed-haar-christoffel-app}
\end{equation}
$(R+1)^{-1}$ times the
reciprocal of the standard Christoffel function of the window, which is the same object whose growth at
the seed roots governs the preparation response in section~\ref{sec:rigidity}. Averaging over pure
preparations therefore does not return the reference problem. It returns the reference measure
reweighted by this normalized kernel diagonal.

To see explicitly how off-diagonal coefficient-space coherences enter the
scalar measure, consider the minimal nontrivial window $R=1$.  In the
ordered basis $(\ket{K_0},\ket{K_1})$, write
\begin{equation}
   \sigma_1=
   \begin{pmatrix}
      p & \alpha\\
      \bar\alpha & 1-p
   \end{pmatrix},
   \qquad
   |\alpha|^2\le p(1-p).
\end{equation}
Then
\begin{equation}
   W_{\sigma_1}(E)=
   p+(1-p)\frac{P_1(E)^2}{h_1}
   +2\operatorname{Re}\!\left(
      \alpha\frac{P_1(E)}{\sqrt{h_1}}
   \right).
   \label{eq:mixed-two-site-weight-app}
\end{equation}
For $0<p<1$, the choice $\alpha=0$ is a full-rank diagonal state,
whereas saturation $|\alpha|^2=p(1-p)$ gives a rank-one extremal state
on the coefficient-space state cone.  %
Although these coefficient-space preparations have different ranks, restriction to the
commutative algebra generated by $H$ retains only the multiplier $W_\sigma(E)$.  Physically
distinct preparations can therefore be measure-equivalent and generate the same scalar
energy-distribution Lanczos chain.

Indeed, eq.~\eqref{eq:mixed-window-weight-app} shows that $W_\sigma$ is
a globally nonnegative real polynomial, not merely a polynomial that is
nonnegative on the spectral support. Every real zero therefore has even
algebraic multiplicity, and the nonreal zeros occur in conjugate pairs.
Taking half the algebraic multiplicity of each real zero and the full
algebraic multiplicity of one member of every nonreal conjugate pair
gives a polynomial $Q_\sigma$ such that
\begin{equation}
   W_\sigma(E)=Q_\sigma^\sharp(E)Q_\sigma(E)
   =|Q_\sigma(E)|^2,\qquad E\in\RR\,,
   \qquad
   \deg Q_\sigma\le R\,,
   \label{eq:mixed-spectral-factor-app}
\end{equation}
where
$Q_\sigma^\sharp(z)=\overline{Q_\sigma(\bar z)}$.
The normalization of $W_\sigma$ gives
$\int|Q_\sigma|^2\dd\mu=1$, so $Q_\sigma$ expands with a unit
coefficient vector in the orthonormal basis
$P_r/\sqrt{h_r}$, $0\le r\le R$. Consequently, for the
Krylov-window parent measure in eq.~\eqref{eq:mixed-parent-measure-app},
\begin{equation}
\begin{aligned}
   \left\{
      \operatorname{Tr}(\sigma\,\dd M):
      \sigma\ge0,\ \operatorname{Tr}\sigma=1
   \right\}
   =
   \left\{
      |Q(E)|^2\dd\mu(E):
      \deg Q\le R,\ \int|Q|^2\dd\mu=1
   \right\}.
\end{aligned}
   \label{eq:mixed-pure-filter-equivalence-app}
\end{equation}
The forward inclusion follows from the factorization above, while the
reverse inclusion is the rank-one construction in
eq.~\eqref{eq:mixed-rank-one-app}.  For example, with
$x=P_1/\sqrt{h_1}$,
\begin{equation}
   p+(1-p)x^2
   =\left|\sqrt p+\ii\sqrt{1-p}\,x\right|^2.
\end{equation}
Thus even a full-rank diagonal state on the first two Krylov directions
has the same scalar measure as an effective pure polynomial seed.  This
equivalence concerns scalar measures and Jacobi problems, not physical
preparations.  The map $\sigma\mapsto W_\sigma$ is
many-to-one and forgets the ensemble decomposition as well as part of
the coefficient-space coherence data.  In particular, the complexity
of the contracted measure is not an ensemble average of the pure-state
complexities.  The reduction in
eq.~\eqref{eq:mixed-pure-filter-equivalence-app} uses the rank-one
polynomial density
$v_R(E)v_R(E)^\dagger\dd\mu(E)$ of
eq.~\eqref{eq:mixed-parent-measure-app}. It need not hold for a general
seed subspace whose compressed measure has matrix density of rank
greater than one relative to a scalar measure dominating its entries.

The scalar transfer now follows directly.  Let $R_n^{[\sigma]}$ be the
monic orthogonal polynomials for $W_\sigma\dd\mu$, with norms
$\thh_n^{[\sigma]}$, and put $L_\sigma=\deg W_\sigma$.  Before a terminal
edge,
\begin{equation}
\begin{aligned}
   W_\sigma(E)R_n^{[\sigma]}(E)
   &=\sum_{m=n}^{n+L_\sigma}
      \Gamma_{n,m}^{[\sigma]}P_m(E),\\
   \tphi_n^{[\sigma]}(t)
   &=\frac{1}{\sqrt{\thh_n^{[\sigma]}}}
      \sum_{m=n}^{n+L_\sigma}
      \Gamma_{n,m}^{[\sigma]}I_m(t).
\end{aligned}
   \label{eq:mixed-connector-amplitude-app}
\end{equation}
The Hankel and Gram constructions may be applied directly to the positive multiplier $W_\sigma$, while the projected-kernel formula uses any scalar factor $Q_\sigma$ in eq.~\eqref{eq:mixed-spectral-factor-app}.  Different factor choices give the same scalar Jacobi data and complexity, both depending on $W_\sigma$ alone; appendix~\ref{app:numerics} collects the corresponding Hankel, Gram and projected-kernel routes.

In finite support, the product in
eq.~\eqref{eq:mixed-connector-amplitude-app} is reduced modulo the
terminal polynomial and expanded over the physical range
$n\le m\le d-1$.  An atom $E_\alpha$ is deleted precisely when
\begin{equation}
   W_\sigma(E_\alpha)=0
   \quad\Longleftrightarrow\quad
   v_R(E_\alpha)\in\ker\sigma\,.
\end{equation}
Because the first component of $v_R$ is
$P_0/\sqrt{h_0}=1$, a positive-definite coefficient state cannot delete
an atom.  Support loss is possible only for rank-deficient $\sigma$,
including rank-one polynomial seeds, and the shifted chain then
terminates at the surviving dimension prescribed by
eq.~\eqref{eq:support-loss-dimension}.

\paragraph{Cyclic realization and the distinction from Liouville dynamics.}
The scalar functional $\omega_\rho$ has a cyclic GNS representation
\cite{BratteliRobinson1987}.  In finite dimension, it can be realized by
introducing the conjugate auxiliary Hilbert space and the vector
\begin{equation}
   \ket{\Omega_\rho}
   =\bigl(\sqrt\rho\otimes I\bigr)
      \sum_a\ket{e_a}\otimes\ket{\bar e_a},
   \qquad
   H_L=H\otimes I.
   \label{eq:mixed-left-purification-app}
\end{equation}
Here $\{\ket{e_a}\}$ is any orthonormal basis and
$\{\ket{\bar e_a}\}$ is its conjugate auxiliary basis.  The sum in
eq.~\eqref{eq:mixed-left-purification-app} is the canonical vector
representing the identity and is invariant under the corresponding
conjugate change of basis.  Moreover,
\begin{equation}
   \bra{\Omega_\rho}f(H_L)\ket{\Omega_\rho}
   =\operatorname{Tr}\!\left(\rho f(H)\right).
   \label{eq:mixed-purification-functional-app}
\end{equation}
Lanczos for $(H_L,\ket{\Omega_\rho})$ therefore produces the scalar
Jacobi matrix of $\mu_\rho$.  More generally, the GNS cyclic space is the
closure of the polynomial subspace in $L^2(\mu_\rho)$, subject to the
standing moment and domain assumptions.

This left-action realization encodes the energy functional; it is not
the conjugation dynamics of the density matrix.  Physical evolution
retains transition-sector matrix elements and is described by an
operator-space Krylov problem for the commutator Liouvillian.  Its
spectral variable is an energy gap, with the zero-gap sector containing
both diagonal data and coherences within degenerate eigenspaces.  The
next appendix develops this distinct Liouville-space construction.

\section{Polynomial seed transfer in Liouville space}
\label{app:operator-krylov}

The spectral transfer developed for state Krylov complexity depends on a
self-adjoint cyclic generator and its scalar spectral measure.  The same
construction therefore applies in operator space once an operator inner
product has been fixed and the commutator Liouvillian admits the required
self-adjoint realization.  Polynomial changes of operator seed then
filter Liouvillian frequencies rather than energies.  This appendix
establishes the fixed-inner-product dictionary and separates it from
changes of thermal inner product or temperature, for which the relevant
spectral data may be more extensive than a single gap measure.

\subsection{Polynomial Liouvillian seeds}
\label{app:operator-polynomial-seeds}

Let $\mathcal A$ be an operator space with positive sesquilinear inner
product $(A|B)_\star$.  The symbol $\star$ denotes the chosen
representation, such as a ground-state GNS product, the
infinite-temperature Hilbert--Schmidt product, or a thermal product.
Unless stated otherwise, the reference operator is normalized by
$(O|O)_\star=1$; an unnormalized spectral measure is identified
explicitly.  On the closure of the cyclic subspace generated by $O$, we
assume the domain conditions under which
$
   \mathcal L B=[H,B]
$ 
admits a self-adjoint realization with respect to
$(\cdot|\cdot)_\star$.  Its Heisenberg evolution and autocorrelation are
\begin{equation}
   |O(t))=\ee^{\ii t\mathcal L}|O),
   \qquad
   C_O(t)=(O|\ee^{\ii t\mathcal L}|O)_\star\,.
\end{equation}
These are the standard Heisenberg-space conventions used in operator
Krylov complexity \cite{Parker2019,Nandy2024,Rabinovici2025}.
The associated cyclic space is
\begin{equation}
   \mathcal K(O,\mathcal L)
   =
   \operatorname{cl}\!\Span\{|O),\mathcal L|O),
      \mathcal L^2|O),\ldots\},
   \label{eq:operator-krylov-space-app}
\end{equation}
where the closure is taken in the operator Hilbert space defined by
$(\cdot|\cdot)_\star$ and may be omitted at finite Krylov dimension.

The sign differs from the Schr\"odinger convention used for state
amplitudes.  Operator Fourier--OP moments are obtained from the state
formulas by $t\mapsto-t$, and hence by complex conjugation for a real
spectral measure.  If the operator seed is a density matrix, its physical
conjugation is
$\rho(t)=\ee^{-\ii t\mathcal L}\rho$ rather than
$\ee^{\ii t\mathcal L}\rho$.  Time reversal leaves Jacobi data,
probabilities, cumulative probabilities and Krylov complexity unchanged.
Subject to the same moment, domain and terminal assumptions as in the
state problem, the spectral dictionary is therefore
\begin{equation}
\begin{gathered}
   H\longmapsto\mathcal L\,,
   \qquad
   \ket{K_0}\longmapsto |O),
   \qquad
   \dd\mu(E)\longmapsto\dd\mu_O(\omega),\\
   \dd\mu_O(\omega)
   =(O|\Pi_{\mathcal L}(\dd\omega)|O)_\star\,.
\end{gathered}
   \label{eq:operator-dictionary-app}
\end{equation}

A polynomial operator seed is the finite nested-commutator descendant
\begin{equation}
\begin{gathered}
   |O_Q)=\frac{Q(\mathcal L)|O)}{\sqrt{N_Q}}\,,\\
   Q(\mathcal L)O
   =\sum_{j=0}^{r}q_j\operatorname{ad}_H^j(O),
   \qquad
   \operatorname{ad}_H(O)=[H,O],\\
   N_Q=(O|\Qsh(\mathcal L)Q(\mathcal L)|O)_\star\,.
\end{gathered}
   \label{eq:operator-polynomial-seed-app}
\end{equation}
This is a deformation in Liouvillian frequency.  It is not, in general,
a spatial translation $O_x\mapsto O_{x+1}$ of a local operator; such an
interpretation is available only when the translated operator belongs to
the same cyclic Liouvillian subspace with the corresponding polynomial
representative.  The descendant measure is
\begin{equation}
   \dd\nu_Q(\omega)
   =\frac{|Q(\omega)|^2}{N_Q}\dd\mu_O(\omega),
   \qquad
   N_Q=\int_{\RR}|Q(\omega)|^2\dd\mu_O(\omega).
   \label{eq:operator-christoffel-measure-app}
\end{equation}
It is therefore exactly the polynomial modification treated in
sections~\ref{sec:finiteband} and~\ref{sec:kernels}.

Let $P_n$ and $R_n^{[Q]}$ be the reference and shifted monic orthogonal
polynomials, respectively, and set
\begin{equation}
   W_Q=\frac{\Qsh Q}{N_Q}\,,
   \qquad
   L_Q=\deg W_Q=2\deg Q\,.
\end{equation}
Before a terminal edge, their connector is
\begin{equation}
   W_Q(\omega)R_n^{[Q]}(\omega)
   =\sum_{m=n}^{n+L_Q}
      \Gamma_{n,m}^{[Q]}P_m(\omega),
   \label{eq:operator-connector-app}
\end{equation}
with the finite-support reduction and support-loss rules of
sections~\ref{sec:setup} and~\ref{sec:finiteband}.  Shifted amplitudes,
cumulative probabilities and Jacobi data follow from the same connector
and projected-kernel formulas.  Thus a fixed-degree descendant can be
reconstructed algebraically at every level for which the reference
operator problem is known through the required window
$n\le m\le n+L_Q$, without rerunning Lanczos in the full operator space.

\subsection{Ground-state and gap-spectrum realizations}
\label{app:operator-realizations}

For the ground-state product, let 
$H\ket0=E_0\ket0$ and define
\begin{equation}
   (A|B)_0=\bra0A^\dagger B\ket0.
\end{equation}
After quotienting operators that annihilate $\ket0$, the map
$[A]\mapsto A\ket0$ identifies the commutator Liouvillian with
$H-E_0$ on the cyclic subspace.  This gives its self-adjoint realization
under the usual domain assumptions and shows directly why the spectral
measure is one-sided.  Inserting energy eigenstates gives
\begin{align}
   C_O^{(0)}(t)
   &=\bra0O^\dagger\ee^{\ii t\mathcal L}O\ket0
   =\sum_n|\bra nO\ket0|^2\ee^{\ii(E_n-E_0)t}\,, 
   \label{eq:zero-temp-op-corr-app} \\
   \dd\mu_O^{(0)}(\omega)
   &=\sum_n|\bra nO\ket0|^2
      \delta\!\left(\omega-(E_n-E_0)\right)\dd\omega\,.
   \label{eq:zero-temp-op-measure-app}
\end{align}

For normalized $O$, this is a probability measure (unit total
spectral weight). A local or few-body descendant $Q(\mathcal L)O$ is a finite linear combination of time
derivatives at the insertion point, up to powers of $\ii$, and multiplies
the zero-temperature spectral function by $|Q(\omega)|^2$.  Because the
measure in eq.~\eqref{eq:zero-temp-op-measure-app} need not be invariant
under $\omega\mapsto-\omega$, its Jacobi chain may have nonzero diagonal
coefficients.  A symmetrized correlator or a doubled channel containing
both $O$ and $O^\dagger$ defines a different, two-sided operator
Hilbert space.

\paragraph{Hilbert--Schmidt product.}
For a Hilbert space of dimension $d$, the infinite-temperature product is
\begin{equation}
   (A|B)_\infty
   =\frac{1}{d}\operatorname{Tr}(A^\dagger B).
   \label{eq:operator-HS-inner-product-app}
\end{equation}
If $H\ket a=E_a\ket a$ and $O_{ab}=\bra aO\ket b$, then
\begin{equation}
   \dd\mu_O^{(\infty)}(\omega)
   =\frac{1}{d}\sum_{a,b}|O_{ab}|^2
      \delta\!\bigl(\omega-(E_a-E_b)\bigr)\dd\omega\,.
   \label{eq:operator-gap-measure-app}
\end{equation}
The normalized density-matrix operator
$
   O_\rho
   =\frac{\sqrt d\,\rho_{\rm in}}
          {\sqrt{\operatorname{Tr}\rho_{\rm in}^2}}
$ 
has unit Hilbert--Schmidt norm and gives the density-matrix operator
problem studied in ref.~\cite{CaputaDensityMatrix2024},
\begin{equation}
   \dd\mu_{\rho_{\rm in}}^{(\infty)}(\omega)
   =\frac{1}{\operatorname{Tr}\rho_{\rm in}^2}
      \sum_{a,b}|(\rho_{\rm in})_{ab}|^2
      \delta\!\bigl(\omega-(E_a-E_b)\bigr)\dd\omega\,.
   \label{eq:density-operator-gap-measure-app}
\end{equation}
The diagonal entries contribute at $\omega=0$, as do coherences within
degenerate energy eigenspaces.  More generally, if
$[\rho_{\rm in},H]=0$, the entire measure is supported at zero frequency
and no nonstationary Liouville sector is present.  A descendant
$Q(\mathcal L)\rho_{\rm in}$ multiplies the transition matrix element
$(\rho_{\rm in})_{ab}$ by $Q(E_a-E_b)$.  Thus the scalar mixed-state
functional of appendix~\ref{app:mixed-seeds} retains an induced energy
   distribution, whereas the operator measure resolves transition-sector
   matrix elements.  State jumps filter energies whereas operator jumps filter
energy gaps.  A zero of $Q$ at a gap of positive spectral weight removes
that entire gap sector and can reduce the finite operator Krylov
dimension.

\paragraph{Zero-temperature support loss.}
As an example, let
\begin{equation}
\begin{gathered}
   H\ket0=0\,,
   \qquad
   H\ket1=\omega_1\ket1,
   \qquad
   H\ket2=\omega_2\ket2,\\
   \omega_1,\omega_2>0\,,
   \qquad
   \omega_1\ne\omega_2\,.
\end{gathered}
   \label{eq:three-level-zero-temp-spectrum-app}
\end{equation}
Choose the normalized operator
\begin{equation}
   O=c_1\ket1\bra0+c_2\ket2\bra0,
   \qquad
   c_1c_2\ne0\,,
   \qquad
   |c_1|^2+|c_2|^2=1\,.
   \label{eq:three-level-zero-temp-operator-app}
\end{equation}
The ground-state product gives
\begin{equation}
   \dd\mu_O^{(0)}(\omega)
   =|c_1|^2\delta(\omega-\omega_1)\dd\omega
    +|c_2|^2\delta(\omega-\omega_2)\dd\omega\,.
   \label{eq:three-level-zero-temp-measure-app}
\end{equation}
For the descendant $Q(\mathcal L)O$,
\begin{equation}
   \dd\nu_Q^{(0)}(\omega)
   =
   \frac{
      |Q(\omega_1)|^2|c_1|^2\delta(\omega-\omega_1)
      +|Q(\omega_2)|^2|c_2|^2\delta(\omega-\omega_2)}
      {|Q(\omega_1)|^2|c_1|^2+|Q(\omega_2)|^2|c_2|^2}\dd\omega\,,
   \label{eq:three-level-zero-temp-shifted-measure-app}
\end{equation}
provided the denominator is nonzero.  If $Q(\omega_1)=0$ and
$Q(\omega_2)\ne0$, the atom at $\omega_1$ is removed and the shifted cyclic dimension drops from
two to one.  This is the operator-space counterpart of eq.~\eqref{eq:support-loss-dimension} and
shows that support loss is not tied to the Hilbert--Schmidt limit.

\subsection{Thermal reweighting and transition-resolved gap data}
\label{app:operator-thermal}

The operator inner product is additional spectral data.  A useful
finite-temperature family is parametrized by a nonzero finite
nonnegative Borel measure $\gamma_\beta$ on $[0,\beta]$,
\begin{equation}
   (A|B)^\gamma_\beta
   =\frac{1}{Z_\beta}
      \int_{[0,\beta]}\gamma_\beta(\dd\lambda)\,
      \operatorname{Tr}\!\left(
         \ee^{-(\beta-\lambda)H}A^\dagger
         \ee^{-\lambda H}B
      \right).
   \label{eq:thermal-inner-product-g-app}
\end{equation}

The endpoint measure $\delta_0$ gives the one-sided left Gibbs, or GNS, product.  The symmetric
endpoint measure $\tfrac12(\delta_0+\delta_\beta)$ gives the standard thermal product, while
$\delta_{\beta/2}$ gives the Wightman product \cite{TanWeiZhang2024}.  Normalized Lebesgue measure
gives the Kubo--Mori canonical-correlation product \cite{Kubo1957CanonicalCorrelation}.
We assume that this form
and all moments required at the working Krylov depth are finite.  If the
form is semidefinite, its null space is quotiented before constructing
the cyclic operator space.  The commutator Liouvillian is then
self-adjoint under the same domain assumptions as above.

Define the Laplace transform of the finite positive measure $\gamma_\beta$ on $[0,\beta]$
\begin{equation}
   \widehat\gamma_\beta(\omega)
   =\int_{[0,\beta]}
      \ee^{-\lambda\omega}\gamma_\beta(\dd\lambda).
\end{equation}
The corresponding gap measure of an operator that has not necessarily
been normalized in this product is
\begin{equation}
   \dd\mu^\gamma_{\beta,O}(\omega)
   =\frac{1}{Z_\beta}\sum_{a,b}|O_{ab}|^2
      \ee^{-\beta E_b}
      \widehat\gamma_\beta(E_a-E_b)
      \delta\!\bigl(\omega-(E_a-E_b)\bigr)\dd\omega\,.
   \label{eq:thermal-gap-measure-app}
\end{equation}
At fixed $\beta$, replacing $\gamma$ by $\gamma'$ changes the scalar
Liouvillian measure by
\begin{equation}
   \frac{\dd\mu^{\gamma'}_{\beta,O}}
        {\dd\mu^{\gamma}_{\beta,O}}(\omega)
   =\frac{\widehat\gamma'_\beta(\omega)}
          {\widehat\gamma_\beta(\omega)}\,,
   \label{eq:fixed-beta-relative-density-app}
\end{equation}
where the denominator is nonzero on the relevant support.  This compares
measures with total masses
$\|O\|_\gamma^2=(O|O)^\gamma_\beta$ and
$\|O\|_{\gamma'}^2=(O|O)^{\gamma'}_\beta$.  For the normalized measures
$\dd\bar\mu^\gamma=\dd\mu^\gamma/\|O\|_\gamma^2$,
\begin{equation}
   \frac{\dd\bar\mu^{\gamma'}_{\beta,O}}
        {\dd\bar\mu^{\gamma}_{\beta,O}}(\omega)
   =\frac{\|O\|_\gamma^2}{\|O\|_{\gamma'}^2}
      \frac{\widehat\gamma'_\beta(\omega)}
           {\widehat\gamma_\beta(\omega)}\,.
   \label{eq:fixed-beta-normalized-relative-density-app}
\end{equation}
If the normalized ratio is $|Q(\omega)|^2/N_Q$, the polynomial
finite-band transfer applies directly.  Rational ratios lead to
Christoffel--Geronimus transformations
\cite{BuenoMarcellan2004,Zhedanov1997}; their zeros and poles must be
compatible with the support, and inverse polynomial modifications may
require compensating mass terms.  On infinite support, a genuinely
non-polynomial analytic ratio, such as the standard-to-Wightman thermal
filter, is generically an infinite-band transformation or may be treated
by polynomial or rational approximation \cite{GutlebOlverSlevinsky2024}.  On
finite gap support, by contrast, every nonnegative scalar ratio has an
exact polynomial representative in the terminal quotient, obtained by
interpolating its square root on the surviving atoms.  The degree of this
representative may grow with the number of gaps.

The distinction between changing the thermal kernel at fixed temperature
and changing the temperature itself becomes transparent before the
transition data are projected onto the gap variable.  Each matrix element
$O_{ab}$ describes a transition from energy $F=E_b$ to energy $E=E_a$,
with Liouvillian frequency $\omega=E-F$.  These endpoint data are retained
by the joint transition measure
\begin{equation}
\begin{aligned}
   \dd\mathfrak m_O(E,F)
   &=
   \sum_{a,b}|O_{ab}|^2
   \delta(E-E_a)\delta(F-E_b)\dd E\dd F\,,\\
   \int_{\RR}f(\omega)\dd\mu^\gamma_{\beta,O}(\omega)
   &=
   \frac{1}{Z_\beta}
   \int_{\RR^2}
   f(E-F)\ee^{-\beta F}
   \widehat\gamma_\beta(E-F)
   \dd\mathfrak m_O(E,F)
\end{aligned}
   \label{eq:bivariate-transition-measure-app}
\end{equation}
for every bounded Borel function $f$.  The second line is the weighted
pushforward of $\dd\mathfrak m_O$ under the gap map
$\pi_-(E,F)=E-F$ and reproduces
eq.~\eqref{eq:thermal-gap-measure-app}.

At fixed $\beta$, changing $\gamma$ multiplies all transitions with a common gap by the same
known factor and therefore descends to the scalar measure.  After the known partition-function and gap-kernel
factors have been stripped, changing $\beta$ to $\beta'$ reweights a transition by~
$\ee^{-(\beta'-\beta)F}$.  Transitions with the same gap but different source
energies are then reweighted differently.  One gap measure retains only their sum and cannot in
general determine another temperature.  Under the exponential-moment and known-prefactor hypotheses of
proposition~\ref{prop:temperature-tomography-app}, the full unnormalized stripped temperature family
recovers this missing coordinate.

Push $\dd\mathfrak m_O$ forward under
$(E,F)\mapsto(E-F,F)$ and write
\begin{equation}
   \dd\eta_O(\omega,F)
   =\sum_{a,b}|O_{ab}|^2
     \delta\!\bigl(\omega-(E_a-E_b)\bigr)
     \delta(F-E_b)\dd\omega\dd F.
   \label{eq:gap-source-transition-measure-app}
\end{equation}
\begin{proposition}[Measure-valued temperature tomography]
\label{prop:temperature-tomography-app}
Let $H$ and $O$ be fixed, and let $I\subset(0,\infty)$ be an open interval.  Assume that, for
every $\beta\in I$, the partition function $Z_\beta$ is known and finite, the kernel
$\widehat\gamma_\beta(\omega)$ is known, finite and strictly positive on the gap support under
consideration, and
\begin{equation}
   0<\mathcal N_O(\beta)
   =\int_{\RR^2}\ee^{-\beta F}\dd\eta_O(\omega,F)<\infty\,.
   \label{eq:temperature-exponential-moment-app}
\end{equation}
Define the unnormalized stripped gap measure by
\begin{equation}
\begin{aligned}
   \dd\Lambda_\beta(\omega)
   &=\frac{Z_\beta}{\widehat\gamma_\beta(\omega)}
      \dd\mu^\gamma_{\beta,O}(\omega)
   =\int_\RR \ee^{-\beta F}\dd\eta_O(\omega,F)\,.
\end{aligned}
   \label{eq:temperature-measure-laplace-app}
\end{equation}
Knowledge of the measure-valued map $\beta\mapsto\dd\Lambda_\beta$ on $I$ determines the
positive joint measure $\dd\eta_O$ uniquely on that gap support.
\end{proposition}

\begin{proof}
Fix $\beta_0\in I$ and a Borel gap set $B$.  Introduce the finite positive measure
\begin{equation}
   \rho_B(A)=\int_{B\times A}\ee^{-\beta_0F}\dd\eta_O(\omega,F).
\end{equation}
Its moment-generating function is known on an open interval containing the origin because
\begin{equation}
   \int_\RR\ee^{tF}\rho_B(\dd F)=\Lambda_{\beta_0-t}(B),
   \qquad t\in\beta_0-I.
\end{equation}
Because this moment-generating function is finite on a neighborhood of the origin, it extends holomorphically to the corresponding vertical strip. The identity theorem determines the extension from its real values, and uniqueness of the Fourier transform of a finite measure~\cite{Folland1999} then determines $\rho_B$. Thus the finite tilted joint measure $\ee^{-\beta_0F}\dd\eta_O$ is known on every measurable rectangle $B\times A$ with $B$ in the gap support. Rectangles form a generating $\pi$-system, so the monotone-class theorem determines that measure uniquely; multiplication by the strictly positive density $\ee^{\beta_0F}$ then recovers $\dd\eta_O$ on the gap support.%
\end{proof}

\paragraph{Conditional response and information loss.}
Fix $\beta_0\in I$.  The finite measure
$\ee^{-\beta_0F}\dd\eta_O(\omega,F)$ admits the disintegration
\begin{equation}
   \ee^{-\beta_0F}\dd\eta_O(\omega,F)
   =\dd\Lambda_{\beta_0}(\omega)\,\pi_{\beta_0,\omega}(\dd F),
   \label{eq:temperature-reference-disintegration-app}
\end{equation}
where $\pi_{\beta_0,\omega}$ is a conditional probability measure.  The measures
$\Lambda_\beta$ and $\Lambda_{\beta_0}$ are mutually absolutely continuous, with
\begin{equation}
\begin{aligned}
   L_{\beta\mid\beta_0}(\omega)
   &=\frac{\dd\Lambda_\beta}{\dd\Lambda_{\beta_0}}(\omega)
   =\int_\RR\ee^{-(\beta-\beta_0)F}\pi_{\beta_0,\omega}(\dd F).
\end{aligned}
   \label{eq:temperature-conditional-laplace-app}
\end{equation}
The conditional source-energy distribution at inverse temperature $\beta$ is
\begin{equation}
   \pi_{\beta,\omega}(\dd F)
   =\frac{\ee^{-(\beta-\beta_0)F}\pi_{\beta_0,\omega}(\dd F)}
          {L_{\beta\mid\beta_0}(\omega)}\,.
   \label{eq:temperature-conditional-tilt-app}
\end{equation}
For each interior $\beta\in I$ and almost every $\omega$, differentiation gives
\begin{equation}
\begin{aligned}
   -\partial_\beta\log L_{\beta\mid\beta_0}(\omega)
      &=\mathbb E_\beta[F\mid\omega],\\
   \partial_\beta^2\log L_{\beta\mid\beta_0}(\omega)
      &=\operatorname{Var}_\beta(F\mid\omega)\ge0\,,\\
   (-1)^k\partial_\beta^k\log L_{\beta\mid\beta_0}(\omega)
      &=\kappa_{k,\beta}(F\mid\omega),\qquad k\ge1\,.
\end{aligned}
   \label{eq:temperature-response-cumulants-app}
\end{equation}
Here $\kappa_{k,\beta}(F\mid\omega)$ is the $k^{\text{th}}$ conditional cumulant under
$\pi_{\beta,\omega}$.  Normalize the stripped joint measure and its gap marginal according to
\begin{equation}
\begin{aligned}
   \dd P_\beta(\omega,F)
   &=\frac{\ee^{-\beta F}\dd\eta_O(\omega,F)}{\mathcal N_O(\beta)}\,,\\
   \dd P_\beta^{\rm gap}(\omega)
   &=\frac{\dd\Lambda_\beta(\omega)}{\mathcal N_O(\beta)}\,.
\end{aligned}
   \label{eq:temperature-normalized-laws-app}
\end{equation}
The derivatives of their log densities with respect to inverse temperature, namely the score
functions, are
\begin{equation}
\begin{aligned}
   s_{\rm full}(\omega,F)&=-F+\mathbb E_\beta F\,,\\
   s_{\rm gap}(\omega)&=-\mathbb E_\beta[F\mid\omega]+\mathbb E_\beta F\,.
\end{aligned}
   \label{eq:temperature-scores-app}
\end{equation}
Consequently, the Fisher informations of the full and gap probability distributions satisfy
\begin{equation}
\begin{aligned}
   \mathcal I_{\rm full}(\beta)&=\operatorname{Var}_\beta(F),\\
   \mathcal I_{\rm gap}(\beta)
      &=\operatorname{Var}_\beta\!\left(\mathbb E_\beta[F\mid\omega]\right),
\end{aligned}
   \label{eq:temperature-fisher-information-app}
\end{equation}
and hence
\begin{equation}
   \mathcal I_{\rm full}(\beta)-\mathcal I_{\rm gap}(\beta)
   =\mathbb E_\beta\!\left[\operatorname{Var}_\beta(F\mid\omega)\right].
   \label{eq:temperature-fisher-deficit-app}
\end{equation}
Equation~\eqref{eq:temperature-conditional-laplace-app} follows by applying the
disintegration~\eqref{eq:temperature-reference-disintegration-app} after multiplication by
$\ee^{-(\beta-\beta_0)F}$.  Differentiating its logarithm gives the conditional cumulants.  Because
$\beta$ lies in the interior of $I$, finiteness on the open interval gives the local
exponential moments needed to justify these derivatives.  
A single null set serves for all $\beta$ and all $k$. Indeed,
$\int_\RR L_{\beta\mid\beta_0}\dd\Lambda_{\beta_0}=\mathcal N_O(\beta)<\infty$.
Choose a countable dense set $D\subset I$ and discard the union, over $\beta\in D$, of the null sets on which $L_{\beta\mid\beta_0}$ is infinite. For every remaining $\omega$, log-convexity in $\beta$ propagates finiteness from $D$ to all of $I$. Each $\beta\in I$ is then an interior point of the convergence interval, and local uniform domination on compact subintervals justifies differentiation under the integral to all orders.

Finally, $s_{\rm gap}$ is the conditional expectation of $s_{\rm full}$ at fixed $\omega$.
Equation~\eqref{eq:temperature-fisher-deficit-app} is therefore the conditional-variance
decomposition of $\operatorname{Var}_\beta(F)$.  Thus the temperature information lost by
projection onto the Liouvillian gap is exactly the unresolved conditional source-energy variance.
This is an identifiability statement for exact data, not a stability result.  Bilateral Laplace
inversion remains severely ill-conditioned at finite precision.  The reconstructed measure
contains the aggregated positive transition intensity.  It does not recover phases, eigenvectors,
separate labels inside degenerate endpoint sectors or higher transition correlations.  The proposition 
also requires the unnormalized masses $\mathcal N_O(\beta)$.  A family of individually normalized
gap measures need not determine $\eta_O$, even up to an overall constant.  If raw thermal measures
are used without stripping a $\beta$-dependent kernel, the known contribution
$\partial_\beta\log\widehat\gamma_\beta(\omega)$ must be included in the gap score.

\paragraph{Coarse-grained ETH interpretation.}
The obstruction is especially transparent for a many-body system satisfying the
eigenstate-thermalization hypothesis.  After coarse graining in a microcanonical window, write
\begin{equation}
   O_{ab}=O(\bar E)\delta_{ab}
   +\ee^{-S(\bar E)/2}f(\bar E,\omega)R_{ab}\,,
   \qquad
   \bar E=\tfrac12(E_a+E_b),\quad \omega=E_a-E_b\,,
\end{equation}
with $\overline{|R_{ab}|^2}=1$ in the off-diagonal sector
\cite{Srednicki1999ETH}.  We first restrict to that sector; a nonzero diagonal
term instead gives a separate singular contribution at $\omega=0$.  Thus $\dd\mathfrak m_O$ is the transition-resolved measure
on which the coarse-grained ETH intensity is naturally represented.

Taking $\rho(E)=\ee^{S(E)}$ for the smooth density of states, the off-diagonal part of
eq.~\eqref{eq:bivariate-transition-measure-app} has the coarse-grained intensity
\begin{equation}
   \frac{\dd\mathfrak m_{O,{\rm off}}}{\dd\bar E\,\dd\omega}
   \simeq
   \exp\!\left[
      S\!\left(\bar E+\frac{\omega}{2}\right)
     +S\!\left(\bar E-\frac{\omega}{2}\right)-S(\bar E)
   \right]|f(\bar E,\omega)|^2
   \label{eq:bivariate-eth-envelope}
\end{equation}
and, for smooth $S$ at fixed or subextensive $\omega$,
\begin{equation}
   S\!\left(\bar E+\frac{\omega}{2}\right)
  +S\!\left(\bar E-\frac{\omega}{2}\right)-S(\bar E)
   =S(\bar E)+\frac{\omega^2}{4}S''(\bar E)
    +O\!\left(\omega^4S^{(4)}(\bar E)\right).
   \label{eq:eth-entropy-expansion-app}
\end{equation}
The multiplicative expansion about $\ee^{S(\bar E)}$ is justified only when
$|S''(\bar E)|\omega^2\ll1$ and the higher terms are controlled.
Substituting into
eq.~\eqref{eq:thermal-gap-measure-app} and using
$\ee^{-\beta E_b}=\ee^{\beta\omega/2}\ee^{-\beta\bar E}$,
\begin{equation}
   \frac{\dd\mu^\gamma_{\beta,O,{\rm off}}}{\dd\omega}(\omega)
   \simeq\frac{\widehat\gamma_\beta(\omega)\,\ee^{\beta\omega/2}}{Z_\beta}
    \int\dd\bar E\;\ee^{-\beta\bar E}
    \exp\!\left[
      S\!\left(\bar E+\frac{\omega}{2}\right)
     +S\!\left(\bar E-\frac{\omega}{2}\right)-S(\bar E)
    \right]|f(\bar E,\omega)|^2.
   \label{eq:thermal-gap-eth}
\end{equation}

For a differentiable concave entropy with an interior nondegenerate thermodynamic saddle,
and provided $f$ and the remaining prefactors vary subexponentially across its width, the integral
is concentrated in a finite window around $S'(\bar E_\beta)=\beta$.  Changing $\beta$ moves this
window.  Endpoint saddles, phase coexistence or exponentially rapid energy dependence of $f$
require separate treatment; the scalar measure at one temperature therefore does not determine
the measure at another.

A qualified converse is available at the level of the coarse-grained intensity.  At a fixed
$\omega$ for which $\widehat\gamma_\beta(\omega)\ne0$, divide out the known thermal prefactors and
define the reduced off-diagonal gap density as
\begin{equation}
   \mathcal G_\beta(\omega)
   =\frac{Z_\beta\ee^{-\beta\omega/2}}
           {\widehat\gamma_\beta(\omega)}
      \frac{\dd\mu^\gamma_{\beta,O,{\rm off}}}{\dd\omega}(\omega).
   \label{eq:thermal-gap-Laplace-data-app}
\end{equation}
In terms of the
stripped gap measure introduced in eq.~\eqref{eq:temperature-measure-laplace-app}, wherever
$\Lambda_{\beta,{\rm off}}$ is absolutely continuous with respect to $\dd\omega$,
\begin{equation}
   \mathcal G_\beta(\omega)
   =\ee^{-\beta\omega/2}
     \frac{\dd\Lambda_{\beta,{\rm off}}}{\dd\omega}(\omega).
   \label{eq:eth-exact-temperature-relation-app}
\end{equation}
Under the ETH coarse graining, $\mathcal G_\beta(\omega)$ is the $\bar E$-Laplace transform of
the entropy-dressed off-diagonal transition intensity.

At fixed gap, $\bar E=F+\omega/2$.  The exact conditional cumulants in
eq.~\eqref{eq:temperature-response-cumulants-app}, applied to the normalized off-diagonal
transition measure after the known thermal kernel has been divided out,
therefore imply
\begin{equation}
\begin{aligned}
   -\partial_\beta\log\mathcal G_\beta(\omega)
   &=\mathbb E_{\beta,{\rm off}}[F\mid\omega]+\frac{\omega}{2}
    =\mathbb E_{\beta,{\rm off}}[\bar E\mid\omega],\\
   \partial_\beta^2\log\mathcal G_\beta(\omega)
   &=\operatorname{Var}_{\beta,{\rm off}}(F\mid\omega)
    =\operatorname{Var}_{\beta,{\rm off}}(\bar E\mid\omega).
\end{aligned}
   \label{eq:thermal-gap-response-cumulants-app}
\end{equation}
Higher derivatives give the remaining conditional cumulants.  These identities are exact for
the stripped transition measure.  The ETH hypothesis enters only when its density is interpreted
as the entropy-dressed envelope $|f(\bar E,\omega)|^2$, and it still does not recover phases or the
microscopic variables $R_{ab}$.

\paragraph{Parity and non-Hermitian qualifications.}
Two qualifications delimit this scalar self-adjoint setting.  First, the
infinite-temperature gap measure is even for Hermitian $O$, including a
density matrix, and more generally whenever the transition weights are
invariant under $a\leftrightarrow b$.  A polynomial seed preserves this
evenness whenever
$|Q(-\omega)|=|Q(\omega)|$ on the support; definite parity is a
sufficient, but not necessary, condition. A parity-mixing seed may therefore produce a general Jacobi chain with nonzero diagonal
coefficients.  Second, generic non-Hermitian generators, including
dissipative Lindbladians, require Arnoldi, bi-Lanczos or biorthogonal
polynomial methods \cite{BhattacharyaNandyNathSahu2023}.  Quantum
detailed balance can sometimes give a self-adjoint realization in a
weighted inner product, but the physical dissipative evolution remains a
nonunitary semigroup; normalized Krylov probabilities and spread
complexity then require an additional convention and do not follow
unchanged from the unitary construction above.

\end{document}